\documentclass[10pt]{iopart}

\usepackage{iopams}  

\expandafter\let\csname equation*\endcsname\relax 
\expandafter\let\csname endequation*\endcsname\relax 

\usepackage{overpic}

\usepackage{bm}
\usepackage{mathrsfs}
\usepackage{verbatim}
\usepackage[all,cmtip]{xy}
\usepackage{amssymb}
\usepackage{amsmath}
\usepackage{amsthm}
\usepackage{color}
\usepackage{dsfont}
\usepackage{amscd}
\usepackage{cases}
\usepackage{amsfonts}
\usepackage{amssymb}
 \usepackage{mathdots}
\usepackage{bbm}
\usepackage{epsfig}
\usepackage{color}
\usepackage[figuresright]{rotating}
\usepackage{makecell}
\usepackage{tabularx}
\usepackage{footmisc}
\usepackage{epsfig,pstricks,graphicx}
\usepackage[mathscr]{eucal}
\usepackage{comment}
\usepackage[normalem]{ulem}
\usepackage{pstricks,graphicx}
\definecolor{darkred}  {rgb}{0.5,0,0}
\definecolor{darkblue} {rgb}{0,0,0.5}
\definecolor{darkgreen}{rgb}{0,0.5,0}
\usepackage{hyperref}

\hypersetup{
	pdftitle = {QRT Proposal},
	pdfauthor = {},
	colorlinks = true,
	urlcolor  = blue,         
	linkcolor = red,     
	citecolor = blue,    
	filecolor = darkred       
}

\def\ra{\rangle}
\def\la{\langle}
\def\bb{\mathbb}
\def\ot{\otimes}
\def\CC{{\rm\kern.24em \vrule width.04em height1.46ex depth-.07ex
		\kern-.30em C}}

\newtheorem{theorem}{Theorem}

\newtheorem{definition}{Definition}
\theoremstyle{remark}

\newcommand{\beax}{\begin{eqnarray*}}
\newcommand{\eeax}{\end{eqnarray*}}

\def\be{\begin{eqnarray}}
\def\ee{\end{eqnarray}}
\newcommand{\bea}{\begin{eqnarray}}
\newcommand{\eea}{\end{eqnarray}}

\newcommand{\mF}{\mathcal{F}}
\newcommand{\mE}{\mathcal{E}}
\newcommand{\mC}{\mathcal{C}}
\newcommand{\mQ}{\mathcal{Q}}

\newcommand{\mH}{\mathcal{H}}

\newcommand{\mT}{\mathcal{T}}

\newcommand{\mW}{\mathcal{W}}

\newcommand{\mP}{\mathcal{P}}

\newcommand{\mS}{\mathcal{S}}

\newcommand{\lr}{\rangle\langle}

\newcommand{\mbR}{\mathbb{R}}
\newcommand{\C}{\mathbb{C}}
\newcommand{\B}{\rm{B}}
\newcommand{\G}{\rm{G}}
\newcommand{\GHZ}{\rm{GHZ}}
\newcommand{\GWV}{\rm{GWV}}   
\newcommand{\W}{\rm{W}} 

\def\>{\rangle}
\def\<{\langle}
\def\tr{ \mathrm{Tr}}
\def\diag{ \mathrm{diag}}
\def\supp{ \mathrm{supp}}
\def\fh{\mathfrak{h}}

\makeatletter
\def\@mkboth#1#2{}
\makeatother

\begin{document}

\review[Measure of entanglement and the monogamy relation]
{Measure of entanglement and the monogamy relation: a topical review}

\author{Yu Guo}

\address{School of Mathematical Sciences, Inner Mongolia University, Hohhot, Inner Mongolia 010021, People's Republic of China}
\address{Inner Mongolia Key Laboratory of Mathematical Modeling and Scientific Computing, Inner Mongolia University, Hohhot, Inner Mongolia 010021, People's Republic of China}
\ead{guoyu3@aliyun.com}

\author{Zhi-Xiang Jin}

\address{School of Computer Science and Technology, Dongguan University of Technology, Dongguan 523808, People's Republic of China}
\ead{jzxjinzhixiang@126.com}

\begin{abstract}
Characterizing entanglement, including quantifying and distribution of entanglement, which lies at heart of the quantum resource theory, have been investigated extensively ever since Bennett \etal proposed three seminal measures of entanglement in 1996. Up to now, there are numerous measures of entanglement that have been proposed from different point of view and plenty of monogamy relations have been explored which make the distribution of entanglement became more and more clear. While this is relatively easy in the case of pure states, it is much more intricate for the case of mixed quantum states especially with higher dimension and more particles in the system. We present here an overview of the theory along this line. We outline most of the results in this field historically and focus on the finite-dimensional systems. In particular we emphasize the point of view that (i) which yardsticks haven been applied in quantifying entanglement and its distribution, (ii) what are the substantive characteristics and interrelations of these measures and their monogamy relations mathematically by comparing, and (iii) which concepts should be improved or revised and how they were developed accordingly.

\vspace{2pc}
\end{abstract}

\maketitle

\makeatletter
\@mkboth{Measure of entanglement and the monogamy relation: a topical review}{Measure of entanglement and the monogamy relation: a topical review}
\makeatother



\tableofcontents


\section{Introduction}


Although the phenomenon of quantum entanglement was firstly discovered by Einstein, Podolsky, and Rosen (EPR) and Schr\"{o}dinger 
as a ``spook'' feature of quantum machinery in 1935~\cite{Einstein1935,Schrodinger1935}, its quantitative description started with Bell's inequalities in 1964~\cite{Bell1964}. Since then many experimental tests of violation of the local hidden variable model have been performed~\cite{Horodecki2009}. Consequently, the entanglement was proved to be a crucial resource in many quantum information processing tasks~\cite{Bennett1996prl,Burkhart2021prx,DiVincenzo1995,Gisin2002rmp,Gisin2007np,Guhne2009,Horodecki2009,Nielsenbook,Raussendorf2003pra,Trenyi2024njp,Wangdongsheng2024pra,Yuxiaodong2021pral}. The second milestone from the theory point of view was Werner's 
mathematical definition of entanglement in 1989~\cite{Werner1989pra}. It gave rise to numerous studies on characterizing entanglement which is called the resource theory of entanglement that including the entanglement criteria, measure of entanglement, and the distributions of entanglement, etc.

Historically, the first quantity that quantifies the degree of entanglement was proposed by Shimony in 1995~\cite{Shimony95}, which is called the geometric measure of entanglement in literature (up to a whole factor).
The first three well-known seminal entanglement measures are the entanglement of formation, the entanglement cost and the distillable entanglement proposed by Bennett \etal simultaneously in 1996~\cite{Bennett1996pra2046,Bennett1996pra3824}.
Soon after, Vedral, Plenio, Rippin and Knight established the axiomatic definition of a bipartite measure of entanglement in 1997~\cite{Vedral1997} which keep the leading idea that entanglement should not increase under local operation and classical communication (LOCC)~\cite{Bennett1996pra3824}. They pointed out that a measure of bipartite entanglement need satisfies three conditions: (i) vanishing on separable states, (ii) invariant under local unitary operation, and (iii) nonincreasing under LOCC.  
In the next year, Vedral and Plenio improved item (iii) as nonincreasing on average under LOCC on account of the LOCC is always stochastic~\cite{Vedral1998pra}. From then, a large number of different approaches of quantifying bipartite entanglement were proposed~\cite{Eltschka2014jpa,Guhne2009,Horodecki2009} according to the axiomatic definition, such as the concurrence~\cite{Hill,Wootters} and its various generalizations~\cite{Albeverio2001,Audenaert2001pra,Fan2003jpa,Gour2005pra,Rungta2001pra,Rungta2003pra,Uhlmann2000pra,Wei2022jpa,Yangxue2021pra}, negativity~\cite{Lee2003pra,VidalWerner,Zyczkowski1998pra} and the associated generalizations~\cite{Carrasco2021qip,Plenio2005prl,Wang2020pra}, entropy induced entanglement measures~\cite{Christandl2004jmp,Datta2009ieee,Datta2009ijqi,Gour2007jmp,Kim2010pra,Kim2010jpa,Kim2011jpa,Vedral1997,Vedral1998pra,Vidal2000,Yang2008prl,Yangxue2023epj}, fidelity based entanglement measures~\cite{Biham2002pra,Guo2020qip,Shapira2006pra,Vedral1998pra}, etc. These entanglement measures are always shown to be convex but not all~\cite{Plenio2005prl}. In 2000, Vidal~\cite{Vidal2000} put forward a concept that termed entanglement monotone which is an entanglement measure that not only nonincreases on average under LOCC but also is convex additionally${\footnotemark{}}$. $\footnotetext{~In Ref.~\cite{Vidal2000}, entanglement monotone is defined to be a magnitude that is non-increasing on average under LOCC. Later, Refs.~\cite{Christandl2004jmp,Eisert2001pra,Eisert2003jpa,Horodecki2000prl} stated that entanglement monotone also need to be convex. We adopt the later case throughout this paper.}$ Up to now there have been more than 36 bipartite entanglement measures presented in literature and most of them are shown to be entanglement monotones. In the next section, we review these measures in detail.

Yet, concerning multipartite cases, the situation is much more complicated due to the exponential increase in the complexity of states with increasing dimension of the state space and the number of the particles. There are different types of entanglement in multipartite system, such as the genuine entanglement, the $k$-entanglement (when $k=2$, it is just the genuine entanglement), the $k$-partite entanglement~\cite{Guhne2005njp,Guo2025qip,Hong2023epjp,Lihui2024aqt}, and the partitewise entanglement~\cite{Dong2024pra,Guo2025pra}. While the $k$-entanglement reflects how many splitted subsystems are entangled under partitions of the systems, the $k$-partite entanglement focus on at most how many particles in the global system are entangled but separable from other particles~\cite{Guhne2005njp}. They are complementary to each other in some sense. Partially motivated by the pairwise entanglement put forward in Ref.~\cite{Dong2024pra}, Guo \etal established the framework of the partitewise entanglement~\cite{Guo2025pra} which refers to the entanglement of the given subsystems in the view of the global system. In such a sense, two subsystems $AB$ may live in a genuine entangled system $ABC$ but $AB$ may be separable when it is regarded as an independent state. So $AB$ shared entanglement with part $C$ whenever $ABC$ is genuinely entangled, this is just the motivation of the pairwise entanglement of $A$ and $B$ in $ABC$.

Some axiomatic bipartite entanglement measures can be immediately extended to the multipartite case, e.g., the relative entropy of entanglement~\cite{Vedral1997,Vedral1998pra}, the geometric measure of entanglement~\cite{Das2016pra,Sen2010pra,Shimony95}, and the measure that induced by the fidelity~\cite{Cao2007jpa,Guo2020qip}. Another simple extension of bipartite measures to mulitipartite case is the one induced from the bipartite partitions~\cite{Guo2022jpa,Guo2024rip,Ma2011pra}, in which the sum over bipartite partitions can quantify the ``global entanglement'' contained in the state~\cite{Guo2020pra,Guo2022jpa,Guo2024rip} while the minimum one over bipartite partitions exports genuine entanglement measures (GEMs)~\cite{Guo2024rip,Ma2011pra}. Of course, there are also other approaches such as the SLOCC invariants, concurrence vector, Ent, Hyperdeterminant, the $q$- and $c$-squashed entanglement, and the informationally complete entanglement measure~\cite{Jin2023pra}. In general, based on the $k$-partitions, we can derive the $k$-entanglement measures, and from the $k$-fineness partitions, we can obtain the $k$-partite entanglement measures. But the quantification of partitewise entanglement seems a little hard.

So far, a series of different approaches to quantifying multipartite entanglement which includes the genuine entanglement, the global entanglement, the $k$-entanglement, the $k$-partite entanglement, and the partitewise entanglement, have been put forward. The first effort in quantifying genuine entanglement is the ``three tangle'' which reports the genuine three-qubit
entanglement by Coffman \etal in 2000~\cite{Coffman2000pra}.
In 2011, Ma \etal~\cite{Ma2011pra} established some conditions for a quantity to be
a GEM and proposed the so-called genuinely multipartite concurrence (GMC), according to the origin bipartite concurrence.
The $k$-entanglement measure was firstly discussed in Ref.~\cite{Hong2012pra,Li2024pra} and latter was discussed systematically by Guo \etal in Ref.~\cite{Guo2024pra}. Recently, the $k$-partite entanglement measures were explored~\cite{Guo2025qip,Hong2023epjp,Lihui2024aqt}. Very recently, Guo~\etal investigated the partitewise entanglement and give three classes of measures in Ref.~\cite{Guo2025pra}. We review all these measures in detail in Sec.~\ref{sec-9}-Sec.\ref{sec-14}.

From the information-theoretical point of view, another crucial issue for characterizing multiparticle entagnlement is the distribution of the entanglement up to the given measure.
This is the monogamy law of entanglement~\cite{Terhal2004}, which states
that, unlike classical correlations, if two parties $A$ and $B$ are
maximally entangled, then neither of them can share entanglement
with a third party $C$. Entanglement monogamy has many applications not only in quantum physics~\cite{Bennett2014,Seevinck,Toner} but also
in other area of physics, such as no-signaling theories~\cite{Augusiak2014pra,Streltsov}, 
condensed matter physics~\cite{Brandao2013,Garcia,Ma2011}, 
statistical mechanics~\cite{Bennett2014}, and even black-hole physics~\cite{Lloyd}.
Particularly, it is the key feature 
that guarantees quantum key distribution secure~\cite{Pawlowski,Terhal2004}.

The fundamental matter in this context is to
determine whether a given measure of quantum correlation is monogamous.
Quantitatively, the monogamy of entanglement is always described by an inequality~\cite{Bai2014prl,Coffman2000pra,Dhar,Hehuan,Osborne,Streltsov}. However, it seems very hard to verify whether a given entanglement masure is monogamous by means of the inequality especailly for the higher dimensional case since it need to check all the states of the system. Guo and Gour improved the definition of monogamy relation by an 
equality between the states and the reduced states up to the given measure~\cite{GG2018q}. This improved definition captures the spirit of monogamy of entanglement, and perhaps not surprisingly,
can yield a family of monogamy relations similar to the origin monogamy relation in terms of inequality by replacing the measure $E$ with $E^\alpha$ for 
some $\alpha>0$~\cite{GG2018q}. The smallest value of $\alpha$ is called the monogamy exponent of $E$~\cite{GG2018q}. Shortly after this improved definition, Guo and Gour proved in Ref.~\cite{GG2019pra} that a bipartite entanglement measures of an arbitrary dimensional system that defined by the convex-roof extension is monogamous whenever the reduced function is strictly concave. This class of measures include almost all of the convex-roof extened measures so far, which indicates that the improved definition is more precise and effecitve than the origin one. In addition, it was proved in Ref.~\cite{Guo2023njp} that, the entanglement monotone termed partial-norm of entanglement is not monogamous. Notice that the corresponding reduced function is concave but not strictly concave. This fact suggest the axiomatic postulates of the entanglement should be improved by adding the requirement that the reduced function is strictly concave additionally. This additional condition is also consistent with the strict-LOCC entanglement monotone argued in Ref.~\cite{Guo2019pra}. Along the line of Ref.~\cite{Lan16}, Jin \etal suggested another equivalent definition of the monogamy relation in Ref.~\cite{Jin2022aqt}. Monogamy is believed to be an intrinsic feature of entanglement. In the absence of the measure of entanglement, the monogamy of entanglement can also be verified from a qualitative aspect~\cite{Doherty2005pra,Terhal2004,Werner1989lmp,Yang2006pla}.

Although we can prove the monogamy of almost all the entanglement monotones, the monogamy exponent is still unknown except for some special multi-qubit system with some special entanglement monotones. Considerable efforts have been directed to evaluating the monogamy exponent for a given entanglement monotones which are mainly focus on the $n$-qubit system~\cite{Allen,Audenaert,Bai2014prl,Chengshuming,Coffman2000pra,Eltschka2015prl,Gao2021rp,Gao2021qip,Hehuan,Kim2009,Kim2010pra,Kim2010jpa,Koashi,Lancien,Liang2017qip,Luo2015ap,Luo2016pra,Osborne,Ouyongcheng,Ouyongcheng2007pra2,Shi2020pra,Song,Wang2020pra,Xuan2025aqt,Yangxue2023epj,Yu2005pra,Zhu2014pra,Zhu2017qip2}. Moreover, the tighter monogamy relation, which contributes to finer characterization of the entanglement distribution, are also attracted wide attention~\cite{Cao2024lpl,Gao2020ctp,Gao2020qip,Gao2020ijtp,Gu2022qip,Jin2017qip,Jin2018pra,Jin2019qip,Jin2020qip,Kim2018pra1,Lai2021jpa,Li2022qip,Li2023lpl,Qi2022ijtp,Ren2021lpl,Shen2023epjp,Shi2019arxiv,Shi2020pra,Tao2023m,Yang2019ctp,Yang2021qip,Zhang2022ijtp,Zhang2022lpl,Zhang2022qip,Zhang2023ps,Zhu2019qip}. The strong monogamy relation indicates more accurate monogamy relation in another way~\cite{Regula2014prl}. As a dual notion to the convex-roof extended entanglement monotone (or called entanglement of formation sometimes~\cite{Guo2019pra}), the entanglement of assistance can be calculated as the maximum over all possible pure states ensembles of the given state, which is origin defined as the most entanglement of $AB$ derived by Charlie performs local measurement on part $C$ from a tripartite pure state that is the purification of $AB$~\cite{DiVincenzo,GourSpekkens,Hughston}. Accordingly, the polygamy relation for entanglement of assistance were investigated~\cite{Bai2014prl,Choi2015pra,Coffman2000pra,Gour2005pra2,Gour2007jmp,Kim2009,Kumar2016pla,Luoshunlong,Osborne,Ouyongcheng2007pra2,Zhu2014pra}. Ref.~\cite{Guo2018qip} improved the definition of polygamy relation from which a unified way of checking the polygamy of the entanglement of assistance was obtained. This results in qualitative conclusion that any entanglement of assistance is polygamous and any faithful entanglement measures can not be polygamous. Likewise the monogamy exponent, the polygamy exponent is also hard to calculate in general. For some particular systems, the tighter polygamy relations were explored~\cite{Cao2024lpl,Chen2019ijtp,Gao2020qip,Jin2019qip,Jin2020qip,Kim2018pra1,Kim2018pra2,Lai2021jpa,Ren2021lpl,Shi2020pla,Yang2019ctp,Zhang2022ijtp,Zhang2023ps}.

The monogamy relations discussed via the bipartite measures (e.g., the bipartite entanglement measures) display certain drawback: Only the relation between $A|BC$, $AB$ and $AC$ are revealed, the global correlation in $ABC$ and the correlation contained in part $BC$ is missed, where the vertical bar indicates 
the bipartite split across which we will measure the (bipartite) 
correlation. In addition, the correlation between any partitions or any subsystems with the coarsening relation can not be compared with each other thoroughly by means of the bipartite measure.
To address such a subject, by adding additional axiomatic postulates, Guo \etal developed a framework of complete measure of multipartite quantum correlation and the associated complete monogamy relation~\cite{Guo2020pra,Guo2021qst,Guo2022entropy,Guo2023pra,Guo2024pra,Guo2024rip}, especially for the global multipartite entanglement, the genuine entanglement and the $k$-entanglement. The complete monogamy relation is shown to be complementary to the traditional monogamy relation~\cite{Guo2023njp,Guo2023pra,Guo2024pra}.
In such a context, a measure of multipartite correlation should be complete
in the sense that the distribution of the correlation could be depicted exhaustively.
It has been showed that many complete multipartite entanglement measures are completely monogamous~\cite{Guo2020pra,Guo2022entropy,Guo2024rip}.

Note that, in general, we cannot discuss the monogamy of a multipartite entanglement measure (MEM), and similarly the complete monogamy of a bipartite entanglement measure since they are incompatible with each other. However, $k$-entanglement corresponds to the entanglement of the split state over the subsystems with a fixed number of partition ``$k$'', which is similar to both the bipartite entanglement measure and the general multipartite ones to some extent. We thus need take into account both the monogamy and the complete monogamy for a given $k$-entanglement measure ($k$-EM). Very recently, such an issue was addressed by Guo in Ref.~\cite{Guo2024pra}. Moreover, several unfined way of quantifying the $k$-entanglement have been explored in terms of sum and product of the reduced functions~\cite{Guo2024pra}.

Consequently, all these multipartite entanglement measures could be classified into two classes: the complete multipartite entanglement measures and the previous ones that are not involved with the completeness. We call the original one the multipartite entanglement measure 1.0., and the complete one is the multipartite entanglement measure 2.0. Version 1.0 only requires basic conditions in order to quantify entanglement while the version 2.0 is equipped with better performance that allow the amount of the entanglement can be compared compatibly under any partition of the system and its subsystems. For the version 1.0, we can only discuss the monogamy relation (e.g., the monogamy of the $k$-entanglement), but the complete monogamy relation and the tightly complete relation are compatible with version 2.0 while the version 1.0 fails. It is worth mentioning that the difference between the two versions can also be depicted from another perspective---coarsening relation of multipartite partitions of the multipartite system--- which was proposed by Guo \etal~\cite{Guo2020pra,Guo2023pra,Guo2024pra,Guo2024rip}. The coarsening relation of multipartite partitions are divided into three basic relations~\cite{Guo2023pra,Guo2024pra,Guo2024rip}: coarsening relation of type (a) means discarding some subsystem(s) of a given partition, type (b) refers to combining some subsystem(s) of a given partition, and type (c) indicates discarding some subsystem(s) of some subsystem(s) in a given partition. Then the monogamy relation is related with the coarsening relation of type (c), the complete monogamy relation is related with that of type (a), while the tightly complete monogamy relation is related with that of type (b). Any partition or partition of the subysystem can be derived from the origin system by these coarsening operations. So this scenario can also be applied on other multipartite quantum correlations~\cite{Guo2021qst,Guo2023pra}.

Apart from the polygamy relation of the entanglement of assistance, the triangle relation also exhibits the polygamy of entanglement. The first triangle relation is based on concurrence for the three qubit~\cite{Qian2018njp,Zhu2014pra}. Consequently, it was shown to be true for the general $n$-partite pure state via concurrence. In 2022, Guo \etal demonstrated a general conclusion that any continuous entanglement measure could induce a triangle relation. Very recently, Ge~\etal~\cite{Ge2024pra} found a class of concrete triangle relations for measure that associated with subadditive reduced functions. All these relations indeed capture the `polygamy' of entanglement better than that of the entanglement of assistance since entanglement of assistance is not a well-defined entanglement measure. We thus suggest that, when the polygamy of entanglement is concerned, we consider the triangle relation rather than the polygamy of the entanglement assistance.

According to the framework of the complete multipartite entanglement measure and the corresponding complete monogamy relation, some previous problems such as the additivity of the entanglement measure and the maximally bipartite entangled state are more clear. Recall that, the additivity of the entanglement formation was a long standing open problem which wae conjectured to be
true~\cite{Plenio2007qic} and then disproved by Hastings in 2009~\cite{Hastings}. In fact most of the entanglement measures are not additive. But as an axiomatic condition, any unified mulitpartite entanglement measure should be additive in another sense~\cite{Guo2020pra,Guo2024pra,Guo2024rip} that different from the additivity of the biparite measure. Namely, the additivity seems suitable for mulitpartite entanglement measure other than the biparite case. The origin definition of maximally bipartite entangled mixes state~\cite{Li2012qic} contradicts the monogamy law of entanglement~\cite{Guo2020pra}. But this issue can be addressed by the complete monogamy relation, which indicates that a bipartite state is maximally entangled if and only if its any extension have the same entanglement quantified by the complete tripartite entanglement measure as the origin state quantified by the associated bipartite entanglement measure~\cite{Guo2020pra}.

In this review, we mainly focus on the mathematical characterization of the various entanglement measures, monogamy relations, complete monogamy relations, and the related mathematical structure. We list as detail as possible these properties for readers' convenience under the outline of the framework of the resource theory. We thus can get as complete as possible the whole story for the quantifying entanglement in the finite dimensional systems by now. We unify all the associated notations and the terminologies, but some figures are borrowed from the original literature. Throughout this paper, we denote by $\mH^{A_1A_2\cdots A_n}=\mathcal{H}^{A_1}\otimes
\mathcal{H}^{A_2}\otimes\cdots\otimes\mathcal{H}^{A_n}$
an $n$-partite Hilbert space with finite dimension and by $\mS^{X}$ we denote the set of all density
operators (or called states) acting on $\mH^{X}$. The superscript or subscript $X$ always denotes the corresponding system. For example, the state in $\mS^{X}$ is denoted by $\rho^X$ (or $\rho_X$ sometimes), and is also denoted by $\rho$ for simplicity whenever the associated system $X$ is clear from the context. 
For any operator acting on $\mH$ ($\dim\mH<\infty$ here), $\|A\|_{\tr}$ denotes the trace norm of $A$, i.e.,
$\|A\|_{\tr}=\tr|A|$, where $|A|=(A^\dag A)^{1/2}$, $\|A\|_2$ is the Hilbert-Schmidt norm of $A$, i.e., $\|A\|_2=\left[\tr(A^\dag A)\right]^{1/2}$, $\|A\|$ is the operator norm, i.e.,
$\|A\|=\sup_{|\psi\ra}\|A|\psi\ra\|=\lambda_{\max}$, where $\lambda_{\max}$ denotes the maximal eigenvaule of $|A|$, and $r(A)$ is the rank of $A$. $I^X$ (sometimes $I_X$ or $I$) denotes the identity operator on $\mH^X$. 
Throughout this paper, for any system that is split into two parts $XY$, $\overline{X}$ denotes the complementary part of $X$, i.e., $\overline{X}=Y$. For example, in the system $ABCD$, $\overline{A}=BCD$, $\overline{AB}=CD$, etc.


\section{Bipartite entanglement measure}


\subsection{Basic concepts}

\subsubsection{Bipartite entanglement}

A bipartite state $\rho^{AB}$ defined on $\mH^{AB}$ is separable if and only if it can be represented by
\bea
\rho^{AB}=\sum\limits_{i=1}^mp_i\rho_i^A\ot\rho_i^B,
\eea
where $p_i>0$, $\sum_ip_i=1$. In fact, $\rho_i^A$ and $\rho_i^B$ can be chosen to be pure.
Then, from the Caratheodory theorem, the number
$m$ in the convex combination admits $m
\leq  d_{AB}^2$, where $d_{AB}=\dim \mH^{AB}$~\cite{Horodeki1997pla,Vedral1998pra}. For two qubits the number of states (or
called the cardinality) needed in the separable decomposition
is always 4, which corresponds to the dimension of
the Hilbert space itself~\cite{Sanpera1998pra,Wootters}. There exist, however, $d\ot d$ states that for $d\geq3$ have cardinality of order of $d^4/2$~\cite{DiVincenzo}.

\begin{figure}[htbp]
	\centerline{
		\includegraphics[width=7cm]{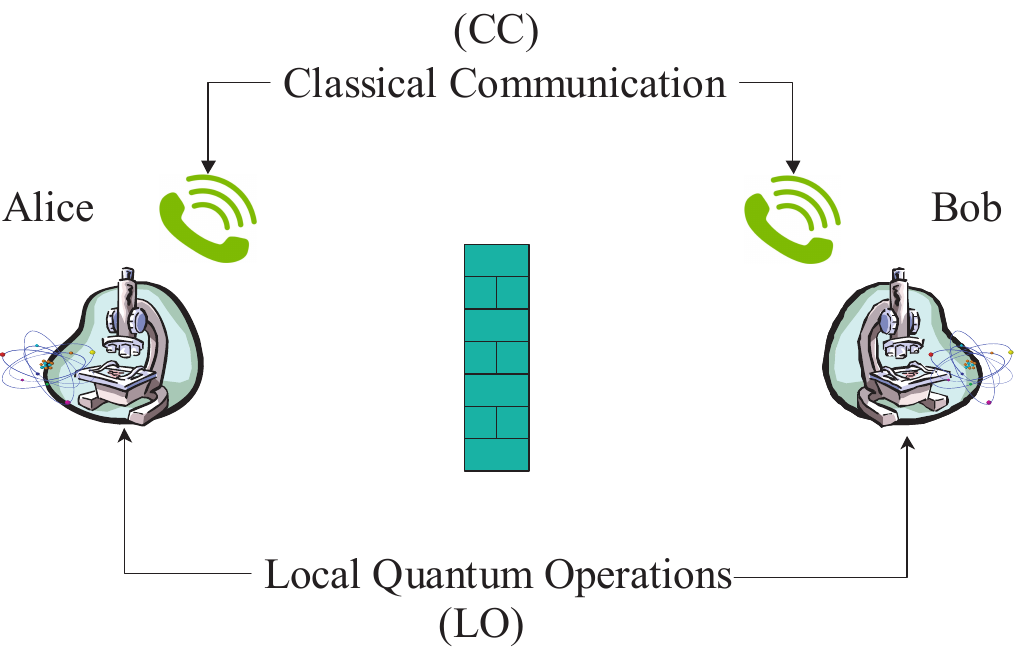}}
	\vspace*{0.15cm} \caption{\label{fig2-1} In a standard quantum
		communication setting two parties Alice and Bob may perform any
		generalized measurement that is localized to their laboratory and
		communicate classically. The brick wall indicates that no quantum
		particles may be exchanged coherently between Alice and Bob. This
		set of operations is generally referred to as LOCC.}
\end{figure}

\subsubsection{Bipartite entanglement measure}

A function $E: \mS^{AB}\to\bb{R}^{+}$ is called a bipartite entanglement measure if
it satisfies~\cite{Vedral1997}: (E-1) $E(\rho)=0$ if $\rho$ is separable; (E-2) $E$ cannot increase under LOCC, i.e.,
$E[\varepsilon(\rho)]\leq  E(\rho)$ for any LOCC $\varepsilon$. (E-2) implies that $E$ is invariant under local unitary
operations, i.e., $E(\rho)=E(U_A\otimes U_B\rho U_A^{\dag}\otimes U_B^{\dag})$ for any local unitaries $U_A$ and $U_B$.
LOCC $\varepsilon$ is a completely positive and trace preserving (CPTP) map that can be presented as
\bea \label{locc}
\varepsilon(\cdot)=\sum_iA_i\ot B_i(\cdot)A_i^\dag\ot B_i^\dag,
\eea
where $A_i$'s are operators from $\mH^A$ into $\mH^{A'}$ and $B_i$'s are operators from $\mH^B$ into $\mH^{B'}$, respectively, 
$\sum_iA_i^\dag A_i\ot B_i^\dag B_i=I^{AB}$. LOCC is the class of operations that has both fundamental and technological motivations~\cite{Plenio2007qic}. It can be implemented by separated parties acting locally through CPTP maps upon the individual
subsystems, while the only communication channel connecting the subsystems is classical, such as a telephone [see Fig.~\ref{fig2-1} (borrowed from Ref.~\cite{Plenio2007qic}) for the bipartite case, the multiparty case follows as a consequence].

LOCC represents the free operations in the quantum resource theory (QRT) of entanglement, while all separable states are the ``free state'' which constitute a closed convex set and this leads to a convex QRT~\cite{Chitambar2019rmp}.
The general structure of an LOCC map is quite complex~\cite{Donald2002jmp} although it can always be represented as Eq.~\eref{locc} mathematically. Eq.~\eref{locc} is indeed the form of the separable measurement which including LOCC as a subset. The multiparty case is similar.

\subsubsection{Bipartite entanglement monotone}

\begin{figure}[th]
	\begin{center}
		\includegraphics[width=7cm]{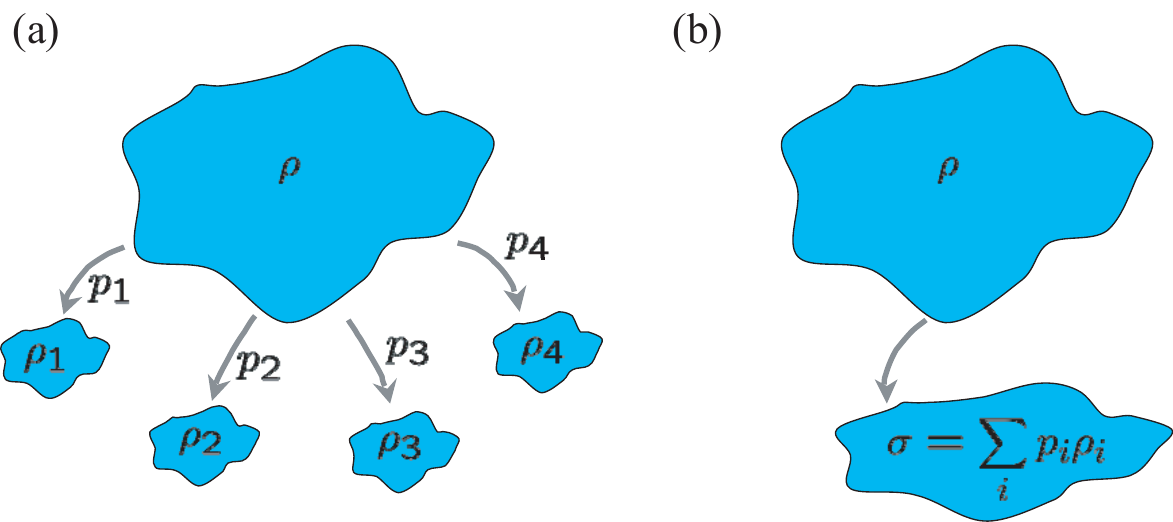}
	\end{center}
	\caption{\label{subselect} Schematic picture of
		the action of quantum operations with and without sub-selection
		shown in part (a) and part (b) respectively.  }
\end{figure}

In general, LOCC can be stochastic, in the sense that $\rho$ can be converted to $\rho_{j}$ with some probability $p_j$ [see part (a) of fig.~\ref{subselect} (borrowed from Ref.~\cite{Plenio2005prl})]. We thus require the measure $E$ additionally obeys~\cite{Plenio2005prl,Vedral1997pra,Vedral1998pra,Vidal2000}
\bea\label{average}
\sum_{j}p_jE(\rho_{j})\leq  E\left(\rho\right).
\eea 
Note that Eq.~\eref{average} is more restrictive than (E-2), i.e., $E(\rho)\geq E[\varepsilon(\rho)]=E(\sum_jp_j\rho_j)$ with $\varepsilon(\rho)=\sigma=\sum_jp_j\rho_j$ (see part (b) of fig.~\ref{subselect}). Namely, item (E-2) means that we are unable to select sub-ensembles according to a measurement outcome. This in fact lead to an additional constraint on LOCC. But such a restriction on LOCC is not directly related to the non-local structure of quantum mechanics and would obscure key features of entanglement~\cite{Plenio2005prl}. So, as in most of the literature, Eq.~\eref{average} was considered instead of (E-2).

On the other hand, the map from $\rho$ to $\rho_{j}$ can not be described in general by a CPTP map. However, by introducing a ``flag'' system $A'$, we can view the ensemble $\{\rho_{j},p_j\}$ as a classical quantum state $\sigma'=\sum_{j}p_j|j\lr j|^{A'}\otimes\rho_{j}$. Hence, if $\rho$ can be converted by LOCC to $\rho_{j}$ with probability $p_j$, then there exists a CPTP LOCC map $\varepsilon$ such that $\varepsilon(\rho)=\sigma'$. Therefore, the definition above of a measure of entanglement captures also probabilistic transformations in general. In particular, $E$ must satisfy $E\left(\sigma'\right)\leq  E\left(\rho\right)$. Most measures of entanglement studied in literature so far satisfies
\bea\label{flag}
E\left(\sigma'\right)=\sum_{j}p_jE(\rho_{j})\;,
\eea
which is very intuitive since $A'$ is just a classical system encoding the value of $j$. In this case the condition $E\left(\sigma'\right)\leq  E\left(\rho\right)$ becomes Eq.~\eref{average}. That is, LOCC can not increase entanglement on average (it is possible that $E(\rho_{j_0})>E(\rho)$ for some $j_0$) in general.

If a measure of bipartite entanglement $E$ does not increase {on average} under LOCC and is convex additionally, it is called a bipartite {entanglement monotone}~\cite{Christandl2004jmp,Eisert2001pra,Eisert2003jpa,Horodecki2000prl,Vidal2000}. So a convex bipartite entanglement measure $E$ is a bipartite entanglement monotone if and only if it satisfies Eq.~(\ref{flag})~\cite{Horodecki2005osid}. Obviously, any entanglement monotone is an entanglement measure but not vice versa (although most of entanglement measures are entanglement monotone, there exist some counterexamples, e.g., if it is not convex~\cite{Datta2009ieee,Datta2009ijqi,Plenio2005prl,Wang2020prl,Wang2023pra}, or it is nonincreasing under LOCC but can increase on average under LOCC~\cite{Datta2009ijqi,Gittsovich2010pra}).

\subsubsection{Reduced function}

For any bipartite entanglement measure $E$ on $\mS^{AB}$, if there exists a nonnegative function $h: \mS^A\rightarrow\mbR^+$ such that 
\bea\label{h}
h\left( \rho^A\right) = E\left( |\psi\lr\psi|^{AB}\right), \quad \rho^A=\tr_B|\psi\lr\psi|^{AB}
\eea 
when it is evaluated for the pure states, we call $h$ the {reduced function} of $E$~\cite{Guo2023njp,Guo2024pra,Guo2024rip}. 
It is clear that, any reduced function $h$ is invariant under local unitaries, i.e.,
\beax
h\left( U\rho^AU^\dag\right) =h\left( \rho^A\right)
\eeax 
for any unitary operator $U$ acting on $\mH^A$. Let $E$ be a bipartite entanglement measure, then the convex-roof extension of $E$ that defined by
\bea\label{EOF}
E_F\left(\rho^{AB}\right)=\min\sum_{j=1}^{n}p_jE\left(|\psi_j\lr\psi_j|^{AB}\right),\quad \rho^{AB}\in\mS^{AB},
\eea
where the minimum is taken over all pure state decompositions of $\rho^{AB}=\sum_{j=1}^{n}p_j|\psi_j\lr\psi_j|^{AB}$, is an entanglement monotone if and only if the reduced function $h$ is concave, i.e.,
\beax 
h[\lambda\rho_1+(1-\lambda)\rho_2]\geq\lambda h(\rho_1)+(1-\lambda)h(\rho_2)
\eeax
for any states $\rho_1$, $\rho_2$, and any $0\leq\lambda\leq1$~\cite{Vidal2000}. Conversely, if $E$ is an entanglement monotone (not necessarily defined as $E_F$ for mixed states), $h$ is concave~\cite{Guo2023njp,Vidal2000}. It is proved by Vidal~\cite{Vidal2000} that if $E$ is an entanglement monotone on pure states, then $E_F$ is also an entanglement monotone on $\mS^{AB}$. Hereafter, $E_F$ is abbreviated as $E$ sometimes. The Carath\'{e}odory theorem ensures that, for the decompositions of a mixed state into a pure states ensemble, we need only finite pure states, or, to be more precise, the number of pure states $m$ satisfies $m \leq [r(\rho)]^2 \leq (\dim\mH^{AB})^2$~\cite{Uhlmann1998}. This leads to the existence of the minimum in~Eq.~\eref{EOF}. A method of evaluating such convex-roof entanglement measures was proposed in Ref.~\cite{Toth2015prl}.

\subsubsection{Majorization relation: condition for LOCC transformation}

Majorization is a topic of much interest in varies of mathematics and statistics~\cite{Alberti1982book,Bhatia1997book,Marshall1979book}. In quantum information, it is a basic tool that has been used to reflect the disorder of the information contained in the state.
Suppose  $\textbf{x}$ and $\textbf{y}$ are vectors in $\mathbb{R}^n$ with their components in decreasing order, i.e., $x_1\geq x_2\geq\dots\geq x_n$ and $y_1\geq y_2\geq\dots\geq y_n$. We usually suppose in addition that  
$\textbf{x}$ and $\textbf{y}$ are probability distributions. If $\textbf{x}$ and $\textbf{y}$ satisfy
\bea 
\sum_{i=1}^kx_i\leq\sum_{i=1}^ky_i
\eea  
for $k=1,\dots,n$ with equality at $k=n$, then we say that $\textbf{x}$ is majorized by $\textbf{y}$ and denote by
\bea \label{majorization}
\textbf{x}\preceq\textbf{y}.
\eea 
The relation $\textbf{x}\preceq\textbf{y}$ is intended to capture the notion that $\textbf{x}$ is more``chaotic'' (i.e., disordered) than $\textbf{y}$. The majorization relation is connected with the double stochastic matrix: $\textbf{x}\preceq\textbf{y}$ if and only if $\textbf{x}=D\textbf{y}$, where $D$ is a double stochastic matrix. Namely, $\textbf{x}\preceq\textbf{y}$ means that $\textbf{y}$ is the input probability distribution to a noisy channel described by the doubly stochastic matrix $D$, inducing a more disordered output probability distribution, $\textbf{x}$.

Recall that, for any $|\psi\ra^{AB}\in\mH^{AB}$ with $\dim\mH^A=m\leq\dim\mH^B=n$, there exist orthonormal sets $\{|e_i\ra^A\}$ and $\{|e_i\ra^B\}$ such that
\bea \label{Schmidt}
|\psi\ra^{AB}=\sum\limits_{i=1}^{t}\lambda_i|e_i\ra^A|e_i\ra^B
\eea
with $\lambda_i>0$, $i=1$, 2, $\dots$, $t$, $t\leq m$. Eq.~\eref{Schmidt} is call the Schmidt decomposition of $|\psi\ra^{AB}$, $t$ is called the Schmidt rank (or Schmidt number) and denoted by $S_r(|\psi\ra^{AB})$ hereafter, $\{\lambda_i\}$ is the Schmidt coefficents.
Let $|\psi\ra^{AB}$ and $|\phi\ra^{AB}$ be two pure states in $\mH^{AB}$ with Schmidt coefficents $\{\lambda_i\}$ and $\{\delta_i\}$, respectively. We write $\tr_B|\psi\ra\la\psi|^{AB}=\rho^A$ and $\tr_B|\phi\ra\la\phi|^{AB}=\varrho^A$,
${\vec{\sigma}(\rho^A)}=(\lambda_1^2, \lambda_2^2, \dots, \lambda_m^2)$ and  ${\vec{\sigma}(\varrho^A)}=(\delta_1^2, \delta_2^2, \dots, \delta_m^2)$. We assume with noloss of generality that $\lambda_1\geq\lambda_2\geq\cdots\lambda_m$ and $\delta_1\geq\delta_2\geq\cdots\geq\delta_m$. Zeros are appended in these two inequalities in order to make $\vec{\sigma}(\rho^A)$ and $\vec{\sigma}(\varrho^A)$ are $m$-dimensional vectors. In particular, we take $|\psi\ra^{AB}=|\Phi^+\ra$ to be the maximally entangled state, i.e., $\lambda_i=\frac{1}{\sqrt{m}}$, $i=1$, $2$, $\dots$, $m$, and $|\phi\ra^{AB}$ is separable. Then $\vec{\sigma}(\rho^A)\preceq\vec{\sigma}(\varrho^A)$ and for any bipartite entanglement measure $E$ with reduced function $h$, we 
always expect that $h({\vec{\sigma}(\rho^A)})>h({\vec{\sigma}(\varrho^A)})$. In general, we expect that  
\bea\label{Schur-concave}
\vec{\sigma}(\rho^A)\preceq\vec{\sigma}(\varrho^A)\Leftrightarrow h[\vec{\sigma}(\rho^A)]\geq h[\vec{\sigma}(\varrho^A)]
\eea
for any $|\psi\ra^{AB}$ and $|\phi\ra^{AB}$ that satisfies $\vec{\sigma}(\rho^A)\preceq\vec{\sigma}(\varrho^A)$. If $h$ satisfies Eq.~\eref{Schur-concave}, we call $h$ is Schur concave~\cite{Marshall1979book}. According to Nielsen's Theorem in Ref.~\cite{Nielsen1999prl}, there exists some LOCC $\varepsilon$ such that $\varepsilon(|\psi\ra\la\psi|^{AB})=|\phi\ra\la\phi|^{AB}$ iff $\vec{\sigma}(\rho^A)\preceq\vec{\sigma}(\varrho^A)$. Namely, the reduced function is also Schur concave.

\subsubsection{Equivalent entanglement measures}	

Two entanglement measures $E$ and $E'$ are called equivalent to each other if for any states $\rho^{AB}$ and $\sigma^{AB}$ in $\mS^{AB}$, $E(\rho^{AB})\geq E(\sigma^{AB})$ impiles $E'(\rho^{AB})\geq E'(\sigma^{AB})$ and vice versa~\cite{Virmani2000pla}. Namely, equivalent measures exports the same order of the degree of the entanglement contained in the states. There are entanglement measures which export opposite ordering (see Sec.~\ref{sec-2.2}) since there are incomparable states under LOCC, i.e., states $\rho^{AB}$, $\varrho^{AB}$, that neither $\varepsilon(\rho^{AB})=\varrho^{AB}$ nor $\varepsilon(\varrho^{AB})=\rho^{AB}$ is possible for any LOCC $\varepsilon$. The equivalent measures of multipartite entanglement can be argued similar.

\subsection{Examples of bipartite entanglement measures}\label{sec-2.2}

In this subsection, we list bipartite entanglement measures in literature up to now (some of them may not be well-defined entanglement measures). We fix some notations and introduce some basic concepts first. We denote the set of all separable states in $\mS^{AB}$ by $\mS_{sep}$. $S(\rho)=-\tr(\rho\log_2\rho)$ is the von Neumann entropy of the state $\rho$~\cite{Bennett1996pra2046,Bennett1996pra3824,Bennett1996prl,HaydenJozaPetsWinter,Nielsenbook,Rungta2003pra,Wootters2001qic}.
The quantum relative entropy (or called the von neumann relative entropy) between the two states $\rho$ and $\sigma$ is defined as $S(\rho\|\sigma)=\tr\rho(\log_2\rho-\log_2\sigma)$~\cite{Fawzi2015cmp,HaydenJozaPetsWinter,Ibinson2007cmp,Nielsenbook}. $S(\rho\|\sigma)$ quantifies how difficult it is to distinguish the state $\rho$ form the state $\sigma$~\cite{Hiai1991cmp}.
Another measure of distance between two quantum states is the fidelity. Although it is not a metric on $\mS^X$, it does give rise to a useful quantity~\cite{Nielsenbook}. The most widely-employed fidelity that has been proposed in the literatures is the Uhlmann-Jozsa fidelity $F$~\cite{Jozsa1994,Uhlmann1976}, which is defined as
\bea\label{fidelity}
F(\rho,\sigma)=\left(  \tr\sqrt{\sqrt{\rho}\sigma\sqrt{\rho}}\right) ^2.
\eea
It is shown that $F$ is monotonic under any quantum operation and it admits many desirable properties~\cite{liang2019}. The Uhlmann-Jozsa fidelity can also be defined as~\cite{Fawzi2015cmp,Luo2004pra,Zhanglin2015qip}
\bea
\sqrt{F}(\rho,\sigma)=\sqrt{F(\rho,\sigma)}.
\eea
Another alternative fidelity measure that also non-decreases under quantum operation is the square of the quantum affinity $A(\rho,\sigma)$ proposed in~\cite{Ma2008pra,Raggio1984}, by the name of {A-fidelity}:
\bea
F_A(\rho,\sigma)=[\tr(\sqrt{\rho}\sqrt{\sigma})]^2.
\eea

There are two classes of bipartite states that are discussed very often when we explore the measure of entanglement. One is the Werner state and the other is the isotripic state, which are highly symmetric and the amount of the entanglement contained in them can be calculated analytically for several entanglement measures. Recall that, the Werner state is just the one invairant under local unitary operation $U\ot U$ acting on $\mH^{AB}$ with $\dim\mH^A=\dim\mH^B$, i.e., the state $\rho$ that admits $U\ot U\rho U^\dag\ot U^\dag=\rho$ for any unitary operator $U$~\cite{Werner1989pra}. For the $m\ot m$ system, it can be written as 
\bea\label{rhox}
\rho_x=\frac{2(1-x)}{m(m+1)}\Pi^++\frac{2x}{m(m-1)}\Pi^-,\quad 0\leq x\leq 1,
\eea
where $\Pi^+=\frac12(I+\check{F})$, $\Pi^-=\frac12(I-\check{F})$, $I$ is the identity operator on $\mH^{AB}$, $\check{F}$ is the flip operator or called the swap operator, i.e., $\check{F}|\psi\ra|\phi\ra=|\phi\ra|\psi\ra$ for any $|\psi\ra\in\mH^A$ and $|\phi\ra\in\mH^B$. Another expression of the Werner state is
\bea\label{rhoc}
\rho_c=\frac{1}{m^3-m}[(m-c)I+(mc-1)\check{F}], \quad -1\leq c\leq 1.
\eea
The isotropic sate is the one that invariant under the local unitary operation $U\ot U^*$ acting on $\mH^{AB}$ with $\dim\mH^A=\dim\mH^B$, where $U^*$ refers to the conjugate of $U$ up to some given orthonormal basis of the state space~\cite{Horodecki1999pra}. The states of this class can be written as
\bea\label{rhot}
\rho_t=\frac{1-t}{m^2-1}I+\frac{tm^2-1}{m^2-1}P^+,\quad 0\leq t\leq 1,
\eea
where $P^+=|\Phi^+\ra\la\Phi^+|$. The isotropic state can also be expressed as
\bea\label{rhof} 
\rho_f=\dfrac{1-f}{m^2-1}(I-P^+)+fP^+,
\eea
where $f=\la\Phi^+|\rho_f|\Phi^+\ra$. $\rho_f$ is separable if $f\leqslant 1/m$.

Hereafter in this section, we always assume that $\rho\in\mS^{AB}$, $|\psi\ra\in\mH^{AB}$ with $\dim\mH^{AB}<\infty$, and $\rho^{A,B}=\tr_{B,A}\rho$ ($\rho_{A,B}=\tr_{B,A}\rho$ sometimes) unless otherwise specified.

\subsubsection{Distillable entanglement}

Let $\rho$ be a state shared between Alice and Bob. For $n$ copies of state $\rho$, they apply an LOCC operation $\varepsilon$ and the final state is $\varepsilon(\rho^{\ot n})$. We expect that for large $n$ the final state approaches $n'$ copies of the singlet state, i.e., $|\phi^+\ra^{\ot n'}$, where $|\phi^+\ra=\frac{1}{\sqrt{2}}(|00\ra+|11\ra)$. If it is impossible for any $n$, then we call $\rho$ is not distillable. Otherwise we say that the LOCC operations constitute a distillation protocol $\varOmega$ and the rate of distillation is $R_{\varOmega}=\lim_n\frac{n'}{n}$. The distillable entanglement, denoted by $E_d$, is defined as the supremum of such singlet distillation rates over all possible distillation protocols~\cite{Bennett1996pra2046,Bennett1996pra3824,Plenio2007qic}, i.e.,
\bea\label{E_d}
E_d(\rho)=\sup\left\lbrace \frac{n'}{n}:\lim_{n\rightarrow\infty}\left[\inf_\varepsilon\left\|\varepsilon\left( {\rho}^{\otimes n}\right) -\Phi^+_{2^{n'}}\right\|_{\tr} \right]=0\right\rbrace,
\eea
where $\Phi^+_{2^{n'}}=\left(|\phi^+\ra\la\phi^+| \right)^{\otimes n'}$. The distillable entanglement is just the entanglement of communication suggested in Ref.~\cite{Nielsen2001jpa}.

\subsubsection{Entanglement cost}

As a quantity dual to $E_d$, entanglement cost of a given state $\rho$, denoted by $E_c(\rho)$, is defined as the infimum rate of singlets that have to be used to create many copies of $\rho$ via LOCC~\cite{Bennett1996pra2046,Bennett1996pra3824,Hayden2001jpa}:
\bea\label{E_c}
E_c(\rho)=\inf\left\lbrace \frac{n'}{n}:\lim_{n\rightarrow\infty}\left[\inf_\varepsilon\left\| {\rho}^{\otimes n} -\varepsilon\left(\Phi^+_{2^{n'}}\right)\right\|_{\tr} \right]=0\right\rbrace.
\eea
$E_c$ is faithful~\cite{Yangdong2005prl}.

\subsubsection{Entanglement of formation}\label{2.3.3}

The first convex-roof extended measure is the entanglement of formation (EoF)~\cite{Bennett1996pra2046,Bennett1996pra3824,Wootters2001qic},	which is defined by
\bea\label{EoF}
E_f(|\psi\ra)=E(|\psi\ra)=S(\rho^A),
\eea
for pure state, and
\bea
E_f(\rho)=\min_{\{p_i,|\psi_i\ra\}}\sum_ip_iE(|\psi_i\ra)
\eea
for mixed state, where the minimum is taken over all pure-state decomposition $\{p_i,|\psi_i\ra\}$ of $\rho$. For pure state $|\psi\ra$, both $E_c$ and $E_d$ are equal to the entropy of its reduced state~\cite{Bennett1996pra2046,Bennett1996pra3824} 
\beax 
E_f(|\psi\ra)=E_d(|\psi\ra)=E_c(|\psi\ra).
\eeax
$E_c\leq E_f$~\cite{Christandl2003,Hayden2001jpa}. EoF is the minimum value that may be achieved for the ratio $m/n$ in the protocol that creating $n$ ``pretty good'' copies of $\rho$ from $m$ shared singlet state $|\phi^+\ra$ between Alice and Bob using only LOCC~\cite{Bennett1996pra2046,Nielsen2001jpa,Wootters2001qic}. Hayden \etal proved that~\cite{Hayden2001jpa} 
\beax 
E_c(\rho)=\lim_{n\rightarrow\infty}E_f(\rho^{\ot n})/n.
\eeax  
EoF coincides with the so-called entanglement of creation discussed in Reff.~\cite{Nielsen2001jpa}.

Throughout this paper, we identify the original bipartite entanglement of formation with $E_f$, the notation $E_F$ with capital $F$ in the subscription denotes the convex-roof extended measure of $E$ execpt for $E_f$ (in some literature, all the convex-roof extended measures of entanglement are called entanglement of formation uniformly~\cite{Guo2019pra}). In what follows, if we only give the entanglement measures (bipartite or multipartite) for pure states, then for the case of mixed states, they are all defined by the convex-roof extension with no further statement. Obviously, any entanglement measure that defined in this way is convex straightforwardly.

	$E_f(\rho)$ is hard to be calculated for mixed state $\rho$ since the minimizing process is needed. But for some highly sysmmetric states such as the Werner state and the isotropic state, they have analytical expressions~\cite{Terhal2000prl,Vollbrecht}:
	\beax 
	E_f(\rho_x)&=&1-\left[ \frac12-\sqrt{x(1-x)}\right] \log_2\left[ 1-2\sqrt{x(1-x)}\right] \\
	&&-\left[ \frac12+\sqrt{x(1-x)}\right] \log_2\left[ 1+2\sqrt{x(1-x)}\right] ,\quad x>\frac12
	\eeax
	and 
	\beax
	E_f(\rho_t)=\left\lbrace\begin{array}{ll}
		0, &t\leq\frac{1}{m},\\
		h(\gamma)+(1-\gamma)\log_2(m-1), &\frac{1}{m}<t<\frac{4(m-1)}{m^2},\\
		\dfrac{(t-1)m\log_2(m-1)}{m-2}+\log_2m,&\frac{4(m-1)}{m^2}\leq t\leq 1, 
	\end{array}\right.
	\eeax
	where $\gamma=\dfrac{1}{m}\left[ \sqrt{t}+\sqrt{(m-1)(1-t)}\right]^2$.

\subsubsection{Concurrence}

	For the two-qubit case, Hill and Wootters found in 1997 that~\cite{Hill}, 
	\begin{eqnarray}\label{eof}
		E_f(|\psi\rangle)=H_2\left(\frac{1+\sqrt{1-C(|\psi\ra)^2}}{2}\right), 
	\end{eqnarray}
	where
	\bea \label{H_2}
	H_2(x)=-x\log_2x-(1-x)\log_2(1-x)
	\eea
	is the binary entropy function,
	\beax 
	C(|\psi\rangle)=2|a_{00}a_{11}-a_{01}a_{10}|=2\delta_0\delta_1=\sqrt{2(1-\tr\rho_A^2)}
	\eeax
	whenever
	$|\psi\rangle=\sum_{i,j}a_{ij}|i\rangle^A|j\rangle^B=\sum\limits_{i}\delta_i|e_i\ra^A|e_i\ra^B$ with $\sum\limits_{i}\delta_i|e_i\ra^A|e_i\ra^B$ is the Schmidt decomposition.
	$E_f$ is a monotonic increasing function of $C$, so $C$ can be regarded as an entanglement measure,
	which is called concurrence~\cite{Hill,Wootters}.

	Concurrence seems the one that has been investigated the most widely among numerous entanglement measures.
	For two-qubit pure state $|\psi\ra$, $C(|\psi\rangle)$ can also be represented as
	\bea\label{concurrence0}
	C(|\psi\ra)=\la\psi|\sigma_y\ot\sigma_y|\psi^*\ra,
	\eea
	where $\sigma_y$ is the Pauli operator, $|\psi^*\ra$ denotes the complex conjugation of $|\psi\ra$ in the eigenbasis of $\sigma_y$.
	Soon afterwards, Wootters
	extended it to the $2\otimes2$ mixed states by~\cite{Wootters}
	\bea\label{concurrence1}
	C(\rho)=\max\{0,\lambda_1-\lambda_2-\lambda_3-\lambda_4\}
	\eea
	in light of Eq.~\eref{concurrence0}, where $\lambda_i$'s ($i=1$, 2, 3, 4) are the eigenvalues, in decreasing order, of
	$R=(\rho^{\frac{1}{2}}\tilde{\rho}\rho^{\frac{1}{2}})^{\frac{1}{2}}$, 
	$\lambda_1\geq\lambda_2\geq\lambda_3\geq\lambda_4$, where
	$\tilde{\rho}$ is the result of applying the spin-flip operatioin to $\rho$, i.e.,
	$\tilde{\rho}=(\sigma_y\otimes\sigma_y)\rho^*(\sigma_y\otimes\sigma_y)$, $\rho^*$ denotes the complex conjugation of $\rho$ in the eigenbasis of $\sigma_y$.
	Motivated by Eq.~\eref{concurrence1}, in Ref.~\cite{Uhlmann2000pra}, Uhlmann extended the concurrence to higher-dimensionl systems via the conjugation $\Theta$ acting on the state space (a conjugation is an antiunitary operator satisfying $\Theta=\Theta^{-1}$) and called it $\Theta$-concurrence, which is formulated as $C_{\Theta}(|\psi\ra)=\la\psi|\Theta|\psi\ra$ for pure state and the convex-roof extension for mixed is
	\bea\label{Ctheta}
	C_{\Theta}(\rho)=\max\left\lbrace 0, \lambda_1-\sum\limits_{j>1}\lambda_j \right\rbrace,
	\eea
	where $\lambda_j$'s are the eigenvalues of $(\sqrt{\rho}\tilde{\rho}\sqrt{\rho})^{1/2}$ in decreasing order, $\tilde{\rho}=\Theta\rho\Theta$.
	Later, in 2001, Rungta \etal generalized the spin-flip operator $\sigma_y$ to a universal inverter $S_d$ defined as $S_d(\rho)=v_d(I-\rho)$, and then the concurrence of $|\psi\ra$ in $d_1\ot d_2$ system is defined by~\cite{Rungta2001pra} 
	\beax
	C(|\psi\ra)=\sqrt{\la\psi|(S_{d_1}\ot S_{d_2}|\psi\ra\la\psi|)|\psi\ra}=\sqrt{2v_{d_1}v_{d_2}(1-\tr\rho_A^2)},
	\eeax
	where $v_d>0$, $\rho_A=\tr_B|\psi\ra\la\psi|$. This version of concurrence was called the $I$-concurrence~\cite{Rungta2001pra}.
	In general, the sensible choice of $v_d$ is $v_d=1$, which is consistent with the two-qubit case~\cite{Hill,Rungta2001pra,Rungta2003pra,Wootters}, i.e., 
	\bea\label{concurrence}
	C(|\psi\ra)=\sqrt{2(1-\tr\rho_A^2 }),
	\eea
	for arbitrary dimensional system. Audenaert \etal~\cite{Audenaert2001pra} proposed another generalization of concurrence by defining a concurrence vector in terms of a specific set of antilinear operators. However, the length of the concurrence vector in Ref.~\cite{Audenaert2001pra} turns out to be equal to the definition given in equation ~\eref{concurrence}~\cite{Wootters2001qic}.
	In addition, Albeverio and Fei generalized the notion of concurrence for the $d\ot d$ system by using invariants of local unitary transformations as~\cite{Albeverio2001}
	\bea\label{Albeverio-Fei-concurrence}
	C(|\psi\ra)=\sqrt{\frac{d}{d-1}(1-\tr\rho_A^2)},
	\eea
	which turns out to be the same as that of Rungta \etal up to a whole factor~\cite{Rungta2001pra}.

	Let
	\beax
	|\psi\ra=\sum_{i=1}^{d_A}\sum_{j=1}^{d_B}
	a_{ij}|i\ra^A|j\ra^B,
	\eeax
	where $\dim\mH^{A,B}=d_{A,B}$.
	The concurrence vector ${\bf C}$ with components
	$C_{\alpha\beta}$ was defined as~\cite{Akhtarshenas2005jpa}
	\bea\label{N1N2Ccomp}
	C_{\alpha\beta}=\langle\psi |{\tilde \psi}_{\alpha\beta} \rangle,
	\qquad |{\tilde \psi}_{\alpha\beta} \rangle=(L_{\alpha}\otimes
	L_{\beta})\left|\psi^\ast \right>.
	\eea
	where $L_{\alpha}$, $\alpha=1,..., d_A(d_A-1)/2$ and $L_{\beta}$,
	$\beta=1,..., d_B(d_B-1)/2$ are generators of $SO(d_A)$ and
	$SO(d_B)$ respectively. Recalling that, the generators of $SO(d)$ group, denoted by $L_{\alpha}$ with no loss of generality, are operators that satisfy 
	$(L_{\alpha})_{kl}=(L_{[j_1j_2\cdots
		j_{d-2}]})_{kl}=\epsilon_{[j_1j_2\cdots j_{d-2}]kl}$ where
	$\alpha$ is used to denote the set of $d-2$ indices
	$[j_1j_2\cdots j_{d-2}]$ with $1\le j_1 <j_2<\cdots <j_{d-2}\le
	d$ in order to label $d(d-1)/2$ generators of $SO(d)$, and
	$\epsilon_{j_1j_2\cdots j_{d}}$ is antisymmetric under
	interchange of any two indices with $\epsilon_{12\cdots d}=1$.  
	It follows that the norm of the concurrence vector can
	be defined as a measure of entanglement, i.e.~\cite{Akhtarshenas2005jpa},
	\beax\label{N1N2con1}
	C=\|{\bf C}\|= \sqrt{\sum_{\alpha=1}^{d_A(d_A-1)/2} \;\;
		\sum_{\beta=1}^{d_B(d_B-1)/2}|C_{\alpha\beta}|^2}.
	\eeax
	It can also be expressed as~\cite{Akhtarshenas2005jpa}
	\beax\label{C2}
	C=2\sqrt{\sum_{i<j}^{d_A}\;\;\sum_{k<l}^{d_B}|a_{ik}a_{jl}-a_{il}a_
		{jk}|^2 }.
	\eeax
	Noted that the norm of the concurrence vector here
	is the same as that obtained in Eq.~\eref{Albeverio-Fei-concurrence}, and, up to a whole
	factor, it is also in accordance with Eq.~\eref{concurrence}, in turn, it is equal to the
	trace norm of the concurrence matrix proposed in Ref.~\cite{Badziag2001}. In addition, the definition for concurrence vectors in Eq.~\eref{N1N2Ccomp} is closely related
	to that of Ref.~\cite{li2003}.

	In Ref.~\cite{Androulakis2020q,Gudder2020q}, the quantity $e(|\psi\ra)=\sqrt{1-\tr\rho_A^2}$ is called the entanglement number of $|\psi\ra$, $\rho_A=\tr_B|\psi\ra\la\psi|$. Obviously, $C(|\psi\ra)=\sqrt2e(|\psi\ra)$. In what follows, we adopt Eq.~\eref{concurrence} for the definition of concurrence as in the most of literature otherwise specified.

	For the two-qubit case, any bipartite entanglement measure $E$ is a monotonic increasing function of its concurrence, i.e., $E(|\psi\ra)=g\left[C(|\psi\ra)\right]$ for some monotonic increasing function $g$~\cite{Abouraddy2001pra,Chenjingling2002pra}. $g$ is in fact strictly monotonic increasing~\cite{GG2018q}. Wootters proved in Ref.~\cite{Wootters} that, for any two-qubit state $\rho$, there exsits an ensemble~$\{p_i, |\psi_i\ra\}$ such that $C(|\psi_i\ra)=C(\rho)$ holds for each~$i$.

	$C(\rho)$ is hard to be calculated, but for the Werner state and the isotropic state, they have analytical expressions~\cite{Chen2006rmp,Rungta2003pra}. For general states, the lower bound can be obtained~\cite{Chen2005prl} 
	\beax 
	C(\rho)\geq\sqrt{\frac{a}{m(m-1)}}\left(\max\left\lbrace\left\|\rho^{T_a}\right\|_{\tr}, \|\rho^{R}\|_{\tr} \right\rbrace-1  \right), 
	\eeax 
	where $\rho^{T_a}$ is the partial transposition of $\rho$ up to part $A$~\cite{Horodecki1996pla,Peres1996prl}, $\rho^{R}$ is the realignment of $\rho$~\cite{Chen2003qic,Guo2011csb,Guo2013rmp,Rudolph2004lmp}, that is, if $\rho=\sum_{i,j,k,l}a_{i,j,k,l}|i\ra\la j|^A\ot|k\ra\la l|^B$ up to some basis $\{|i\ra^A|k\ra^B\}$ of $\mH^{AB}$, then 
	\bea
	\rho^{T_a}=\sum_{i,j,k,l}a_{i,j,k,l}\left( |i\ra\la j|^A\right)^T\ot|k\ra\la l|^B,\label{partial-transpose}\\
		\rho^{R}=\sum_{i,j,k,l}a_{i,j,k,l}|j\ra^A|i\ra^A\la l|^B\la k|^B,\label{realignment}
	\eea
	where $X^T$ denotes the transpose of $X$ up to some given orthonormal basis. $\rho^R$ here is in fact the column realignment while the row realignment is formulated as $\rho^{R}=\sum_{i,j,k,l}a_{i,j,k,l}|i\ra^A| j|^A\la k|^B\la l|^B$. The trace norm of these two kinds of realignments coincide with each other.

	In Ref.~\cite{Wanghaofan2025pra}, a symmetric measurement-induced lower bounds of concurrence was proposed.
	A set of $N$ $d$-dimensional positive operator-valued measures (POVMs) $\{E_{\alpha,k}|k=1,2\cdots,M\}$ ($\alpha=1,2,\cdots,N$) constitute an $(N,M)$-POVM if
	\beax 
	\tr(E_{\alpha,k}) = \dfrac{d}{M}, \quad \tr(E_{\alpha,k}^{2}) = x,\\
	\tr(E_{\alpha,k}E_{\alpha,l}) = \dfrac{d-Mx}{M(M-1)},~~ l\neq k\\
	\tr(E_{\alpha,k}E_{\beta,l}) = \dfrac{d}{M^{2}},~~ \beta\neq\alpha
	\eeax
	where the parameter $x$ satisfies ${d}/{M^{2}}<x\leq \min\left\{{d^{2}}/{M^{2}},{d}/{M}\right\}$. When $N(M-1)=d^{2}-1$, the $(N,M)$-POVM is called an informationally complete $(N,M)$-POVM.
	Let $\{E_{\alpha,k}|\alpha=1,\cdots, N;\,k=1,\cdots,M\}$ be an informationally complete $(N, M)$-POVM with the free parameter $x$ on the $d$ dimensional Hilbert space $\mH$, $\rho$ be a bipartite state acting on $\mH\otimes\mH$, $\{|w_{\alpha,k}\ra|\alpha=1,\cdots,N;\,k=1,\cdots,M\}$ be an orthonormal basis of $\mathbb{C}^{NM}$. Define $\mP(\rho)=\sum\limits_{\alpha,\beta=1}^{N}\sum\limits_{k,l=1}^{M}{\tr\left[\rho\left(E_{\alpha,k}\otimes E_{\beta,l}\right)\right]|w_{\alpha,k}}\ra\la w_{\beta,l}|$. The concurrence $C(\rho)$ satisfies
	\beax\label{thm}
	C(\rho)\geq \dfrac{M(M-1)}{xM^{2}-d}\sqrt{\dfrac{2}{d(d-1)}}\left[\|\mP(\rho)\|_{\tr}-\dfrac{(d-1)(xM^{2}+d^{2})}{dM(M-1)}\right].
	\eeax
	Let $\{P_1,\cdots,P_{d^2}\}$ be a general sysmmetric informationally complete (SIC)\cite{Kalev2014jpa,Renes2004jmp} POVM with the free parameter $x$ on $\mH$, $\rho$ be a bipartite state on $\mH\otimes\mH$. Denote $E_k=\sqrt{\dfrac{d(d+1)}{xd^{2}+1}}P_k$($k=1,2,\cdots,d^{2}$), $p_{kl}=\tr\left[\rho\left(E_k\otimes E_l\right)\right]$ and $\mP(\rho)=\left[p_{kl}\right]_{d^{2}\times d^{2}}$. We have
	\beax
\fl \qquad \qquad	C(\rho)\geq \dfrac{(d-1)(xd^{2}+1)}{xd^{3}-1}\sqrt{\dfrac{2}{d(d-1)}}\left[ \|\mP(\rho)\|_{\tr}-\dfrac{xd^{3}-1+d(1-xd)}{(d-1)(xd^{2}+1)}\right].
	\eeax
	Let $\Pi_1,\cdots,\Pi_{d^2}$ be SIC projectors. Denote $E_k=\sqrt{\dfrac{d+1}{2d}}\Pi_k$ ($k=1,2,\cdots,d^{2}$), $p_{kl}=\tr\left[\rho\left(E_k\otimes E_l\right)\right]$ and $\mP(\rho)=\left[ p_{kl}\right]_{d^{2}\times d^{2}}$. Then
	\beax
	C(\rho)\geq 2\sqrt{\dfrac{2}{d(d-1)}}\left(\|\mP(\rho)\|_{\tr}-1\right).
	\eeax

	For a bipartite state $\rho$, Ref.~\cite{Luyu2025qip} defined
	\beax\label{eq:m1}
		\mathcal{Q}_{\mu, \nu}(\rho)=\left(\begin{array}{cc}
			\mu\nu^T & \mu \operatorname{vec}(\rho_B)^T \\
			\operatorname{vec}(\rho_A)\nu^T& \rho^{R}
		\end{array}\right),
	\eeax
	where $\mu=(u_1,...,u_{n})^T$ and $\nu=(v_1,...,v_{m})^T$,
	$u_i$ $(i=1,...,n)$ and $v_j$ $(j=1,...,m)$ are given real numbers, $m$ and $n$ are positive integers. 
	From $\mathcal{Q}_{\mu, \nu}(\rho)$, a lower bound of concurrence was obtained~\cite{Luyu2025qip}, i.e.,
	the concurrence of $\rho$ satisfies that
	\begin{eqnarray}\notag
		C \!\left(\rho\right)\geq \frac{\sqrt{2}}{\sqrt{d(d-1)}}\!\!\left[\left\|\mathcal{Q}_{\mu, \nu}\left(\rho\right)\right\|_\mathrm{T r}\!\! - \!\sqrt{\left(|\mu|^{2}+1\!\right)\left(|\nu|^{2}+1\!\right)}\right],
	\end{eqnarray}
	where $d=\min\dim\mH^{A,B}$.

\subsubsection{Tangle}

The tangle of a pure state $|\psi\ra$ is defined as the square of the concurrence~\cite{Rungta2003pra}, i.e.,
	\bea\label{tangle}
	\tau(|\psi\ra)=C^2(|\psi\ra)=2(1-\tr\rho_A^2).
	\eea
	For any $\rho\in\mS^{AB}$, we have~\cite{Guo2013qip,Mintert2007prl,Osborne,Zhang2008pra}
	\beax 
	C^2(\rho)\leq \tau(\rho)\leq 2(1-\tr\rho_A^2).
	\eeax 
	If $\rho$ is a two-qubit state, $C^2(\rho)=\tau(\rho)$~\cite{Osborne}.

\subsubsection{Concurrence hierarchy}

	Fan \etal defined concurrence hierarchy for $d\otimes d$ systems in Ref.~\cite{Fan2003jpa}, and later,  it was explored in detail in Ref.~\cite{Gour2005pra} wherein it is called concurrence monotones by Gour.
	Let $|\psi\ra$ be a pure state with Schmidt decomposition
	$|\psi\ra=\sum_j\lambda_j|e_j\ra^A|e_j\ra^B$ and let
	\beax
	\eqalign{
		S_1(\lambda)=\sum_j\lambda_j^2,\quad S_2(\lambda)=\sum_{i<j}\lambda_i^2\lambda_j^2,\cr
		S_3(\lambda)=\sum_{i<j<k}\lambda_i^2\lambda_j^2\lambda_k^2,\quad, \dots,\quad  S_d(\lambda)=\prod_{j=0}^{d-1}\lambda_j^2.}
	\eeax
	The concurrence hierarchy (or called concurrence monotones) were defined as~\cite{Fan2003jpa,Gour2005pra}
	\bea\label{Concurrence hierarchy}
	C_k(|\psi\ra)=\left[\frac{S_k(\lambda)}{S_k(1/d, 1/d,\dots, 1/d)}\right]^{1/k}, \quad k\geq 2.
	\eea
	The corresponding reduced function is
	\beax\label{h_k}
	h_k(\rho)=[S_k(\lambda(\rho))/S_k(1/d, 1/d,\dots, 1/d)]^{1/k},
	\eeax
	where $\lambda(\rho)$ denotes the vector of eigenvalues of $\rho$.
	$C_k$ is entanglement monotone since $h_k$ is concave whenever $1\leq k\leq d$ and it is strictly concave whenever $k>1~\cite{Marshall1979book}$.

	The case of $k=d$ is called $G$-concurrence which is denoted by $G_d$~\cite{Gour2005pra}, i.e.,
	\bea\label{G-concurrence}
	G_d(|\psi\ra)=C_{d}(|\psi\ra)=d(\lambda_0\lambda_1\cdots\lambda_{d-1})^{1/d}
	\eea
	It is coincides with the original concurrence for $d=2$.
	$C_k$'s have the following properties~\cite{Gour2005pra}: (i) $C_2$ is reduced to the expression for the concurrence given in~\cite{Albeverio2001,Rungta2001pra},
	\beax
	C_2(|\psi\ra)=\sqrt{\frac{d}{d-1}(1-\tr\rho_A^2)};
	\eeax
	(ii) $C_k(|\psi\ra)=0$ if $k$ is greater than the Schmidt rank of $|\psi\ra$;
	(iii) $[C_k(|\psi\ra)]^k\geq[C_l(|\psi\ra)]^l$ whenever $2\leq  k\leq  l\leq  d$;
	(iv) $G_d(\rho)\leq  C_k(\rho)$ for any bipartite state $\rho$, $k\leq  d$;
	(v) $G_{dd'}(|\psi\ra|\phi\ra)=G_d(|\psi\ra)G_d(|\phi\ra)$ for any $d\ot d$ pure state $|\psi\ra$ and $d'\ot d'$ pure state $|\phi\ra\in\mH^{A'B'}$, where $G_{dd'}(|\psi\ra|\phi\ra)$ is measured up to the partition $AA'|BB'$;
	(vi) $G_d(c|\psi\ra)=|c|^2G_d(|\psi\ra)$ holds for any complex number $c$, and 
	$G_d(A\ot B|\psi\ra)=|{\rm Det}(A)|^{2/d}|{\rm Det}(B)|^{2/d}G_d(|\psi\ra)$ for any operators $A$ and $B$ that acting on
	$\mH^A$ and $\mH^B$, respectively.
	A similar idea was also proposed by Sinolecka, Zyczkowski and Kus~\cite{Sino2002app}.

\subsubsection{$q$-concurrence}

	Another generalization of concurrence is the $q$-concurrence~\cite{Yangxue2021pra},
	\bea\label{C_q}
	C_q(|\psi\ra)=1-\tr\rho_A^q,\quad q\geq 2,~ \rho_A=\tr_B|\psi\ra\la\psi|.
	\eea
	Clearly, the reduced function is concave, so it is an entanglement monotone. In fact, it can be defined for $q>1$~\cite{Guo2024rip}.
	Suppose a pure state $|\psi\rangle$ has the Schmidt decomposition
	\begin{eqnarray*}
		|\psi\rangle=\sum^m_{i=1}{\lambda_i}|e_i\rangle^A|e_i\rangle^B
	\end{eqnarray*}
	Then one can get the expression of $q$-concurrence as following
	\beax
		C_q(|\psi\rangle)=1-\sum^m_{i=1} {\lambda^{2q}_i},
		\label{eqni}
	\eeax
	and $0\leq C_q(|\psi\rangle)\leq 1-m^{1-q}$. The lower bound is obtained for product states while the upper bound is achieved for $|\Phi^+\ra$.
	
	For any mixed entanglement state $\rho$ with the dimension of $\dim\mH^A=d_A$ and $\dim\mH^B=d_B$ ($d_A \leq d_B$), the $q$-concurrence $C_q(\rho)$ satisfies the following inequality~\cite{Yangxue2021pra}
	\beax
		C_q(\rho)\geq\frac{\left( \max\left\lbrace \left\| \rho^{T_a}\right\|_{\tr}^{q-1},\left\|\rho^R\right\|_{\tr}^{q-1}\right\rbrace -1\right) ^2}{d_A^{2q-2}-d_A^{q-1}}.
	\eeax

\subsubsection{$\alpha$-concurrence}

	In Ref.~\cite{Wei2022jpa}, a parameterized bipartite entanglement monotone called $\alpha$-concurrence is proposed:
	\bea\label{C_alpha}
	C_\alpha(|\psi\ra)=\tr\rho_A^\alpha-1,\quad 0\leq  \alpha\leq  1/2
	\eea
	for any pure state $|\psi\ra$. Indeed, $C_\alpha$ is valid for $0<\alpha<1$.
	$C_\alpha$ has a lower bound~\cite{Wei2022jpa}
	\beax
	C_\alpha(\rho)\geq \frac{d^{1-\alpha}}{d-1}\left[\max\left(\left\|\rho^{T_a}\right\|_{\tr}, \left\|\rho^{R}\right\|_{\tr}\right)-1\right].
	\eeax

\subsubsection{Total concurrence}

	The total concurrence $C_q^{to}$ ($q\geqslant 2$) is defined by~\cite{Xuan2025aqt}
	\bea \label{C_q^{to}}
	C_q^{to}(|\psi\ra)=\frac{1}{\mu}\left[  1-\tr\rho_A^q+\tr(I_A-\rho_A)-\tr(I_A-\rho_A)^q\right],
	\eea 
	where $\mu=d-d^{1-q}[1+(d-1)^q]$ serves as the normalization factor to ensure the scaling of the measure,
	$d=\dim\mH^{A,B}$. It is an entanglement monotone since its reduced function is strictly concave. If $|\psi\ra=\sum_{i}\lambda_i|e_i\ra^A|e_i\ra^B$ is the Schmidt decomposition of $|\psi\ra$, then
	\beax 
	C_q^{to}(|\psi\ra)=\frac{1}{\mu}\left[ d-\sum_i\lambda_i^{2q}-\sum_i\left(1-\lambda_i^2\right)^q\right].
	\eeax 
	Let $d=\dim\mH^{A,B}$, then
	\beax 
	C_q^{to}(\rho)\geqslant\dfrac{d-d^{1-q}[1+(d-1)^q]}{(d-1)^2}\left[\max\left(\left\|\rho^{T_a}\right\|_{\tr}, \left\|\rho^{R}\right\|_{\tr}\right)-1\right]^2
	\eeax 
	for $q\geqslant 2$ with $d\geqslant 3$ or $q\geqslant 4$ with $d=2$, and 
	\beax 
	C_q^{to}(\rho)>\dfrac{1-2^{1-q}}{1-2^{1-s}}\left[\max\left(\left\|\rho^{T_a}\right\|_{\tr}, \left\|\rho^{R}\right\|_{\tr}\right)-1\right]^2
	\eeax 
	for $3.33902\leqslant q<4$ with $d=2$. $C_3^{to}$ of the Werner state and the isotropic state when $d=2$ and $d=2, 3$ are calculated in Ref.~\cite{Xuan2025aqt}, respectively.

\subsubsection{Tsallis-$q$ entanglement}

	Quantum Tsallis-$q$ entropy is defined as~\cite{Landsberg,Tsallis1988jsp}
	\begin{eqnarray}\label{Quantum Tsallis-entropy}
		S_q(\rho)=\frac{1-{\tr}\rho^q}{q-1}
	\end{eqnarray}
	for $q>0$, $q\neq 1$.
	The Tsallis-$q$ entanglement is defined by~\cite{Kim2010pra}
	\begin{eqnarray}\label{E_q}
		E_q\left( |\psi\ra\right) =S_q\left( \rho^A\right) 
	\end{eqnarray}
	for pure state. It is an entanglement monotone since $S_q$ is strictly concave~\cite{GG2019pra,Landsberg,Tsallis1988jsp}.

	For any $2\otimes d$ pure state
	$|\psi\rangle$ with Schmidt coefficients $\{\sqrt{\lambda}, \sqrt{1-\lambda}\}$, its Tsallis-$q$ entanglement is \cite{Kim2010pra}
	\begin{eqnarray*}
		E_q\left(|\psi\rangle \right)=S_q(\rho^A)
		=\frac{1}{1-q}\left[\lambda^{q}+\left(1-\lambda\right)^{q}-1 \right],
	\end{eqnarray*} 
	and it can also be expressed as~\cite{Kim2011jpa}
	\beax 
	E_q\left(|\psi\rangle\right)=f_q\left(C(|\psi\rangle) \right),
	\eeax
	where 
	\beax
	f_q(x)=\frac{\left(1+\sqrt{1-x^2}\right)^{q}
		+\left(1-\sqrt{1-x^2}\right)^{q}-2^{q}}{(1-q)2^{q}},\quad 0 \leq x \leq 1.
	\label{f}
	\eeax
	Specially, for any two-qubit mixed state $\rho$,
	$E_q(\rho)=f_q(C(\rho) )$ whenever $1 \leq q \leq4$ and $0 \leq x \leq 1$.

\subsubsection{R\'{e}nyi-$\alpha$ entanglement}

	The R\'{e}nyi-$\alpha$ entropy is defined as~\cite{Dur2000pra2,Greenberger,Renyi}
	\begin{eqnarray}\label{Salpha}
		S_\alpha(\rho)=\frac{1}{1-\alpha}\log_2{\tr}\rho^\alpha, \quad \alpha>0, \alpha\neq 1.
	\end{eqnarray}
	The R\'{e}nyi-$\alpha$ entanglement for pure state reads as~\cite{Gour2007jmp,Kim2010jpa,Vidal2000}
	\begin{eqnarray}\label{E_alpha}
		E_\alpha\!\left( |\psi\ra\right) =S_\alpha\!\left(\rho^A\right),\quad 0<\alpha<1.
	\end{eqnarray}
	It is an entanglement monotone as $S_\alpha$ is strictly concave whenever $0\leq  \alpha\leq  1$~\cite{Vidal2000}. 
	For the case where $\alpha\leq  2$, R\'{e}nyi-$\alpha$ entropy is
	concave if the dimension of the quantum system is 2~\cite{Ben1978}. That is, $E_\alpha$ is still an entanglement monotone of $2\ot n$ quantum system. However, for
	$\alpha>2$ or for $\alpha>1$ in higher dimensional systems, $S_\alpha$ is not even concave. So $E_\alpha$ may not be an entanglement monotone in such cases.
	
	For any $2\otimes d$ pure state
	$|\psi\rangle$ with Schmidt coefficients $\{\sqrt{\lambda}, \sqrt{1-\lambda}\}$, its R\'{e}nyi-$\alpha$ entanglement is \cite{Kim2010jpa}
	\begin{eqnarray*}
		E_\alpha\!\left(|\psi\rangle \right)=S_\alpha(\rho^A)
		=\frac{1}{1-\alpha}\log_2\left[\lambda^{\alpha}+\left(1-\lambda\right)^{\alpha} \right].
	\end{eqnarray*} 
	Let \begin{eqnarray*}
		g_\alpha(x)=\frac{1}{1-\alpha}\log_2\left[\left(\frac{1+\sqrt{1-x^2}}{2}\right)^\alpha+\left(\frac{1-\sqrt{1-x^2}}{2}\right)^\alpha\right],
	\end{eqnarray*}
	then~\cite{Kim2010jpa}
	\beax
	E_\alpha\!\left(|\psi\rangle\right)=g_\alpha\!\left[C(|\psi\rangle] \right)
	\eeax 
	whenever $0 \leq x \leq 1$, $\alpha\geq0$ and $\alpha\neq1$.

	Here, we note that $g_\alpha(x)$ converges to the function $H_2\left(\frac{1+\sqrt{1-x^2}}{2}\right)$ with $H_2$ as in Eq.~\eref{H_2} whenever $\alpha\to1$.
	Specially, for any two-qubit mixed state $\rho$,
	\beax 
	E_\alpha\!\left(\rho\right)=g_\alpha\!\left[C(\rho)\right]
	\eeax
	 with $\alpha\geq1$ and $0 \leq x \leq 1$ \cite{Kim2010jpa}.

	Interestingly, the R\'{e}nyi entropies are Schur concave but not concave when $\alpha>1$, thus $E_\alpha$ is an entanglement measure on pure states, but do
	not satisfy Eq.~\eref{average} in such a case. As pointed in Ref.~\cite{Horodecki2009}, it still remains unknown how we can extend such measure to mixed states.

\subsubsection{Unified-$(q,s)$ entanglement }

	Hu \etal~\cite{Hu2006jmp} introduced in 2006 the notion of unified-$(q,s)$ entropy
	\bea\label{unified entropy}
	S_{q,s}(\rho)=\frac{1}{(1-q)s}\left[\left(\mathrm{Tr}\rho^q\right)^s-1\right],\quad q>0, q\neq 1, s\neq0.
	\eea 
	In 2011, Rastegin~\cite{Rastegin2011jsp} showed that $S_{q,s}$ are concave for $0<q<1$ and $s\leq 1$, and that it is subadditive for $q>1$ and $s>1/q$.
	Consequently, Kim and Sanders introduced the unified-$(q,s)$ entanglement in Ref.~\cite{Kim2011jpa}:
	\bea\label{uep}
	E_{q,s}(|\psi\rangle)=S_{q,s}(\rho^A),\quad 0<q<1, 0< s\leq 1.
	\eea

	Unified-$(q,s)$ entanglement reduces to R\'{e}nyi-$q$ (i.e., the case of $\alpha=q$) entanglement when $s$ tends to 0, and converges to Tsallis-$q$ entanglement when $s$ tends to 1. With any nonnegative $s$, unified-$(q,s)$ entanglement converges to EoF when $q$ tends to 1.
	Namely, $E_{q,s}$ is one of the most general classes of bipartite entanglement measures, including $E_\alpha$, $E_q$ and $E_f$ as special classes~\cite{Kim2011jpa}.

	For any $2\otimes d$ pure state
	$|\psi\rangle$ with Schmidt coefficients $\{\sqrt{\lambda}, \sqrt{1-\lambda}\}$, we have~\cite{Kim2011jpa}
	\beax
	E_{q,s}\!\left(|\psi\rangle \right)=S_{q,s}(\rho^A)
	=\frac{1}{(1-q)s}\left\lbrace \left[\lambda^{q}+\left(1-\lambda\right)^{q}\right]^s-1 \right\rbrace 
	\eeax
	and
	\beax 
	E_{q,s}\!\left(|\psi\rangle\right)=f_{q,s}\!\left[C(|\psi\rangle) \right],
	\eeax
	where 
	\beax
	f_{q,s}(x)=\frac{\left[ \left(1+\sqrt{1-x^2}\right)^{q}
		+\left(1-\sqrt{1-x^2}\right)^{q}\right]^s-2^{qs}}{(1-q)s2^{qs}}, \quad 0 \leq x \leq 1.
	\label{f2}
	\eeax

	Furthermore, $f_{q,s}(x)$ converges to the function $H_2\left(\frac{1+\sqrt{1-x^2}}{2}\right)$ with $H_2$ as in Eq.~\eref{H_2} whenever $q\to1$, and
	it reduces to $g_\alpha$ and $f_q$ in the previous subsections when $s$ tends to $0$ and $1$ respectively~\cite{Kim2010pra,Kim2010jpa}.
	Especially, for any two-qubit mixed state $\rho$,
	\beax 
	E_{q,s}\!\left(\rho\right)=f_{q,s}\left[C(\rho) \right]
	\eeax 
	whenever $q\geq1$, $0 \leq s \leq1$, and $qs\leq 3$.

\subsubsection{The dual entropy of entanglement}

	In Ref.~\cite{Yangxue2023epj}, the quantum dual entropy was proposed, which is defined as
	\bea \label{quantum dual entropy}
	S^{to}(\rho)=-\tr[\rho\log_2\rho+(I-\rho)\log_2(I-\rho)].
	\eea 
	$-\tr(I-\rho)\log_2(I-\rho)$ is regarded as the quantum version of the ``extropy'' , where ``extropy'' of a given distribution $\{p_i\}$ is defined by $H(\{p_i\})=-\sum\limits_{i}(1-p_i)\log_2(1-p_i)$~\cite{Lad}.
	$S^{to}(\rho)$ has the following properties~\cite{Lad}: (i) $0\leq S^{to}(\rho)\leq d\log_2d-(d-1)\log_2(d-1)$ (assume that $\dim\mH=d$), (ii) $S^{to}(\rho)$ is concave, (iii) 
	\beax 
	\max\{S^{to}(\rho^A), S^{to}(\rho^B)\}\leq S^{to}(\rho^A\ot\rho^B)<S^{to}(\rho^A)+S^{to}(\rho^B).
	\eeax 
	Then the $S^{to}$-entropy entanglement
	\bea \label{E_t}
	E_t(|\psi\ra)=\frac{1}{r}S^{to}(\rho^A)
	\eea 
	is a well-defined entanglement monotone, $r=d\log_2d-(d-1)\log_2(d-1)$, $\dim\mH^{A,B}=d$. For the two-qubit case, there is an analytic formula
	\bea 
	E_t(\rho)=h(C(\rho))
	\eea
	with 
	\beax 
\fl \qquad \quad	h(x)=-\frac{1+\sqrt{1-x^2}}{2}\log_2\!\frac{1+\sqrt{1-x^2}}{2}
	-\frac{1-\sqrt{1-x^2}}{2}\log_2\!\left(\frac{1-\sqrt{1-x^2}}{2}\right).
	\eeax

\subsubsection{Negativity}

	Negativity is a computable entanglement monotone via the well-known positive partial transpose (PPT) criterion of entanglement~\cite{Horodecki1996pla,Horodeki1997pla,Peres1996prl}. It was firstly introduced by \.{Z}yczkowski \etal in 1998~\cite{Zyczkowski1998pra}, and then Vidal and Werner showed that it was a well-defined entanglement monotone~\cite{VidalWerner}. Recall that, the negativity is defined as
	\bea \label{negativity}
	N(\rho)=\left|\sum_i\mu_i\right| =\frac{1}{2}\left(  \left\| \rho^{T_a}\right\|_\tr-1\right) ,
	\eea
	where $\mu_i$'s are the negative eigenvalues of $\rho^{T_a}$. Namely, $N$ is a quantitative version of the PPT criterion  of entanglement. $N$ is shown to be convex and is non-increasing on average under LOCC, so it is an entanglement monotone.
	$N(\rho)$ is not faithful since $N(\rho)=0$ if $\rho$ is an entangled state with positive partial transpose.
	Later, Lee \etal discussed the convex-roof extension of the negativity, denoted by $N_F$, and showed that $N_F$ is an entanglement monotone~\cite{Lee2003pra}. Note that, there is a gap in the proof. In fact, the reduced function of $N$ (or $N_F$) is strictly concave~\cite{Guo2023njp}. In Ref.~\cite{Lee2003pra}, $N$ is defined as 
	\bea \label{2-3-14}
	N(\rho)=\frac{1}{d-1}\left(\left\|\rho^{T_a}\right\|_\tr-1\right) ,
	\eea 
	where $d=\min\dim\mH^{A,B}$. In what follows, we use definition~\eref{negativity} unless otherwise specified.

	Let $\rho^{T_a}=A^+-A^-$ be the Jordan decomposition of $\rho^{T_a}$, then
	$A^{\pm}\geq 0$, $A^+A^-=A^-A^+=0$. It follows that $N(\rho)=\|A^-\|_\tr$.
	If we take 
	\bea 
	\tilde{N}(\rho)=\|A^-\|_2,
	\eea
	then $\tilde{N}_F$ is still an entanglement monotone since
	$C(|\psi\ra)=2\tilde{N}(|\psi\ra)$.

	For pure state $|\psi\ra$ with the Schmidt coefficients $\{\lambda_j\}$,
	\beax
	N(|\psi\ra)=\frac12\left[\left( \sum_j\lambda_j\right)^2 -1\right]=\frac12\left[\left(\tr\sqrt{\rho_A}\right)^2 -1\right].
	\eeax
	For the $d\ot d$ state $\rho$ that commutes with all unitaries of the form $U\ot U$, where $U$ is real orthogonal, i.e.,
	\beax 
	\rho=adP^+ +b\check{F}+cI
	\eeax 
	with $a, b, c$ are suitable real numbers, we have~\cite{VidalWerner}
	\beax 
	N(\rho)=\frac14|1-f|+\frac14|1+f-2g/d|+\frac12|g/d|-\frac12,
	\eeax 
	where $f=d\tr(\rho P^+)$ and $g=\tr(\rho\check{F})$.
	Clearly,
	\beax 
	2\tilde{N}(\rho)\leq  C(\rho). 
	\eeax
	In Ref.~\cite{Verst2001jpa}, the authors proved that for any 2-qubit state $\rho$, 
	\beax \label{N-C}
	\sqrt{(1-C(\rho))^2+C^2(\rho)}-1+C(\rho)\leq  2N(\rho)\leq  C(\rho).
	\eeax 
	In the first inequality, the equality happens iff $\rho$ is a rank 2 quasi-distillable state. 
	For any 2-qubit state $\rho$~\cite{Kim2009}, 
	\beax 
	2N(\rho)= C(\rho).
	\eeax 
	In addition, 
	\beax 
	2N(|\psi\ra^{A|BC})= C(|\psi\ra^{A|BC})
	\eeax 
	holds for any 3-qubit pure state $|\psi\ra^{ABC}$~\cite{Ouyongcheng2007pra2}.

	For any state $\rho\in\mS^{AB}$, it was proved that~\cite{Luyu2025qip}	
	\begin{eqnarray}\notag
		N_F\left(\rho\right) \geq \frac{\left\|\mathcal{Q}_{\mu, \nu}\left(\rho\right)\right\|_{\mathrm{T r}}-\sqrt{\left(|\mu|^{2}+1\right)\left(| \nu|^{2}+1\right)}}{d-1},
	\end{eqnarray}
	where $d=\min\dim\mH^{A,B}$, $N_F$ here is defined from Eq.~\eref{2-3-14}, $\mathcal{Q}_{\mu, \nu}\left(\rho\right)$ is defined as in Eq.~\eref{eq:m1}.

Another quantifier of entanglement, similar to negativity, that based on the realignment criterion~\cite{Chen2003qic,Rudolph2004lmp}, was defined as~\cite{Yinchao2023prl}
\bea\label{E_R}
E_R(\rho)=\ln\left\|\rho^R\right\|_\tr 
\eea
in order to calculate the entanglement shared by two intervals in the ground state of a $(1+1)$-dimensional conformal field theory (CFT). Wherein it was called CCNR negativity. $E_R(\rho)>0$ implies $\rho$ is entangled, but it is not a well-defined entanglement measure. The advantage of $E_R$ is that the Riemann surface for any number of replicas always has genus 1, which leads to connection with CFT on the torus.

\subsubsection{Logarithmic negativity}

	The logarithmic negativity $E_N$ is defined
	as~\cite{Plenio2005prl,VidalWerner}
	\bea\label{E_N}
	E_N(\rho)=\log_2\left\|\rho^{T_a}\right\|_\tr\,.
	\eea
	It is non-increasing on average under LOCC, but it is not convex, so it is not an entanglement monotone~\cite{Plenio2005prl}.
	By definition, $E_N$ is additive.
	$E_N$ was always at least as great as the entanglement of distillation $E_d^\epsilon(\rho)$, where $\epsilon$ denotes the degree of imperfection allowed in the distilled singlets~\cite{VidalWerner}.

	The logarithmic negativity was extended into a more general case in Ref.~\cite{Carrasco2021qip}, which was called trace-norm group negativities, including the original logarithmic negativity as a special case.
	We review some preliminary notations introduced in Ref.~\cite{Carrasco2021qip}. 
	The group logarithms $\log_G$ is the function of the form
	\beax\label{eq:GL}
	\log_{G}(x)= G(\ln x), \quad x>0
	\eeax
	where $G$ is a suitable, strictly increasing function (vanishing at $0$), assuring concavity of $\log_{G}(x)$.
	A group logarithm $\log_{G}(x)$ is said to be subadditive if
	\beax 
	\log_G(xy)\le\log_Gx+\log_G y
	\eeax
	for any $x>0$ and $y>0$. 
	The trace-norm group negativity is defined as~\cite{Carrasco2021qip}
	\bea \label{eq:tngn}
	E_{N,G}(\rho)=\log_{G} \left\|   \rho^{T_b}\right\| _{\tr}, 
	\eea
	Then, the original logarithmic negativity is just the case of $G(x)=x$.
	A non-trivial example is provided by the use of the $q$-logarithm form, i.e.,
	\bea\label{E_{N-q}}
	E_{N\text{-}q}(\rho)=\frac{\left\| \rho^{T_b}\right\| _\tr^{1-q}-1}{1-q}, \quad q>1  \,.
	\eea
	For $q\to 1$, it reduces to the logarithmic negativity, i.e., $\lim_{q\to 1}E_{N\text{-}q}(\rho)=E_N(\rho)$. 
	$E_{N,G}$ was shown to be an entanglement measure since it does not increase on average under LOCC~\cite{Carrasco2021qip}.

\subsubsection{The relative entropy of entanglement} 

	Another important entanglement monotone that
	is not derived from the convex-roof extension
	is the relative entropy of entanglement~\cite{Vedral1997,Vedral1998pra}:
	\bea\label{E_r}
	E_r(\rho)= \min\limits_{\sigma}S\left( \rho||\sigma\right) ,
	\eea
	where the minimum is taken over all separable states $\sigma$ in $\mS_{sep}$. 
	For pure state, it is equal to EoF, and $E_r(\rho)\leq E_f(\rho)$ for any $\rho$~\cite{Vedral1998pra}.
	$E_r$ was shown to be strictly subadditive; i.e., $E_r(\rho^{\ot2})<2E_r(\rho)$ for some $\rho$~\cite{Vollbrecht}.  where $E_r(\rho^{\ot2})$ is measured under the split $AA|BB$.
	
	The ``standard'' regularization of $E_r$ was defined as~\cite{Brandao2010cmp,Horodecki2000prl}
	\bea 
	E_r^\infty(\rho)=\lim\limits_{n\rightarrow\infty}\frac{1}{n}E_r(\rho^{\ot n}).
	\eea 
	$E_r^\infty\neq E_r$ in general. $E_r^\infty$ plays an analogous role to the entropy in thermodynamics~\cite{Brandao2010cmp} and it is faithful~\cite{Piani2009prl}. $E_r^\infty(\rho)$ satisfies $E_c\geq E_r^\infty\geq E_d$~\cite{Brandao2008np}.

For any give $\rho\in\mS^{AB}$, let
\beax
\mS_\rho^{AB}=\{\sigma\in\mS_{sep}^{AB}: \sigma^A=\rho^A, \sigma^B=\rho^B\}.  
\eeax 
Eisert \etal~\cite{Eisert2003jpa} defined 
\bea
E_A(\rho)&=&\inf\limits_{\sigma\in\mS_\rho^{AB}}S(\sigma\|\rho),\label{E_A}\\
E_M(\rho)&=&\inf\limits_{\sigma\in\mS_\rho^{AB}}S(\rho\|\sigma).\label{E_M}
\eea
$E_A$ and $E_M$ are shown to be entanglement monotones, $E_M(|\psi\ra)=S(\rho^A)$, and $E_A$ is additive, i.e.,
\beax 
E_A(\rho\ot\varrho)=E_A(\rho)+E_A(\varrho),
\eeax
where $\varrho\in\mS^{A'B'}$, $E_A(\rho\ot\varrho)$ is measured under the split $AA'|BB'$.

\subsubsection{Squashed entanglement}

	The squashed entanglement $E_{sq}$~\cite{Christandl2004jmp}
	is an additive entanglement monotone and has a nice 
	operational meaning.
	For any state $\rho\in\mS^{AB}$, $E_{sq}$ is defined by~\cite{Christandl2004jmp}
	\bea\label{E_{sq}}
	E_{sq}(\rho)=\inf\limits_{E}\left\lbrace \frac{1}{2}I(A;B|E)\bigg| {\rm Tr}_E\rho^{ABE}=\rho\right\rbrace,
	\eea
	where $I(A;B|E)=S(\rho^{AE})+S(\rho^{BE})-S(\rho^{ABE})-S(\rho^{E})$, and the infimum is taken over all extensions of 
	$\rho^{ABE}$ of $\rho$. It was shown that~\cite{Christandl2004jmp}: (i) For pure state, it equals to the entanglement of formation $E_f$; (ii) 
	\beax 
	E_d\leq  E_{sq}\leq  E_c, \quad E_{sq}\leq  E_f;
	\eeax  
	(iii) $E_{sq}(\rho)=0$ for any separable $\rho$, and conversely, if there exists an extension $\rho^{ABE}$ with $\dim \mH^E<\infty$, then $\rho$ is separable;
	(iv) $E_{sq}$ is additive, i.e.,
	\beax
	E_{sq}(\rho\ot\varrho)=E_{sq}(\rho)+E_{sq}(\varrho),
	\eeax
	where $\varrho\in\mS^{A'B'}$, $E_{sq}(\rho\ot\varrho)$ is measured under the split $AA'|BB'$;
	(v) It is superadditive, namely,
	\beax 
	E_{sq}(\rho^{AA'|BB'})\geq E_{sq}(\rho)+E_{sq}(\varrho)
	\eeax
	for any $\rho^{AA'BB'}$, where $\rho$ and $\varrho$ are marginal states of $\rho^{AA'BB'}$.

\subsubsection{Conditional entanglement of mutual information}

	Yang \etal. introduced an additive and operational entanglement monotone called conditional entanglement of mutual information, which was defined as~\cite{Yang2008prl}
	\bea\label{E_I}
	E_I(\rho^{AB})=\frac12\inf\{I(AA':BB')\},
	\eea
	where the infimum is taken over all extensions of $\rho^{AB}$, i.e., over all states that satisfying 
	$\tr_{A'B'}\rho^{AA'BB'}=\rho^{AB}$.
	$E_I$ is additive, i.e., 
	\beax 
	E_I(\rho^{AB}\ot\sigma^{CD})=E_I(\rho^{AB})+E_I(\sigma^{CD}),
	\eeax  
	where
	$E_I(\rho^{AB}\ot\sigma^{CD})$ is measured under the cutting $AC|BD$.

	For pure state, it reduces to EoF. For mixed states, $E_I\leq E_f$.
	Its operational
	meaning is elaborated to be the minimal net ``flow of qubits'' in the process of partial state merging~\cite{Yang2008prl}.
	It can also be extended into multipartite case (see Subsec.~\ref{sec-Conditional entanglement for multipartite mutual information}.)

\subsubsection{The geometric measure}

	We denote by $\mP^{AB}$ the set of all separable pure states in $\mH^{AB}$. The geometric measure of entanglement was firstly introduced by Shimony in 1995~\cite{Shimony95}:
	\beax 
	E_G(|\psi\ra)=\frac12\left(  1-\max _{|\phi\rangle \in \mP^{AB}}|\langle\phi \vert \psi\rangle|^2\right),
	\eeax
	where the maximum runs over all separable pure states in $\mP^{AB}$. Later, the factor $\frac12$ was omitted in literaure~\cite{Cao2007jpa,Das2016pra,Sen2010pra,Wei2003pra,Wei2003arxiv,Wei2004pra}, i.e.,
	\bea\label{geometric-measure-1}
	E_G(|\psi\ra)=  1-\max_{|\phi\rangle \in \mP^{AB}}|\langle\phi \vert \psi\rangle|^2.
	\eea
	This approach is valid for both bipartite and multipartite cases. In this subsubsection, we only consider the bipartite case and the multipartite case will be discussed in Sec.~\ref{sec-14.1}.

	$E_G$ is an entanglement monotone since it is nonincreasing on average under LOCC~\cite{Wei2003pra}. In fact, for any pure state $|\psi\ra$ with Schmidt decomposition $|\psi\ra=\sum_j\lambda_j|e_j\ra^A|e_j\ra^B$, one can easily get (see also in~\cite{Weinbrenner2025}) 
	\beax 
	\Lambda(|\psi\ra)=\max_{|\phi\rangle \in \mP^{AB}}|\langle\phi \vert \psi\rangle|=\lambda_1.
	\eeax
	Namely, the reduced function of $E_G$ is $1-\|\rho^A\|$, which is a concave function.
	 By definition, $E_G$ is hard to compute. But for the two-qubit state, it was shown in Ref.~\cite{Wei2003pra,Wei2003arxiv} that
	\beax
E_{G}(\rho)=\frac{1}{2}\left[{1-\sqrt{1-C^2\left(\rho\right)}}\right] 
	=\frac{1}{2}\left[{1-\sqrt{1-4N^2\left(\rho\right)}}\right].
	\eeax

	In Ref.~\cite{Cao2007jpa}, the authors gave the definition of revised geometric measure of entanglement (RGME) 
	\begin{eqnarray}\label{b-revised geometric-measure}
		\check{E}_{G} ( \rho ) = \min\limits_{\sigma\in \mathcal{S}_{sep}} \left[  {1 -
			F( {\rho ,\sigma } )} \right].
	\end{eqnarray}
	$\check{E}_{G}(\rho)$ is non-increasing under LOCC, thus it is an entanglement measure. $\check{E}_{G}$ can be expressed in fidelity form, namely, it does not need the convex hull for the case of mixed state. 
	In addition,
	\beax
	1 - \sqrt{F}\left( {\rho ,\sigma } \right) \le \sqrt{\check{E}_{G} \left( \rho \right)}.
	\eeax
	It was proved that~\cite{Cao2007jpa}
	\beax
	\check{E}_{G}(|\psi\ra) \leq E_f(|\psi\ra)=E_{r}(|\psi\ra)
	\eeax
	and the equality is valid only when $|\psi\ra$ is separable.
	Streltsov \etal proved in Ref.~\cite{Streltsov2010njp} that
	\beax 
	E_G= \min\limits_{\sigma\in \mathcal{S}_{sep}} \left[  {1 -
		F( {\rho ,\sigma } )} \right],
	\eeax 
	namely $\check{E}_{G}(\rho)=E_G(\rho)$ for any state and thus $\check{E}_{G}$ is an entanglement monotone as well.
	They also proved that
	\beax
	S(\rho\lVert\sigma)\leq \tr(\rho\log_2\rho)-\log_2F(\rho, \sigma).
	\eeax

\subsubsection{Entanglement measure induced by the Hilbert-Schmidt norm}

Witte and Trucks introduced an entanglement measure induced by the Hilbert-Schmidt norm (or called the Hilbert-Schmidt entanglement)~\cite{Witte1999pla}:
\bea \label{H-S-entanglement}
E_{\rm HS}(\rho)=\min\limits_{\sigma\in\mS_{sep}}\|\rho-\sigma\|_2^2,
\eea
where the minimum is taken over all separable states in $\mS^{AB}$. It is unknown whether it is nonincreasing on average under LOCC, but numerical investigations have not give any counterexamples~\cite{Witte1999pla}.

\subsubsection{Trace distance entanglement measure}

Eisert \etal presented a trace-distance entanglement monotone which is defined by~\cite{Eisert2003jpa}
\bea\label{E_tr}
E_{\tr}(\rho)=\min\limits_{\sigma\in\mS_\rho^{AB}}\|\rho-\sigma\|_\tr. 
\eea 
But 
\bea\label{E'_tr}
E'_{\tr}(\rho)=\min\limits_{\sigma\in\mS_{sep}}\|\rho-\sigma\|_\tr 
\eea 
is an entanglement measure but it may be not an entanglement monotone.

\subsubsection{Fidelity based entanglement measure}

Three kinds of measures were defined in Ref.~\cite{Guo2020qip}:
	\bea
	E_{\mF}(|\psi\rangle):&=&
	1-F\left( |\psi\rangle\la\psi|,\rho^A\otimes\rho^B\right),\label{definition-purestate1}\\
	E_{{\mF}'}(|\psi\rangle):&=&
	1-\sqrt{F}\left( |\psi\rangle\la\psi|,\rho^A\otimes\rho^B\right),\label{definition-purestate2}\\
	E_{A\mF}(|\psi\rangle):
	&=&1-F_A(|\psi\rangle\la\psi|,\rho^A\otimes\rho^B).\label{definition-purestate3}
	\eea
	Let
	$|\psi\rangle=\sum_k\lambda_k|k\rangle^A|k\rangle^B$ be the Schmidt decomposition of $|\psi\ra$, then
	\beax
	\eqalign{
		E_{\mF}\left( |\psi\ra\right)
		=1-\tr  \rho_A^3 =1-\sum_k\lambda_k^6,\cr
		E_{A\mF}\left( |\psi\ra\right)
		=1-\left( \tr \rho_A^2 \right) ^2
		=1-\left( \sum_k\lambda_k^4\right)^2.}
	\eeax
	The reduced functions of these measures are strictly concave, so they are entanglement
	monotones~\cite{Guo2020qip}.

\subsubsection{Bures measure of entanglement}

	Although not a metric, the fidelity can easily be turned
	into a metric such as the Bures angle~\cite{Gilchrist,liang2019} $A(\rho,\sigma)=\arccos\sqrt{F(\rho,\sigma)}$,
	the modified Bures metric~\cite{Gilchrist,Hubner,liang2019} $B(\rho,\sigma)=\sqrt{1-\sqrt{F(\rho,\sigma)}}$,
	and the sine distance~\cite{Gilchrist,liang2019,Rastegin} $C(\rho,\sigma)=\sqrt{1-{F(\rho,\sigma)}}$.
	The Bures metric can induce an entanglement measure~\cite{Vedral1998pra},
	\begin{eqnarray}\label{Bures measure}
		E_{\B}(\rho)= \min_{\sigma\in \mS_{sep}}\left[2-2\sqrt{F(\rho,\sigma)}\right],
	\end{eqnarray}
	where the minimum is taken over all the separable states in $\mS_{sep}$.
	According to Eq.~(3.2) in Ref.~\cite{liang2019}
	we can obtain that $B(\rho,\sigma)$
	is convex, therefore $E_{\B}$ is an
	entanglement monotone indeed.
	For the two-qubit pure state $|\psi\ra$ with Schmidt coefficients $\{\lambda_1, \lambda_2\}$, $E_{\B}(|\psi\ra)=4\lambda_1^2\lambda_2^2$~\cite{Vedral1998pra}.

	$E_{\B}$ can be represented by the fidelity of separability $F_s$, 
	\beax
	E_{\B}(\rho)=2-2\sqrt{F_s(\rho)},
	\eeax
	where $F_s$
	is defined by~\cite{Streltsov2010njp} 
	\bea\label{Fs}
	F_s(\rho)=\max_{\sigma\in\mS_{sep}} F(\rho,\sigma),
	\eea
	where the maximum is taken over all separable states of the the set $\mS_{sep}$.
	$F_s$ is closely related to tangle for the $2\ot d$ states~\cite{Gao2021qip}:
	\beax 
	\eqalign{
		F_s(|\psi\ra)=\frac{1+\sqrt{1-\tau(|\psi\ra)}}{2},\cr
		1-\frac{\tau(\rho)}{2}\leq  F_s(\rho)\leq \frac{1+\sqrt{1-\tau(\rho)}}{2}.}
	\eeax
	It was shown in Ref.~\cite{Streltsov2010njp} that
	\beax 
	F_s(\rho)=\max\limits_{\{p_i, |\psi_i\ra\}}\sum_ip_iF_s(|\psi_i\ra),
	\eeax 
	where the maximization is done over all pure state ensembles $\{p_i, |\psi_i\ra\}$ of $\rho$.

\subsubsection{Groverian measure of entanglement}

	Very similar to the Bures measure of entanglement, the Groverian measure of entanglement was defined as~\cite{Biham2002pra,Shapira2006pra}
	\begin{eqnarray}\label{f4}
		E_{\G r}(\rho)= \min_{\sigma\in \mS_{sep}}\sqrt{1-F(\rho,\sigma)},
	\end{eqnarray}
	where the minimum is taken over all the separable states in $\mS_{sep}$.
	It is interesting that $E_{\G r}(\rho)=E_{{\G r},F}(\rho)$~\cite{Gao2021qip} for any state $\rho$,
	where $E_{{\G r},F}(\rho)$ denotes the convex-roof extension of $E_{\G r}$. $E_{\G r}$ was proven to be an entanglement monotone~\cite{Biham2002pra,Shapira2006pra}. Obviously, $E_G=E_{\G r}^2$. $E_{\G r}$ can also be represented by the fidelity of separability $F_s$, 
	\beax
	E_{\G r}(\rho)=\sqrt{1-{F_s(\rho)}}.
	\eeax

\subsubsection{A sharp geometric measure of entanglement}

In Ref.~\cite{Ramachandran2025qip}, an entanglement measure is called a sharp entanglement measure if it is sensitive to smooth variations in any Schmidt coefficients for any pure state. In such a sense, the geometric measure $E_G$ is not sharp. So they improved the geometric measure in terms of the distance to the closest maximally entangled state~\cite{Ramachandran2025qip}, i.e.,
\bea\label{E_SG}
E_{SG}(|\psi\ra)=1-\min\limits_{|\Phi\ra}\frac{1}{\sqrt{1-{1}/{\sqrt{d_A}}}}\sqrt{1-|\la\psi|\Phi\ra|},
\eea 	
where $d_A=\dim\mH^A\leq\dim\mH^B$, $|\Phi\ra$ is the maximally entangled state in $\mH^{AB}$, and the minimum runs over all  maximally entangled states in $\mH^{AB}$. It was shown that~\cite{Ramachandran2025qip}
\beax
E_{SG}(|\psi\ra)=1-\frac{1}{\sqrt{1-{1}/{\sqrt{d_A}}}}\left( 1-\frac{\sum_i\sqrt{\lambda_i}}{\sqrt{d_A}}\right)^{\frac12} ,
\eeax 	
where $\{\lambda_i\}$ are the Schmdit coefficients of $|\psi\ra$. The reduced function is concave~\cite{Ramachandran2025qip}, which implies that $E_{SG}$ is an entanglement monotone.

\subsubsection{Robustness of entanglement}

	The robustness of entanglement was introduced by Vidal and Tarrach in 1999~\cite{Vidal1999pra}.
	It is defined for both the bipartite case and the multipartite case. We consider here only the bipartite case, and the multipartite case is discussed in Sec.~\ref{Robustness of multipartite entanglement}.
	For a given state $\rho\in\mS^{AB}$ and a separable
	state $\sigma\in\mS_{sep}$. Let 
	\beax
	R(\rho|\sigma)=\min\left\lbrace t \Big|  \frac{\rho+t\sigma}{1+t}\in\mS_{sep}\right\rbrace. 
	\eeax
	The robustness of entanglement is defined
	as~\cite{Vidal1999pra} 
	\bea \label{Robustness of entanglement}
	R(\rho)=\inf\limits_{\sigma\in\mS^{sep}}R(\rho|\sigma).
	\eea 
	$R(\rho)$ reports what minimal level of mixing of $\rho$ with separable state will produce separable state again. $R$ was shown to be convex and nonincreasing on average under LOCC, so it is an entanglement monotone. 
	For pure state $|\psi\ra$,
	\bea
	R(|\psi\ra)=\left( \tr\sqrt{\rho_A}\right)^2-1, \quad \rho^A=\tr_B|\psi\ra\la\psi|.
	\eea 
	That is, $R(|\psi\ra)=2N(|\psi\ra)$ for pure state $|\psi\ra$.
	
	Harrow and Nielsen~\cite{Harrow2003pra} and Steiner~\cite{Steiner2003pra}
	generalized robustness as
	\bea\label{generalized robustness}
	\check{R}(\rho)=\inf\limits_{\sigma\in\mS}R(\rho|\sigma),
	\eea 
	where the infimum is taken over all states rather than just the separable ones. 
	$\check{R}\leq R$ immediately, but for pure
	states it does not make a difference. $\check{R}$ is an entanglement monotone
	too. Brand\~{a}o~\cite{Brandao2005pra} showed that the generalized robustness $\check{R}$ has operational interpretation: it is equal to the 
	measure $E^{(d)}$ quantifying the activation power.

\subsubsection{Vidal's entanglement monotone}

	Let $|\psi\ra$ be a pure state with Schmidt coefficients $\{\lambda_i\}$ and $S_r(|\psi\ra)=r$.
	We assume with no loss of generality that $\lambda_1\geq\lambda_2\geq\cdots\geq\lambda_r$.
	In 1999, Vidal proposed an entanglement monotone in Ref.~\cite{Vidal1999pral}, i.e.,
	\bea\label{E_k}
	E_{V\text{-}k}\left( |\psi\ra \right) =\sum\limits_{i=k}^r\lambda_i^2,\quad k\geq2.
	\eea 
	In particular, Guo developed the case of $k=2$, i.e.,
	\bea\label{E_2}
	E_2\left( |\psi\ra \right) =\sum\limits_{i=2}^r\lambda_i^2=1-\lambda_1^2=1-\|\rho^A\|,\quad \rho^A=\tr_B|\psi\ra\la\psi| 
	\eea 
	in Ref.~\cite{Guo2023njp} and called it {partial-norm of entanglement} in the sense that $1-\|\rho^A\|$ counts for only a portion of the norm $\|\rho^A\|$ for the qubit case. By definition, $E_2=E_G$.
	Obviously, $E_2\geq0$ for any $|\psi\ra \in\mH^{AB}$ and
	$E_2(|\psi\ra)=0$ if and only if $|\psi\ra$ is separable.
	The reduced function $h(\rho)=1-\|\rho\|$ is concave, so $E_2$ is a well-defined entanglement monotone ($E_{V\text{-}k}$ is entanglement monotone since the reduced funciton of $E_{V\text{-}k}$ is concave for any $k$).
	But it is not strictly concave, since,
	generally, 
	$\|A+B\|=\|A\|+\|B\|$ does not guarantee $A=\alpha B$ for hermitian operators $A$ and $B$. 
	In addition, $E_2\geq\tau$ and both of them are monotonically increasing with $0\leq p\leq1/2$, and
	\beax
	E_2(\rho)\leq\frac{d}{d-1}\min\{1-\|\rho^A\|, 1-\|\rho^B\|\}.
	\eeax
	This is the first entanglment monotone whose redeced function is not strictly concave, and it is shown by counterexample that $E_2$ is not monogaous~\cite{Guo2023njp} (also see Sec.~\ref{sec-3.7} for detail). It is also the first example of entanglement monotone that is not monogamous except for the Schmidt number.

\subsubsection{Minimal partial-norm of entanglement}

	Let $\lambda_{\min}$ be the minimal positive Schmidt coefficient of $|\psi\ra $.
	We define
	\bea\label{E_min}
	E_{\min}(|\psi\ra )=\left\lbrace \begin{array}{ll}\!\!\lambda_{\min}^2,& \lambda_{\min}<1,\\
		\!\!0, &\lambda_{\min}=1.
	\end{array}\right. 
	\eea 
	Denoting by 
	\bea\label{min norm}
	\|\rho\|_{\min}=\left\lbrace \begin{array}{ll}\!\!\lambda_{\min}^2,& \lambda_{\min}<1,\\
		\!\!0, &\lambda_{\min}=1,
	\end{array}\right. 
	\eea  
	it turns out that
	\beax
	E_{\min}(|\psi\ra )=h(\rho^A)=\|\rho^A\|_{\min}.
	\eeax
	The reduced function $h(\rho^A)=\|\rho^A\|_{\min}$ is concave, so $E_{\min}$ is an entanglement monotone, and it is called the {minimal partial-norm of entanglement}, which reflects as the minimal case of the partial-norm~\cite{Guo2023njp}. 
	It is clear that $E_{\min}(\rho)=0$ iff $\rho$ is separable.

	However, $E_{\min}$ does not achieve the maximal value for the maximally entangled state.
	For making up the disadvantages, we can define~\cite{Guo2023njp}
	\bea\label{E'_min}
	E_{\min'}(|\psi\ra)=\left\lbrace \begin{array}{ll}\!\!\lambda_{\min}^2S_r(|\psi\ra),& \lambda_{\min}<1,\\
		\!\!0, &\lambda_{\min}=1,
	\end{array}\right. 
	\eea 
	where $S_r(|\psi\ra)$ denotes the Schmidt rank of $|\psi\ra$. $E_{\min'}$ is also an entanglement monotone and it is called the {reinforced minimal partial-norm of entanglement}~\cite{Guo2023njp}.
	$E_{\min'}$ is equal to $2E_{\min}$
	for any $2\ot n$ state.
	In such a case, $E_{\min'}$ reaches the maximal quantity for the maximally entangled state but not only for these states.
	$E_2$, $E_{\min}$, and $E_{\min'}$ have the following properties: (i) They are not equivalent to each other.
	(ii) $E_{\min'}$ is equal to $2E_{\min}$
	for any $2\ot n$ state. 
	(iii) The maximal value of $E_2$ is $(d-1)/d$.
	We thus, in order to get a normalized measure, replace $E_2$ by $dE_2/(d-1)$. Hereafter the notation $E_2$ refers to the normalized one.
	For the $2\ot n$ system, $E_2$ coincides with $E_{\min'}$ but not for $m\ot n$ system with $2<m\leq n$.
	For any pure state $|\psi\ra$ with Schmidt coefficients are $\sqrt{p}$ and $\sqrt{1-p}$ in $2\ot n$ system, $p\leq1/2$, it is immediate that 
	$E_2(|\psi\ra)=2E_{\min}(|\psi\ra)=E_{\min'}(|\psi\ra)=2p$.
	(iv) $E_{\min}(\rho)\leq \min\{\|\rho^A\|_{\min}, \|\rho^B\|_{\min}\}$
	and 
	$E_{\min'}(\rho)\leq \min\{r_A\|\rho^A\|_{\min}, r_B\|\rho^B\|_{\min}\}$,
	where $r_{A,B}$ is the rank of $\rho^{A,B}$. (v) $E_2$, $E_{\min}$, and $E_{\min'}$ are not equivalent to each other.
	Moreover, it was proved in Ref.~\cite{Guo2023njp} that $E_{\min}$, and $E_{\min'}$ are not monogamous (also see in Sec.~\ref{sec-3.7}).

\subsubsection{Kaniadakis entropy of entanglement}

	Kaniadakis entropy, which was introduced primarily to deal
	with relativistic statistical systems, have been used in a large variety of physical systems~\cite{Ourabah2015pc}.
	Recall that, Kaniadakis proposed a generalization of logarithm which reads~\cite{Kaniadakis2006pa,Kaniadakis2009epjb,Kaniadakis2009epja}
	\bea \label{ln-kappa}
	\ln_{\kappa}(x)=\frac{x^{\kappa}-x^{-\kappa}}{2\kappa},
	\eea
	where $0\leq  \kappa\leq  1$. $\ln_{\kappa}$ reduces to $\ln$ when $\kappa\to 0$. 
	In Ref.~\cite{Ourabah2015pc}, the quantum version of Kaniadakis entropy was proposed~\cite{Ourabah2015pc}:
	\bea\label{Kaniadakis entropy}
	S_{\kappa}=-\tr\rho\ln_{\kappa}(\rho)=\frac{\tr(\rho^{1-\kappa}-\rho^{1+\kappa})}{2\kappa}.
	\eea
	$S_{\kappa}$ is the generalization of the quantum von Neumann entropy, namley, whenever $\kappa\to 0$, it is recovered to $S$. 
	$S_{\kappa}$ was shown to be a concave function~\cite{Beck}, so 
	we can define
	\bea \label{E_kappa}
	E_{\kappa}(|\psi\ra)=S_{\kappa}(\rho^A),
	\eea 
	and it is an entanglement monotone.

\subsubsection{Informationally complete entanglement measure}

	For bipartite state $|\psi\rangle$ with Schmidt rank $S_r(|\psi\rangle)=r+1$, the Informationally complete entanglement measure was defined by~\cite{Jin2023pra}
	\begin{eqnarray}\label{ICEM}
		E_{\rm ic}(|\psi\rangle)=1-\frac{1}{2^r}\sum_{i=0}^rP_r^i\tr\rho^{i+1}_A,
	\end{eqnarray}
	where $P_r^i=\frac{r!}{i!(r-i)!}$. It is an entanglement monotone since the reduced function is strictly concave.

\subsubsection{Schmidt number}

	Almost all entanglement measure for pure states can be
	represented by its Schmidt coefficients~\cite{Donald2002jmp}, such as entanglement formation, concurrence, negativity, etc. (see in previous subsubsections).
	In fact, the {Schmidt number} is indispensable in
	characterization and quantification of entanglement associated with
	pure states~\cite{Buscemi,Donald2002jmp,Fan2003jpa,Fedorov,Guhne2009,Horodecki2009,Sanpera,Sperling2011},
	it can be used to
	characterize and quantify the degree of bipartite entanglement for
	pure state directly without the Schmidt coefficients~\cite{Donald2002jmp,Sperling2011,Terhal2000pra}.
	
	Recall that, the
	Schmidt number (or called Schmidt rank) is
	\bea\label{Schmidt number0}
	S_r(|\psi\rangle)=r(\rho^A)=r(\rho^B).
	\eea
	The Schmidt number can be extended to mixed states by
	means of a convex-roof extension~\cite{Terhal2000pra}
	\begin{eqnarray}\label{Schmidt number}
		S_r(\rho)=\min\limits_{p_i,|\psi_i\ra}\max\limits_{\psi_i}S_r(|\psi_i\rangle),
	\end{eqnarray}
	where the minimum runs over all the pure sates decompositions of $\rho$.
	$S_r$ is non-increasing under LOCC~\cite{Terhal2000pra}, so $S_r-1$ is an entanglement measure. Later, Sperling and Vogel proved in~\cite{Sperling2011} that $S_r(\rho)$ is non-increasing on average under LOCC, so it is in fact an entanglement monotone.

\subsubsection{Min- and max-relative entropy of entanglement}

	Datta introduced an entanglement monotone, called the max-relative entropy of entanglement, in Ref.~\cite{Datta2009ieee}.
	The max-relative entropy of a state $\rho$ and a positive operator $\delta$ is 
	defined as~\cite{Datta2009ieee}
	\bea\label{max-relative entropy}
	D_{\max}(\rho\| \sigma)=\log_2\min\{\lambda: \,\rho\leq \lambda \sigma \}.
	\label{dmax}
	\eea
	Namely, $D_{\max}(\rho||\sigma)$ is well-defined when $\mathrm{supp}\,\rho \subseteq \mathrm{supp}\,\sigma$, where $\mathrm{supp}\,\rho$ is the support of $\rho$, i.e., the subspace spanned by eigenvectors of $\rho$. 
	The min-relative entropy of a state $\rho$ and a positive operator $\delta$ is defined by~\cite{Datta2009ieee}
	\bea\label{dmin}
	D_{\min}(\rho\| \sigma)
	=    - \log_2 \tr\bigl(\pi_\rho\sigma\bigr) \ ,
	\eea 
	where $\pi_\rho$ denotes the projector onto $ \mathrm{supp}\,\rho$. It is well-defined if 
	$\mathrm{supp}\,\rho\cap\mathrm{supp}\,\sigma\neq\emptyset$.
	In particular,
	\beax 
	0\leq D_{\min}(\rho\| \sigma)\leq S(\rho\|\sigma)\leq D_{\max}(\rho\| \sigma)
	\eeax 
	when $\rho$ and $\sigma$ are states. $D_{\min}(\rho\| \sigma)=D_{\max}(\rho\| \sigma)=0$ when $\rho=\sigma$,
	moreover, $D_{\min}(\rho\| \sigma)=0$ if $\mathrm{supp}\,\rho= \mathrm{supp}\,\sigma$.

	In Ref.~\cite{Datta2009ieee,Datta2009ijqi}, two entanglement measures were defined via $D_{\min}$ and $D_{\max}$.
	The max- and min-relative entropy of entanglement were defined by
	\bea 
	E_{\max\text{-}r}(\rho)&=&\min\limits_{\sigma\in\mS_{sep}}D_{\max}(\rho\|\sigma),\label{max-relative}\\
	E_{\min\text{-}r}(\rho)&=&\min\limits_{\sigma\in\mS_{sep}}D_{\min}(\rho\|\sigma). \label{min-relative}
	\eea
	It was shown in Ref.~\cite{Datta2009ieee} that $E_{\max\text{-}r}$ is an entanglement measure and it is in fact non-increasing on average under LOCC, but it is not convex in general so it may be not an entanglement monotone. In fact $E_{\max\text{-}r}$ was show to be quasiconvex~\cite{Datta2009ijqi}, i.e., for a mixture of state $\rho=\sum_{i}p_i\rho_i$, $E_{\max\text{-}r}(\rho)\leq \max_i E_{\max\text{-}r}(\rho_i)$.
	We can prove that $E_{\min\text{-}r}$ is convex since $D_{\min}$ is jointly convex. $E_{\min\text{-}r}$ is nonincreasing under LOCC but it can increase on average under LOCC~\cite{Datta2009ijqi}, so it is still an entanglement measure but not an entanglement monotone. Obviously, this approach can be extended into the multipartite system directly.

	It is immediate that 
	\beax 
	E_{\min\text{-}r}(\rho)\leq E_r(\rho)\leq E_{\max\text{-}r}(\rho)
	\eeax 
	for any $\rho$.
	It turns out that, $E_{\max\text{-}r}$ is equal to the logarithmic version of the global robustness of entanglement in Eq.~\eref{generalized robustness}~\cite{Datta2009ijqi}, 
	\bea\label{E_{check{R}}}
	E_{\check{R}}(\rho)=\log_2\left[1+\check{R}(\rho)\right].
	\eea
	Thus
	\beax 
	E_{\max\text{-}r}(|\psi\ra)=2\log_2\tr\sqrt{\rho^A}
	\eeax
	does not reduce to $S(\rho^A)$ in general. But for the maximally entangled state $|\Phi^+\ra$, $E_{\max\text{-}r}(|\Phi^+\ra)=S(\rho^A)$.
	Interestingly, $E_{\max\text{-}r}(|\psi\ra)=2E_N(|\psi\ra)$ for pure state.
	So $E_{\max\text{-}r}$ is also additive since $E_N$ is additive on pure states.
	In addition, $E_{\max\text{-}r}(\rho\ot P^+)=E_{\max\text{-}r}(\rho)+E_{\max\text{-}r}(P^+)$ for any state $\rho$~\cite{Brandao2011ieee}.

\subsubsection{$\alpha$-logarithmic negativity}

	Wang and Wilde proposed another additive entanglement measure termed $\alpha$-logarithmic negativity in Ref.~\cite{Wang2020pra}. We recall some preliminary notations first.
	For $\alpha\geq1$, a Hermitian operator $X\neq0$, and a positive operator $\sigma\neq0$~\cite{Wang2020pra}%
	\begin{align}
		\mu_{\alpha}(X\Vert\sigma)  &  =\left\{
		\begin{array}
			[c]{cl}%
			\left\Vert \sigma^{\frac{1-\alpha}{2\alpha}}X\sigma^{\frac{1-\alpha}{2\alpha}%
			}\right\Vert _{\alpha} & \text{if}\ \operatorname{supp}(X)\subseteq
			\operatorname{supp}(\sigma)\\
			+\infty & \text{else}%
		\end{array}
		\right.  ,\label{eq:mu-def}\\
		\nu_{\alpha}(X\Vert\sigma)  &  =\log_{2}\mu_{\alpha}(X\Vert\sigma),\nonumber
	\end{align}
	where the $\alpha$-norm of an operator $Y$ is defined for $\alpha\geq1$ as%
	\bea\label{alphanorm}
	\left\Vert Y\right\Vert _{\alpha}    =\left(  \operatorname{Tr}%
	\left\vert Y\right\vert ^{\alpha}\right)  ^{1/\alpha},
	\eea
	and the inverse $\sigma^{\left(  1-\alpha\right)  /2\alpha}$ is understood in
	the generalized sense (i.e., taken on the support of $\sigma$). The definition
	in Eq.~\eqref{eq:mu-def} is consistent with the following limit~\cite{Wang2020pra}:%
	\beax
	\mu_{\alpha}(X\Vert\sigma)=\lim_{\varepsilon\rightarrow0}\mu_{\alpha}%
	(X\Vert\left(  1-\varepsilon\right)  \sigma+\varepsilon\delta),
	\label{eq:limit-mu-alpha-support}%
	\eeax
	where $\delta$ is a state and $\delta>0$.

	$\mu_{\infty}(X\Vert\sigma)$ is defined by
	\beax
	\mu_{\infty}(X\Vert\sigma)  &=&\lim_{\alpha\rightarrow\infty}\mu
	_{\alpha}(X\Vert\sigma)
	=\left\Vert \sigma^{-1/2}X\sigma^{-1/2}\right\Vert _{\infty}\\
	& ~ =&\inf\left\{  \lambda:-\lambda\sigma\leq X\leq\lambda\sigma\right\}  ,
	\eeax
	and%
	\beax
	\nu_{\infty}(X\Vert\sigma)=\log_{2}\mu_{\infty}(X\Vert\sigma).
	\label{eq:nu-infty}%
	\eeax
	Both of the above formulas are defined as above in the case that
	$\operatorname{supp}(X)\subseteq\operatorname{supp}(\sigma)$, and $\mu
	_{\infty}(X\Vert\sigma)$\ and $\nu_{\infty}(X\Vert\sigma)$ are set to
	$+\infty$ otherwise.

	The function $\nu_{\alpha}(X\Vert\sigma)$ is related to the sandwiched R\'enyi
	relative entropy $\widetilde{D}_{\alpha}(X\Vert\sigma)$%
	\ \cite{Muller2013jmp,Wilde2014cmp} of a Hermitian operator $X\neq0$ and a positive
	operator $\sigma\neq0$ as follows:%
	\bea\label{SR relative entropy}
	\widetilde{D}_{\alpha}(X\Vert\sigma)=\frac{\alpha}{\alpha-1}\nu_{\alpha
	}(X\Vert\sigma). \label{eq:sandwiched-Renyi}%
	\eea
	Analogously,
	\beax
	\nu_{\infty}(X\Vert\sigma)  &  =&D_{\max}(X\Vert\sigma
	)
	=\log_{2}\left\Vert \sigma^{-1/2}X\sigma^{-1/2}\right\Vert _{\infty}\\
	&  =&\log\inf\left\{  \lambda:-\lambda\sigma\leq X\leq\lambda\sigma\right\}  ,
	\eeax 
	where $D_{\max}(X\Vert\sigma)$ is the max-relative entropy \cite{Datta2009ieee}\ of a
	Hermitian operator $X\neq0$ and a positive semi-definite operator $\sigma
	\neq0$. Note that $D_{\max}(X\Vert\sigma)=+\infty$ if $\operatorname{supp}%
	(X)\not \subseteq \operatorname{supp}(\sigma)$.

	The $\alpha$-logarithmic negativity of $\rho$ was defined as~\cite{Wang2020pra}
	\bea\label{alpha-logarithmic negativity}
	E_{N}^{\alpha}(\rho)=\inf_{\sigma\in\mS_{\rm PPT}}%
	\nu_{\alpha}(\rho^{T_b}\Vert\sigma),
	\eea
	where $\mS_{\rm PPT}$ is the set
	of positive partial transpose states in $\mS^{AB}$.
	It was proved that~\cite{Wang2020pra}: (i) $E_{N}^{\alpha}(\rho)=0$ iff $\rho$ is a PPT state,
	(ii) it does not increase on average under compeltely PPT preserving and trace preserving operations which include LOCC operations as a special case, (iii) it is not convex, (iv) subadditive, and (v) for the tripartite state $|\psi\ra^{ABC}=\frac12\left(|000\ra^{ABC}+|011\ra^{ABC}+\sqrt2|110\ra^{ABC} \right)$,
	$E_{N}^{\alpha}(A|BC)<E_{N}^{\alpha}(AB)+E_{N}^{\alpha}(AC)$.
	So $E_{N}^{\alpha}$ is an entanglement measure but not an entanglement monotone.
	Comparing with logarithmic negativity $E_N$, it was shown that~\cite{Wang2020pra}:
	(i) $E_N(\rho)\leq E_N^{\alpha}(\rho)\leq E_N^{\beta}(\rho)$ whenever $1\leq\alpha\leq\beta$ for any $\rho$, (ii) $\lim_{\alpha\rightarrow1}E_{N}^{\alpha}(\rho)=E_{N}(\rho)$ for any $\rho$,
	and (iii) if $\rho$ satisfies $|\rho^{T_b}|^{T_b}\geq0$, then
	all $\alpha$-logarithmic negativities are equal, i.e., the following equality
	holds for all $\alpha\geq1$:%
	\beax
	E_{N}^{\alpha}(\rho)=E_{N}(\rho). 
	\eeax

\subsubsection{$\kappa$ entanglement}

	In Ref.~\cite{Wang2020prl,Wang2023pra}, in order to determine the PPT exact entanglement cost, Wang and Wild introduced an efficiently computable entanglement, called $\kappa$ entanglement, defined as
	\bea \label{Kappa-entanglement}
	\check{E}_{\kappa}(\rho)=\log_2\inf\limits_{S_{AB}\geq 0}\left\lbrace \tr S_{AB}:-S_{AB}^{T_b}\leq\rho_{AB}^{T_b}\leq S^{T_b}_{AB}\right\rbrace.
	\eea
	$\check{E}_{\kappa}$ can be calculated by means of a semi-definite program~\cite{Vandenberghe1996}.
	It was shown in Ref.~\cite{Wang2020prl,Wang2023pra} that
	(i) $\check{E}_{\kappa}(\rho)=0$ iff $\rho$ is a PPT state, (ii) it does not increase on average under compeltely PPT preserving and trace preserving operations which include LOCC operations as a special case, (iii) it is not convex, (iv) additive, and (v) for the tripartite state $|\psi\ra^{ABC}=\frac12\left(|000\ra^{ABC}+|011\ra^{ABC}+\sqrt2|110\ra^{ABC} \right)$,
	$\check{E}_{\kappa}(A|BC)<\check{E}_{\kappa}(AB)+\check{E}_{\kappa}(AC)$.
	So $\check{E}_{\kappa}$ is an entanglement measure but not an entanglement monotone.
	Interestingly, $\check{E}_{\kappa}$ coincides with the PPT exact entanglement cost for any bipartite state~~\cite{Wang2020prl,Wang2023pra}.
	
	In addition, it was proved in Ref.~\cite{Wang2023pra} that
	\beax 
	\check{E}_{\kappa}(\rho)\geq E_N(\rho), 
	\eeax
	and if $\rho$ satisfies the binegativity condition
	\beax 
	\left| \rho^{T_b}\right| ^{T_b}\geq 0,
	\eeax 
	then
	\beax 
	\check{E}_{\kappa}(\rho)= E_N(\rho).
	\eeax 
	For the $m\ot m$ maximally entangled state $|\Phi^+\ra$, $\check{E}_{\kappa}(|\Phi^+\ra)=\log_2m$.
	Furthermore, for any $\rho$ with $\dim\mH^A=d_A$ and $\dim\mH^B=d_B$, $\check{E}_{\kappa}(\rho)\leq\log_2\min\{d_A, d_B\}$.
	$\kappa$ entanglement has a relation to the $\alpha$-logarithmic negativity, i.e.,
	$E_{\kappa}(\rho) =\lim_{\alpha\rightarrow\infty
	}E_{N}^{\alpha}(\rho)$ for any $\rho$~\cite{Wang2020pra}.

\subsubsection{Entanglement monotone from complementarity relations}

	For any given pure state $|\psi\ra$, let $P(\rho^A) + C_v(\rho^A) \leq \alpha$ be a complementarity relation for the state $\rho_A$ such that it saturates only if $\rho^A$ is pure, with $P(\rho^A)$ and $C_v(\rho^A)$ being bone-fide measures of predictability and visibility, respectively, and $\alpha\in \mathbb{R}$ with $\alpha>0$. The quantity~\cite{Basso2022jpa}
	\bea\label{E_{cr}}
	E_{cr}(|\psi\ra)= \alpha - P(\rho^A) - C_v(\rho^A) \label{eq:entmon_}
	\eea
	was shown to be an entanglement monotone 
	since $P(\rho^A)$ and $C_v(\rho^A)$ are convex functions of $\rho^A$.
	Recall that, the quantitative formulation of the
	wave-particle duality was devised by D\"{u}rr~\cite{Durr2001pra} and Englert \etal~\cite{Englert2008ijqi} for the first time, for which they proposed reasonable criteria for checking
	the reliability of newly defined predictability and visibility measures.

	For example, we take $P_{\rm vn}(\rho^{A})=\log_2d_A-S(\rho^A_\diag)$~\cite{Basso2020qip},
	$C_r(\rho^A)=S(\rho^A_\diag)-S(\rho^A)$ the bone-fide measure of visibility, $\rho^A_\diag$ denotes the diagonal matrix up to the reference basis of the coherence~\cite{Baumgratz2014prl}.
	Then 
	\beax 
	E_{cr}(|\psi\ra)= \log_2d_A - P_{\rm vn}(\rho^A) - C_r(\rho^A) 
	\eeax
	is an entanglement monotone of this kind, where $d_A=\dim\mH^A$.

\subsubsection{Entanglement measure in terms of LOCC on pure states}

	Gour and Tomamichel proposed a quantity which called optimal extension of given resource measure by means of the minimal/maximal extension. Consequently, the optimal extension of bipartite entanglement measure was discussed~\cite{Shi2021adp,Yu2021cpb}.
	Let $E$ be a bipartite entanglement measure for pure states and it does not increase on average under LOCC. In Ref.~\cite{Shi2021adp,Yu2021cpb}, the authors extended $E$ to the entanglement measure $\check{E}$ for mixed states as
	\bea\label{Ev}
	\check{E}(\rho)=\inf_{|\psi\ra,\varepsilon} E(|\psi\ra), 
	\eea
	where the infimum takes over all the pure states $|\psi\ra$'s and all LOCC $\varepsilon$ such that $\rho=\varepsilon(|\psi\ra\la\psi|)$.
	$\check{E}$ is non-increasing on average under LOCC as well~\cite{Shi2021adp,Yu2021cpb}. 
	In general, $\check{E}\geq E$~\cite{Gour2020pra,Yu2021cpb}. But, sometimes, $\check{E}$ coincides with $E$. It was proved that, (i) if $E=E_G$, $\check{E}_G=E_G$, and (ii) when $E=C$, $\check{C}=C$ for the 2-qubit states. In general, if $\check{E}$ is convex additionally, i.e., it is an entanglement monotone, then $\check{E}= E_F$~\cite{Shi2021adp,Yu2021cpb}.

\subsubsection{Passive-state energy as an entanglement quantifier}

	For a given state $\rho$, the lowest energetic state with the same spectrum is called the passive state, represented by $\rho_p$. Namely, if the system is governed by a fixed Hamiltonian $H=\sum_j\epsilon_j|\epsilon_j\ra\la\epsilon_j|$ and the spectrum of $\rho$ is $\{\lambda_j\}$, then $\rho_p=\sum_j\lambda_j|\epsilon_j\ra\la\epsilon_j|$, where $\epsilon_j\leq\epsilon_{j+1}$, $j=1$, $2$, $\dots$, $d-1$, and $\lambda_j\geq\lambda_{j+1}$.

	In Ref.~\cite{Alimuddin2020pre}, the authors presented passive-state energy acts as an entanglement monotone.
	For the given Hamiltonian $H^A$, the passive-state energy was defined as~\cite{Alimuddin2020pre} 
	\bea\label{pasivemonotone}
	E_p(|\psi\ra)=\tr(\rho_p^AH^A),
	\eea
	where $\rho_p^A$ is the passive state corresponding to $\rho_A $ governed by the Hamiltonian $H^A$. From definition, the passive-state energy is not only a function of the spectrum of the reduced state, but also a function of the energy eigenvalues. 
	It was shown to be a well-defined faithful entanglement monotone and not additive.

	Passive-state energy as an entanglement measure has an
	important physical interpretation in the context of thermodynamics, and it was also called thermodynamic
	measure of entanglement.

\subsubsection{Ergotropic gap entanglement}\label{Ergotropic gap}

With the notations as above, the ergotropy, i.e., the maximal extractable work from the initial state $\rho$, is defined by~\cite{Allahverdyan2004epl}
\beax 
W_e(\rho)=\tr(\rho H)-\tr(\rho_p H).
\eeax 
The antiergotropy as the minimal extractable work from the system cyclic processes is defined as
\beax 
W_{ae}(\rho)=\tr(\rho H)-\tr(\rho_{ac}H),
\eeax
where $\rho_{ac}$ is the active state of $\rho$ that maximizes the mean energy $\la H\ra$. The battery capacity of the system is defined as~\cite{Yang2023prl}
\beax 
\mC(\rho)=W_e(\rho)-W_{ae}(\rho)=\tr(\rho_{ac}H)-\tr(\rho_pH).
\eeax
For a bipartite state $\rho^{AB}$, the global Hamiltonian is $H^{AB}=H^A\ot I^B+I^A\ot H^B$. The global ergotropy is
\beax 
W_e^g(\rho^{AB})=\tr(\rho^{AB}H^{AB})-\tr(\rho_p^{AB}H^{AB}),
\eeax 
and the local ergotropy is
\beax 
W_e^l(\rho^{AB})&=&W_e^A(\rho^{AB})+W_e^B(\rho^{AB})\\
&=&\tr(\rho^{AB}H^{AB})-\tr(\rho_p^{A}H^{A})-\tr(\rho_p^{B}H^{B}).
\eeax
The bipartite ergotropy gap is~\cite{Alimuddin2019pra}
\beax
\Delta_{A|B}(\rho^{AB})&=& W_e^g(\rho^{AB})-W_e^l(\rho^{AB})\\
&=&\tr(\rho_p^{A}H^{A})+\tr(\rho_p^{B}H^{B})-\tr(\rho_p^{AB}H^{AB})
\eeax 
which refers to the total extractable energy. If $\rho^{AB}=|\psi\ra\la\psi|^{AB}$, it follows that
\beax 
\Delta_{A|B}(|\psi\ra^{AB})=\tr(\rho_p^{A}H^{A})+\tr(\rho_p^{B}H^{B}).
\eeax 
The ergotropic gap entanglement is defined by~\cite{Yang2024qip}
\bea \label{E_{e-g}}
E_{e\text{-}g}(|\psi\ra)=\tr(\rho_p^AH^A)+\tr(\rho_p^BH^B).
\eea 	
It was shown to be an entanglement monotone.

\subsubsection{Entanglement parameter}

	We first denote the local orthogonal observables as the complete set of orthogonal observables $A_i$ on $\mathcal{H}^A$ with
	$i=1,...,d_A^2$ and $\tr({A_i A_j})=\delta_{ij}$ and a similar 
	set $B_j$ for $\mathcal{H}^B.$  For example, $d_A=2$ are the Pauli operators and the identity. Then, we can consider observables on $\mathcal{H}^{AB}$ defined by
	\beax
	\{M_{\alpha}\}&=&\{A_i\otimes I, I\otimes B_j\},\;\;\;\;i=1,\dots, d_A^2,
	~j=d_A^2+1,\dots,d_A^2+d_B^2,\label{LOOs}
	\eeax
	which then also obey $\tr({M_{\alpha}M_{\beta}}) =\delta_{\alpha\beta}$.

	A covariance matrice (CM) of a given bipartite state $\rho$ is defined by the following entries
	\begin{eqnarray*}
		\gamma(\rho)_{\alpha\beta} = 
		\frac{1}{2}\langle M_{\alpha} M_{\beta} + M_{\beta} M_{\alpha}\rangle_{\rho} 
		- \langle M_{\alpha}\rangle_{\rho}\langle M_{\beta}\rangle_{\rho}.
	\end{eqnarray*}
	Choosing the observables as in Eq.~(\ref{LOOs}) one can write the CM 
	in a handy block form
	\begin{eqnarray*}
		\gamma = \begin{pmatrix} A & C\\
			C^T & B
		\end{pmatrix},
		\label{blockcm}
	\end{eqnarray*}
	where $A = \gamma(\rho_A,\{A_i\})$, 
	$B = \gamma(\rho_B,\{B_i\})$ are CMs of reduced states and $c_{ij}=\langle A_i\otimes B_j\rangle_{\rho} - \langle {A_i}\rangle_{\rho}\langle{B_j}\rangle_{\rho}$ 
	denote correlations between the two parties, $C=[c_{ij}]$.
	
	Let $\rho$ be a bipartite quantum state with CM $\gamma(\rho)$. We 
	define a function $V(\rho)$ as \cite{Gittsovich2010pra}
	\begin{eqnarray*}
		V(\rho)= \max_{t,\kappa_A,\kappa_B}\{t\leq 1 : \gamma(\rho)-
		t\kappa_A\oplus\kappa_B\geq 0\},
		\label{covmass}
	\end{eqnarray*}
	where $\kappa_A=\sum_k p_k \gamma(|\psi_k\rangle\langle\psi_k|)$ 
	and $\kappa_B= \sum_k p_k\gamma(|\phi_k\rangle\langle\phi_k|)$
	with $|\psi_k\rangle\langle\psi_k|$ 
	and $|\phi_k\rangle\langle\phi_k|$ are pure states in $\mathcal{H}^A$ and $\mathcal{H}^B$ respectively.

	The entanglement parameter $\tilde{\mE}(\rho)$ is then defined as~\cite{Gittsovich2010pra}
	\begin{eqnarray}\label{entanglement-parameter}
	\tilde{\mE}(\rho) = 1 - V(\rho).
		\label{entmeas}
	\end{eqnarray}
	It has not been proved that it is an entanglement measure although it is convex and invariant under local unitary operation. It is certain that $\mathcal{\tilde{E}}$ is not an entanglement monotone since there
	exists a two-qubit state and a LOCC protocol such that $\mathcal{\tilde{E}}$
	increases on average from zero to a positive value under this protocol~\cite{Gittsovich2010pra}.
	Also, a similar function is presented in Ref.~\cite{Giedke2002pra} to 
	quantify entanglement in infinite dimensional systems.
	Specially, assuming that $d=d_A =d_B$ we have the following lower bounds on $\mathcal{\tilde{E}}(\rho)$ ~\cite{Gittsovich2010pra}
	\begin{eqnarray*}
		\tilde{\mE}(\rho)\geq \frac{\mathrm{Tr}{\rho_A^2} + \mathrm{Tr}{\rho_B^2} + 2\mathrm{Tr}{|C|} - 2}{2 d-2}
		\label{ubofe}
	\end{eqnarray*}
	and
	\begin{eqnarray*}
		\tilde{\mE}(\rho)\geq \frac{1}{d-1}
		\left[  
		\frac{\mathrm{Tr}{\rho_A^2} + \mathrm{Tr}{\rho_B^2}-2}{2} + 
		\sqrt{\frac{1}{4}\left( \mathrm{Tr}{\rho_A^2} - \mathrm{Tr}{\rho_B^2}\right) ^2+\Vert C \Vert_{\mathrm{tr}}^2}
		\right]  .
		\label{ubofe2}
	\end{eqnarray*}

\subsubsection{Entanglement coherence}

	For a bipartite pure state $|\psi\rangle$, then its entanglement coherence is given by~\cite{Pathania2022ijtp}
	\begin{eqnarray}\label{ecm}
		C_E(|\psi\ra) = \frac{1}{d-1}\left[(\tr\sqrt{\rho_A})^2 - 1\right].
		\label{ECbi}
	\end{eqnarray}
	where $d=\min\dim\mH^{A,B}$. By definition, $C_E$ coincides with $N_F$ up to a whole factor, so $C_E$ is also an entanglement monotone.
	In fact, $C_E$ can be regarded as the normalized coherence of the entangled state in its Schmidt basis, and the maximally entangled states are the ones which are maximally coherent in their Schmidt basis. 
	For a bipartite pure state $|\psi\rangle$ in its Schmidt decomposed form
	$|\psi\rangle = \sum_{i=1}^d \lambda_i |e_i\rangle^A|e_i\rangle^B$,
	One can find coherence in the Schmidt basis is equal to the $C_E$ since
	\beax
\fl\quad \quad 	\frac{1}{d-1} \sum_{j\neq k} {\lambda_j\lambda_k}
	&=& \frac{1}{d-1} \left[\left(\sum_{j} {\lambda_j}\right)^2 - \sum_{j} \lambda_j^2\right] 
	= \frac{1}{d-1}\left[\left( \mathrm{Tr}{\sqrt{\rho_A}}\right)^2 - \mathrm{Tr}{\rho_A} \right]\\
	&= &\frac{1}{d-1}\left[\left( \mathrm{Tr}{\sqrt{\rho_A}}\right)^2 - 1 \right]
	=C_E (|\psi\ra).
	\eeax
	The expression for $C_E$ in an arbitrary basis of the subsystem is defined as~\cite{Pathania2022ijtp}
	\begin{eqnarray*}
		C_E(|\psi\ra) = \frac{1}{d-1} \sum_{j\neq k} \left( \sqrt{\rho}_{jj}\sqrt{\rho}_{kk}
		- |\sqrt{\rho}_{jk}|^2\right).
		\label{EC}
	\end{eqnarray*}
	where $\sqrt{\rho}_{jk}$ are the elements of the square-root of the
	reduced state in the chosen
	basis. The entanglement coherence $C_E$ also has a connection with concurrence, unified entropy and skew information~\cite{Pathania2022ijtp}.

\subsubsection{Entanglement measure based on observable correlations}

	In 2008, Luo proposed an entanglement measure in terms of the observable correlations~\cite{Luo2008tmp}.
	For a given state $\rho\in\mS^{AB}$, the observable correlations is defined by
	\bea \label{Ioc}
	I_{oc}(\rho)=\max\limits_{A,B}I(\rho|A;B)
	\eea  
	with $I(\rho|A;B)=S(\rho^{A'})+S(\rho^{B'})-S(\rho')$, where $\rho'$ denotes the output state under the local measurements $\{A_i\}$ on part $A$ and  $\{B_j\}$ on part $B$, respectively, $\sum_iA_i^\dag A_i=I^A$,  $\sum_jB_j^\dag B_j=I^B$, $\rho^{A',B'}=\tr_{B,A}\rho'$. It follows that
	\bea \label{E_oc}
	E_{oc}(\rho)=\min\sum_kp_kI_{oc}(\rho_k)
	\eea 
	with the minimum runs over all the convex combinations $\sum_kp_k\rho_k$ of $\rho$ has the following properties: (i) $E_{oc}(|\psi\ra\la\psi|)=S(\rho^A)$, (ii) it is convex, (iii) $E_{oc}(\rho)\geq 0$ and $E_{oc}(\rho)= 0$ if and only if $\rho$ is separable, and (iv) it cannot increase under local operations in the sense that $E_{oc}(\varepsilon(\rho))\leq E_{oc}(\rho)$, where $\varepsilon(\rho)=\sum_kM_k\ot I^B\rho M_k^\dag\ot I^B$, $\sum_kM_k^\dag M_k=I^A$. In addition,
	\beax 
	E_{oc}(\rho)\leq \max\limits_{A,B}I(\rho|A;B)\leq I(\rho)
	\eeax 
	for any $\rho$ while 
	\beax 
	E_f(\rho)>I(\rho)
	\eeax
	for some $\rho$, namely, $E_{oc}$ is different from $E_f$. Notice here that whether $E_{oc}$ is nonincreasing under LOCC is unknown, so it may be not a well-defined entanglement measure in fact.

\subsection{Entanglement of assistance: the dual quantity of entanglement measure}

	In 1998, DiVincenzo~\etal put forward the entanglement of assistance~\cite{DiVincenzo,GourSpekkens}.
	Let $E$ be a bipartite entanglement measure, for any bipartite state $\rho^{AB}$, if~$|\Psi\ra^{ABC}$ is a purification of $\rho^{AB}$, i.e., $\rho^{AB}=\tr_C|\Psi\ra\la\Psi|^{ABC}$, then the entanglement of assistance of $E$ is defined as~\cite{DiVincenzo,GourSpekkens}
	\bea\label{eoa}
	E_a(\rho^{AB})=\max\sum_{j=1}^{n}p_jE\left(\rho^{AB}_{j}\right),
	\eea
	where the maximum is taken over all the local measurements on part $C$, $\{p_j, \rho^{AB}_{j}\}$ is the post states of $AB$ part when~$|\Psi\ra^{ABC}$ underling the local measurement on part $C$.

	According to Ref.~\cite{Hughston},
	\bea\label{eoa2}
	E_a(\rho^{AB})=\max\sum_{j=1}^{n}p_jE\left(|\psi\ra^{AB}_{j}\right),
	\eea
	where the maximum is taken over all the ensembles $\left\lbrace p_j, |\psi_j\ra^{AB}\right\rbrace$ of $\rho^{AB}$.

	By definition, the entanglement of assistance is dual to the convex-roof extended entanglement monotone. The entanglement of assistance can increase under the three-partite LOCC~\cite{GourSpekkens}, and it is not convex, so it is not entanglement measure any more. The first entanglement of assistance is the concurrence of assistance $C_a$~\cite{Laustsen2003qic}.
	For any pure state $|\psi\ra^{AB}$ with Schmidt~number~2, $C(|\psi\ra^{AB})=2N(|\psi\ra^{AB})$,
	and $C_a=2N_a$ holds for any 2-qubit state~\cite{Kim2009}. $C_a$ also satisfies $C_a^2(\rho^{AB})\leq 2(1-\tr\rho_A^2)$~\cite{Lizongguo}. Later, several entanglement of assistance are discussed ~\cite{Bai2014prl,Choi2015pra,Coffman2000pra,Gour2005pra2,Gour2007jmp,Kim2009,Kumar2016pla,Luoshunlong,Osborne,Ouyongcheng2007pra2,Zhu2014pra} (also see in Subsec.~\ref{sec-5.2}).

\subsection{Negative entanglement measure}

For a biparite separable state $\rho$, its negative entanglement is defined as~\cite{Zhang2010prl}
\bea\label{E_S} 
E_S(\rho)=-\inf\limits_{t, \sigma}\left\lbrace  tC(\sigma): C\left( \frac{\rho+t\sigma}{1+t}\right) >0 \right\rbrace,
\eea 
where the infimum is taken over all possible $t>0$ and entangled state $\sigma$, $C$ is concurrence or the $I$-concurrence. $E_S$ adimits the following properites~\cite{Zhang2010prl}: (i) $E_S(\rho)=0$ for any edge state of separable states, where $\rho$ is called an edge state of separable states (ESS) if for any $\epsilon>0$, there exists an entangled state $\sigma$ such that $(\rho+\epsilon\sigma)/(1+\epsilon)$ is entangled; (ii) $E_S$ is invariant under local unitary operations; (iii) $E_S(\sum_kp_k\rho_k)\leq\sum_kp_kE_S(\rho_k)$, where $\rho_k$'s are separable states and $\sum_kp_k=1$, (iv) $E_S(|\psi\ra^A|\psi\ra^B)=0$. In addition, for the two-qubit case, (i) $E_S(\rho)=\lambda_1-\sum\limits_{j>1}\lambda_j$, where $\lambda_j$'s are the eigenvalues of $(\sqrt{\rho}\tilde{\rho}\sqrt{\rho})^{1/2}$ in decreasing order as in Eq.~\eqref{Ctheta}; (ii) $E_S(\rho)=0$ if and only if $\rho$ is an ESS state; (iii) $|E_S(\rho)|$ does not increase on average under LOCC for any separable states. Fot the isotropic state $\rho_f$, $E_S(\rho_f)\geq\sqrt{2m/{m-1}}(f-1/m)$ whenever $0\leq f\leq 1/m$.


\section{Monogamy of the bipartite entanglement measure} \label{sec-3}


In 2000, Coffman, Kundu, and Wootters (CKW) presented the first quantitative monogamy relation, which was also called CKW inequality, for
three-qubit state $\rho$, in the seminal paper~\cite{Coffman2000pra}, i.e.,
\begin{eqnarray}\label{Coffman}
	C^2(\rho^{A|BC})\geq C^2(\rho^{AB})+C^2(\rho^{AC}),
\end{eqnarray}
where $\rho^{AB}=\tr_C\rho$, $\rho^{AC}=\tr_B\rho$, the vertical bar indicates the split 
across which the entanglement is measured. Henceforth, considerable efforts have been devoted to this task in the last two
decades~\cite{Allen,Bai2014prl,Camalet,Chengshuming,Coffman2000pra,Deng,Dhar,Eltschka2009pra,Eltschka2015prl,Eltschka2018pra,Eltschka2019quantum,Gour2010jmp,GG2018q,GG2019pra,Guo2019pra,Guo2022entropy,Guo2023njp,Guo2023pra,Guo2024pra,Guo2024rip,Hehuan,Karczewski,Kim2009,kim2012limitations,Koashi,Kumar2016pla,Lan16,Luo2016pra,Oliveira2014pra,Osborne,Ouyongcheng2007pra2,Regula2014prl,Regula2016pra,Streltsov,Zhu2014pra}.

In this section, we review the monogamy relation of bipartite entanglement measure so far in detail. We review the definition of monogamy relation first.


\subsection{Definition of the monogamy relation}


For an bipartite entanglement measure $E$, $E$ is said to be monogamous if~\cite{Coffman2000pra,Koashi}
\bea\label{monogamy1}
E(\rho^{A|BC})\geq E(\rho^{AB})+E(\rho^{AC})
\eea
for any $\rho\in\mS^{ABC}$. However, Eq.~\eref{monogamy1} is not valid for many entanglement measures~\cite{Coffman2000pra,Dhar,GG2018q} but some power function of $E$ admits the monogamy relation (i.e., $E^\alpha(\rho^{A|BC})\geq E^\alpha(\rho^{AB})+E^\alpha(\rho^{AC})$ for some $\alpha>0$). In Ref.~\cite{GG2018q}, we improved the definition of monogamy as: 

\noindent
\begin{definition}[\cite{GG2018q}]\label{main}
	A bipartite measure of entanglement $E$ is monogamous if for any $\rho^{ABC}\in\mS^{ABC}$ that satisfies the {disentangling condition},~i.e.,    
	\bea\label{cond}
	E( \rho^{A|BC}) =E( \rho^{AB}),
	\eea
	we have that $E(\rho^{AC})=0$, where $\rho^{AB}=\tr_C\rho^{ABC}$. 
\end{definition}

With respect to this definition, a continuous measure $E$ is monogamous according to this definition if and only if there exists $0<\alpha<\infty$ such that
\bea\label{power}
E^\alpha( \rho^{A|BC}) \geq E^\alpha( \rho^{AB}) +E^\alpha( \rho^{AC})	
\eea
for all $\rho$ acting on the state space $\mH^{ABC}$ with fixed $\dim\mH^{ABC}=d<\infty$ (see Theorem 1 in Ref.~\cite{GG2018q}).
In 2016, Lancien \etal proposed an approach to depict the monogamous measure of entanglement by a family
of monogamy relations of the form,
\begin{eqnarray}\label{lan}
	E(\rho^{A|BC})\geq f\left[E\left(\rho^{AB}\right), E\left(\rho^{AC}\right)\right] 
\end{eqnarray}
that satisfied by any state $\rho^{ABC}$, where $f$ is some function of two variables that satisfy certain conditions~\cite{Lan16}. For the particular choice of the function $f(x,y)=\sqrt{x^2+y^2}$, one recovers the CKW inequality~Eq.~\eref{Coffman}.
If $f(x,y)=x+y$, one gets the monogamy relation satisfied by $E=\tau$. This supports that one can obtain a family of monogamy inequalities based on a proper choice of the function $f$. Motivated by this point, Jin \etal proposed the following definition of monogamy relation.

\begin{definition}[\cite{Jin2022aqt}]\label{main2}
	Let $E$ be a bipartite measure of entanglement. $E$ is said to be monogamous if there exists $0<\mu\leq 1$ such that for any state $\rho\in\mS^{ABC}$,   
	\bea\label{m4}
	E(\rho^{A|BC})= \mu E(\rho^{AB})+E(\rho^{AC})
	\eea
	holds provided that $E(\rho^{AB})\leq E(\rho^{AC})$. 
\end{definition}

Definition~\ref{main2} is equivalent to Definition~\ref{main}, and it is equivalent to Eq.~\eref{power} for any continuous bipartite entanglement measure $E$ as well~\cite{Jin2022aqt}. $\mu$ in Eq.~\eref{m4} is called the monogamy weight with respect to the  entanglement measure $E$.
The parameter $\mu$ has an operational interpretation as the ability to be monogamous for an entanglement measure $E$.

For any given $y>0$, let~\cite{Zhu2023pra} 
\beax \fl \qquad\qquad~~
X_y&=&\left\lbrace x({\rho^{ABC}})\mid  x({\rho^{ABC}})\left[ E^{y}(\rho^{A|BC})-\max\{E^{y}(\rho^{AB}),E^{y}(\rho^{AC})\}\right] \right. \nonumber\\
&&~=\left. \min\{E^{y}(\rho^{AB}),E^{y}(\rho^{AC})\}\right\rbrace,
\eeax 
where $x({\rho^{ABC}})$ is a nonnegative function on $\mS^{ABC}$, and
we take $x({\rho^{ABC}})=0$ for any $y$ when $E(\rho^{A|BC})=\max\left\lbrace E(\rho^{AB}),E(\rho^{AC})\right\rbrace$ and $\min\left\lbrace E(\rho^{AB}),E(\rho^{AC})\right\rbrace=0$.
It was proved in Ref.~\cite{Zhu2023pra} that a bipartite entanglement measure $E$ (not necessarily continuous) satisfies Eq.~\eref{power} if and only if there exists $y_0>0$ such that $X_{y_0}$ is a bounded set. For the general case, see in Ref.~\cite{Chenwei2025epjp} for detail.

By definition, if $E$ is monogamous on tripartite states, it is also monogamous on any $n$-partite systems ($n>3$). So we only need to consider the tripartite case.


\subsection{Monogamy \& $n$-extension}


Apart from the measure of entanglement, the monogamy of entanglement can also be verified from a qualitative aspect~\cite{Doherty2005pra,Terhal2004,Werner1989lmp,Yang2006pla}.
It was shown that $\rho^{AB}$ is separable if and only if it is $n$-sharable for any $n$~\cite{Doherty2005pra,Terhal2004,Werner1989lmp,Yang2006pla}. Conversely, for any entangled state, there always exists a finite $N$ such that no valid $n$-extension can be found for any $n>N$~\cite{Yang2006pla}.
Recall that, a bipartite state $\rho^{AB}$ is called $n$-sharable if there exists an extension $\rho^{AB_1B_2\cdots B_n}$ such that 
\bea\label{n-extension}
\rho^{AB}=\rho^{AB_1}=\rho^{AB_2}=\cdots\rho^{AB_n},
\eea
where $\rho^{AB_k}$ is the reduced state of $\rho^{AB_1B_2\cdots B_n}$ with respect to the subsystem $AB_k$.
$\rho^{AB_1B_2\cdots B_n}$ is called an $n$-extension of $\rho^{AB}$ if it satisfies Eq.~\eref{n-extension} (The $n$-extension is the bipartite case of the locally symmetric extension (LSE) in Ref.~\cite{Doherty2005pra}).
Indeed, the upper bound $N$ can be calculated by~\cite{Yang2006pla}
\beax 
N= \left\lfloor \frac{S(\rho^A)}{G_{\leftarrow}(\rho^{AB})}\right\rfloor, 
\eeax 
where $\lfloor x \rfloor$ denotes the maximal integer not larger than $x$,
$G_{\leftarrow}(\rho^{AB})=\min\sum_ip_iC_\leftarrow(\rho_i^{AB})$~\cite{Yangdong2005prl},
$C_\leftarrow(\rho^{AB})=\max\limits_{B_i}\left[S(\rho^A)-\sum\limits_ip_iS(\rho^A_i) \right] $
is the classical correlation of $\rho^{AB}$~\cite{Henderson2001jpa},
$\{B_i^\dag B_i\}$ is a POVM performed on part $B$,
$p_i\rho_i^A=\tr_B(I\ot B_i\rho^{AB} I\ot B_i^\dag)$, 
where the minimum is taken over all possible mixed ensembles $\{p_i, \rho_i^{AB}\}$ of $\rho^{AB}$.
Namely, entanglement in a given state cannot be sharable in arbitrarily many parties, which demonstrate the monogamy of the entanglement qualitativly.
In Ref.~\cite{Guo2020pra}, it was proved that, the maximally entangled state is not extendible for any $n$~\cite{Cavalcanti2005pra,Guo2020pra} (see Sec.~\ref{sec-9.2} for detail).
It seems that, as one may expect, the more the state is entangled, the less sharability it has.


\subsection{Monogamy of entanglement: pure vs mixed tripartite states }


In general, it is typically hard to check if a measure of entanglement is 
monogamous since the condition in Eq.~\eref{cond} involves mixed tripartite states. 
On the other hand, it is significantly simpler to check the disentangling condition 
if $\rho^{ABC}$ that appears in Eq.~\eref{cond} is pure. 
$E$ is called monogamous on pure states if for any pure tripartite state $\rho^{ABC}$ that satisfies~\eref{cond}, $E(\rho^{AC})=0$~\cite{GG2018q}.
In Ref.~\cite{GG2019pra}, it was shown that for any entanglement monotone $E$, if $E_F$ is monogamous on pure states 
it is also monogamous on mixed states (that is, it is monogamous according to Def.~\ref{main}).


\subsection{Quantum Markov states and monogamy of entanglement}


Recall that quantum Markov states are states that saturate the strong subadditivity 
of the von-Neumann entropy. That is, they saturate the inequality:
\bea
S(\rho^{AB})+S(\rho^{BC})\geq S(\rho^{ABC})+S(\rho^B).
\eea
In Ref.~\cite{HaydenJozaPetsWinter} it was shown that the inequality above 
is saturated if and only if the Hilbert space of system $B$, $\mH^B$, can be decomposed into a direct sum of tensor products
\beax
\mH^B=\bigoplus_{j}\mH^{B_{j}^{L}}\otimes\mH^{B_{j}^{R}}
\eeax
such that the state $\rho^{ABC}$ has the form
\bea\label{aMarkov}
\rho^{ABC}=\bigoplus_{j}q_j\rho^{AB_{j}^{L}}\otimes\rho^{B_{j}^{R}C},
\eea
where $q_j$ is a probability distribution.
It was proved in Ref.~\cite{GG2018q} that, for all entangled Markov quantum states $\rho^{ABC}$ of the form~\eref{aMarkov}, any entanglement monotone $E$ satisfies the disentangling 
condition~\eref{cond}.


\subsection{Entanglement of collaboration and monogamy of entanglement}


\subsubsection{Entanglement of collaboration}

Monogamy of entanglement is closely related to entanglement of collaboration. 
Given a measure of bipartite entanglement $E$, its corresponding entanglement of 
collaboration, $E_{co}^{AB|C}$, is a measure of entanglement on tripartite mixed states, $\rho^{ABC}$, given by~\cite{GourSpekkens}:
\bea\label{EoC}
E_{co}^{AB|C}\left(\rho^{ABC}\right)=\max\sum_{j=1}^{n}p_jE\left(\rho^{AB}_{j}\right),\label{aeocmixmax}
\eea
where the maximum is taken over all tripartite LOCC protocols yielding the bipartite 
state $\rho^{AB}_{j}$ with probability $p_j$.
For any entanglement monotone $E$, if $\rho^{ABC}$ (possibly mixed) satisfies the disentangling condition~\eref{cond}, then.~\cite{GG2018q} 
\bea\label{astrong}
E\left(\rho^{AB}\right)= E_{co}^{AB|C}\left(\rho^{ABC}\right).
\eea

Note that $E_{co}^{AB|C}$ can be strictly larger than $E_a$, and in general, $E_{co}^{AB|C}\geq E_a$~\cite{GourSpekkens,GG2018q}. 
However, if $\rho^{ABC}$ satisfies the disentangling condition then we must have 
$E_{co}^{AB|C}=E_a$. Indeed, if $\rho^{ABC}$ satisfies Eq.~\eref{cond} we get~\cite{GG2018q}
\beax
E_a(\rho^{ABC})\geq E(\rho^{AB})=E\left(\rho^{A|BC}\right)=E_{co}^{AB|C}\left(\rho^{ABC}\right).
\eeax
Therefore, one can replace $E_{co}^{AB|C}$ in Eq.~\eref{astrong} with $E_a$, which may be 
convenient since $E_a$ is somewhat a simpler measure than $E_{co}^{AB|C}$. Note however 
that we left $E_{co}^{AB|C}$ in Eq.~\eref{astrong} since 
$E\left(\rho^{A|BC}\right)=E_{co}^{AB|C}\left(\rho^{ABC}\right)$ 
implies $E\left(\rho^{A|BC}\right)=E_{a}\left(\rho^{ABC}\right)$ but not vice versa.

\subsubsection{When entanglement of formation equals entanglement of assistance?}

Let $E$ be a bipartite entanglement monotone, and let $\rho$ 
be a pure tripartite state satisfying the disentangling condition~\eref{cond}. By Eq.~\eref{EoC}, one can immediately get~\cite{GG2018q}
\bea\label{aeq}
E\left(\rho^{AB}\right)=E_F(\rho^{AB})=E_a(\rho^{AB}).
\eea
Moreover, 
for and bipartite entanglement measure $E$, if 
$E_F(\rho^{AB})=E_a(\rho^{AB})$ then for any $\sigma^{AB}\in\mS^{AB}$ 
with $\supp(\sigma^{AB})\subseteq\supp(\rho^{AB})$ we have $E_F(\sigma^{AB})=E_a(\sigma^{AB})$,
where $\supp(\rho^{AB})$ denotes the support subspace of $\rho^{AB}$.

In addition, for the two-qubit state $\rho^{AB}$, Eq.~\eref{aeq} holds for any injective (up to local unitaries) bipartite entanglement measure $E$ if and only if~\cite{GG2018q}  
$\rho^{AB}=\tr_C|\check{\W}\lr \check{\W}|$, where $|\check{\W}\ra$
is the 3-qubit W-class state
\bea\label{wstate}
|\check{\W}\ra=\lambda _{0}|000\rangle +\lambda _{1}e^{\rmi\varphi }|100\rangle
+\lambda _{2}|101\rangle +\lambda _{3}|110\rangle,  
\eea
where $\lambda_i\geq 0$, $i=0$, $1$, 2, 3, $\sum_{i=0}^{3}\lambda_i^2=1$, $0\leq \varphi\leq \pi$, $\lambda _{0}\lambda _{2}\lambda_{3}\neq 0$.


\subsection{Monogamy criterion of the convex-roof extended bipartite entanglement monotone}


In Ref.~\cite{GG2019pra}, we proposed the following Monogamy criterion: 
\begin{theorem}[Monogamy criterion in~\cite{GG2019pra}]\label{2-monogamy criterion}
	Let $E$ be an entanglement monotone with the reduced function $h$ is strictly concave. Then,
	\begin{itemize}
		\item[(i)] If  $\rho^{ABC}=|\psi\lr\psi|^{ABC}$ is pure and~\eref{cond} holds then 
		$\mH^B$ has a subspace isomorphic to $\mH^{B_1}\otimes\mH^{B_2}$ and up to local unitary on system $B_1B_2$,
		\bea\label{productdecomposition}
		|\psi\ra^{ABC}=|\phi\ra^{AB_1}|\eta\ra^{B_2C}\,,
		\eea
		where $|\phi\ra^{AB_1}\in\mH^{AB_1}$ and $|\eta\ra^{B_2C}\in\mH^{B_2C}$ are pure states.
		In particular, $\rho^{AC}$ is a product state [and consequently $E(\rho^{AC})=0$], so that $E$ is monogamous on pure tripartite states.
		
		\item[(ii)] If  $\rho^{ABC}$ is a mixed tripartite state and 
		$E_F(\rho^{A|BC})=E_F(\rho^{AB})$,
		then 
		\beax
		\rho^{ABC}=\sum_{x}p_x|\psi_x\ra\la\psi_x|^{ABC},
		\eeax 
		where $\{p_x\}$ is some probability distribution, and for each $x$ the Hilbert space
		$\mH^B$ has a subspace isomorphic to $\mH^{B_1^{(x)}}\otimes\mH^{B_2^{(x)}}$ such that up to local unitary on system $B$, each pure state $|\psi_x\ra^{ABC}$ is  given by
		\beax
		|\psi_x\ra^{ABC}=|\phi_x\ra^{AB_1^{(x)}}|\eta_x\ra^{B_2^{(x)}C}\,.
		\eeax
		In particular, the marginal state $\rho^{AC}$ is separable so that $E_F$ is monogamous (on mixed tripartite states).
	\end{itemize}
\end{theorem}

The condition that $E$ in the conclusion above is an entanglement monotone can be replaced with a weaker condition that the measure of entanglement $E$ satisfies $E\leq E_F$ on all bipartite states. The item (i) above generalizes a similar result that was proved in~\cite[the disentagling theorem]{Hehuan} for the special case in which $E$ is taken to be the negativity.

If system $B$ in the second part of the result above has dimension not greater than 3, then we must have for each $x$ that either $\mH^{B_1^{(x)}}$ or ${\mH}^{B_{2}^{(x)}}$ are one dimensional. Thus, if $E_F(\rho^{A|BC})=E_F(\rho^{AB})$ and $\dim\mH^B\leq 3$ then $\rho^{ABC}$ is bi-separable, and in particular it admits the form~\cite{GG2019pra}
\beax
\rho^{ABC}=t\sigma^{A|BC}+(1-t)\gamma^{AB|C},
\eeax
where $\sigma^{A|BC}$ is $A|BC$ separable, $\gamma^{AB|C}$ is $AB|C$ separable, and $t\in[0,1]$.
In particular, if $\rho^{ABC}=|\psi\ra\la\psi|^{ABC}$ is a pure state, then $|\psi\ra^{ABC}$ has the form $|\phi\ra^{AB}|\eta\ra^{C}$ or $|\phi\ra^{A}|\eta\ra^{BC}$.

For any entanglement monotone $E$, and any monotonically increasing function $f:\mbR^{+}\to\mbR^+$ with the property that $f(0)=0$, denote~\cite{GG2019pra}
\bea
E_{\rm f}(\rho^{AB})=\min\sum_{j}p_jf\left[E\left(|\psi_j\ra^{AB}\right)\right] ,
\eea
where the minimum is taken over all pure state decompositions of $\rho^{AB}=\sum_{j=1}^{n}p_j|\psi_j\lr\psi_j|^{AB}$. Observe that if in addition $f$ is convex then we must have
\beax
E_{\rm f}\left( \rho^{AB}\right)\geq f\left[E\left(\rho^{AB}\right)\right].
\eeax
Therefore, if $E_{\rm f}$ is monogamous on pure tripartite states then $E$ is also monogamous on pure tripartite states~\cite{GG2019pra}. 
As a simple example of this, consider the function $f(x)=x^2$ and take $E=C$ be the concurrence. Then,  $E_{\rm f}=C^2$ is the tangle which is monogamous (the reduced function of tangle is the linear entropy, which is strictly concave). Hence, the concurrence $C$ is also monogamous.


\subsection{Entanglement monotones that are not monogamous}\label{sec-3.7}


Let 
\beax
|\psi_0\ra^{AB}&=&a_0|0\ra^A|0\ra^B+a_1|1\ra^A|1\ra^B+a_2|2\ra^A|2\ra^B,\\
|\psi_1\ra^{AB}&=&a'_0|0\ra^A|3\ra^B+a'_1|1\ra^A|2\ra^B+a'_2|2\ra^A|1\ra^B
\eeax
with $a_0^2={a'}_0^2\geq\frac12$, $a_1'a_2\neq a_1a_2'$, $\sum_i a_i^2=\sum_i {a'}_i^2=1$, $a_0>a_1\geq a_2$, $a'_0>a'_1\geq a'_2$, 
and
\bea\label{counter-eg1}
|\Phi\ra=\frac{1}{\sqrt{2}}\left(|\psi_0\ra^{AB}|0\ra^C+|\psi_1\ra^{AB}|1\ra^C\right). 
\eea
It follows that~\cite{Guo2023njp} 
\beax 
E_2(\rho^{AB})=E_2(|\Phi\ra^{A|BC})=1-a_0^2
\eeax but $\rho_{AC}^{T_a}$ is not positive whenever $a_1'a_2\neq a_1a_2'$, 
and thus $E_2(\rho^{AC})>0$. If the reduced subsystem is two-dimensional, there also exists pure state $|\psi\rangle$ in $\C^{2}\otimes \C^{2}\otimes \C^{2} $ that violates the disentangling condition~\eref{cond}~\cite{Guo2023njp}. Therefore
$E_2$ is still not monogamous.
$E_{\min}$ and $E_{\min'}$ are not monogamous either~\cite{Guo2023njp}.

In addition, it can be easily checked that the Schmidt number is not monogamous~\cite{Guo2024jsnu}. For example, we consider the $W$ state $|\W\ra=\frac{1}{\sqrt{3}}(|100\ra+|010\ra+|001\ra)$.
$S_r(|\W\ra)=2$, and all the Schmidt number of its bipartite reduced states are 2, so
$S_r(A|BC)=S_r(AB)$ can not lead to $\rho^{AC}$ is separable. In the next subsection, we discuss the strict concavity of the reduced functions so far in literature.


\subsection{Strict concavity of the reduced function}\label{sec-3.8}


The reduced functions of the entanglement of formation $E_f$, tangle $\tau$, concurrence $C$, negativity $N$,
the Tsallis $q$-entropy of entanglement $E_q$, and the R\'{e}nyi $\alpha$-entropy of entanglement $E_\alpha$ are
\beax
h(\rho)&=&S(\rho),\\
h_{\tau}(\rho)&=&h_C^2(\rho)=2(1-\tr\rho^2),\\
h_N(\rho)&=&\frac12[(\tr\sqrt{\rho})^2-1],\\
h_q(\rho)&=&\frac{1-\tr\rho^q}{q-1},\quad q>0,\\
h_\alpha(\rho)&=&(1-\alpha)^{-1}\ln(\tr\rho^\alpha),\quad 0<\alpha<1,
\eeax
respectively, where $S$ is the von Neumann entropy. It has been shown that $h$, $h_{\tau}$, $h_C$, $h_N$, $h_q$, and $h_\alpha$ are not only concave but also strictly concave~\cite{GG2019pra,Guo2023njp,Wehrl1978,Vidal2000} (where the strict concavity of $h_N$ is proved very recently in Ref.~\cite{Guo2023njp}). The reduced functions of $q$-concurrence and the $\alpha$-concurrence are
\beax
h_{q'}(\rho)=1-\tr\rho^q, \quad q>1, 
\eeax
and
\beax h_{\alpha'}(\rho)=\tr\rho^\alpha-1,\quad  0<\alpha<1,
\eeax 
respectively. Obviously, the properties of these two functions above are the same as that of $h_q$.

The reduced functions of the entanglement monotones induced by the fidelity-based distances $E_{\mF}$,  $E_{{\mF}'}$, and $E_{A\mF}$ are
\beax
h_{\mathcal{F}}(\rho)&=&1-\tr\rho^3,\\
h_{\mathcal{F}'}(\rho)&=&1-\left( \tr\rho^2\right)^2 ,\\
h_{A\mathcal{F}}(\rho)&=&1-\sqrt{\tr\rho^3},
\eeax
respectively. They are strictly concave~\cite{Guo2022entropy}.

The reduced functions of $E_2$, $E_{\min}$, and $E_{\min'}$ are
\beax
h_2(\rho)&=&1-\|\rho\|,\\
h_{\min}(\rho)&=&\|\rho\|_{\min},\\
h_{\min'}(\rho)&=&r(\rho)\|\rho\|_{\min}.
\eeax
All of them are concave but not strictly concave. The reduced function of $E_{v\text{-}k}$ is $1-\|\rho\|_{(k-1)}$, where $\|\rho\|_{(k)}=\sum_{i=1}^k\delta_i$ is the Fy Fan norm, $\delta_i$'s are eigenvalues of $\rho$ in decreasing order.

Any function that can be expressed as
\bea
H_g(\rho)=\tr g(\rho)=\sum_jg(p_j)\,,
\eea
where $p_j$'s are the eigenvalues of $\rho$, is strictly concave if $g''(p)<0$ for all $0<p<1$. This includes (i) the quantum Tsallis $q$-entropy~\cite{Landsberg,Tsallis1988jsp}
with $q>0$ (the case of $q=2$ is the linear entropy $1-\tr\rho^2$), (ii) the R\'enyi $\alpha$-entropy for
$\alpha\in[0,1]$, (iii) the reduced function of $E_{N_F}$, i.e., $2\log_2\tr\sqrt{\rho}$, which is strictly concave since $(\log_2x^{1/2})''<0$, so is that of $E_{\max-r}$, (iv) the reduced function of $C_q^{to}$ and (v) the reduced function of $E_\kappa$ since $\tr\rho^{1-\kappa}-\tr\rho^{1+\kappa}$ is strictly cave whenever $0\leq\kappa\leq 1$. 

The reduced function of $C_k$ is strictly concave~\cite{Guo2024jsnu}. The reduced functions of $R$ and $C_E$ are multiple of $h_N$, therefore they are also strictly concave. The other reduced functions are not straightforward, which need further research.
We list these reduced functions that are already known in Table~\ref{tab:table3-6} for convenience.

\begin{table}[h]
	\caption{\label{tab:table3-6} The strict concavity of the reduced functions.
	}	
	\begin{indented}
		\item[]		
		\begin{tabular}{@{}lllll}
			\br
			$E$     & Reduced function & Concave & Strictly concave  \\ 
			\mr
			$E_f$  &$S$                          &$\checkmark$         &$\checkmark$      \\
			$C$    &$\sqrt{2(1-\tr\rho^2)}$      &$\checkmark$         &$\checkmark$       \\
			$\tau$ &$2(1-\tr\rho^2)$             &$\checkmark$         &$\checkmark$        \\
			$C_k$&   $\left[\frac{S_k(\lambda)}{S_k(1/d, 1/d,\dots, 1/d)}\right]^{1/k}$           &  $\checkmark $       &$\checkmark$  \\
			$C_q$  &$1-\tr\rho^q$    &$\checkmark$($q>0$)  &$\checkmark$($q>1$)\\
			$C_\alpha$&$\tr\rho^\alpha-1$, $\alpha\in(0,1)$&$\checkmark$       &$\checkmark$        \\
			$C_q^{to}$ & $\frac{1}{\mu}[1-\tr\rho^q+\tr(I-\rho)-\tr(I-\rho)^q]$~${\color{red}^{\rm a}}$& $\checkmark$ &$\checkmark$ \\
			$E_q$  &$\frac{1-\tr\rho^q}{q-1}$    &$\checkmark$($q>0$)  &$\checkmark$($q>1$)\\
			$E_\alpha$&$\frac{\log_2(\tr\rho^\alpha)}{1-\alpha}$, $\alpha\in(0,1)$
			&$\checkmark$       &$\checkmark$        \\
			$E_{q,s}$ & $\frac{1}{(1-q)s}\left[(\mathrm{Tr}\rho^q)^s-1\right]$ &$\checkmark$  & ?${\color{red}^{\rm b}}$\\
			$E_t$ & $-\frac{1}{r}\tr[\rho\log_2\rho+(I-\rho)\log_2(I-\rho)]~{\color{red}^{\rm c}}$& $\checkmark$& $\checkmark$\\
			$N$  &$\frac{(\tr\sqrt{\rho})^2-1}{2}$   &$\checkmark$         &$\checkmark$       \\
			$N_F$  &$\frac{(\tr\sqrt{\rho})^2-1}{2}$   &$\checkmark$         &$\checkmark$       \\
			$E_N$  &$2\log_2\tr\sqrt{\rho}$   &$\checkmark$         &$\checkmark$       \\
			$E_{\mF}$&$1-\tr\rho^3$                &$\checkmark$         &$\checkmark$       \\
			$E_{{\mF}'}$&$1-(\tr\rho^2)^2$            &$\checkmark$         &$\checkmark$       \\
			$E_{A\mF}$&$1-\sqrt{\tr\rho^3}$         &$\checkmark$         &$\checkmark$       \\
			$E_{SG}$&$1-\frac{1}{\sqrt{1-{1}/{\sqrt{d}}}}\left( 1-\frac{\tr\sqrt{\rho}}{\sqrt{d}}\right)^{\frac12}$         &$\checkmark$         &$?$       \\
			$R$&$\left( \tr\sqrt{\rho}\right)^2-1$         &$\checkmark$         &$\checkmark$       \\
			$E_2$, $E_G$	   &$1-\|\rho\|$                 &$\checkmark$         &$\times$           \\
			$E_{\min}$&$\|\rho\|_{\min}$            &$\checkmark$         &$\times$           \\
			$E_{\min'}$&$r(\rho)\|\rho\|_{\min}$     &$\checkmark$         &$\times$          \\
			$E_r$ &$S$                          &$\checkmark$         &$\checkmark$      \\
			$E_M$ &$S$                          &$\checkmark$         &$\checkmark$      \\
			$E_{sq}$ &$S$                          &$\checkmark$         &$\checkmark$      \\
			$E_I$&$S$                          &$\checkmark$         &$\checkmark$      \\
			$R$ &$(\tr\sqrt{\rho})^2-1$ &$\checkmark$  &$\checkmark$ \\
			$E_{V\text{-}k}$ &$1-\|\rho\|_{(k-1)}$ &$\checkmark$  & $\times$${\color{red}^{\rm d}}$ \\
			$E_\kappa$ &$\frac{\tr(\rho^{1-\kappa}-\rho^{1+\kappa})}{2\kappa}$ &$\checkmark$ & $\checkmark$\\
			$E_{\rm ic}$ & $1-\frac{1}{2^r}\sum_{i=0}^rP_r^i\tr\rho^{i+1}$, $P_r^i=\frac{r!}{i!(r-i)!}$${\color{red}^{\rm e}}$ &$\checkmark$ &$\checkmark$ \\
			$S_r$ &$r(\rho)-1$ &$\times$ & $\times$\\
			$E_{\max\text{-}r}$ &$2\log_2\tr\sqrt{\rho}$ &$\checkmark$ & $\checkmark$\\
			$C_E$ &$\frac{1}{d}[(\tr\sqrt{\rho})^2-1]$ &$\checkmark$ & $\checkmark$\\
			$E_{oc}$ & $S$& $\checkmark$& $\checkmark$\\
			\br
		\end{tabular}
		\item[] ${\color{red}^{\rm a}}$ {``?'' means it is unknown.}	
		\item[] ${\color{red}^{\rm b}}$ $\mu=d-d^{1-q}[1+(d-1)^q]$.	
		\item[] ${\color{red}^{\rm c}}$ $r=d\log_2d-(d-1)\log_2(d-1)$, $d=\dim\mH^{A,B}$.	
		\item[] ${\color{red}^{\rm d}}$ It is not strictly concave when $k=2$. We conjecture that they are not strictly cave for all $k\geq 2$.
		\item[] ${\color{red}^{\rm e}}$ $r(\rho)=r+1$.
	\end{indented}
\end{table} 

In light of the monogamy criterion, we know that except for the partial-norm of entanglement, the geometric measure, the minimal partial-norm of entanglement, and the Schmidt number, almost all other convex-roof extended bipartite entanglement monotones so far are monogamous. It was showed that many measures of entanglement, such as the entanglement of formation, that were believed not to be monogamous (irrespective of the specific monogamy relation~\cite{Lan16}), are in fact monogamous according to the improved definition of monogamy without inequalities. Therefore, the results presented here support this improved definition of monogamy. The fact that so many important measures of entanglement are not universally monogamous~\cite{Lan16} may give the impression that monogamy of entanglement cannot be attributed to entanglement itself but rather is a property of the particular measure that is used to quantify entanglement. Furthermore, as was shown in~\cite{Lan16}, measures of entanglement cannot be simultaneously faithful (as defined in~\cite{Lan16}) and universally monogamous. 
In Ref.~\cite{GG2019pra}, all these issues were avoided by adopting the improved definition of monogamy that allows for non-universal monogamy relations, while at the same time maintaining a quantitative way (as in~\eref{power}) to express monogamy relations.

While we were not able to show that all measures of entanglement are monogamous (according to the Definition~\ref{main}), we are also not aware of any continuous measures of entanglement that are not monogamous. It may be the case that all continuous measures of entanglement are monogamous, which will support our assertion that monogamy is a property of entanglement and not of some particular functions quantifying entanglement. Moreover, many measures of entanglement, are not defined in terms of convex-roof extensions, such as the negativity, the relative entropy of entanglement, the trace distance entanglement, the Groverian measure of entanglement, the robustness of entanglement, the generalized robustness of entanglement, ect. For such measures, the monogamy criterion does not provide any information regarding their monogamy on mixed tripartite states.


\subsection{Strict entanglement monotone}


Intuitively, the discussion in previous subsections supports that, 
for an entanglement monotone $E_F$ with reduced function $h$,
$E_F$ is monogamous if and only if $h$ is strictly concave. On the other hand, the monogamy is an inherent feature of entanglement.
So the axiomatic definition of an entanglement monotone should be improved as follows~\cite{Guo2023njp}. 

\begin{definition}\label{strict entanglement monotone}
Let $E$ be a nonnegative function on $\mS^{AB}$ with $E(|\psi\ra)=h(\rho^A)$ for pure state. We call $E$ a {strict entanglement monotone (SEM)} if (i) $E(\sigma^{AB})=0$ for any 
separable state $\sigma^{AB}\in\mS^{AB}$, (ii) 
$E$ behaves monotonically decreasing under LOCC on average, (iii) $E$ is convex, and (iv) the reduced function $h$ is strictly concave.
\end{definition}

With such a spirit,  except for $E_2$, $E_G$, $E_{\rm min}$, $E'_{\rm min}$and the Schmidt number, all the previous entanglement monotones that are shown to be monogamous or monogamous on pure states are strict entanglement monotones (see in Sec.~\ref{sec-2.2}, especially Table~\ref{tab:table3-6}). However, it still remains unknown that whether or not the non convex-roof extended strict entanglement monotones in literature are monogamous in addition to the squashed entanglement and the one-way distillable entanglement. We conjecture that all the strict entanglement monotones are monogamous.


\subsection{Strict LOCC entanglement monotone}


Any entanglement measure is non-increasing under LOCC and it is non-increasing on average under LOCC in general.
As we discussed above, the reduced function is strictly concave in most cases, so we think that the action of LOCC should be strictly decreasing. Consequently, the strict LOCC entanglement measure was proposed in Ref.~\cite{Guo2019pra}.
An entanglement measure $E$ is said to be strictly decreasing on average
under LOCC if 
for any stochastic LOCC~\cite{Guo2019pra}
\beax
&&\left\lbrace \varepsilon_j:{\rm Tr}\varepsilon_j(\rho)=p_j,\sum_jp_j=1, \varepsilon_j(\rho)\neq p_jU_j^A\otimes U_j^B\rho {(U_j^A)}^\dag\otimes {(U_j^B)}^\dag,\right.  \\
&&\quad\quad \left. ~~~{(U_j^X)}^\dag U_j^X=I_j^X,~ j=1,~ 2,~\dots,~ d.\right\rbrace
\eeax 
there exists $\rho\in\mS^{AB}$ such that 
\bea\label{strict1}
E\left( \rho\right)>\sum_jp_jE\left( \sigma_j\right),
\eea
where $p_j\sigma_j=\varepsilon_j(\rho)$.
Equivalently, an entanglement measure $E$ decreases strictly on average under LOCC if and only if~\cite{Guo2019pra}
\beax\label{strict2}
E\left( \rho\right)=\sum_jp_jE\left( \sigma_j\right)
\eeax
holds for all states $\rho\in\mS^{AB}$ implies that
the LOCC is either a local unitary operation
(if the LOCC is a map from system $AB$ to $A'B'$, then it is a local isometric operation; here we always assume with
no loss of generality that the LOCCs are acting on $AB$ to itself) or a convex mixture of local unitary operations.
If an entanglement monotone $E$ is strictly decreasing on average under LOCC, we call it is a strict-LOCC
entanglement monotone (S-LEM)~\cite{Guo2019pra} (note here that, in Ref.~\cite{Guo2019pra}, S-LEM is called strict entanglement monotone).
If an entanglement monotone $E$ is strictly decreasing under LOCC for pure states,
we call it is an S-LEM on pure states.

It is shown that, the S-LEM is similar to that of strict entanglement monotone~\cite{Guo2019pra}:
(i) If $E$ is an S-LEM on pure states, then $E_F$ is an S-LEM as well;
(ii) $E_F$ as defined in~(\ref{EOF}) is a S-LEM if the reduced function $h$ is strictly concave; Let $E$ be a measure of entanglement that satisfies $E\leq E_F$ on all bipartite states. If the reduced function is strictly concave, then it is an S-LEM on pure states.

\begin{table}[htp]
	\caption{\label{tab:table3-1} A list of  entanglement measures.}		
\footnotesize
\hspace{5mm}\begin{tabular}{@{}lllllll}\br
	$E$          &Additivity       & Eq.~\eref{average}      &Convex      &  SEM            & Monogamy           & S-LEM  \\ \mr
	$E_d$                         &$\times$\cite{Horodecki1999prl,Shor2001} &$\checkmark$&$\times$\cite{Shor2001} & $\times$& $\checkmark$${\color{red}^{\rm a}}$      & pure states\\
	$E_c$                      &  $\checkmark$ &$\checkmark$                    &   $\checkmark$\cite{Donald2002jmp}         &$\checkmark$& pure states   &pure states\\
	$E_f$ &$\times$ &$\checkmark$               &$\checkmark$&$\checkmark$& $\checkmark$& $\checkmark$\\
	$C$              &? &$\checkmark$                      &$\checkmark$&$\checkmark$& $\checkmark$&$\checkmark$\\
	$\tau$           & ?    &$\checkmark$         &$\checkmark$&$\checkmark$& $\checkmark$&$\checkmark$\\
	$C_k$              & ?&$\checkmark$                     &$\checkmark$& $\checkmark$& $\checkmark$&$\checkmark$\\
	$G_d$               & ?&$\checkmark$                      &$\checkmark$& $\times$ & $\checkmark$&$\checkmark$\\
	$C_q$ & ?&$\checkmark$&$\checkmark$ & $\checkmark$& $\checkmark$& $\checkmark$\\
	$C_\alpha$ & ?&$\checkmark$&$\checkmark$ & $\checkmark$&$\checkmark$ & $\checkmark$\\	
	$E_{q}$, $ q>0$     & ?    &$\checkmark$    &$\checkmark$&$\checkmark$ & $\checkmark$&$\checkmark$\\
	$E_{\alpha}$, $0 \leq\alpha\leq1$ & ?  &$\checkmark$   &$\checkmark$&$\checkmark$& $\checkmark$&$\checkmark$\\
	$E_{q,s}$, $0<q, s<1$ & ?  &$\checkmark$   &$\checkmark$&?& ?&?\\
	$E_t$ & ?&$\checkmark$& $\checkmark$&$\checkmark$ & $\checkmark$& $\checkmark$\\
	$N$              & ? &$\checkmark$     &$\checkmark$&$\checkmark$& pure states     &$\checkmark$\\
	$N_F$             &?  &$\checkmark$      &$\checkmark$&$\checkmark$ & $\checkmark$&$\checkmark$ \\
	$E_N$            &$\checkmark$\cite{VidalWerner} &$\checkmark$ &$\times$    &$\times$  & pure states&$\checkmark$ \\
	$E_r$               &$\times$\cite{Vollbrecht} &$\checkmark$   &$\checkmark$& $\checkmark$& pure states     &$\checkmark$\\
	$E_A$               &$\checkmark$\cite{Eisert2003jpa} &$\checkmark$              &$\checkmark$&? &   ?   &?\\
	$E_M$               &? &$\checkmark$              &$\checkmark$&? & pure states     & pure states \\
	$E_{sq}$\cite{Christandl2004jmp} &$\checkmark$  &$\checkmark$         &$\checkmark$&$\checkmark$&$\checkmark$&pure states${\color{red}^{\rm b}}$ \\
	$E_I$\cite{Yang2008prl}&$\checkmark$   &$\checkmark$        &$\checkmark$& $\checkmark$&pure states      &pure states\\
	$E_{\rm HS}$&?   &?        &?& ?&?      &?\\
	$E_{\tr}$&?   &$\checkmark$        &?& ?&?      &?\\
	$E_{\mF}$ & ?&$\checkmark$& $\checkmark$& $\checkmark$& $\checkmark$& $\checkmark$\\
	$E_{{\mF}'}$ & ?&$\checkmark$&$\checkmark$ & $\checkmark$& $\checkmark$& $\checkmark$\\
	$E_{A\mF}$ &? &$\checkmark$&$\checkmark$ &$\checkmark$ &$\checkmark$ & $\checkmark$\\
	$E_{\B}$ & ?& $\checkmark$ & $\checkmark$&? & ?&? \\
	$E_{\rm Gr}$ &? & $\checkmark$ & $\checkmark$&? &? &? \\
	$E_{SG}$ &? & $\checkmark$ & $\checkmark$&? &? &? \\
	$R$ &? &$\checkmark$ & $\checkmark$& $\checkmark$&pure states  &pure states \\
	$\check{R}$ &? &$\checkmark$ & $\checkmark$& $\checkmark$&pure states  &pure states \\
	$E_{V\text{-}k}$ &? &$\checkmark$&$\checkmark$ & ?& ?& ?\\
	$E_2$, $E_G$, $\check{E}_G$ &? &$\checkmark$&$\checkmark$ & $\times$& $\times$& $\times$\\
	$E_{\rm min}$ & ?&$\checkmark$&$\checkmark$ & $\times$&$\times$ & $\times$\\
	$E_{\rm min'}$ & ?&$\checkmark$&$\checkmark$ & $\times$&$\times$ & $\times$\\
	$E_\kappa$ &? &$\checkmark$&$\checkmark$ &$\checkmark$ &$\checkmark$ & $\checkmark$\\
	$E_{\rm ic}$ &? &$\checkmark$ &$\checkmark$ & $\checkmark$& $\checkmark$& $\checkmark$\\
	$S_r$ & $\times$ &$\checkmark$&$\times$  &$\times$  &$\times$  & $\times$ \\
	$E_{\max\text{-}r}$ &pure states&$\checkmark$ & $\times$&$\times$ &? &$\checkmark$ \\
	$E_{\min\text{-}r}$ & ?&$\times$& $\checkmark$& $\times$& ?&  $\times$\\
	$E_{N}^\alpha$ &?&$\checkmark$ & $\times$&$\times$ &? &  $\times$\\
	$\check{E}_{\kappa}$ &$\checkmark$ & $\checkmark$ &$\times$ &$\times$ & ?& $\times$\\
	$E_{cr}$& ?& $\checkmark$ & $\checkmark$ & ?&? &?\\
	$\check{E}$ & ?  & $\checkmark$ & ? &$\times$  & ? & $\times$\\
	$E_p$ & $\times$   & $\checkmark$ & $\checkmark$ & ? &?&?\\		
	$E_{e\text{-}g}$ & ?  & $\checkmark$ & $\checkmark$ & ? &?&?\\		
	$\tilde{\mE}$&? & $\times$ & $\checkmark$ & $\times$& ?& $\times$\\
	$C_E$ &? &$\checkmark$&$\checkmark$ & $\checkmark$& $\checkmark$& $\checkmark$\\
	$E_{oc}$ &? &? &$\checkmark$ & ?& ? &? \\
	\br
\end{tabular}\\
\indent\hspace{5mm}${\color{red}^{\rm a}}$ {The one-way distillable entanglement is monogamous~\cite{Koashi}.}\\
\indent\hspace{5mm}${\color{red}^{\rm b}}$ {For mixed states, see~\cite[Theorem~7]{Guo2019pra}.}\\
\end{table}
\normalsize

Note that, many entanglement measures, such as entanglement of distillation $E_d$, entanglement cost $E_c$, the squashed entanglement $E_{sq}$~\cite{Christandl2004jmp} and the relative entropy of entanglement $E_r$ coincide with the entanglement of formation $E_f$ for pure states. Thus $E_f$, $E_d$, $E_c$, $E_{sq}$ and $E_r$ are S-LEMs on pure states.

The strict concavity of $h$ is closely related to the strict-LOCC~\cite{Guo2019pra}:
For any entanglement monotone, the reduced function $h$ is strictly concave if and only if for any
stochastic LOCC $\left\lbrace\varepsilon_j: 
\varepsilon_j(\cdot)=I_A\otimes M_j(\cdot)I_A\otimes M_j^\dag\right\rbrace$ and any pure state $\rho\in\mS^{AB}$ that satisfies
\beax\label{monogamycond}
\sigma_{j_0}^A\neq\rho^A\,
\eeax
for some $j_0$ we have~(\ref{strict1}) holds,
where $p_j\sigma_j=\varepsilon_j(\rho)$, $X^A={\rm Tr}_BX$.

The negativity $N$ is an S-LEM on pure states since the reduced function is strictly concave and thus
$N_F$ is also an S-LEM.
In fact, many entanglement measures, such as negativity, the relative entropy of entanglement, the squashed entanglement, etc., that are not derived via the convex-roof structure are S-LEMs~\cite{Guo2019pra} (for the squashed entanglement, we assume that $E_{sq}(\rho^{AB})$ can be attained by optimal extension for any state $\rho^{AB}\in\mS^{AB}$).
For readers' convenience, we present a list of the related properties of all entanglement measures by now (see in Table~\ref{tab:table3-1}).


\subsection{Monogamy exponent}\label{sec-3.11}


As discussed in Ref.~\cite{GG2018q}, the expression for $\alpha$ is optimal 
in the sense that it provides the smallest possible value for $\alpha$ that 
satisfies Eq.~\eref{power}, and we call it monogamy exponent. 
If Eq.~\eref{power} holds, then 
\bea \label{monogamy-beta}
E^{\beta}(\rho^{A|BC})\geq E^{\beta}(\rho^{AB})+E^{\beta}(\rho^{AC})
\eea for any 
$\beta\geq\alpha$. Namely, the monogamy exponent is a function of the measure 
$E$, and we denote it by $\alpha(E)$. 
Not that, according to the monogamy criterion and the equivalent monogamy relation with the monogamy exponent in Eq.~\eref{power}, many monogamy relations discussed in literature are in fact special cases of Eq.~\eref{monogamy-beta}, which indicate some bounds of the  monogamy exponent $E(\alpha)$.
We list all the monogamy exponents discussed in literature so far in Table~\ref{tab:table3-7}. 
$E(\alpha)$ may depend also on the dimension 
$d=\dim\mH^{ABC}$ (see in Table~\ref{tab:table3-7}), and, in general, 
is hard to compute especially in higher dimensions~\cite{Allen,Audenaert,Chengshuming,Eltschka2015prl,Hehuan,Lancien}. By definition, $\alpha(E)$ 
can only increase with $d$ (e.g.
see Table~\ref{tab:table3-7}).

We need explain some abbreviations in Table~\ref{tab:table3-7}.
The generalized $n$-qubit W states is a set of all pure states that admit the form of
\bea \label{W-state}
|\widetilde{\W}_{n,2}\ra=b_1|10...0\rangle+b_2|01...0\rangle+...+b_n|00...1\rangle
\eea
with $\sum_{i=1}^{n}|b_i|^2=1$. A pure state in $d^{\ot n}$ is called a generalized W (GW) state if it can be written as
\begin{eqnarray}\label{GW}
	|\widetilde{\W}_{n,d}\ra=\sum_{i=1}^da_{1i}|i00\cdots 0\rangle+a_{2i} |0i0\cdots 0\rangle+\cdots+a_{ni}|000\cdots i\rangle
\end{eqnarray}
with $\sum_{ij}|a_{ij}|^2=1$.

\begin{table}
	\caption{\label{tab:table3-7} {The monogamy exponent}}
		\footnotesize
	\hspace{5mm}	\begin{tabular}{@{}llll}\br
			$E$& $\alpha(E)$ & System & Reference \\
			\mr
			$E_d$ & $\leq1$& any system & \cite{Koashi}\\
			$E_{sq}$& $\leq 1$& any system &\cite{Koashi}\\
			$C$ & $2$& $2^{\otimes n}$, pure state & \cite{Coffman2000pra,Osborne,Zhu2014pra}\\
			$C$ & $\leq 2$& any system & \cite{Yu2005pra}\\
			$C$ & $\leq 2$& reduced states of $|\widetilde{\W}_{n,2}\ra$ & \cite{Zhu2017qip2}\\
			$\mC^{to}_q$, $2 \leq q \leq 3$ & $\leq 1$& $2^{\ot n}$ & \cite{Xuan2025aqt}\\
			$\tau$ &   $1$    & $2\otimes 2\otimes2$ &\cite{Coffman2000pra}\\
			$\tau$       & $1$    & $2\otimes 2\otimes2^m$, $2^{\otimes n}$ &\cite{Osborne}\\
			$\tau$      & $>1$    & $3\otimes 3\otimes3$ &\cite{Ouyongcheng}\\
			$E_f$& $\leq \sqrt{2}$ & $2^{\otimes n}$ & \cite{Zhu2014pra}\\
			$E_f$& $\leq \sqrt{2}$ & $2\otimes 2\otimes2^m$ & \cite{Luo2015ap}\\
			$E_f$	& $>1$ & $2\otimes 2\otimes 2$ & \cite{Bai2014prl}\\
			$E_t$& $\leq \sqrt{2}$ & $2^{\otimes n}$ & \cite{Yangxue2023epj}\\
			$N$& $\leq 2$ & $2^{\otimes n}$ &\cite{Ouyongcheng2007pra2}\\
			$N$& $\leq 2$ & $2\otimes2\otimes 3$,  pure state &\cite{Gao2021rp}\\
			$N_{F}$ & $\leq 2$ & $2\otimes 2\otimes2^m$ & \cite{Luo2015ap}\\
			$N_{F}$& $\leq 2$ & $2^{\otimes n}$,  pure state & \cite{Kim2009}\\
			$N_{F}$& $\leq 2$ & $2\otimes2\otimes 3$ & \cite{Gao2021rp}\\
			$N_{F}$ & $\leq 2$& reduced states of $|\widetilde{\W}_{n,2}\ra$ & \cite{Liang2017qip}\\
			$E_{q}$, $2\leq q\leq 3$& $\leq 1$& $2^{\otimes n}$ &\cite{Kim2010pra}\\
			$E_{q}$, $q\in[\frac{5-\sqrt{13}}{2},\frac{5-\sqrt{13}}{2}]$& $\leq 2$& $2\otimes 2\otimes2^m$,
			$2^{\otimes n} $ &\cite{Luo2016pra}\\
			$E_{q}$, $q\in[\frac{5-\sqrt{13}}{2},\frac{5-\sqrt{13}}{2}]$& $>\sqrt{2}$& $2\otimes 2\otimes2$ &\cite{Luo2016pra}\\
			$E_{q}$, $q\in[\frac{5-\sqrt{13}}{2},\frac{5-\sqrt{13}}{2}]$& $\leq 2$& reduced states of $|\widetilde{\W}_{n,d}\ra$ &\cite{Shi2020pra}\\
			$E_{\alpha}$, $\alpha\geq2$& $\leq 1$& $2^{\otimes n}$ &\cite{Kim2010jpa}\\
			$E_{\alpha}$, $\alpha\geq\frac{\sqrt{7}-1}{2}$& $\leq 2$& $2^{\otimes n}$ &\cite{Song}\\
			$E_{\B}$& $\leq 1$& $2^{\otimes n}$ &\cite{Gao2021qip}\\
			$E_{G}$& $\leq 1$& $2^{\otimes n}$ &\cite{Gao2021qip}\\
			$E_N$& $\leq 4\ln2$ & $2^{\otimes n}$,  pure state & \cite{Gao2021rp}\\
			$E_N$& $\leq 4\ln2$ & $2\otimes2\otimes 3$,  $2\otimes2\otimes 2^n$, pure state& \cite{Gao2021rp}\\
			$E_{N_F}$& $\leq 4\ln2$ & $2^{\otimes n}$ & \cite{Gao2021rp}\\
			$E_{N_F}$& $\leq 4\ln2$ & $2\otimes2\otimes 3$,  $2\otimes2\otimes 2^n$& \cite{Gao2021rp}\\
			$E_N$   & $>1$            & $2\otimes2\otimes 2$ & \cite{Wang2020pra}\\
			\br
		\end{tabular}
\end{table}
\normalsize


\subsection{Monogamy of the CHSH inequality}


The monogamy of entanglement can also be exhibited by the the CHSH inequality.
In 2009, Toner proved that the CHSH inequality is
monogamous~\cite{Toner} in light of the earlier results concerning a
link between the security of quantum communication
protocols and violation of Bell's inequalities~\cite{Scarani2001prl} and theory of nonlocal games~\cite{Cleve2004ieee}.
For any $\rho^{ABC}\in\mS^{ABC}$, it was proved that~\cite{Toner}
\bea \label{monogamyofchsh}
\left| \tr\left( {B}^{AB}\rho^{AB}\right) \right| + \left| \tr\left( {B}^{AC}\rho^{AC}\right)\right| \leq 4,
\eea
where ${B}^{AX}=\check{A}_1\ot(\check{X}_1+\check{X}_2)+\check{A}_2\ot (\check{X}_1-\check{X}_2)$ are the CHSH operators, $\check{A}_1$, $\check{A}_2$, $\check{X}_1$, and $\check{X}_2$ are Hermitiaon operators on the corresponding system $A$ and $X$ respectively, with spectrum in $[-1, 1]$, $X=B$ or $C$, $\rho^{AB}=\tr_C\rho^{ABC}$ and $\rho^{AC}=\tr_B\rho^{ABC}$.  
Eq.~\eref{monogamyofchsh} reveals that, if $\rho^{AB}$ violates the CHSH inequality, then $\rho^{AC}$ cannot.
In general, for any $\rho^{AB_1B_2\cdots B_n}$, part $A$ violates the CHSH inequality with at most one of the part $B_i$s~\cite{Toner}.


\section{Tighter monogamy relation of the bipartite entanglement measure}


Since the monogamy property has emerged as the ingredient in the security analysis of quantum key distribution, tighter monogamy relations imply finer characterizations of the quantum correlation distributions, which tightens the security bounds in quantum cryptography~\cite{Cao2024qip,Jin2020cpb,Tao2023m}.  
The aim of this section is to outline the tighter monogamy relations proposed in literature so far.

As we discussed in previous section, the monogamy exponent $\alpha(E)$ of a given bipartite measure $E$ is difficult to calculated in general, even for the multiqubit case. So the monogamy relations as in Eq.~\eref{power} we obtained so far are always not the tightest one. Namely, the inequality we obtained are always 
\bea\label{suppower}
E^\beta( \rho^{A|B_1B_2\cdots B_n}) \geq \sum_{i=1}^nE^\beta( \rho^{AB_i})
\eea
with $\beta>\alpha(E)$, and only for some special system and particular measure we can obtain Eq.~\eref{suppower} with $\beta=\alpha(E)$ (see Tab.~\ref{tab:table3-7}). This leads to some tighter monogamy relation of Eq.~\eref{suppower} by adjusting the items $E^\beta( \rho^{AB_i})$, such as adding some coefficients $x_i>1$ to the some items $E^\beta( \rho^{AB_i})$, or together with adding new items $y$, i.e.,
\bea\label{tighter-suppower}
E^\beta( \rho^{A|B_1B_2\cdots B_n}) \geq\sum_{i=1}^n x_iE^\beta( \rho^{AB_i})+y,
\eea
where $x_i$ and $y$ are function of $\rho^{A|B_1B_2\cdots B_n}$ or some of $\rho^{AB_i}$, in general. It is clear that
Eq.~\eref{tighter-suppower} is tighter than Eq.~\eref{suppower}.
Since Eq.~\eref{suppower} implies 
\bea\label{suppower2}
E^\gamma( \rho^{A|B_1B_2\cdots B_n}) \geq \sum_{i=1}^nE^\gamma( \rho^{AB_i})
\eea
for any $\gamma\geq \beta$, the tighter monogamy relation were always discussed with power $\gamma$ instead of $\beta$ in Eq.~\eref{tighter-suppower}.
In addition, for some special states in some subset of $\mS^{AB_1B_2\cdots B_n}$, one can get other tighter monogamy relation, in contrast with the one in Eq.~\eref{suppower} with $\beta\geq\alpha(E)$.

These coefficients $x_i$ and $y$ are always obtained by some class of inequality which can always be written in the form of
\bea \label{tighter-inquality1}
(1+t)^x\geq 1+\eta(t,x)+\theta(t,x)t^x,\quad 0\leq t\leq 1,~ x\geq 1
\eea
with some functions $\eta(t,x)\geq 0$ and $\theta(t,x)\geq0$,
or in the from of
\bea \label{tighter-inquality3}
(1+t)^x\geq 1+t^x+\varphi(t,x),\quad x\geq 1,~ t\geq 1
\eea
for some $\varphi(t,x)\geq 0$.
For example, for the case of $(1+t)^x \geq 1+xt^x$ (i.e,, $\eta(t,x)=0$ and $\theta(t,x)=x$),
if $E^2({A|BC})\geq E^2({AB})+E^2({AC})$ holds for any $\rho\in\mS^{ABC}$ for some measure $E$, $\gamma\geq 2$, it turns out that
\begin{eqnarray*}\label{method-I}
	E^\gamma(A|BC)&&\geq \left[ E^2(AB)+E^2(AC)\right] ^{\frac{\gamma}{2}}=E^\gamma(AB)\left[1+\frac{E^2(AC)}{E^2(AB)}\right]^{\frac{\gamma}{2}}\nonumber \\
	&& \geq E^\gamma(AB)\left[1+\frac{\gamma}{2}\left(\frac{E^2(AC)}{E^2(AB)}\right)^{\frac{\gamma}{2}}\right]\nonumber\\
	&&=E^\gamma(AB)+\frac{\gamma}{2}E^\gamma(AC).
\end{eqnarray*}
If $\gamma>2$, $E^\gamma(A|BC)\geq E^\gamma(AB)+\frac{\gamma}{2}E^\gamma(AC)$
is tighter than $E^\gamma(A|BC)\geq E^\gamma(AB)+E^\gamma(AC)$. Hereafter, we denote $E(\rho^X)$ by $E(X)$ sometimes for simplicity if one can know the state associated from the context.

On the other hand, for any $0<\zeta<\alpha(E)$, 
\beax 
E^\zeta( \rho^{A|B_1B_2\cdots B_n}) \ngeq E^\zeta( \rho^{AB_1}) +E^\zeta( \rho^{AB_2})+\cdots +E^\zeta( \rho^{AB_n})
\eeax 	
for some $\rho^{A|B_1B_2\cdots B_n}$.
However, with some factors $a_i~\!(a_i<1)$ on some of the items $E^\zeta(\rho^{AB_i})$s, and some $b>0$, we can obtain 
	\bea\label{tighter-subpower}
	E^\zeta( \rho^{A|B_1B_2\cdots B_n}) \geq\sum_i a_iE^\zeta( \rho^{AB_i})+b
	\eea
	holds still.
	But $\alpha(E)$ is unknown in general. So we always derive Eq.~\eref{tighter-subpower} based on Eq.~\eref{suppower} for some $0<\zeta<\beta$ with $\beta\geq \alpha(E)$. Similar to that of $x_i$ got from Eq.~\eref{tighter-suppower}, $a_i$'s are always derived from the inequality with the form of
\bea \label{tighter-inquality2}
(1+t)^x\geq 1+\iota(t,x)+\varsigma(t,x)t^x,\quad 0\leq x\leq 1,~ t\geq 1,
\eea
where $\iota(t,x)\geq0$, $\varsigma(t,x)\geq 0$.

As we see in Tab.~\ref{tab:table3-7}, the monogamy relation as Eq.~\eref{power} so far are only known for some special systems or some particular class of states. Namely, it also very hard to get the power $\beta$ that satisfies Eq.~\eref{power}. So when we investigate the tighter monogamy relation, we always discuss the case of adding some constraints on the states. 
For convenience, we list the constraints on states concerning with the tighter monogamy relation associated  
with bipartite entanglement measure $E$ up to now as follows. We assume $\rho\in\mS^{AB_1B_2\cdots B_n}$ throughout this section unless otherwise specified.
$$
\eqalign{
	\mS^{(1, \delta)}(E)=\left\lbrace \rho\left| E^\delta(AB_i)\geq E^\delta(A|B_{i+1}\cdots B_{n}), 1\leq i\leq n-1 \right. \right\rbrace.\cr
	\mS^{(2,\delta)}(E)=\left\lbrace \rho\in\mS^{A_1B_2\cdots B_n}\left|E^\delta({AB_i})\geq {E^\delta({A|B_{i+1}\cdots B_{n}})},\right. \right.\cr 
	\quad\quad\quad \quad \quad ~ E^\delta({AB_j})\leq E^\delta({A|B_{j+1}\cdots B_n}),  \cr
	\quad\quad\quad \quad \quad ~ \left. 1\leq i\leq m,  m+1\leq j\leq  n-1,
	1\leq m\leq n-2, n\geq 3\right\rbrace.\cr 
	\mS^{(2',\delta)}(E)=\left\lbrace \rho\in\mS^{A_1B_2\cdots B_n}\left|E^\delta({AB_i})\leq {E^\delta({A|B_{i+1}\cdots B_{n}})},\right. \right.\cr 
	\quad\quad\quad \quad \quad ~ E^\delta({AB_j})\geq E^\delta({A|B_{j+1}\cdots B_n}),  \cr
	\quad\quad\quad \quad \quad ~ \left. 1\leq i\leq m,  m+1\leq j\leq  n-1,
	1\leq m\leq n-2, n\geq 3\right\rbrace.\cr 
	\mS^{(3,\delta)}(E)=\left\lbrace \rho\left| E^\delta(AB_j)\geq E^\delta(AB_{j+1})\geq 0,\right. 1\leq j\leq  n-1 \right\rbrace.\cr
	\mS^{(4,\delta)}(E)=\left\lbrace \rho\left| E^\delta(AB_i)\geq \sum_{j=i+1}^{n}E^\delta(AB_j),\right. ~i=1, 2, \cdots, n-1\right\rbrace.\cr
	\mS^{(5,\delta)}(E)= \left\lbrace \rho\left| k E^\delta(AB_i) \geq E^\delta(A|B_{i+1}\cdots B_n),\right. ~~0<k\leq 1, i=1,2,\cdots n-1 \right\rbrace.\cr 
	\mS^{(6,\delta)}(E)=\left\lbrace \rho\left| k E^\delta(AB_i) \geq E^\delta(A|B_{i+1}\cdots B_n),\right. ~~i=1,2,\cdots,m, \right. \cr  
	\quad\quad\quad \quad \quad \left. ~~ E^\delta(AB_j)\leq k E^\delta(A|B_{j+1}\cdots B_n),~~ j=m+1,\cdots,n-1, \right. \cr  
	\quad\quad\quad \quad \quad \left. ~~1\leq m\leq n-2, n\geq 3, 0<k\leq 1\right\rbrace.\cr
	\mS^{(6',\delta)}(E)=\left\lbrace \rho\left| k E^\delta(AB_i) \leq E^\delta(A|B_{i+1}\cdots B_n),\right. ~~i=1,2,\cdots,m, \right. \cr  
	\quad\quad\quad \quad \quad \left. ~~ E^\delta(AB_j)\geq k E^\delta(A|B_{j+1}\cdots B_n),~~ j=m+1,\cdots,n-1, \right. \cr  
	\quad\quad\quad \quad \quad \left. ~~1\leq m\leq n-2, n\geq 3, k\geq 1\right\rbrace.\cr
	\mS^{(7,\delta)}(E)= \left\lbrace \rho\left| k E^\delta(AB_i)\geq E^\delta(AB_{i+1})\geq 0,\right.  ~~i=1,2,\cdots n-1, 0<k\leq 1\right\rbrace.\cr
	\mS^{(8,\delta)}(E)= \left\lbrace \rho\left|  E^\delta(AB_i)\geq k \sum_{j=i+1}^{n}E^\delta(A|B_j),\right.  ~~i=1, 2, \cdots, n-1, k\geq 1\right\rbrace.\cr
	\mS^{(9,\delta)}(E)=\left\lbrace \rho\left|  E^\delta(AB_i) \geq k
	\sum_{l=i+1}^nE^\delta(AB_l), k'E^\delta(AB_j)\leq \sum_{l=j+1}^nE^\delta(AB_l),\right. \right. \cr  
	\quad\quad\quad \quad \quad~~\left. ~~ i=1,2,\cdots,m, j=m+1,\cdots,n-1, \right. \cr  
	\quad\quad\quad \quad \quad \left. ~~~~1\leq m\leq n-2, n\geq 3, k\geq 1, k'\geq 1\right\rbrace.\cr
}
$$

Here, $\delta>0$ corresponds to exponent $\gamma$ in the tighter monogamy relation. Namely, the tighter monogamy relations in literature are not valid universally in general, but they are true on some set $\mS^{(i,\delta)}(E)$ under the associated measure $E$.

In general, for any given tighter monogamy relation, it admits either the form of~Eq.~\eref{tighter-suppower}, or the form of~Eq.~\eref{tighter-subpower}. We thus classify the tighter monogamy relations in literature so far into three classes: (i) supper-power tighter monogamy relation [as in~Eq.~\eref{tighter-suppower}], (ii) sub-power tighter monogamy relation [as in~Eq.~\eref{tighter-subpower}], and (iii) other relations that are not about any of the former two cases.


\subsection{Supper-power tighter monogamy relation}


For more clarity, we list these tighter monogamy relations in the Tables~\ref{tab:table4-1-1}-\ref{tab:table4-1-4}.
We need explain some abbreviations for more convenience. 
We denote by $|\GWV\ra$ a pure state in a
superposition of an $n$-qudit generalized W state and vacuum~\cite{Choi2015pra}, i.e.,
\bea\label{GWV}
|{\GWV}\ra=\sqrt{p}\big|\widetilde{\W}_{n,d} \big\rangle+\sqrt{1-p}|0\cdots 0\rangle,\quad 0\leqslant p \leqslant 1.
\eea
In the following tables we denote by $\rho_{A_{i_1}A_{i_2}\cdots A_{i_{m}}}$ the reduced density matrices of a $|\widetilde{\W}_{n,2}\ra$, $|\widetilde{\W}_{n,d}\ra$ or $|\GWV\ra$, and $P|P_1|\cdots|P_{r}$ denotes some partition of $A_{i_1}A_{i_2}\cdots A_{i_m}$. In table~\ref{tab:table4-1-3}, we 
let $h_1=2^{\frac{\gamma}{2}}-1$, $h_2=\frac{(1+k)^{\frac{\gamma}{2}}-1}{k^{\frac{\gamma}{2}}}$ and $h_3=2^\gamma-1$.

\begin{table}[tp]
	\caption{\label{tab:table4-1-1} {The inequality used in the supper-power tighter monogamy relations in literature.}}
	\footnotesize
		\begin{tabular}{@{}lllll}\br
			SN	&$E$& $\beta$ & Inequality&$\gamma$, System   \\
			\mr
			(1)	&$C$  & $2$    &\eref{tighter-inquality1}: \cite{Jin2017qip} $(1+t)^x\geq 1+xt^x$  & $\geq 2$, $2^{\ot (n+1)}$ \\
			(2)	&  $C$  &$2$ &  \eref{tighter-inquality1}: \cite{Jin2017qip} $(1+t)^x\geq 1+xt^x$                   &            $\geq 2$,       $2^{\ot (n+1)}$     \\
			(3)&$C$    &$2$ &\eref{tighter-inquality1}: \cite{Jin2018pra} $(1+t)^x\geq 1+\left(2^x-1\right)t^x$                  & $\geq 2$,$2^{\ot(n+1)}$ \\
			(4)&  $C$  &$2$ &  \eref{tighter-inquality1}: \cite{Jin2018pra} $(1+t)^x\geq 1+(2^x-1)t^x$                    &            $\geq 2$,       $2^{\ot (n+1)}$        \\
			(5)& $\tau$${\color{red}^{\rm a}}$  &$1$ &    \eref{tighter-inquality1}: \cite{Kim2018pra1} $(1+t)^x\geq 1+xt^x$                &            $\geq 1$,           $2^{\ot (n+1)}$                \\
			(6)& $\tau$${\color{red}^{\rm a}}$  &$1$ &    \eref{tighter-inquality1}: \cite{Kim2018pra1} $(1+t)^x\geq 1+xt^x$                 &            $\geq 1$,           $2^{\ot (n+1)}$                \\
			(7)& $C$   &$2$ &  \eref{tighter-inquality1}: \cite{Jin2019qip}  $(1+t)^x\geq 1+(2^{x}-1)t^x$                &  $\geq 2$,           $2^{\ot(n+1)}$         \\
			(8)& $C$   &$2$ &  \eref{tighter-inquality1}: \cite{Jin2019qip} $(1+t)^x\geq 1+(2^{x}-1)t^x$                 &  $\geq 2$,  $2^{\ot(n+1)}$      \\
			(9)&  $C$  &$2$& \eref{tighter-inquality1}: \cite{Yang2019ctp} $(1+t)^x\geq 1+\frac{(1+k)^x-1}{k^x}t^x$${\color{red}^{\rm, m}}$      &     $\geq 2$,    $2^{\ot (n+1)}$   \\
			(10)&  $C$  &$2$& \eref{tighter-inquality1}: \cite{Yang2019ctp} $(1+t)^x\geq 1+\frac{(1+k)^x-1}{k^x}t^x$${\color{red}^{\rm, m}}$      &    $\geq 2$,   $2^{\ot (n+1)}$     \\
			(11)	&  $E_\alpha$${\color{red}^{\rm b}}$  &$1$&  \eref{tighter-inquality1}:  \cite{Gao2020ctp} $(1+t)^{x}\geq1+\frac{x}{2}t+(2^{x}-\frac{x}{2}-1)t^{x}$      &     $\geq 1$,    $2^{\ot (n+1)}$     \\
			(12)&  $E_\alpha$${\color{red}^{\rm c}}$  &$1$& \eref{tighter-inquality1}:  \cite{Gao2020ctp} $(1+t)^{x}\geq1+\frac{x}{2}t+(2^{x}-\frac{x}{2}-1)t^{x}$      &     $\geq 2$,   $2^{\ot (n+1)}$    \\
			(13)&  $E_\alpha$${\color{red}^{\rm b}}$  &$1$&  \eref{tighter-inquality1}: \cite{Gao2020ctp} $(1+t)^{x}\geq1+\frac{x}{2}t+(2^{x}-\frac{x}{2}-1)t^{x}$      &     $\geq 1$,    $2^{\ot (n+1)}$     \\
			(14)&  $E_\alpha$${\color{red}^{\rm c}}$  &$1$& \eref{tighter-inquality1}: \cite{Gao2020ctp} $(1+t)^{x}\geq1+\frac{x}{2}t+(2^{x}-\frac{x}{2}-1)t^{x}$     &     $\geq 2$,    $2^{\ot (n+1)}$     \\
			(15)& $E_X$${\color{red}^{\rm a,}}$${\color{red}^{\rm d}}$   &$1$ &    \eref{tighter-inquality1}: \cite{Gao2020ijtp} $(1+t)^x\geq 1+\frac{kx}{k+1}t+[(k+1)^x$ &\\
			&&&\quad\qquad\qquad\qquad~$-\left. \left( 1+\frac{x}{k+1}\right) k^x\right] t^x$${\color{red}^{\rm, n}}$           &           $\geq 1$,    $2^{\ot (n+1)}$         \\
			(16) & $E_X$${\color{red}^{\rm a,}}$${\color{red}^{\rm d}}$   &$1$ &    \eref{tighter-inquality1}: \cite{Gao2020ijtp}  $(1+t)^x\geq 1+\frac{kx}{k+1}t+[(k+1)^x$ &\\
			&&&\quad\qquad\qquad\qquad~$-\left. \left( 1+\frac{x}{k+1}\right) k^x\right] t^x$${\color{red}^{\rm, n}}$        &              $\geq 1$,          $2^{\ot (n+1)}$         \\
			(17)& $E$   &$\alpha(E)$ &     \eref{tighter-inquality1}: \cite{Gao2020qip} $(1+t)^x\geq 1+\frac{kx}{k+1}t+[(k+1)^x$ &\\
			&&&\quad\qquad\qquad\qquad~$-\left. \left( 1+\frac{x}{k+1}\right) k^x\right] t^x$${\color{red}^{\rm, n}}$           &         $\geq\alpha(E)$,   $d^{\ot (n+1)}$         \\				
			(18) &$E$  &    $\alpha(E)$           & \eref{tighter-inquality3}: \cite{Gao2020qip} $(1+t)^{x}\geq t^{x}+(m+1)^{x}-m^{x}$, &\\
			&&&\quad\qquad\qquad\qquad~ $t\geq m\geq1$, $x\geq1$                 &       $\geq \alpha(E)$,      $d^{\ot(n+1)}$       \\
			(19)& $E_q$${\color{red}^{\rm e}}$ &$2$ &   \eref{tighter-inquality1}: \cite{Lai2021jpa} $(1+t)^x\geqslant 1+\frac{(1+k)^x -1}{k^x} t^x$${\color{red}^{\rm, m}}$     &   $\geq 2$,                   $|\GWV\ra$          \\
			(20)& $E_q$${\color{red}^{\rm e}}$   &$2$ & \eref{tighter-inquality1}:  \cite{Lai2021jpa} $(1+t)^x\geqslant 1+\frac{(1+k)^x-1}{k^x} t^x$${\color{red}^{\rm, m}}$      &        $\geq 2$,     $|\GWV\ra$             \\
			(21) & $E_q$${\color{red}^{\rm e}}$   &$2$ & \eref{tighter-inquality1}: \cite{Lai2021jpa} $(1+t)^x\geqslant 1+\frac{(1+k)^x-1}{k^x} t^x$${\color{red}^{\rm, m}}$       &        $\geq 2$                    $|\GWV\ra$            \\
			(22)  & $E_q$${\color{red}^{\rm e}}$   &$2$ & \eref{tighter-inquality1}: \cite{Lai2021jpa} $(1+t)^x\geqslant 1+\frac{(1+k)^x-1}{k^x} t^x$${\color{red}^{\rm, m}}$   &     $\geq 2$,              $|\GWV\ra$  \\
			(23)&$E_{q,s}$${\color{red}^{\rm f}}$  & $1$    &\eref{tighter-inquality1}: \cite{Jin2020cpb} $(1+t)^x\geq 1+\left(2^x-1\right)t^x$ & $\geq 1$, $2^{\ot(n+1)}$ \\
			(24)&$E_{q,s}$${\color{red}^{\rm f}}$  & $1$    &\eref{tighter-inquality1}: \cite{Ren2021lpl} $(1+t)^{x}\geq 1+\frac{(1+k^{\delta})^{x}-1}{k^{\delta x}}t^{x}$,&\\
			&&&\quad\qquad\qquad\qquad~ $0<k \leq 1$, $\delta\geq 1$, $0\leq t \leq k^{\delta}$   & $\geq 1$, $2^{\ot(n+1)}$ \\
			(25) &$E_{q,s}$${\color{red}^{\rm f}}$  & $1$    &\eref{tighter-inquality1}: \cite{Ren2021lpl} $(1+t)^{x}\geq 1+\frac{(1+k^{\delta})^{x}-1}{k^{\delta x}}t^{x}$,&\\
			&&&\quad\qquad\qquad\qquad~ $0<k \leq 1$, $\delta\geq 1$, $0\leq t \leq k^{\delta}$  & $\geq 1$, $2^{\ot (n+1)}$ \\	    
			(26) & $E$     &  $\alpha(E)$           & \eref{tighter-inquality1}: \cite{Yang2021qip}  $(1+t)^x\geq1+[(k+1)^x-k^x]t^x$${\color{red}^{\rm, n}}$         &  $\geq \alpha(E)$, $2^{\ot (n+1)}$     \\
			(27)  & $E$     &  $\alpha(E)$           & \eref{tighter-inquality1}: \cite{Yang2021qip} $(1+t)^x\geq1+[(k+1)^x-k^x]t^x$${\color{red}^{\rm, n}}$        &  $\geq \alpha(E)$, $2^{\ot (n+1)}$     \\
			(28)	&  $E_q$${\color{red}^{\rm g}}$  &$1$ & \eref{tighter-inquality1}: \cite{Qi2022ijtp} $(1+t)^{x}\geq 1+\frac{x^2}{x+1}t$ $+(2^{x}-\frac{x^2}{x+1}-1)t^{x}$                   &   $\geq 1$,         $2^{\ot (n+1)}$          \\
			(29)	&  $E_q$${\color{red}^{\rm g}}$  &$1$ & \eref{tighter-inquality1}: \cite{Qi2022ijtp}  $(1+t)^{x}\geq 1+\frac{x^2}{x+1}t$ $+(2^{x}-\frac{x^2}{x+1}-1)t^{x}$                   &   $\geq 1$,         $2^{\ot (n+1)}$          \\
			(30)	&  $C$  &$2$ & \eref{tighter-inquality1}: \cite{Gu2022qip}  $(1+t)^x\geq1+\frac{x}{2}t+d(x,t)$${\color{red}^{\rm h}}$               &         $\geq 2$,   $2^{\ot (n+1)}$           \\
			(31)&  $C$  &$2$ & \eref{tighter-inquality1}: \cite{Gu2022qip} $(1+t)^x\geq1+\frac{x}{2}t+d(x,t)$${\color{red}^{\rm h}}$         &     $\geq 2$,         $2^{\ot (n+1)}$           \\
			(32) &  $E$    &  $\beta$           &    \eref{tighter-inquality1}: \cite{Li2022qip} $(1+t)^x\geq1+x t+d_1(x,k,t)$${\color{red}^{\rm i}}$.        &   $\geq 2\beta$,   $d^{\ot (n+1)}$     \\
			(33)  &  $E$    &  $\beta$           &    \eref{tighter-inquality1}: \cite{Li2022qip} $(1+t)^x\geq1+x t+d_1(x,k,t)$${\color{red}^{\rm i}}$.        &   $\geq 2\beta$,    $d^{\ot (n+1)}$     \\
			(34)	&  $C$  &$2$&  \eref{tighter-inquality1}: \cite{Tao2023m} $(1+t)^x\geq 1+\left(2^x-t^x\right)t^x$, $x\geq 2$    &     $\geq 4$, $2^{\ot (n+1)}$              \\
			(35)	&  $C$  &$2$&  \eref{tighter-inquality1}: \cite{Tao2023m} $(1+t)^x\geq 1+\left(2^x-t^x\right)t^x$, $x\geq 2$    &     $\geq 4$, $2^{\ot (n+1)}$              \\
			(36)& $C$   &$2$ &  \eref{tighter-inquality1}: \cite{Li2023lpl}  $(1+t)^x\geq 1+d_2(x,k,t)$${\color{red}^{\rm j}}$     &    $\geq 4$, $2^{\ot (n+1)}$                 \\
			(37)& $C$   &$2$ &  \eref{tighter-inquality1}: \cite{Li2023lpl} $(1+t)^x\geq 1+d_2(x,k,t)$${\color{red}^{\rm j}}$  &    $\geq 4$,      $2^{\ot (n+1)}$                 \\
			\br
		\end{tabular}\\
		 ${\color{red}^{\rm a}}$ pure state;
		${\color{red}^{\rm b}}$ $\alpha\geq2$;
		 ${\color{red}^{\rm c}}$ $\frac{\sqrt{7}-1}{2}\leq\alpha< 2$;
		 ${\color{red}^{\rm d}}$ $E_X=\{E_{\B}, \check{E}_{G}\}$;
		${\color{red}^{\rm e}}$	$\frac{5-\sqrt{13}}{2}\leq q\leq\frac{5+\sqrt{13}}{2}$;
	${\color{red}^{\rm f}}$ $q\geq2$, $0\leq s \leq1$, $qs\leq3$;
		${\color{red}^{\rm g}}$ $2\leq q\leq 3$;
		${\color{red}^{\rm m}}$ $0\leq t\leq k~,0<k\leq1$;
		${\color{red}^{\rm n}}$ $0\leq t\leq \frac{1}{k}, k\geq1$;
		${\color{red}^{\rm h}}$ $d(x,t)=\frac{(x-1)^2}{4}t^2+\left[ 2^x-\frac{x}{2}+\frac{(x-1)^2}{4}-1\right] t^x-{\frac{(x-1)^2}{2}}t^{x+1}$;
		${\color{red}^{\rm i}}$ $d_1(x,k,t)=[(k+1)^x$ $-x k^{x-1}-k^x]t^x$, $0\leq t\leq \frac{1}{k}, k\geq1$, $x\geq2$;
			${\color{red}^{\rm j}}$ $d_2(x,k,t)=\left[ \frac{(1+k)^x-1}{k^x}+k^x-t^x\right] t^x$,  $0\leq t\leq k~,0<k\leq1$, $x\geq2$.
\end{table}
\normalsize

\begin{sidewaystable}[htbp]
	\vspace{130mm}
	\caption{\label{tab:table4-1-3} Supper-power tighter monogamy relations in literature corresponding to Table~\ref{tab:table4-1-1}. CS is the abbreviation for conditional states. }
		\lineup
		\footnotesize
		\begin{tabular}{@{}lll}\br
			SN	& Tighter monogamy relation  &CS \\
			\mr
			(1)	&\cite{Jin2017qip}: $C^\gamma(A|B_1\cdots B_{n})\geq \sum_{i=1}^{n}(\frac{\gamma}{2})^{i-1}C^\gamma(AB_i)$  & $\mS^{(1,1)}(C)$\\
			(2)	&\cite{Jin2017qip}: $C^\gamma(A|B_1B_2\cdots B_{n})\geq \sum_{i=1}^m (\frac{\gamma}{2})^{i-1}C^\gamma(AB_i)$
			$+(\frac{\gamma}{2})^{m+1}\sum_{j=m+1}^{n-1}C^\gamma(AB_j)+(\frac{\gamma}{2})^{m}C^\gamma(AB_n)$    &$\mS^{(2,1)}(C)$\\
			(3)		&\cite{Jin2018pra}: $C^\gamma(A|B_1\cdots B_{n})\geq\sum_{i=1}^{n} h_1^{i-1}C^\gamma(AB_i)$ &$\mS^{(1,1)}(C)$\\
			(4)&\cite{Jin2018pra}: $C^\gamma(A|B_1B_2\cdots B_{n})\geq \sum_{i=1}^m h_1^{i-1}C^\gamma(AB_i)$
			$+h_1^{m+1}\sum_{j=m+1}^{n-1}C^\gamma(AB_j)+h_1^{m}C^\gamma(AB_n)$ &$\mS^{(2,1)}(C)$\\
			(5)&\cite{Kim2018pra1}: $ \tau^\gamma(A|B_1\cdots B_{n})\geq \sum_{j=1}^{n} \gamma^{w_H(\overrightarrow{j-1})}\tau^\gamma(AB_j)  $            &    $\mS^{(3,1)}(\tau)$       \\
			(6)&\cite{Kim2018pra1}: $ \tau^\gamma(A|B_1\cdots B_{n})\geq \sum_{j=1}^{n} \gamma^{j-1} \tau ^\gamma(AB_j)  $            &    $\mS^{(4,1)}(\tau)$       \\
			(7)&   \cite{Jin2019qip}: $C^\gamma(A|B_1\cdots B_{n})\geq \sum_{j=1}^{n} h_1^{w_H(\overrightarrow{j-1})}C^\gamma(AB_j)$              &   $\mS^{(3,1)}(C)$     \\
			(8)& \cite{Jin2019qip}: $C^\gamma(A|B_1\cdots B_{n})\geq\sum_{j=1}^{n} h_1^{j-1}C^\gamma(AB_j)$            &    $\mS^{(4,2)}(C)$         \\
			(9)	&\cite{Yang2019ctp}: $C^\gamma(A|B_1\cdots B_{n})\geq\sum_{i=1}^{n}h_2^{i-1}C^\gamma(AB_i)$ &$\mS^{(5,2)}(C)$\\ 
			(10)&\cite{Yang2019ctp}: $C^\gamma(A|B_1\cdots B_{n})\geq \sum_{i=1}^m h_2^{i-1}C^\gamma(AB_i)+h_2^{m+1}\sum_{j=m+1}^{n-1}C^\gamma(AB_j)+h_2^m C^\gamma(AB_n)$       & $\mS^{(6,2)}(C)$          \\
			(11)&\cite{Gao2020ctp}: $E_{\alpha}^{\gamma}({A|B_{1}\cdots B_{n}})\geq\sum_{i=1}^{n-2}h_3^{i-1}E_{\alpha}^{\gamma}({AB_{i}})+h_3^{n-2}J(AB)$${\color{red}^{\rm a}}$  &	$\mS^{(1,1)}(C)$\\
			(12)		&\cite{Gao2020ctp}: $E_{\alpha}^{\gamma}({A|B_{1}\cdots B_{n}})\geq\sum_{i=1}^{n-2}h_1^{i-1}E_{\alpha}^{\gamma}({AB_{i}})+h_1^{n-2}Q(AB)$${\color{red}^{\rm b}}$  &	$\mS^{(1,1)}(C)$\\
			(13)	&\cite{Gao2020ctp}: $E_{\alpha}^{\gamma}({A|B_{1}\cdots B_{n}})\geq\sum_{i=1}^{m}h_3^{i-1}E_{\alpha}^{\gamma}({AB_{i}})+h_3^{m+1}\sum_{j=m+1}^{n-2}E_{\alpha}^{\gamma}({AB_j}) + h_3^{m} H(AB)$${\color{red}^{\rm c}}$   &	$\mS^{(2,1)}(C)$\\
			(14)	&\cite{Gao2020ctp}: $E_{\alpha}^{\gamma}({AB_{1}\cdots B_{n}})\geq\sum_{i=1}^{m}h_1^{i-1}E_{\alpha}^{\gamma}({AB_{i}})+h_1^{m+1}\sum_{j=m+1}^{n-2}E_{\alpha}^{\gamma}({AB_j}) + h_1^{m} G(AB)$${\color{red}^{\rm d}}$  &	$\mS^{(2,1)}(C)$\\
			(15)& \cite{Gao2020ijtp}: $E_X^{\gamma}({A|B_{1}\cdots B_{n}}) \geq E_X^{\gamma}({AB_{1}})+(2^{\gamma}-1)E_X^{\gamma}({AB_{2}})+\cdots+[n^{\gamma}-(n-1)^{\gamma}]E_X^{\gamma}({AB_{n}})$   &   $\mS^{(3,1)}(E_X)$       \\
			(16)&\cite{Gao2020ijtp}: $E_X^{\gamma}({A|B_{1}\cdots B_{n}}) \geq  \sum_{i=1}^m h^{i-1}E_X^{\gamma}(AB_i)+h^{m}[(k'+1)^{\gamma}-k'^{\gamma}]\sum_{j=m+1}^{n-2}E_X^{\gamma}({AB_j})+h^{m}P(AB)$${\color{red}^{\rm f}}$        &   $\mS^{(9,1)}(E_X)$          \\			
			(17)&\cite{Gao2020qip}: $E^{\gamma}({A|B_{1}\cdots B_{n}}) \geq \sum_{i=1}^m h^{i-1}E^{\gamma}({AB_i}) +h^{m}[(\eta'+1)^{t}-\eta'^{t}]\sum_{j=m+1}^{n-2}E^{\gamma}({AB_j})+h^{m}N_{AB}$ ${\color{red}^{\rm g}}$, &\\
			&\qquad\qquad \qquad\qquad$h=(\eta+1)^{t}-\eta^{t}$, $t=\frac{\gamma}{\alpha(E)}$, $\eta=k^{\alpha(E)}$, $\eta'=k'^{\alpha(E)}$    &  $\mS^{(9,1)}(E)$        \\
				(18) &\cite{Gao2020qip}: $E^{\gamma}({A|B_{1}\cdots B_{n}}) \geq
			E^{\gamma}({AB_{1}})+(2^{t}-1)E^{\gamma}({AB_{2}})+\cdots
			+[n^{t}-(n-1)^{t}]E^{\gamma}({AB_{n}})$, $t=\frac{\gamma}{\alpha(E)}$   &$\mS^{(3,1)}(E)$           \\	
			(19)& \cite{Lai2021jpa}: $E_q^\gamma(\rho_{P|P_1\cdots P_{r}})\geqslant \sum_{j=1}^{r}h_2^{j-1}E_q^\gamma(\rho_{PP_j})$        &    $\mS^{(5,2)}(E_q)$        \\
			(20)& \cite{Lai2021jpa}: $E_q^\gamma(\rho_{P|P_1\cdots P_{r}})\geqslant \sum_{j=1}^{m}h_2^{j-1}E_q^\gamma(\rho_{PP_j})+h_2^{m+1}\sum_{j=m+1}^{r-1}E_q^\gamma(\rho_{PP_j}) +h_2^{m}E_q^\gamma(\rho_{PP_{r}})$, $1 \leqslant m\leqslant r-2$, $r\geqslant3$  &     $\mS^{(6,2)}(E_q)$           \\
			(21)& \cite{Lai2021jpa}: $E_q^\gamma(\rho_{P|P_1\cdots P_{r}})\geqslant\sum\limits_{j=1}^{r} h_2^{\omega_H(\overrightarrow{j-1})}E_q^\gamma(\rho_{ PP_j})$ &      $\mS^{(7,2)}(E_q)$     \\
			(22)& \cite{Lai2021jpa}: $E_q^\gamma(\rho_{P|P_1\cdots P_{r}})\geqslant\sum\limits_{j=1}^{r} h_2^{j-1} E_q^\gamma(\rho_{ PP_j})$       &      $\mS^{(8,2)}(E_q)$     \\ 
			(23)& \cite{Jin2020cpb}: $E_{q, s}^\gamma({A|B_1\ldots B_{n}})
			\geq\sum\limits_{j=1}^{n} h_3^{j-1}E_{q, s}^\gamma({AB_j})$          &     $\mS^{(4,1)}(E_{q,s})$       \\
				(24)& \cite{Ren2021lpl}: $E_{q, s}^\gamma({A|B_1\ldots B_{n}})
			\geq\sum\limits_{j=1}^{n}\left[ \frac{(1+k^\delta)^\gamma-1}{k^{\delta\gamma}}\right] ^{\omega_H(\overrightarrow{j-1})}E_{q, s}^\gamma({AB_j})$          &     $\mS^{(7,1)}(E_{q,s})$       \\
			(25)& \cite{Ren2021lpl}: $E_{q, s}^\gamma({A|B_1\ldots B_{n}})
			\geq\sum\limits_{j=1}^{n}\left[ \frac{(1+k^\delta)^\gamma-1}{k^{\delta\gamma}}\right] ^{j-1} E_{q, s}^\gamma({AB_j})$          &     $\mS^{(8,1)}(E_{q,s})$       \\
			(26)& \cite{Yang2021qip}: $E^\gamma(A|B_1\cdots{}B_{n})\geq\sum_{j=1}^{n} [(k+1)^t-k^t]^{\omega_H(\overrightarrow{j-1})} E^\gamma({AB_j})$, ~$t=\gamma/{\alpha(E)}$          &      $\mS^{(7,1)}(E)$        \\
			\br
		\end{tabular}\\
\end{sidewaystable}
\normalsize

\begin{sidewaystable}[htbp]
	\vspace{130mm}
	\caption{\label{tab:table4-1-4} Supper-power tighter monogamy relations in literature corresponding to Table~\ref{tab:table4-1-1}. }
		\footnotesize
		\begin{tabular}{@{}lll}\br
			SN	& Tighter monogamy relation  &CS \\
			\mr
			(27)& \cite{Yang2021qip}: $E^\gamma(A|B_1\cdots{}B_{n})\geq\sum_{j=1}^{n} [(k+1)^t-k^t]^j E^\gamma({AB_j})$,~ $t=\gamma/{\alpha(E)}$          &      $\mS^{(8,1)}(E)$        \\
			(28)	&\cite{Qi2022ijtp}: $E_{q}^{\gamma}({A|B_{1}\cdots B_{n}})  \geq \sum_{i=1}^{n-2}h_3^{i-1}E_{q}^{\gamma}({AB_i})+h_3^{n-2}T(AB)$${\color{red}^{\rm h}}$,& $\mS^{(1,1)}(C)$\\ 
			(29) &\cite{Qi2022ijtp}: $E_{q}^{\gamma}({A|B_{1}\cdots B_{n}}) \geq \sum_{i=1}^{m}h_3^{i-1}E_{q}^{\gamma}({AB_i})+h_3^{m+1}\sum_{j=m+1}^{n-2}E_{q}^{\gamma}({AB_j})+h_3^mK(AB)$${\color{red}^{\rm k}}$	& $\mS^{(2,1)}(C)$\\
			(30)	&\cite{Gu2022qip}:	$C^\gamma(AB_1\cdots B_{n})\geq\sum\limits_{i=1}^{n-1}h_1^{i-1}[C^\gamma(AB_i)+L(AB)$${\color{red}^{\rm l}}]+h_1^{n-1} C^\gamma(AB_{n})$ &$\mS^{(1,1)}(C)$\\
			(31)	& \cite{Gu2022qip}: $C^\gamma(A|B_1\cdots B_{n})\geq
			\sum\limits_{i=1}^{m}h_1^{i-1}[C^\gamma(AB_i)+L(AB){\color{red}^{\rm l}}]+h_1^m\sum\limits_{j=m+1}^{n-1}[h_1C^\gamma(AB_j)+R(AB)]{\color{red}^{\rm m}}+h_1^m C^\gamma(AB_{n})$  &$\mS^{(2,1)}(C)$ \\
			(32)& \cite{Li2022qip}: $E^\gamma({A|B_1\cdots B_{n}})\geq\sum\limits_{i=1}^{n-1} \Bigg\{h^{i-1}\Bigg[E^\gamma({AB_i})+\mu E^{\gamma-\beta}({AB_i})\Bigg(\sum\limits_{j=i+1}^{n}E^{\beta}({AB_j})\Bigg)\Bigg]\Bigg\}$&\\
			&\quad \quad \quad \quad \quad \quad \quad \quad \qquad~~~
			$+h^{n-1} E^\gamma({AB_{n}})$,
			$\gamma\geq 2\beta$, $\mu =\frac{\gamma}{\beta}~(\geq2)$, $h=(k+1)^\mu-\mu k^{\mu-1}-k^\mu$        &      $\mS^{(8,\beta)}(E)$       \\
		(33)&\cite{Li2022qip}: $E^\gamma({A|B_1\cdots B_{n}})\geq\sum\limits_{i=1}^{m} \Bigg\{h^{i-1}\Bigg[E^\gamma({AB_i})+\mu E^{\gamma-\beta}({AB_i})\Bigg(\sum\limits_{l=i+1}^{n}E^{\beta}({AB_l})\Bigg)\Bigg]\Bigg\}$&\\
		&\quad \quad  \quad \quad \quad \quad  \quad \quad \quad \quad \quad
		$+h^{m}[(k'+1)^\mu-k'^\mu][E^{\gamma}({AB_{m+1}})+\cdots+E^{\gamma}({AB_{n-2}})]$&\\
		&\quad \quad \quad \quad \quad \quad  \quad \quad \quad \quad \quad
		$h^{m} {\Bigg\{[(k'+1)^\mu-\mu k'^{\mu-1}-k'^\mu]E^\gamma({AB_{n-1}})+\mu E^{\beta}({AB_{n-1}})E^{\gamma-\beta}({AB_{n}})+E^{\gamma}({AB_{n}})\Bigg\}}$ &\\
		&\quad \quad \quad\quad \quad \quad \quad  \quad \quad \quad \quad \quad
		$\gamma\geq 2\beta$, $\mu =\frac{\gamma}{\beta}~(\geq2)$, $h=(k+1)^\mu-\mu k^{\mu-1}-k^\mu$       &   $\mS^{(9,\beta)}(E)$           \\
		(34)& \cite{Tao2023m}: $C^{\gamma}(A|B_{1}\cdots B_{n})
		\geq \sum_{i=1}^{n-1}\left(\prod_{j=1}^{i}M_{j}\right)C^{\gamma}(AB_{i})+(\prod_{i=1}^{n-1}M_{i})C^{\gamma}(AB_{n})$&\\
		&\quad \quad \quad $M_{1}=1$, $ M_{i+1}=2^{\frac{\gamma}{2}}-\frac{C^{\gamma}(A|B_{i+1}\cdots B_{n})}{C^{\gamma}(AB_{i})}$, $i=2,\cdots,n-1$  &$\mS^{(1,1)}(C)$\\
		(35)	&\cite{Tao2023m}: $C^{\gamma}(A|B_{1}\cdots B_{n})
		\geq\sum_{i=1}^{m}(\prod_{j=1}^{i}M_{j})C^{\gamma}(AB_{i})+(\prod_{i=1}^{m+1}M_{i})[\sum_{j=m+1}^{n-1}S_{j}$${\color{red}^{\rm n}}$$C^{\gamma}(AB_{j})+C^{\gamma}AB_{n}]$   &\\
		&\quad \quad \quad $M_{1}=1$, $M_{i+1}=2^{\frac{\gamma}{2}}-\frac{C^{\gamma}(A|B_{i+1}\cdots B_{n})}{C^{\gamma}(AB_{i})}$, $i=1,2,\cdots,m$     &$\mS^{(2,1)}(C)$ \\
		(36)& \cite{Li2023lpl}: $C^\gamma(A|B_{i+1}\cdots B_{n}) \geq \sum\limits_{i=0}^{n-1} \Big( \prod\limits_{j=0}^i \mathcal{M}_j \Big) C^{\gamma}(AB_{i+1})$&\\
		&\quad \quad\quad $M=\frac{(1+k)^{\frac{\gamma}{2}}-1}{k^{\frac{\gamma}{2}}} + k^{\frac{\gamma}{2}}$, $\mathcal{M}_0=1$, $\mathcal{M}_j=M- \frac{C^\gamma(A|B_{j+1}\cdots B_{n})}{C^\gamma(AB_j)}$, $j=1,2, \cdots, n-1$         &      $\mS^{(5,2)}(C)$     \\	
		(37)& \cite{Li2023lpl}: $C ^{\gamma}(A|B_1\cdots B_{n}) \geq \sum\limits_{i=0}^{m-1} \Big( \prod\limits_{j=0}^i \mathcal{M}_j \Big) C^{\gamma}(AB_{i+1}) +\Big( \prod\limits_{i=1}^{m} \mathcal{M}_i \Big) \left[  \prod\limits_{j=m+1}^{n-1} \mathcal{Q}_j C^\gamma(AB_j) + C^{\gamma}(AB_{n}) \right] $&\\
		&\quad \quad \quad $M=\frac{(1+k)^{\frac{\gamma}{2}}-1}{k^{\frac{\gamma}{2}}} + k^{\frac{\gamma}{2}}$, $\mathcal{M}_0=1$,
		$\mathcal{M}_i=M- \frac{C^\gamma(A|B_{i+1}\cdots B_{n})}{C^\gamma(AB_i)}$,  &\\ & \quad \quad \quad $i=1,2, \cdots, m$, $\mathcal{Q}_j=M-\frac{C^\gamma(AB_j)}{C ^\gamma(A|B_{j+1}\cdots B_{n})} $, $j=m+1, \cdots , n-1$       &      $\mS^{(6,2)}(C)$   \\  		
		\br
		\end{tabular}\\		
\end{sidewaystable}
\normalsize


\subsection{Sub-power tighter monogamy relation}


For more clarity, we list these tighter monogamy relations in the Tables~\ref{tab:table4-2-1}-\ref{tab:table4-2-2}.

We fix some notations in Tables~\ref{tab:table4-2-1}-\ref{tab:table4-2-2} for simplicity:
$$
\eqalign{
	{\color{red}^{\rm a}} J(AB)=E_{\alpha}^{\gamma}({AB_{n-1}})+\frac{\gamma}{2}E_{\alpha}^{\gamma-1}({AB_{n-1}})E_{\alpha}({AB_{n}})
	+\left( 2^{\gamma}-\frac{\gamma}{2}-1\right) E_{\alpha}^{\gamma}({AB_{n}}),\cr
	{\color{red}^{\rm b}} Q(AB)=E_{\alpha}^{\gamma}({AB_{n-1}})+\frac{\gamma}{4}E_{\alpha}^{2}({AB_{n-1}})E_{\alpha}^{\gamma-2}({AB_{n}})
	+\left( 2^{\frac{\gamma}{2}}-\frac{\gamma}{4}-1\right) E_{\alpha}^{\gamma}({AB_{n}}),\cr
		{\color{red}^{\rm 
		c}} H(AB)=\left( 2^{\gamma}-\frac{\gamma}{2}-1\right) [E_{\alpha}({AB_{n-1}})]^{\gamma}+
	\frac{\gamma}{2}E_{\alpha}({AB_{n-1}})E_{\alpha}^{\gamma-1}({AB_{n}}) +[E_{\alpha}({AB_{n}})]^{\gamma},\cr
	{\color{red}^{\rm 
		d}} G(AB)=\left( 2^{\frac{\gamma}{2}}-\frac{\gamma}{4}-1\right) [E_{\alpha}({AB_{n-1}})]^{\gamma}+
	\frac{\gamma}{4}E_{\alpha}^2({AB_{n-1}})E_{\alpha}^{\gamma-2}({AB_{n}}) +[E_{\alpha}({AB_{n}})]^{\gamma},\cr
	{\color{red}^{\rm 
		f}} P(AB)=
	\left[(k'+1)^{\gamma}-\left(1+\frac{\gamma}{k'+1}\right) k'^{\gamma}\right]E_X^{\gamma}({AB_{n-1}})+\frac{k' \gamma}{k'+1}E_X({AB_{n-1}})\cr
	\quad\qquad ~~~~~\cdot E_X^{\gamma-1}({AB_{n}})+E_X^{\gamma}({AB_{n}}), ~~ h=(k+1)^{\gamma}-k^{\gamma}, k\geq1,~k'\geq1,\cr
	{\color{red}^{\rm 
		g}} N_{AB}=\left[ (\eta'+1)^{t}-\left( 1+\frac{t}{\eta'+1}\right) \eta'^{t}\right] E^{\gamma}({AB_{n-1}})+\frac{\eta' t}{\eta'+1}E^{\alpha(E)}({AB_{n-1}}) \cr
		\qquad ~~~~~ \cdot E^{\gamma-\alpha(E)}({AB_{n}})+E^{\gamma}({AB_{n}}),\cr
		{{\color{red}^{\rm h}}} T(AB)=E_{q}^{\gamma}({AB_{n-1}})+\frac{\gamma^2}{\gamma+1}E_{q}^{\gamma-1}({AB_{n-1}})E_{q}({AB_{n}})+\left( 2^{\gamma}-\frac{\gamma^2}{\gamma+1}-1\right) E_{q}^{\gamma}({AB_{n}}),\cr
			{\color{red}^{\rm 
				k}} K(AB)=E_{q}^{\gamma}({AB_{n}})+\frac{\gamma^2}{\gamma+1}E_{q}^{\gamma-1}({AB_{n}})E_{q}({AB_{n-1}})+\left( 2^{\gamma}-\frac{\gamma^2}{\gamma+1}-1\right) E_{q}^{\gamma}({AB_{n-1}}), \cr 
				{\color{red}^{\rm 
					l}} L_{AB}=\frac{\gamma}{4}C^2(A|B_{i+1}\ldots B_{n})[ C^{\gamma-2}(AB_i)-C^{\gamma-2}(A|B_{i+1}\ldots B_{n})]+\cr
				\quad\qquad	\frac{(\gamma-2)^2}{16}C^4(A|B_{i+1}\ldots B_{n})[C^{\gamma-4}(AB_i)+C^{\gamma-4}(A|B_{i+1}\ldots B_{n})\cr
				\quad \qquad-2C^{\gamma-2}(A|B_{i+1}\ldots B_{n})C^{-2}(AB_i)],
}
$$	
$$
\eqalign{
	{{\color{red}^{\rm h}}}	
		{\color{red}^{\rm 
			m}} R(AB)=\frac{\gamma}{4}C^2(AB_j)[C^{\gamma-2}(A|B_{j+1}\ldots B_{n})-C^{\gamma-2}(AB_j)]\cr
		\quad~~\qquad \quad+\frac{(\gamma-2)^2}{16}C^4(AB_j)[C^{\gamma-4}(A|B_{j+1}\ldots B_{n})+C^{\gamma-4}(AB_j)\cr
		\quad ~~\quad \qquad-2C^{\gamma-2}(AB_j)C^{-2}(A|B_{j+1}\ldots B_{n})],\cr
		{\color{red}^{\rm 
			n}}  S_j=2^{\frac{\gamma}{2}}-\frac{C^{\gamma}(AB_{j})}{C^{\gamma}(A|B_{j+1}\cdots B_{n})},~~ j= m + 1,\cdots,n-1.
}
$$

	\begin{table}[htbp]
	\caption{\label{tab:table4-2-1} {The inequality used in the sub-power tighter monogamy relations in literature.}}
		\footnotesize
\hspace{5mm}		\begin{tabular}{@{}lllll}\br
		SN	&$E$& $\beta$ & Inequality&$\zeta$, System   \\
		\mr
		(1) &  $C$    &   $2$  &\eref{tighter-inquality2}: \cite{Zhu2019qip} $(1+t)^x\geq 1+\left(2^x-1\right)t^x$    &     $\leq \beta$,      $2^{\ot (n+1)}$      \\
		(2) &  $C_a$    &   $2$  &\eref{tighter-inquality2}: \cite{Shi2019arxiv} $(1+t)^x\geq 1+\left(2^x-1\right)t^x$    &     $\leq \beta$,     $|\widetilde{\W}_{n,d}\ra$     \\
		(3) &      $C$    &   $2$          &  \eref{tighter-inquality2}: \cite{Jin2020qip} $(1+t)^x\geq 1+\frac{(1+k)^x-1}{k^x}t^x$,  $t\geq k\geq 1$          &  $\leq \beta$,      $2^{\ot (n+1)}$    \\
		(4) &  $E_q$${\color{red}^{\rm e}}$    &     $2$        &  \eref{tighter-inquality2}: \cite{Shi2020pra} $(1+t)^x\geq1+(2^x-1)t^x$        &  $\leq\beta$,  $|\widetilde{\W}_{n,d}\ra$      \\
		(5) &  $E_q$${\color{red}^{\rm e}}$    &   $2$          &  \eref{tighter-inquality2}: \cite{Lai2021jpa} $(1+t)^x\geqslant 1+\frac{(1+k)^x-1}{k^x} t^x$, $t\geqslant k\geqslant 1$    &   $\leq \beta$,  $|\GWV\ra$      \\		
		(6) &  $E_X$${\color{red}^{\rm d}}$    &   $1$          & \eref{tighter-inquality2}: \cite{Zhang2022lpl} $(1+t)^x\geq(\frac{1}{2})^x+\frac{(1+k^\omega)^x-(\frac{1}{2})^x}{k^{\omega x}}t^x$ & \\ 
		& & & \quad \quad \quad \quad\quad 
		$t\geq k^\omega\geq k\geq1$, $\omega\geq1$, $0\leq x\leq\frac{1}{2}$        &  $\leq \frac{\beta}{2}$,    $2^{\ot (n+1)}$     \\
		(7) &  $E_{X}$${\color{red}^{\rm d}}$    &   $1$          & \eref{tighter-inquality2}: \cite{Zhang2022lpl} $(1+t)^x\geq(\frac{1}{2})^x+\frac{(1+k^\omega)^x-(\frac{1}{2})^x}{k^{\omega x}}t^x$ & \\ 
		& & & \quad \quad \quad \quad\quad 
		$t\geq k^\omega\geq k\geq1$, $\omega\geq1$, $0\leq x\leq\frac{1}{2}$        &  $\leq \frac{\beta}{2}$,   $2^{\ot (n+1)}$     \\
		(8) &  $E_f$    &   $\sqrt{2}$  &\eref{tighter-inquality2}: \cite{Zhang2022qip} $(1+t)^x\geq(\frac{1}{2})^x+\frac{(1+k)^x-(\frac{1}{2})^x}{k^x}t^x$& \\ 
		& & & \quad \quad \quad \quad\quad  $t\geq k\geq 1$, $0\leq x\leq\frac{1}{2}$          &    $\leq\frac{\beta}{2} $,    $2^{\ot (n+1)}$     \\
		(9) &  $E_f$    &    $\sqrt{2}$   &  \eref{tighter-inquality2}: \cite{Zhang2022qip} $(1+t)^x\geq(\frac{1}{2})^x+\frac{(1+k)^x-(\frac{1}{2})^x}{k^x}t^x$, \\ 
		& & & \quad \quad \quad \quad\quad $t\geq k\geq 1$, $0\leq x\leq\frac{1}{2}$          &    $\leq \frac{\beta}{2}$,   $2^{\ot (n+1)}$       \\
		(10) &  $C$    &   $2$          &  \eref{tighter-inquality2}:  \cite{Zhang2022ijtp}$(1+t)^x\geq p^x+\frac{(1+k)^x-p^x}{k^x}t^x$  \\ 
		& & & \quad \quad \quad \quad\quad   $t\geq k\geq 1$, $\frac{1}{2}\leq p\leq1$, $0\leq x\leq\frac{1}{2}$         &       $\leq\frac{\beta}{2} $,    $2^{\ot (n+1)}$        \\
		(11) & $C$     &    $2$         & \eref{tighter-inquality2}: \cite{Zhang2022ijtp} $(1+t)^x\geq p^x+\frac{(1+k)^x-p^x}{k^x}t^x$,  \\ 
		& & & \quad \quad \quad \quad\quad $t\geq k\geq 1$, $\frac{1}{2}\leq p\leq1$, $0\leq x\leq\frac{1}{2}$          &    $\leq \frac{\beta}{2}$,   $2^{\ot (n+1)}$          \\
		(12) &  $E$    &  $\geq2$           &  \eref{tighter-inquality2}: \cite{Shen2023epjp} $(1+t)^x\geq q^{x-1}t^x+(1+k)^x-q^{x-1}k^x$,  \\ 
		& & & \quad \quad \quad \quad\quad $t\geq k\geq 1$, $1+\frac{1}{t}\leq q\leq 1+\frac{1}{k}$           &   $\leq \beta$, $2^{\ot (n+1)}$        \\
		(13) &  $E$    &     $\geq2$          &  \eref{tighter-inquality2}:  \cite{Cao2024lpl} $(1+t)^{x}\geq\left(1+\frac{a}{s}\right)^{x-1}+\left(1+\frac{s}{a}\right)^{x-1}t^{x}$, \\ 
		& & & \quad \quad \quad \quad\quad  $t\geq a\geq 1$, $s>0$         &        $\leq \beta$, $2^{\ot(n+1)}$         \\
		(14) &     $E$     &    $\geq 2$         & \eref{tighter-inquality2}: 	\cite{Zhang2023ps} $(1+t)^x\geq (1+a)^{x-1}+(1+\frac{1}{a})^{x-1}t^x$,      \\ 
		& & & \quad \quad \quad \quad\quad  $t\geq a\geq 1$   &     $\leq \beta$ ,   $d^{\ot (n+1)}$         \\
		\br
		\end{tabular} \\
	\indent\hspace{5mm}	${\color{red}^{\rm d}}$ $E_X\in \{E_{\B}, \check{E}_{G}\}$;   ${\color{red}^{\rm e}}$ $\frac{5-\sqrt{13}}{2}\leq q\leq\frac{5+\sqrt{13}}{2}$.
\end{table}
\normalsize

In Table~\ref{tab:table4-2-2}, we let $l_1=2^{\frac{\zeta}{\beta}}-1$, $l_2=\frac{(1+k)^\frac{\zeta}{\beta}-1}{k^\frac{\zeta}{\beta}}$, $l_3=\frac{(1+k^\omega)^{\frac{\zeta}{\beta}}-(\frac{1}{2})^{\frac{\zeta}{\beta}}}{k^{\frac{\omega\zeta}{\beta}}}$, $l_4=\frac{(1+k)^{\frac{\zeta}{\beta}}-(\frac{1}{2})^{\frac{\zeta}{\beta}}}{k^{\frac{\zeta}{\beta}}}$ and $l_5=\frac{(1+k)^{\frac{\zeta}{\beta}}-p^{\frac{\zeta}{\beta}}}{k^{\frac{\zeta}{\beta}}}$.

\begin{sidewaystable}[htbp]
	\vspace{130mm}
	\caption{\label{tab:table4-2-2} Sub-power tighter monogamy relations in literature corresponding to Table~\ref{tab:table4-2-1}. }
	\footnotesize
		\begin{tabular}{@{}lll}\br
			SN	& Tighter monogamy relation  &CS \\
			\mr
			(1)&\cite{Zhu2019qip}: $C^{\zeta}({A|B_1B_2...B_{n}})\geq
			\sum_{i=1}^{m}l_1^{i-1}C^{\zeta}({AB_i})+l_1^{m+1}\sum_{i=m+1}^{n-1}C^{\zeta}({AB_i})+l_1^{m}C^{\zeta}({AB_{n}})$         &    $\mS^{(2',1)}(C)$         \\
			(2)&\cite{Shi2019arxiv}: $C_a^{\zeta}(P|P_1P_2\cdots P_{r})\geq \sum_{i=1}^m l_1^{i-1}C_a^{\zeta}(PP_i)+l_1^{m+1}\sum_{i=m+1}^{r-1}C_a^{\zeta}(PP_i)+l_1^{m}C_a^{\zeta}(P_1P_{r})$     &     $\mS^{(2',1)}(C_a)$        \\
			(3)&\cite{Jin2020qip}: $C^\zeta(A|B_1B_2\cdots B_{n})\geq \sum_{i=1}^m
			l_2^{i-1}C^\zeta(AB_i)+l_2^{m+1}\sum_{j=m+1}^{n-1}C^\zeta(AB_j)
			+l_2^{m}C^\zeta(AB_{n})$      &       $\mS^{(6',\beta)}(C)$       \\
			(4)&\cite{Shi2020pra}: $E_q^\zeta({A|B_1B_2\cdots B_n})\geqslant \sum_{j=1}^m l_1^{j-1}E_q^\zeta(\rho_{AB_j})+l_1^{m+1}\sum_{j=m+1}^{n-1}E_q^\zeta(\rho_{AB_j})+l_1^{m}E_q^\zeta(\rho_{AB_n})$  &   $\mS^{(2',1)}(E_q)$           \\
			(5)&\cite{Lai2021jpa}: $E_q^\zeta({P|P_1P_2\cdots P_{r}})\geqslant \sum_{j=1}^{m} l_2^{j-1}E_q^\zeta(\rho_{PP_j})+l_2^{m+1}\sum_{j=m+1}^{r-1}E_q^\zeta(\rho_{PP_j})+l_2^mE_q^\zeta(\rho_{PP_{r}})$        &   $\mS^{(6',\beta)}(E_q)$          \\
			(6)&\cite{Zhang2022lpl}: $E_X^\zeta(A|B_1B_2\cdots B_n)\geq l_3\left[ E_X^\zeta(AB_1)+(\frac{1}{2})^{\frac{\zeta}{\beta}}E_X^\zeta(AB_2)+\cdots +(\frac{1}{2})^{\frac{(n-2)\zeta}{\beta}}E_X^\zeta(AB_{n-1})\right] +(\frac{1}{2})^{\frac{(n-1)\zeta}{\beta}}E_X^\zeta(AB_n)$, $0\leq\zeta\leq\frac{\beta}{2}$, $\beta\geq1$    &    $\mS^{(5,\beta)}(E_X)$        \\
			(7)& \cite{Zhang2022lpl}: $E_X^\zeta(A|B_1B_2\cdots B_n)\geq 	(\frac{1}{2})^{\frac{\zeta}{\beta}}\left[ \sum_{i=1}^{m}l_3^{i-1}E_X^\zeta(AB_i) \right] +l_3^{m+1}\left[ E_X^\zeta(AB_{m+1})+(\frac{1}{2})^{\frac{\zeta}{\beta}}E_X^\zeta(AB_{m+2})+\cdots\right. $&\\
			&\quad \quad \quad \quad \quad \quad \quad \quad \quad\quad \quad \quad ~~~
			$\left. +(\frac{1}{2})^{\frac{(n-m-2)\zeta}{\beta}}E_X^\zeta(AB_{n-1})\right] +l_3^{m}(\frac{1}{2})^{\frac{(n-m-1)\zeta}{\beta}}E_X^\zeta(AB_n)$   &    $\mS^{(6',\beta)}(E_X)$        \\
			(8)&\cite{Zhang2022qip}: $E_f^\zeta(A|B_1\cdots B_n)\geq l_4\left[ E_f^\zeta(AB_1)+(\frac{1}{2})^{\frac{\zeta}{\beta}}E_f^\zeta(AB_2)+\cdots+(\frac{1}{2})^{\frac{(n-2)\zeta}{\beta}}E_f^\zeta(AB_{n-1})\right] +(\frac{1}{2})^{\frac{(n-1)\zeta}{\beta}}E_f^\zeta(AB_n)$       &    $\mS^{(5,\beta)}(E_f)$        \\
			(9)&\cite{Zhang2022qip}: $E_f^\zeta(A|B_1\cdots B_n)\geq\left(\frac{1}{2}\right)^{\frac{\zeta}{\beta}}\sum_{i=1}^m l_4^{i-1}E_f^\zeta(AB_i)+l_4^{m+1}\left[ E_f^\zeta(AB_{m+1})+(\frac{1}{2})^{\frac{\zeta}{\beta}}E_f^\zeta(AB_{m+2})\right. $&\\
			&\quad \quad\quad \quad \quad \quad \quad\quad\quad \quad\quad\quad
			$\left. +\cdots+(\frac{1}{2})^{\frac{(n-m-2)\zeta}{\beta}}E_f^\zeta(AB_{n-1})\right] +l_4^{m}(\frac{1}{2})^{\frac{(n-m-1)\zeta}{\beta}}E_f^\zeta(AB_n)$        &   $\mS^{(6',\beta)}(E_f)$          \\
			(10)&\cite{Zhang2022ijtp}: $C^\zeta(A|B_1\cdots B_n)\geq l_5(C^\zeta(AB_1)+p^{\frac{\zeta}{\beta}}C^\zeta(AB_2)+\cdots+p^{\frac{(n-2)\zeta}{\beta}}C^\zeta(AB_{n-1}))+p^{\frac{(n-1)\zeta}{\beta}}C^\zeta(AB_n)$    &      $\mS^{(5,\beta)}(C)$        \\
			(11)& \cite{Zhang2022ijtp}: $C^\zeta(A|B_1\cdots B_n)\geq p^{\frac{\zeta}{\beta}}\sum_{i=1}^m l_5^{i-1}C^\zeta(AB_i)+l_5^{m+1}\left[ C^\zeta(AB_{m+1})+p^{\frac{\zeta}{\beta}}C^\zeta(AB_{m+2})\right. $&\\
			&\quad \quad  ~~~
			$\left. +\cdots+p^{\frac{(n-m-2)\zeta}{\beta}}C^\zeta(AB_{n-1})\right] 
			+l_5^{m}p^{\frac{(n-m-1)\zeta}{\beta}}C^\zeta(AB_n)$ &     $\mS^{(6',\beta)}(C)$        \\
			(12)&\cite{Shen2023epjp}: $E^\zeta(A|B_1\cdots B_{n})\geq s_{1}E^\zeta(AB_1)+
			\sum\limits_{i=2}^{m}\prod\limits_{k=1}^{i-1}q_{k}^{\frac{\zeta}{\beta}-1}s_{i}E^\zeta(AB_i)+(q_1\cdots q_{m+1})^{\frac{\zeta}{\beta}-1}
			E^\zeta(AB_{m+1})$&\\
			&\quad \quad \quad \quad  ~~~
			$+(q_1\cdots q_{m})^{\frac{\zeta}{\beta}-1}\left[ \sum\limits_{j=m+2}^{n-1}\prod\limits_{k=m+1}^{j-1}s_{k}q_{j}^{\frac{\zeta}{\beta}-1}E^\zeta(AB_{j})\right] +(q_1\cdots q_{m})^{\frac{\zeta}{\beta}-1}s_{m+1}\cdots s_{n-1}E^\zeta(AB_{n})$ &\\
			& 
			\quad \quad \quad $s_{r}=(1+t_{r})^\frac{\zeta}{\beta}
			-q_{r}^{\frac{\zeta}{\beta}-1}t_{r}^\frac{\zeta}{\beta}$, $t_r\geq1$, $1\le r\le n-1$, $1\leq m\leq n-2$, $n\geq3$,  $1+\frac{E^\beta(AB_i)}{E^\beta(A|B_{i+1}\cdots B_{n})}\leq q_{i}\leq1+\frac{1}{k}$, $i=1,2,\cdots,m$,  &\\
			&    \quad \quad \quad 	$1+\frac{E^\beta(A|B_{j+1}\cdots B_{n})}{E^\beta(AB_j)}\leq q_{j}\leq1+\frac{1}{k}$, $j=m+1,
			\cdots,n-1$  &   $\mS^{(6',\beta)}(E)$           \\
			(13)&\cite{Cao2024lpl}: $E^{\zeta}(A|B_{1}\cdots B_{n})\geq\left(1+\frac{a}{s}\right)^{\frac{\zeta}{\beta}-1}\sum_{i=1}^{n}\left[ \left(1+\frac{s}{a}\right)^{\frac{\zeta}{\beta}-1}\right]^{n-i}E^{\zeta}(AB_i)$          &     $\mS^{(7,\beta)}(E)$            \\
			(14)& \cite{Zhang2023ps}: $E^{\zeta}(A|B_1...B_n)\geq(1+{a})^{\frac{\zeta}{\beta}-1}\sum_{i=1}^{n} \left[ (1+\frac{1}{a})^{\frac{\zeta}{\beta}-1}\right] ^{n-i}E^{{\zeta}}(AB_i)$           &   $\mS^{(7,\beta)}(E)$         \\
			\br
		\end{tabular}

\end{sidewaystable}
\normalsize

In Ref. \cite{Cao2024qip}, for any $2^{\ot (n+1)}$ qubit state and bipartite entanglement $E$, the authors gave a tighter parameterized monogamy relations under condition $\mS^{(6,\alpha(E))}(E)$.
For any $2\otimes2\otimes2^{n-2}$ mixed state $\rho^{ABC}\in\mS^{ABC}$, one has $C^2(A|BC)\geq C^2(AB)+C^2(AC)$. Therefore, there exists $\mu\ge 1$ such that $C^2(A|BC)\geq C^2(AB)+\mu C^2(AC)$ \cite{Liu2021ijtp}. Using inequalities (\ref{tighter-inquality1}), (\ref{tighter-inquality3}), (\ref{tighter-inquality2}), the authors obtained some tighter monogamy relations in Ref. \cite{Li2024sc,Liu2021ijtp,Xie2023qip}.


\subsection{Other monogamy relation}


In this subsection, we present the other monogamy relations that are neither the supper-tighter monogamy relations nor the sub-tighter monogamy relations.

For any $N$-qubit pure state $|\psi\rangle^{ABC_1C_2...C_{n-2}}$, it was proved in Ref.~\cite{Zhu2015pra} that
\beax\label{cor1}
\fl\quad \quad \quad	C^2(|\psi\rangle^{AB|C_1C_2...C_{n-2}})
	&\geq&\max\left\lbrace \sum_{i=1}^{n-2}C^2(\rho^{AC_i})-\sum_{i=1}^{n-2}C_a^2(\rho^{BC_i}),\right. \nonumber\\
	&&~~~~~~~\left. \sum_{i=1}^{N-2}C^2(\rho^{BC_i})-\sum_{i=1}^{n-2}C_a^2(\rho^{AC_i})\right\rbrace.
\eeax
Ref.~\cite{Yang2018pra} got a similar results based on negativity.

In 2014, de Oliveira \etal \cite{Oliveira2014pra} presented the following inequality which was termed the linear monogamy relation for the three-qubit system,
\beax
	E_f(\rho^{AB})+E_f(\rho^{AC})\leq c
\eeax
for some constant $c>0$. The authors numerically found that the upper bound $Z=1.1882$ by considering a sampling of $10^6$ random states, while for three-qubit pure state $|\psi\rangle=\frac{1}{\sqrt{2}}|100\rangle+\frac{1}{2}(|010\rangle+|001\rangle)$, they found $c= 1.20175$. Thus it was conjectured that $c= 1.20175$ for EoF and the linear
monogamy relations can be saturated only when the focus qubit $A$ is maximally entangled with the joint
qubits $BC$. In 2015, Liu \etal \cite{Liu2015Serp} analytically proved the above inequality for a three-qubit pure state based on EoF, i.e. $E_f(\rho^{AB})+E_f(\rho^{AC})\leq 1.2018$ with equality if and only if $E_f(\rho^{AB})=E_f(\rho^{AC})=0.6009$.
In addition, they proved
\beax
C(\rho^{AB})+C(\rho^{AC})\leq 1.4142
\eeax 
with equality iff $C(\rho^{AB})=C(\rho^{AC)}=0.7071$ for three-qubit mixed state.
Later, in Ref.~\cite{Shi2021pra}, the authors generalize the linear monogamy relation for three-qubit states as
\begin{eqnarray}\label{multilinear}
	E(\rho^{AB})+E(\rho^{AC})+E(\rho^{BC})\leq c
\end{eqnarray}
for some constant $c>0$. Eq.~\eref{multilinear} was termed multilinear monogamy relation in Ref.~\cite{Shi2021pra} (it is in fact a complete monogamy relation, see in Sec.~\ref{sec-7}).
For EoF, it was fonud that $W$ state reaches the upper bound $c_{max}=3H_2(\frac{1}{2}+\frac{\sqrt{5}}{6})$ in Eq.~\eref{multilinear},
where $H_2(x)$ as in Eq.~\eref{eof}. For concurrence $C$ and EoF, the authors present $W$ state is the unique state that can reach the upper bound up to the local unitary transformations.

Let $E$ be a bipartite entanglement measure with $\alpha(E)>1$. Then, for any three partite state $\rho_{ABC}$, there always exists a positive integer $m$ (related with $\rho^{ABC}$) such that~\cite{Jin2019pra}
\beax
	E(\rho^{\otimes m}_{A|BC})\geq E(\rho^{\otimes m}_{AB}) +E(\rho^{\otimes m}_{AC}),
\eeax
where $E(\rho^{\otimes m}_{A_1|A_2...A_n})=E(\rho_{A_{11}A_{12}...A_{1m}|A_{21}...A_{2m}...A_{n1}...A_{nm}})$ denotes the quantum entanglement between the first party and the rest ones after the $m$ copies of $\rho_{A_1A_2...A_n}$, i.e., the quantum entanglement between $A_1$ and $\overline{A_1}$, view $A_{11}A_{12}...A_{1m}$ and $A_{21}...A_{2m}...A_{n1}...A_{nm}$ as $A_1$ and $\overline{A_1}$, respectively.

There were also other kind of monogamy relations for the $|\widetilde{\W}_{n,d}\ra$ state.
For $q=2$ or $q=3,$ assume $|\psi\ra$ is a $|\widetilde{\W}_{n,d}\ra$ state, then~\cite{Shi2020pra}
\beax
	E_q(|\psi\rangle^{AB|C_1C_2\cdots C_n})\geq \sum_{i=1}^{n}\left|E_q\left(\rho^{AC_i}\right)-E_q\left(\rho^{BC_i}\right)\right|.
\eeax

The tighter monogamy relation for higher dimensional system is very rare. In Ref.~\cite{Zhu2017qip1}, a monogamy relations for any $m\otimes n\otimes p\otimes q$ pure quantum state $|\psi\rangle^{A_1A_2|B_1B_2}$ was put forward~\cite{Zhu2017qip1},
\bea\label{T3}
C^2(|\psi\rangle^{A_1A_2|B_1B_2})
\geq \sum_{i=1,2;j=1,2}t_{ij}C^2(\rho^{A_iB_j}),
\eea
where $\rho^{A_{i}B_{j}}$ is the reduced state, $t_{11}=x_1y_{11}+x_3y_{31}$,
$t_{12}=x_2y_{21}+x_3y_{32}$, $t_{21}=x_1y_{12}+x_4y_{41}$,
$t_{22}=x_2y_{22}+x_4y_{42}$, with $\sum_{i=1}^{4} x_i=1$ and $\sum_{i=1}^{2}y_{ji}=1$, $j\in{\{1,2,3,4\}}$.
Eq.~\eref{T3} is similar to that of the second-order monogamy relation argued in Ref.~\cite{Cornelio2010pra}.


\subsection{Strong monogamy relation }


Regula \etal proposed a sharper version of the monogamy relation of the concurrence for the $n$-qubit states, which was called strong monogamy (SM) relation.
For an $n$-qubit pure state $|\psi\ra$, the SM relation up to concurrence reads~\cite{Regula2014prl}
\bea\label{tau-n}
\tau^{(1)}_{A_1|A_2\cdots A_n}&\geq &\sum_{j=2}^n\tau^{(2)}_{A_1A_j}+\sum\limits_{k>2}\left[ \tau^{(3)}_{A_1|A_2|A_k}\right]^{\mu_3}+\cdots\nonumber\\
&&+\sum\limits_{l=2}^n\left[\tau^{(n-1)}_{A_1|A_2|\cdots|A_{l-1}|A_{l+1}|\cdots|A_n} \right]^{\mu_{n-1}},
\eea
where $\tau^{(1)}$ is the tangle, two-tangle $\tau^{(2)}=C^2$, $\tau^{(3)}(A|BC)=\tau^{(1)}(A|BC)-\tau^{(2)}(AB)-\tau^{(2)}(AC)$, and so on.
$\{\mu_{m=2}^{n-1}\}$ with $\mu_2=1$ is a sequence of rational exponents which can regulate the weight assigned to the different $m$-partite contributions.
Obviously, the SM inequality~\eref{tau-n} is stronger than the corresponding original monogamy relation $\tau^{(1)}_{A_1|A_2\cdots A_n}\geq \sum_{j=2}^n\tau^{(2)}_{A_1A_j}$. This is why we call it strong monogamy relation.
The pure state residual $n$-tangle $\tau^{(n)}_{A_1A_2\cdots A_n}$ is defined as the difference between the left- and right-hand side in Eq.~\eref{tau-n}:
\bea \label{n-residual}
\tau^{(n)}_{A_1A_2\cdots A_n}:&=&\tau^{(1)}_{A_1|A_2\cdots A_n}-\sum_{j=2}^n\tau^{(2)}_{A_1A_j}-\sum\limits_{k>2}\left[ \tau^{(3)}_{A_1|A_2|A_k}\right]^{\mu_3}-\cdots\nonumber\\
&&-\sum\limits_{l=2}^n\left[\tau^{(n-1)}_{A_1|A_2|\cdots|A_{l-1}|A_{l+1}|\cdots|A_n} \right]^{\mu_{n-1}}.
\eea
$\tau^{(n)}$ just gives a rough indicator of all the leftover entanglement not distributed in any pairwise form.
For mixed $n$-qubit state $\rho$, its residual $n$-tangle is defined by the physically motivated convex-roof procedure~\cite{Regula2014prl}:
\bea \label{n-residual-mixed}
\tau^{(n)}_{A_1A_2\cdots A_n}(\rho)=\left[ \min\limits_{p_i,|\psi_i\ra}\sum_ip_i\sqrt{\tau^{(n)}_{A_1A_2\cdots A_n}(|\psi_i\ra)} \right]^2,
\eea 
where the minimization is taken over all possible pure state ensembles $\{p_i,|\psi_i\ra\}$ of $\rho$. For $n=3$,
Eq.~\eref{n-residual-mixed} is just the mixed-state extension of the three-tangle $\tau^{(3)}$ ad defined in Ref.~\cite{Verstraete2002pra}.
Any nontrivial selection of the sequence $\{\mu_{m=2}^{n-1}\}$ with $\mu_2=1$ defines in fact a particular SM inequality, sharpening and generalizing the CKW one. Clearly, the verification of Eq.~\eref{tau-n} given a set $\{\mu_m^\star\}$ implies its validity for all $\{\mu^\star\}\preceq\{\mu_m\}$.
So the sharpest choices of theses parameters are $\mu_m=1$ for any $m$.

In Ref.~\cite{Regula2014prl}, it was proved that the SM inequality with $\mu_m=1$ is valid for the supperpositions of the $n$-qubit W state and the generalized $n$-qubit GHZ states ($n\geq 4$)
\beax 
|\Phi^n_{a,b,c}\ra=a|0\cdots 0\ra+b|{\W}_{n,2}\ra+c|1\cdots 1\ra,
\eeax 
where $a, b, c\in\C$, $|a|^2+|b|^2+|c|^2=1$, and 
\bea\label{wn2}
|{\W}_{n,2}\ra=1/\sqrt{n}(|0\cdots 01\ra+\cdots+|10\cdots 0\ra)
\eea  
is the $n$-qubit W state.
Moreover, with $\mu_m=m/2$ ($m>2$), they also proved the SM inequality for the four-qubit state by numerical test with huge data points. Later, Kim in 2014 showed in Ref.~\cite{Kim2014pra} that the SM inequality is also true for a large class of multiqubit generalized W state,
\beax
|\widetilde{{\W}_{n,2}}\ra=a_1|10\cdots 0\ra+a_2|01\cdots0\ra+\cdots +a_n|00\cdots 1\ra, ~ \sum_i|a_i|^2=1.
\eeax

However, the SM inequality is not universally true. In 2015, Kim \etal~\cite{Choi2015pra} discussed the strong monogamy relation for $N_F$ and showed that the strong monogamy inequality holds well even in a higher dimensional system but the original strong monogamy inequality fails. In 2016, Karmakar \cite{Karmakar2016pra} found a subclass of the four-qubit generic class of states for which the strong monogamy inequality is satisfied by both negativity and squared negativity. They also found a few classes of states for which negativity and squared negativity are not strongly monogamous.

In fact, for any monogamous bipartite measure of quantum correlation, similar to the three-tangle, we can define the $\beta$-th power of the residual quantum correlation~\cite{Jin2019oc}
\beax\label{re}
\mathcal{Q}^{\beta}(ABC)=\mathcal{Q}^{\beta}(A|BC) -\mathcal{Q}^{\beta}(AB)-\mathcal{Q}^{\beta}(AC),~~~\beta\geq\alpha(\mQ),
\eeax
where $\alpha(\mQ)$ is defined as that of monogamy exponent $\alpha(E)$ for the entanglement measure $E$, namely,  $\alpha(\mQ)$ is the smallest value $\alpha$ that satisfies $\mathcal{Q}^{\alpha}(ABC)\geq0$ for any $\rho^{ABC}\in\mS^{ABC}$.
For any $d\otimes d\otimes d\otimes d$ state $\rho^{AB_1B_2B_3}$, it was shown in Ref.~\cite{Jin2019oc}) that
\beax
\mathcal{Q}^{\beta}(A|B_1B_2B_3)\geq \sum_{i=1}^{3}\mathcal{Q}^{\beta}(AB_i)+\max\limits_{i\neq j}\mathcal{Q}^{\beta}(AB_iB_j),
\eeax
for $\beta\geq\alpha(\mQ)$.


\section{Polygamy of the entanglement of assistance} \label{sec-5}


Dually,
the polygamy relation in literature is expressed as 
\bea
Q(A|BC)\leq Q(AB)+Q(AC)\;.
\label{polygamydefinition}
\eea
However, Eq.~\eref{monogamy1} [resp. Eq.~\eref{polygamydefinition}] captures only partially 
the property that $Q$ is monogamous (resp. polygamous). For example, it is well known that if $Q$ does not satisfy 
these relations, it is still possible to find a positive power $\alpha\in\mbR_{+}$, 
such that $Q^\alpha$ satisfies the relation. For example, several EoAs, $E_a$s, satisfy the polygamy relation~\eref{polygamydefinition} if $Q$ is replaced by $E_a^\alpha$ for some $\alpha>1$~\cite{Buscemi,Gao2021rp,Gour2005pra2,Kim2009,Kim2010pra,Lizongguo,Song2019qip}.

In general, it is very hard to verify whether a given EoA $E_a$ is polygamous according to Eq.~\eref{polygamydefinition},
from which only the multi-qubit case seems easy to check~\cite{Buscemi,Gao2021rp,Gour2005pra2,Kim2009,Kim2010pra,Lizongguo,Song2019qip}.
Apart from $E_{f,a}$ and $E_{q,a}$, the polygamy relation for all other EoAs remain open for high dimensional systems~\cite{Buscemi,Kim2010pra}.

Throughout this section, a measure of quantum correlation $Q$ refers to any quantity that characterizes the ``quantumness'' contained in bipartite quantum systems, such as
entanglement, quantum discord~\cite{Henderson2001jpa,Ollivier}, 
measurement-induced nonlocality~\cite{Luoshunlong}, 
quantum deficit~\cite{Oppenheim}, and other quantum correlations introduced in recent years~\cite{Guowu2014srep,Guo2015ijtp,Groisman,Luo2008,Wiseman,WPM2009}, etc.


\subsection{Improved definition of polygamy relation} \label{sec-5.1}


In order to find a unified way of checking the polygamy of the entanglement of assistance,
Guo improved the definition of polygamy relation in Ref.~\cite{Guo2018qip}:
Let $Q$ be 
a measure of quantum correlation.
$Q$ is called polygamous if for any $\rho^{ABC}\in\mathcal{S}^{ABC}$ that satisfies 
\bea\label{defcond}
Q(\rho^{A|BC})>\max\{Q(\rho^{AB}),Q(\rho^{AC})\}>0\;,
\eea
we have 
$\min\{Q(\rho^{AB}),Q(\rho^{AC})\}>0$.
This improved definition does not involve the standard polygamy relation as Eq.~\eref{polygamydefinition}, but it can derive a more quantitative inequality
from Eq.~\eref{defcond}~\cite{Guo2018qip}: A continuous measure of quantum correlation $Q$ is pologamous according to definition~\eref{defcond} if and only if 
there exists $0<\beta<\infty$ such that
\bea\label{polygamy-power}
Q^\beta(\rho^{A|BC})\leq Q^\beta(\rho^{AB})+Q^\beta(\rho^{AC})
\eea
holds for any $\rho^{ABC}\in\mS^{ABC}$ with fixed $\dim\mH^{ABC}=d<\infty$.


\subsection{Entanglement of assistance is polygamous}\label{sec-5.2}


Like that of the entanglement of formation, if $E_a$ is polygamous (according to definition in Eq.~\eref{polygamydefinition}) on pure tripartite states in $\mH^{ABC}$, then it is also polygamous on mixed sates acting on $\mH^{ABC}$~\cite{Guo2018qip}. According to definition in Eq.~\eref{defcond}, it was showed that any faithful entanglement measure is not polygamous, e.g., the generalized $n$-qudit GHZ states 
\bea\label{GHZnd}
|\widetilde{\GHZ}_{n,d}\ra =\sum\limits_{i=0}^{d-1}\lambda_i|i\ra^{\ot n}, \quad \lambda_i>0, \sum_i\lambda_i^2=1
\eea
can not be polygamous for any $E$, while any entanglement of assistance is polygamous but not monogamous~\cite{Guo2018qip}.


\subsection{Polygamy exponent}\label{sec-5.3}


Note that the original polygamy relation of $Q^\beta$ can be preserved when we lower the power~\cite{Kumar2016pla}:
Let $\rho^{ABC}\in\mS^{ABC}$ and $Q$ be a polygamy measure of quantum correlation.
Then $Q^\beta(\rho^{A|BC})\leq Q^\beta(\rho^{AB})+Q^\beta(\rho^{AC})$
implies $Q^\gamma(\rho^{A|BC})\leq Q^\gamma(\rho^{AB})+Q^\gamma(\rho^{AC})$
for any $\gamma\in[0,\beta]$.
(Note that, in~\cite{Kumar2016pla}, the bipartite correlation measure $Q$
is assumed to be normalized. This condition however, is not necessary,
which can be easily checked following the argument therein.)
We denote by $\beta(Q)$ the supremum for $Q^{\beta(Q)}$ satisfies
Eq.~\eref{polygamy-power} for all the states, and call it the {polygamy exponent} of $Q$ (Note here that, in Ref.~\cite{Guo2018qip} it was called polygamy power. We use the term polygamy exponent throughout this paper in order to be in consistent with the monogamy exponent in Subsec.~\ref{sec-3.11}).  
For the case of $Q$ is entanglement assistance $E_a$ associated with some entanglement 
measure $E$, it is always 
continuous~\cite{Guo2018qip}. Therefore $\beta(E_a)$ exists for $E_a$.
We list the polygamy exponents in literature so far in Table~\ref{tab:table5.3}.
Polygamy exponent $\beta(E_a)$ is a dual concept of the monogamy exponent $\alpha(E)$. 
$\beta(E_a)$ may depend also on the dimension $d=\dim\mH^{ABC}$ as that of $\alpha(E)$~\cite{GG2018q}.
Now, the pair $[\alpha(E),\beta(E_a)]$
reflects the share-ability of continuous entanglement measure $E$ more efficiently. 
This pair of exponent indexes advances our understanding of 
multipartite entanglement
although these quantities are difficult to calculate.

\begin{table}
	\caption{\label{tab:table5.3}The polygamy exponents of several 
		assisted entanglement. Hereafter, $C_a$, $N_a$, $E_{f,a}$, $\tau_a$, $E_{q,a}$, $E_{\alpha,a}$ denote the entanglement of assistance of $C$, $N$, $E_f$, $\tau$, $E_q$, and $E_\alpha$, respectively; $E_{N_a}=\log_2(2N_a+1)$.}
	\begin{indented}
		\lineup
		\item[]
		\begin{tabular}{@{}llll}\br
			$E_a$& $\beta(E_a)$ & System &References \\
			\mr
			$C_a$ & $2$& $2^{\otimes 3}$,~pure state & \cite{Gour2005pra2}\\ 
			$N_a$& $\geq 2$ & $2^{\otimes n}$,~pure state &\cite{Kim2009}\\ 
			${E}_{f,a}$& $\geq 1$ & anysystem,~pure state & \cite{Buscemi}\\ 
			$\tau_a$& $\geq 1$ & $2^{\otimes n}$ &\cite{Lizongguo}\\   
			$E_{q,a}$, $q\ge1$ &   $\geq 1$    & any system & \cite{Kim2010pra}\\ 
			$E_{q,a}$, $q=2$ &   $\geq 1$    & $|\GWV\ra$, $d^{\ot n}$ & \cite{Shi2020pla}\\ 
			$E_{\alpha,a}$,~$\frac{\sqrt{7}-1}{2}\leq \alpha\leq \frac{\sqrt{13}-1}{2}$&   $\geq 1$    &$2^{\otimes n}$ & \cite{Song2019qip}\\ 
			$E_{N_a}$&   $\geq 2$    &$2^{\otimes n}$, pure state & \cite{Gao2021rp}\\ 
			\br
		\end{tabular}
	\end{indented}
\end{table}


\subsection{Tighter polygamy relation} \label{sec-5.4}


Apart from the $\beta(C_a)$ for the three-qubit system, all other polygamy exponent of $E_a$ are unknown up to now.
But as shown in Tab.~\ref{tab:table5.3}, for some special $E_a$ or some particular system, we have obtained some polygamy relations with the power $\gamma\leq\beta(E_a)$, i.e.,
\beax\label{polygamy-power2}
E_a^\gamma(\rho^{A|B_1B_2\cdots B_n})\leq \sum_{i=1}^nE_a^\gamma(\rho^{AB_i}),\quad \gamma\leq \beta(E_a).
\eeax
Then the tighter version 
\beax\label{polygamy-power3}
E_a^\gamma(\rho^{A|B_1B_2\cdots B_n})\leq \sum_{i=1}^nu_iE_a^\gamma(\rho^{AB_i})+v
\eeax
have been proposed for some $u_i<1$ in the last two decades, where $v\geq 0$ is some function of $\rho^{AB_1\cdots B_n}$ or its reduces states. These coefficients $u_i$ and $v$ are obtained by some class of inequality which can always be written in the form of 
\bea\label{polygamy-inequality1}
(1+t)^x\leq 1+\phi(t,x)+\xi(t,x)t^x,\quad 0\leq t\leq 1, 0\leq x\leq 1
\eea
with some functions $\phi(t,x)\geq 0$ and $\xi(t,x)\geq 0$, or in the form 
\bea\label{polygamy-inequality2}
(1+t)^x\leq 1+\pi(t,x)+\omega(t,x)t^x,\quad   t\geq 1,  x\geq 1
\eea
with some functions $\pi(t,x)\geq 0$ and $\omega(t,x)\geq 0$. 
Additionally, some tighter polygamy relations are derived via the following form inequality 
\bea\label{polygamy-inequality3}
(1+t)^x\leq 1+\epsilon(t,x)+\upsilon(t,x)t^x,\quad   t\geq 1,  0\leq x\leq 1
\eea
with $\epsilon(t,x)\geq0,~\upsilon(t,x)\geq 0$.
In this subsection, we list these tighter polygamy relations of entanglement of assistance up to now in literature.

\subsubsection{Sub-power tighter polygamy relation}

For more clarity, we list these tighter polygamy relations in the Tables ~\ref{tab:table5-3-1}-\ref{tab:table5-3-2}.
In table~\ref{tab:table5-3-2}, we let  $h_1=\frac{(1+k)^{\frac{\gamma}{2}}-1}{k^{\frac{\gamma}{2}}}$, $h_2=\frac{(1+k)^\gamma-1}{k^\gamma}$ and $h_3=\frac{(1+k^\delta)^\gamma-1}{k^{\delta\gamma}}$.

\begin{table}[htbp]
	\caption{\label{tab:table5-3-1} {The inequality used in the sub-power tighter polygamy relations in literature.}}
	\footnotesize
		\begin{tabular}{@{}lllll}\br
			SN	&$E_a$& $\beta$ & Inequality&$\gamma\geq0$, System   \\
			\mr			
			(1) & $E_{f,a}$     &  $1$           & \eref{polygamy-inequality1}: \cite{Kim2018pra2}  $\left(1+t\right)^{x}\leq 1+x t^{x}$       &  $\leq 1$, $d^{\ot (n+1)}$     \\
			(2)& $E_{f,a}$     &  $1$           & \eref{polygamy-inequality1}: \cite{Kim2018pra2}  $\left(1+t\right)^{x}\leq 1+x t^{x}$       &  $\leq 1$, $d^{\ot (n+1)}$     \\
			(3)& $\tau_{a}$${\color{red}^{\rm a}}$      &  $1$           & \eref{polygamy-inequality1}: \cite{Kim2018pra1}  $\left(1+t\right)^{x}\leq 1+x t^{x}$       &  $\leq 1$, $2^{\ot (n+1)}$     \\
			(4)& $\tau_{a}$${\color{red}^{\rm a}}$      &  $1$           & \eref{polygamy-inequality1}: \cite{Kim2018pra1}  $\left(1+t\right)^{x}\leq 1+x t^{x}$       &  $\leq 1$, $2^{\ot (n+1)}$     \\
			(5) & $C_{a}$     &  $2$           & \eref{polygamy-inequality1}: \cite{Jin2019qip}  
			$(1+t)^x\leq 1+(2^{x}-1)t^x$       &  $\leq 2$, $2^{\ot (n+1)}$     \\
			(6) & $C_{a}$     &  $2$           & \eref{polygamy-inequality1}: \cite{Jin2019qip}  
			$(1+t)^x\leq 1+(2^{x}-1)t^x$       &  $\leq 2$, $2^{\ot (n+1)}$     \\
			(7) & $C_{a}$${\color{red}^{\rm a}}$     &  $2$           & \eref{polygamy-inequality1}: \cite{Yang2019ctp}  
			$(1+t)^x\leq1+\frac{(1+k)^x-1}{k^x}t^x$, $0< k\leq1$       &  $\leq 2$, $2^{\ot (n+1)}$     \\         
			(8) & $C_{a}$${\color{red}^{\rm a}}$     &  $2$           & \eref{polygamy-inequality1}: \cite{Yang2019ctp}  
			$(1+t)^x\leq1+\frac{(1+k)^x-1}{k^x}t^x$, $0< k\leq1$       &  $\leq 2$, $2^{\ot (n+1)}$     \\              
			(9) & $E_{f,a}$     &  $1$           & \eref{polygamy-inequality1}: \cite{Chen2019ijtp}  $(1+t)^x\leq1+\frac{(1+k)^x-1}{k^x}t^x$, $0< k\leq1$      &  $\leq 1$, $d^{\ot (n+1)}$     \\  
			(10) & $E_{f,a}$     &  $1$           & \eref{polygamy-inequality1}: \cite{Chen2019ijtp}  $(1+t)^x\leq1+\frac{(1+k)^x-1}{k^x}t^x$, $0< k\leq1$      &  $\leq 1$, $d^{\ot (n+1)}$     \\   
			(11) & $E_a$     &  $\beta(E_a)$           & \eref{polygamy-inequality3}: \cite{Gao2020qip}  $(1+t)^{x}\leq t^{x}+(m+1)^{x}-m^{x}$, &\\
			&&&\qquad\qquad\qquad\quad~$1\leq m\leq t$     &  $\leq \beta(E_a)$, $d^{\ot (n+1)}$     \\   
			(12)& $E_a$     &  $\beta(E_a)$           & \eref{polygamy-inequality1}: \cite{Gao2020qip} $(1+t)^x\leq 1+\frac{k^2x}{(k+1)^2} t+\left\lbrace (k+1)^{x}-\right. $&\\
			&&&\qquad\qquad\qquad\quad~$\left. \left[\frac{kx}{(k+1)^2}+1\right]k^{x}\right\rbrace t^{x}$,  $k\geq1$
			&  $\leq \beta(E_a)$, $d^{\ot (n+1)}$     \\   
			(13)& $E_{2,a}$ & $1$ &  \eref{polygamy-inequality1}: \cite{Shi2020pla} $(1+t)^x\leq 1+(2^x-1) t^x$ & $\leq 1$, $|\GWV\ra$  \\
			(14) &$E_{q,s,a}$${\color{red}^{\rm b}}$  & $1$  &      \eref{polygamy-inequality1}: \cite{Jin2020cpb} $(1+t)^x\leq 1+(2^x-1) t^x$ &$\leq 1$,  $2^{\ot (n+1)}$     \\ 
			(15) &$E_{q,s,a}$${\color{red}^{\rm b}}$  & $1$  &      \eref{polygamy-inequality1}: \cite{Ren2021lpl} $(1+t)^{x}\leq 1+\frac{(1+k^{\delta})^{x}-1}{k^{\delta x}}t^{x}$,  &\\
			&&&\qquad\qquad\qquad\quad$0<k \leq 1$, $\delta\geq 1$        &$\leq 1$,  $2^{\ot (n+1)}$     \\ 
			(16) &$E_{q,s,a}$${\color{red}^{\rm b}}$  & $1$  &      \eref{polygamy-inequality1}: \cite{Ren2021lpl} $(1+t)^{x}\leq 1+\frac{(1+k^{\delta})^{x}-1}{k^{\delta x}}t^{x}$,  &\\
			&&&\qquad\qquad\qquad\quad$0<k \leq 1$, $\delta\geq 1$        &$\leq 1$,  $2^{\ot (n+1)}$     \\  
			(17)&$E_{q,a}$${\color{red}^{\rm c}}$    & $1$ &   \eref{polygamy-inequality1}: \cite{Lai2021jpa}  $(1+t)^x \leqslant 1+\frac{(1+k)^x -1}{k^x}t^x$, $0 < k\leqslant1$,     &$\leq 1$, $|\GWV\ra$    \\  
			(18)&$E_{q,a}$${\color{red}^{\rm c}}$    & $1$ &   \eref{polygamy-inequality1}: \cite{Lai2021jpa}  $(1+t)^x \leqslant 1+\frac{(1+k)^x -1}{k^x}t^x$, $0 < k\leqslant1$,     &$\leq 1$, $|\GWV\ra$    \\ 
			(19)&$E_{q,a}$${\color{red}^{\rm c}}$    & $1$ &   \eref{polygamy-inequality1}: \cite{Lai2021jpa}  $(1+t)^x \leqslant 1+\frac{(1+k)^x -1}{k^x}t^x$, $0 < k\leqslant1$,     &$\leq 1$, $|\GWV\ra$    \\ 						
			\br
		\end{tabular}\\
		${\color{red}^{\rm a}}$ pure state;\\
	${\color{red}^{\rm b}}$ $1 \leq q \leq 2$ and $-q^2+4q-3 \leq s \leq1$;\\
	${\color{red}^{\rm c}}$ $q\in[\frac{5-\sqrt{13}}{2},2]\cup [3,\frac{5+\sqrt{13}}{2}]$. 
\end{table}
\normalsize

\begin{sidewaystable}[htbp]
	\vspace{125mm}
	\caption{\label{tab:table5-3-2} Sub-power tighter polygamy relations in literature corresponding to Table~\ref{tab:table5-3-1}. }
	\footnotesize
		\begin{tabular}{@{}lll}\br
			SN	& Tighter monogamy relation  &CS \\
			\mr
			(1)	&\cite{Kim2018pra2}: $E^\gamma_{f,a}(A|B_1\cdots B_{n})\leq \sum_{j=1}^{n} \gamma^{w_H(\overrightarrow{j-1})}{E^\gamma_{f,a}}(AB_j)$  & $\mS^{(3,1)}(E_{f,a})$\\
			(2)	&\cite{Kim2018pra2}: $E^\gamma_{f,a}(A|B_1\cdots B_{n})\leq \sum_{j=1}^{n} \gamma^{j-1}{E^\gamma_{f,a}}(AB_j)$  & $\mS^{(4,1)}(E_{f,a})$\\
			(3)	&\cite{Kim2018pra1}: $\tau^\gamma_{a}(A|B_1\cdots B_{n})\leq \sum_{j=1}^{n} \gamma^{w_H(\overrightarrow{j-1})}{\tau^\gamma_{a}}(AB_j)$  & $\mS^{(3,1)}(\tau_{a})$\\
			(4)	&\cite{Kim2018pra1}: $\tau^\gamma_{a}(A|B_1\cdots B_{n})\leq \sum_{j=1}^{n} \gamma^{j-1}{\tau^\gamma_{a}}(AB_j)$  & $\mS^{(4,1)}(\tau_{a})$\\
			(5) &\cite{Jin2019qip}: $C^\gamma_{a}(A|B_1\cdots B_{n})\leq \sum_{j=1}^{n} (2^{\frac{\gamma}{2}}-1)^{w_H(\overrightarrow{j-1})}{C^\gamma_{a}}(AB_j)$
			& $\mS^{(3,1)}(C_{a})$\\
			(6) &\cite{Jin2019qip}: $C^\gamma_{a}(A|B_1\cdots B_{n})\leq \sum_{j=1}^{n} (2^{\frac{\gamma}{2}}-1)^{j-1}{C^\gamma_{a}}(AB_j)$
			& $\mS^{(4,2)}(C_{a})$\\    
			(7) &\cite{Yang2019ctp}: $C_a^\gamma(A|B_1\cdots B_{n})\leq \sum_{i=1}^m h_1^{i-1} C_a^\gamma(AB_i)+h_1^{m+1}\sum_{j=m+1}^{n-1} C_a^\gamma(AB_j)+h_1^mC_a^\gamma(AB_{n})$ 
			& $\mS^{(6,2)}(C_a)$\\  
			(8) &\cite{Yang2019ctp}: $C_a^\gamma(A|B_1\cdots B_{n})\leq \sum_{i=1}^nh_1^{i-1} C_a^\gamma(AB_i)$ 	& $\mS^{(5,2)}(C_a)$\\  
			(9)&\cite{Chen2019ijtp}: $E_{f,a}^{\gamma}(A|B_{1}B_{2}\cdots B_n)\leq \sum_{j=1}^{n}h_2^{\omega_{H}(\overrightarrow{j-1})}E_{f,a}^{\gamma}(AB_{j})$ & $\mS^{(7,1)}(E_{f,a})$\\
			(10)&\cite{Chen2019ijtp}: $E_{f,a}^{\gamma}(A|B_{1}B_{2}\cdots B_n)\leq \sum_{j=1}^{n}h_2^{j-1}E_{f,a}^{\gamma}(AB_j)$ & $\mS^{(8,1)}(E_{f,a})$\\
			(11) &\cite{Gao2020qip}: $E_{a}^{\gamma}(A|B_{1}\cdots B_n) \leq
			E_{a}^{\gamma}(AB_1)+(2^{t}-1)E_{a}^{\gamma}(AB_2)+\cdots
			+[n^t-(n-1)^t]E_{a}^{\gamma}(AB_n)$, $t=\frac{\gamma}{\beta(E_a)}$
			& $\mS^{(3,1)}(E_{a})$\\
			(12)&\cite{Gao2020qip}: $E_a^{\gamma}({A|B_{1}\cdots B_{n}}) \leq \sum_{i=1}^m h^{i-1}E_a^{\gamma}({AB_i})+h^{m}[(\eta'+1)^{t}-\eta'^{t}]\sum_{j=m+1}^{n-2}E_a^{\gamma}({AB_j})+h^{m}F(AB)$ ${\color{red}^{\rm a}}$&\\
			&\quad \quad \quad $h=(\eta+1)^{t}-\eta^{t}$, $t=\frac{\gamma}{\beta(E_a)}$, $\eta=k^{\beta(E_a)}$, $\eta'=k'^{\beta(E_a)}$    &  $\mS^{(9,1)}(E_a)$        \\
			(13)&\cite{Shi2020pla}: $E_{2,a}^{\gamma}(P|P_1\cdots P_r)\leq \sum_{i=1}^{r}(2^{\gamma}-1)^{w_H(\overrightarrow{i-1})}E_{2,a}^{\gamma}(PP_i)$	 & $\mS^{(3,1)}(E_{2,a})$\\
			(14)&\cite{Jin2020cpb}: $E_{q,s,a}^\gamma({A|B_1\ldots B_{n}})
			\leq \sum\nolimits_{j=1}^{n}(2^\gamma-1)^{j-1}E_{q,s,a}^\gamma(AB_j)$ & $\mS^{(4,1)}(E_{q,s,a})$\\
			(15)&\cite{Ren2021lpl}: $E_{q,s,a}^\gamma({A|B_1\ldots B_{n}})
			\leq \sum\nolimits_{j=1}^{n}h_3^{j-1}E_{q,s,a}^\gamma(AB_j)$ & $\mS^{(8,1)}(E_{q,s,a})$\\
			(16)&\cite{Ren2021lpl}: $E_{q,s,a}^\gamma({A|B_1\ldots B_{n}})
			\leq \sum\nolimits_{j=1}^{n}h_3^{w_H(\overrightarrow{j-1})}E_{q,s,a}^\gamma(AB_j)$ & $\mS^{(7,1)}(E_{q,s,a})$\\
			(17)&\cite{Lai2021jpa}: $E_{q,a}^\gamma(P|P_1\cdots P_{r})\leqslant\sum\limits_{j=1}^r h_2^{\omega_H(\overrightarrow{j-1})}E_{q,a}^\gamma(PP_j)$ & $\mS^{(7,1)}(E_{q,a})$\\
			(18)&\cite{Lai2021jpa}: $E_{q,a}^\gamma(P|P_1\cdots P_{r})\leqslant\sum\limits_{j=1}^r h_2^{j-1}E_{q,a}^\gamma(PP_j)$ & $\mS^{(8,1)}(E_{q,a})$\\
			(19)&\cite{Lai2021jpa}: $E_{q,a}^\gamma(P|P_1\cdots P_r) \leqslant \sum_{j=1}^m h_2^{j-1} E_{q,a}^\gamma(PP_j)+	h_2^{m+1}\sum_{j=m+1}^{r-1}E_{q,a}^\gamma(PP_j)+h_2^{m}E_{q,a}^\gamma(PP_{r})$ & $\mS^{(6,1)}(E_{q,a})$\\
			\br
		\end{tabular}\\
	${\color{red}^{\rm 
			a}}$ $F(AB)=\left\lbrace (\eta'+1)^{t}-\left[ 1+\frac{\eta' t}{(\eta'+1)^2}\right] \eta'^{t}\right\rbrace E_a^{\gamma}({AB_{n-1}}) +\frac{{\eta'}^2 t}{(\eta'+1)^2}E_a^{\beta(E_a)}({AB_{n-1}})E_a^{\gamma-\beta(E_a)}({AB_{n}})+E_a^{\gamma}({AB_{n}})$.
\end{sidewaystable}
\normalsize

\subsubsection{Supper-power tighter polygamy relation}

For more clarity, we list these tighter polygamy relations in the Tables ~\ref{tab:table5-3-3}-\ref{tab:table5-3-4}.

	\begin{table}[htbp]
	\caption{\label{tab:table5-3-3} {The inequality used in the supper-power tighter polygamy relations in literature.}}
	\footnotesize
	\hspace{2mm}	\begin{tabular}{@{}lllll}\br
			SN	&$E_a$& $\beta$ & Inequality&$\zeta$, System   \\
			\mr
			
			(1) & $E_{a}$     &  $\beta(E_a)$           & \eref{polygamy-inequality2}: \cite{Jin2020qip}   $(1+t)^x\leq 1+\frac{(1+k)^x-1}{k^x}t^x$,  $k\geq 1$       &  $\geq \beta(E_a)$, $d^{\ot (n+1)}$     \\
			(2)& $N_{a}$& $\beta(N_a)$ &\eref{polygamy-inequality2}: \cite{Zhang2022ijtp} $(1+t)^x\leq q^x+\frac{(1+k)^x-q^x}{k^x}t^x$~${\color{red}^{a}}$      &  $\geq \beta(N_a)$, $2^{\ot (n+1)}$     \\
			(3)& $N_{a}$& $\beta(N_a)$ &\eref{polygamy-inequality2}: \cite{Zhang2022ijtp} $(1+t)^x\leq q^x+\frac{(1+k)^x-q^x}{k^x}t^x$~${\color{red}^{a}}$     &  $\geq \beta(N_a)$, $2^{\ot (n+1)}$     \\
			(4)&$E_a$ &$\beta(E_a)$ &\eref{polygamy-inequality2}: \cite{Zhang2023ps} $(1+t)^x\leq (1+{a})^{x-1}+\left( 1+\frac{1}{a}\right) ^{x-1}t^x$~${\color{red}^{b}}$    &  $\geq \beta(E_a)$, $d^{\ot (n+1)}$     \\
			(5)&$E_a$ &$\beta(E_a)$ &\eref{polygamy-inequality2}: \cite{Cao2024lpl} $(1+t)^{x}\leq\left(1+\frac{a}{s}\right)^{x-1}+\left(1+\frac{s}{a}\right)^{x-1}t^{x}$ ${\color{red}^{c}}$  &  $\geq \beta(E_a)$, $2^{\ot (n+1)}$     \\
			\br
		\end{tabular}\\
\indent	\hspace{2mm}				${\color{red}^{a}}$  $0<q\leq1$, $k\geq1$; ${\color{red}^{b}}$  $a\geq 1$; ${\color{red}^{c}}$ $s>0$, $t\geq a\geq 1$  
\end{table}
\normalsize

	\begin{table}[htbp]
	\caption{\label{tab:table5-3-4} Supper-power tighter polygamy relations in literature corresponding to Table~\ref{tab:table5-3-3}. }
	\footnotesize
		\begin{tabular}{@{}lll}\br
			SN	& Tighter monogamy relation  &CS \\
			\mr
			(1)	&\cite{Jin2020qip}: ${E_a}^\zeta(A|B_1B_2\cdots B_{n})\leq \sum_{i=1}^m h^{i-1}{E_a}^\zeta(AB_i)+h^m{E_a}^\zeta(AB_n)$&\\
			&\quad \quad \quad \quad \quad \quad \quad \quad \quad\qquad~~~~$+h^{m+1}\sum_{j=m+1}^{n-1}{E_a}^\zeta(AB_j)$&\\
			&\quad \qquad\quad \quad \quad \quad \quad \quad \quad \quad\quad~~
			$h=\frac{(1+k)^\frac{\zeta}{\beta(E_a)}-1}{k^\frac{\zeta}{\beta(E_a)}}$, $0\leq\beta\leq1$  & $\mS^{(6',\beta(E_a))}(E_{a})$\\
			(2) &\cite{Zhang2022ijtp}: $N_{a}^{\zeta}(A|B_1B_2\cdots B_n)\leq q^\frac{\zeta}{\beta(N_a)}\sum_{i=1}^m h^{i-1}N_{a}^\zeta(AB_i)$&\\
			&\quad \quad \quad \quad \quad \quad \quad\qquad\qquad$+h^{m+1}[\left[ N_{a}^\zeta(AB_{m+1})\right. $&\\
			&\quad \quad \quad \quad \quad \quad \quad\qquad\qquad$+q^{\frac{\zeta}{\beta(N_a)}}N_{a}^\zeta(AB_{m+2})+\cdots$&\\
			&\quad \quad \quad \quad \quad \quad \quad\qquad\qquad$\left. +q^{\frac{(n-m-2)\zeta}{\beta(N_a)}}N_{a}^\zeta(AB_{n-1})\right]$&\\
			&\quad \quad \quad \quad \quad \quad \quad\qquad\qquad$+h^{m} q^{\frac{(n-m-1)\zeta}{\beta(N_a)}}N_{a}^\zeta(AB_n)$&\\
			&\quad \quad \quad \quad \quad \quad \quad ~~
			\qquad\qquad$h=\frac{(1+k)^\frac{\zeta}{\beta(N_a)}-q^\frac{\zeta}{\beta(N_a)}}{k^\frac{\zeta}{\beta(N_a)}}$,  $0<q\leq 1$  & $\mS^{(6',\beta(N_a))}(N_{a})$\\
			(3)&\cite{Zhang2022ijtp}: $N_{a}^\zeta(A|B_1B_2\cdots B_n)\leq h\left[ N_{a}^\zeta(AB_1)+q^{\frac{\zeta}{\beta(N_a)}}N_{a}^\zeta(AB_2)+\cdots\right. $&\\
			&\quad \quad  \quad \quad \quad \qquad\qquad\quad~~~$\left. +q^{\frac{(n-2)\zeta}{\beta(N_a)}}N_{a}^\zeta(AB_{n-1})\right] +q^{\frac{(n-1)\zeta}{\beta(N_a)}}N_{a}^\zeta(AB_n)$&\\
			&\quad  \quad \quad \quad \quad \quad \quad \qquad\qquad~~~~$h=\frac{(1+k)^\frac{\zeta}{\beta(N_a)}-q^\frac{\zeta}{\beta(N_a)}}{k^\frac{\zeta}{\beta(N_a)}}$, $0<q\leq 1$  & $\mS^{(5,\beta(N_a))}(N_{a})$\\
			(4)&\cite{Zhang2023ps}: $E_a^{\zeta}(A|B_1...B_n)\leq(1+{a})^{\frac{\zeta}{\beta(E_a)}-1}\sum_{i=1}^{n} h^{n-i}E_a^{{\zeta}}(AB_i)$,
			&\\
			&\quad \quad \quad \quad \quad \quad \quad \quad \quad~ \quad~~$h= (1+\frac{1}{a})^{\frac{\zeta}{\beta(E_a)}-1}$  & $\mS^{(7,\beta(E_a))}(E_a)$\\
			(5)&\cite{Cao2024lpl}: $	E_a^{\zeta}(A|B_{1}\cdots B_{n})\leq \left(1+\frac{a}{s}\right)^{\frac{\zeta}{\beta(E_a)}-1}\sum_{i=1}^{n} h^{n-i}E_a^{\zeta}(AB_i)$, 	&\\
		&\quad \quad \quad \quad \quad \quad \quad \qquad \quad~	$h= \left(1+\frac{s}{a}\right)^{\frac{\zeta}{\beta(E_a)}-1}$ & $\mS^{(7,\beta(E_a))}(E_a)$\\			
			\br
		\end{tabular}
		
\end{table}
\normalsize

For an arbitrary dimensional multipartite quantum state $\rho^{ABC}\in\mS^{ABC}$, one has $E_{f,a}(A|BC)\leq E_{f,a}(AB)+E_{f,a}(AC)$ \cite{Kim2012pra}. Therefore, there exists $0<\nu\leq 1$ such that $E_{f,a}(A|BC)\leq E_{f,a}(AB)+\nu E_{f,a}(AC)$ \cite{Liu2021ijtp}. Using inequalities \eref{polygamy-inequality1}, \eref{polygamy-inequality2}, \eref{polygamy-inequality3}, the authors obtained some tighter polygamy relations in Ref. \cite{Li2024sc,Liu2021ijtp,Xie2023qip}.


\subsection{Other related relations} \label{sec-5.5}


Although the faithful entanglement measure is not polygamous, it exhibits polygamy property in some sense for some special states. For example, for any $(n+1)$-qubit quantum state $\rho$, if there are at least two reduced states $\rho^{AB_1}$ and $\rho^{AB_2}$ such that
$C(\rho^{AB_i})C(\rho^{AB_j})>0$, $i\neq j$,
then there exists $\alpha_0\in(0,2]$ such that~\cite{Zhu2019ijtp}
\begin{eqnarray*}\label{thc2}
	C^{\alpha}(\rho^{A|B_1B_2...B_{n}})\leq\sum_{i=1}^{n}C^{\alpha}(\rho^{AB_{i}}),\quad  0\leq\alpha\leq\alpha_0.
\end{eqnarray*}
For any $(m+n)$-qubit pure state $|\psi\rangle^{A_1A_2\cdots A_mB_1B_2\cdots B_n}$, the concurrence satisfies \cite{Yang2021qip}
\beax
C^2(A_1A_2\cdots A_m|B_1B_2\cdots B_n)\leq \sum^m_{i=1}C^2(A_i|\overline{A_i}).
\label{eqntg2}
\eeax

For the generalized W state $|\widetilde{{\W}_{n,2}}\ra$, 
if $C(\rho^{PP_i})>0$ for $1\leq i\leq r$, then~\cite{Zhu2017qip2}
\beax\label{cay}
	C_a^y(\rho^{P|P_1\cdots P_r})<\sum_{i=1}^{r}C^y_a(\rho^{PP_i}), \quad y\leq0.
\eeax
The polygamy relations for the entanglement of formation was also presented for the $|\widetilde{{\W}_{n,2}}\ra$ state in Ref.~\cite{Zhu2017qip2}.
If $N(\rho^{PP_i})>0$ for $1\leq i\leq r$, then~\cite{Liang2017qip}
\begin{eqnarray*}\label{nay}
	N_a^y(\rho^{P|P_1\cdots P_r})<\sum_{i=1}^{r}N^y_a(\rho^{PP_i}), \quad y\leq0.
\end{eqnarray*}
The polygamy relations for the R\'{e}nyi-$\alpha$ entanglement was also presented for the $|\widetilde{W_{n,2}}\ra$ state in Ref.~\cite{Liang2017qip}.


\section{Multipartite entanglement measure: basic concepts}\label{sec-6}


In this section, we at first review the definitions of different types of multipartite entanglement which including genuine entanglement, $k$-entanglement, $k$-partite entanglement, and partitewise entanglement. By recalling the coarsening relation of multipartite partitions, which is a basic tool for discussing the multipartite quantum correlations~\cite{Guo2021qst,Guo2022entropy,Guo2023pra,Guo2024rip}, we then list different kinds of MEMs that consisting of GEM, $k$-EM, $k$-partite entanglement measure ($k$-PEM), and partitewise entanglement measure, for which we mainly focus on the version 2.0 under the framework of complete measure of multipartite quantum correlation we established recently~\cite{Guo2020pra,Guo2021qst,Guo2022entropy,Guo2023pra,Guo2024pra,Guo2024rip}, respectively.


\subsection{Multipartite entanglement}\label{sec-6.1}


We denote by $A_1A_2\cdots A_n$ an $n$-partite quantum system and by $X_1|X_2| \cdots |X_{k}$ the $k$-partition of $A_1A_2\cdots A_n$ (or subsystem of $A_1A_2\cdots A_n$ sometimes), $k\leq  n$. For instance, partition $AB|C|DE$ is a $3$-partition of the 5-particle system $ABCDE$ with $X_1=AB$, $X_2=C$ and $X_3=DE$. The case of $k=n$ is just the original $n$-particle system without any other partition, namely, $A_1A_2\cdots A_n$ means $A_1|A_2|\cdots |A_n$. So $k< n$, $l<n$ in general unless otherwise specified. Hereafter, for simplicity, we let $|\psi\ra=|\psi\ra^{A_1A_2\cdots A_n}$ be a pure state in $\mH^{A_1A_2\cdots A_n}$ with arbitrary dimension, $\rho=\rho^{A_1A_2\cdots A_n}\in\mS^{A_1A_2\cdots A_n}$ and $h$ be some reduced function unless stated otherwise.

\subsubsection{$k$-entanglement}

A pure state $|\psi\ra$ in $\mH^{A_1A_2\cdots A_n}$ is said to be $k$-separable if
\beax 
|\psi\ra=|\psi\ra^{X_1}|\psi\ra^{X_2}\cdots|\psi\ra^{X_k}
\eeax 
for some $k$-partition of $A_1A_2\cdots A_n$, $2\leq  k\leq  n$. 
An $n$-partite mixed state $\rho$ is
$k$-separable if it can be written as a convex combination of
$k$-separable pure states
$\rho=\sum_{i}p_i|\psi_i\rangle \langle\psi_i|$, 
wherein the contained $\{|\psi_i\rangle\}$ can be $k$-separable with respect to different
$k$-partitions (i.e., a mixed $k$-separable state does not need to be separable with respect to any particular $k$-partition). $\rho$ is called fully separable if it is $n$-separable. If $\rho$ is not $2$-separable ($2$-separable is also called biseparable), then it is
called genuinely entangled.

\begin{figure}
	\begin{center}
		\includegraphics[width=70mm]{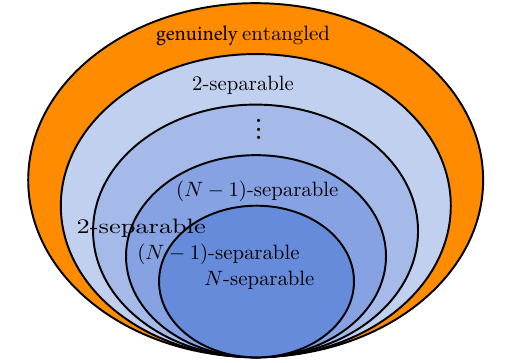}
	\end{center}
	\caption{\label{hierachyofS-k} Schematic picture of the $k$-separable states.
	}
\end{figure}

It is clear that whenever a state $\rho$ is $k$-separable, it is automatically also $l$-separable for all $1<l<k$. We denote the set of all $k$-separable states by $\mS_k$ ($k=2,3,\cdots,n$), then 
\beax
\mS_n\subsetneq\mS_{n-1}\subsetneq\cdots \subsetneq\mS_2\subsetneq\mS
\eeax (also see in Fig.~\ref{hierachyofS-k} which is bowrrowed from~\cite{Hedemann2018qic}. Hereafter we denote $\mS^{A_1A_2\cdots A_n}$ by $\mS$ for simplicity if the system is clear from the context). $\mS\backslash\mS_k$ is the set of all $k$-entangled states~\cite{Akulin2015pra,Gerke2016prl,Guo2022jpa,Guo2024pra} (It is also called $k$-nonseparable or $k$-inseparable in literature~\cite{Dur2000pra,Gao2014prl,Hong2012pra,Seevinck2008pra}. Note that, in Ref.~\cite{Szarek2010jpa}, $k$-entangled state refers to the one with Schmidt number not larger than $k$). $\mS\backslash\mS_2$ is just the set of all genuinely entangled (or equivalently, 2-entangled) states. Hereafter, we denote by $\mS_g$ the set of all genuinely entangled states, i.e., $\mS_g=\mS\backslash\mS_2$. 
If a
state is $k$-entangled, it is automatically also $j$-entangled for all $k<j\leq  n$ but not vice versa. Multipartite entanglement always exhibits counterintuitive phenomenons, e.g., it happens that a multipartite mixed state 
can be separable over any bipartite split but not fully separable~\cite{Bennett1999prl}.  $k$-entanglement here is different from the $\alpha$-entanglement based on the partitions labels of the first kind in Ref.~\cite{Szalay2015pra}, wherein $\alpha$-entangled state refers to the state is not separable under the given partition $\alpha$ of $A_1A_2\cdots A_n$ (In Ref.~\cite{Szalay2015pra}, the partition of $A_1A_2\cdots A_n$ is denoted by $\alpha$).

\subsubsection{$k$-partite entanglement}

Let $X_1|X_2|\cdots |X_m$ be an $m$-partition of $A_1A_2\cdots A_n$. We denote by $\varDelta(X_t)$ the number of subsystems contained in $X_t$, for instance, for the 3-partition $AB|C|DE$ of $ABCDE$, $\varDelta(X_1)=\varDelta(AB)=2$, $\varDelta(X_2)=\varDelta(C)=1$ and $\varDelta(X_3)=\varDelta(DE)=2$. If $\varDelta(X_t)\leq  k$ for any $1\leq  t\leq  m$, we call it a $k$-fineness partition. We denote by $\Gamma_{\!k}^f$ the set of all $k$-fineness partitions of some given system $A_1A_2\cdots A_n$.

\begin{figure}
	\centering
	\includegraphics[width=80mm]{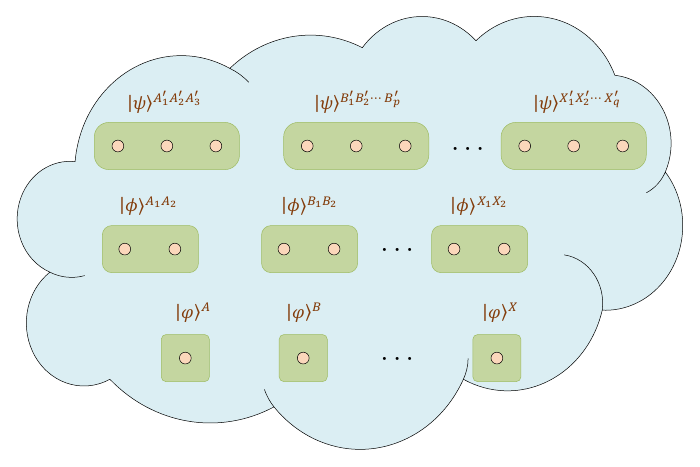}\\
	{\footnotesize (a)}
	\vspace{3mm}\\
	\includegraphics[width=56mm]{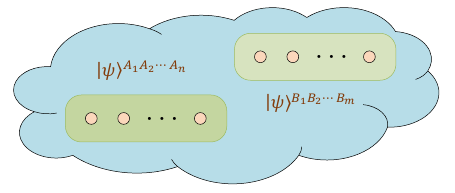}	\\
	{\footnotesize (b)}
	\vspace{2mm}
\fl	\caption{\label{fig12-1}(color online). (a) $3$-partite entangled pure state $|\Psi\ra=|\psi\ra^{A'_1A'_2A'_3}|\psi\ra^{B'_1B'_2\cdots B'_p} \cdots|\psi\ra^{X'_1X'_2\cdots X'_q}|\phi\ra^{A_1A_2}|\phi\ra^{B_1B_2}\cdots|\phi\ra^{X_1X_2}|\varphi\ra^A|\varphi\ra^B\cdots$ $|\varphi\ra^X$, where $|\psi\ra^{A'_1A'_2A'_3}$, $|\psi\ra^{B'_1B'_2\cdots B'_p}$, $\dots$, $|\psi\ra^{X'_1X'_2\cdots X'_q}$ are genuinely entangled states, $3\leqslant p\leqslant q$, $|\phi\ra^{A_1A_2}$, $|\phi\ra^{B_1B_2}$, $\dots$, $|\phi\ra^{X_1X_2}$ are entangled states. In fact, if one of $|\psi\ra^{A'_1A'_2A'_3}$, $|\psi\ra^{B'_1B'_2\cdots B'_p}$, $\dots$, $|\psi\ra^{X'_1X'_2\cdots X'_q}$ is genuinely entangled, $|\Psi\ra$ is also $3$-partite entangled. Here we just take the general form of a $3$-partite entangled pure state.
		(b) $|\Phi\ra=|\psi\ra^{A_1A_2\cdots A_k}|\psi\ra^{B_1B_2\cdots B_l}$ with $k$, $l\geqslant 0$, $k+l\geqslant3$, is a $3$-entangled pure state if one of the following is true: (i) $|\psi\ra^{A_1A_2\cdots A_k}$ and $|\psi\ra^{B_1B_2\cdots B_l}$ are genuinely entangled states, $k$, $l\geqslant3$, (ii)  $|\psi\ra^{A_1A_2\cdots A_k}$ and $|\psi\ra^{B_1B_2\cdots B_l}$ are entangled states, $k=l=2$, (iii) If $k=0$ or $l=0$, $|\Phi\ra$ is genuinely entangled.}
\end{figure}

G\"{u}hne \etal introduced the $k$-partite entanglement in Ref.~\cite{Guhne2005njp}.
A pure state $|\psi\ra$ is called $k$-producible ($1\leq  k\leq  n-1$), if it can be represented as
\beax
|\psi\rangle=|\psi\rangle^{X_1}|\psi\ra^{X_2}\cdots|\psi\ra^{X_m}
\eeax
under some $k$-fineness partition $X_1|X_2|\cdots |X_m$ of $A_1A_2\cdots A_n$.
For mixed state $\rho$, if it can be written as a convex combination of $k$-producible pure states,
i.e., $\rho=\sum\limits_ip_i|\psi_i\rangle\langle\psi_i|$
with $|\psi_i\ra$'s are $k$-producible, 
it is called $k$-producible, where the pure state $|\psi_i\rangle$'s might be $k$-producible in different $k$-fineness partitions. If a quantum state is not $k$-producible, it is termed $(k+1)$-partite entangled.
By definition, the $k$-partite entanglement is different from the $k$-entanglement in general, but they are equivalent only in some special cases. 
For instance, the $n$-partite entangled state is just the genuine
multipartite entangled state and the one-producible states coincide with the fully separable states.
If $|\psi\ra^{ABC}$ is a genuine entangled state, then $|\psi\ra^{ABC}|\psi\ra^{D}|\psi\ra^{E}|\psi\ra^F$ is four-separable and three-partite entangled state. Also note that, a state of which some reduced state of $m$ parties is genuinely entangled, contains $m$-partite entanglement, but not vice versa in general~\cite{Guhne2005njp}. For more clarity, we compare $3$-partite entangled pure state with $3$-entangled pure state in Fig.~\ref{fig12-1}.

A pure state $|\psi\rangle$ is said to be genuinely $k$-producible (or genuinely $k$-partite entangled) \cite{Guhne2005njp} if it is $k$-producible but not $(k\!\!-\!\!1)$-producible. A mixed state $\rho$ is genuinely $k$-producible if it is not only $k$-producible but also for any $k$-producible pure states ensemble of $\rho$, $\rho=\sum_ip_i|\psi_i\ra\la\psi_i|$, there is at least one $|\psi_i\rangle$ is genuinely $k$-producible.

Let $\mS_{P(k)}$ $(k=1,2,\ldots,n-1)$ denote the set of $k$-producible quantum states in $\mS^{A_1A_2\cdots A_n}$ and $\mS_{P(n)}=\mS$. It follows that
\beax 
\mS_{P(1)} \subset \mS_{P(2)}\subset\cdots\subset \mS_{P(n-1)}\subset \mS_{P(n)},
\eeax
$\mS\!\setminus\!\mS_{P(k)}$ is the set consisting of all $(k+1)$-partite entangled states, and $\mS_{P(k)}\!\!\setminus\!\mS_{P(k-1)}$ is the set of all genuinely $k$-producible states.

\subsubsection{Partitewise entanglement}

Very recently, another kind of entanglement was explored, which is termed the pairwise entanglement~\cite{Dong2024pra}.
It is pointed out that although $\rho^{AB}$ as a bipartite reduced state of the three-qubit GHZ state $|\GHZ\ra= 1/\sqrt{2}(|000\ra+|111\ra$ is separable, part $A$ and part $B$ are indeed entangled with each other without tracing out part $C$.

If an $n$-partite pure state $|\psi\ra$ can not be decomposed as $|\psi\ra^X|\psi\ra^Y$ with $A_1$ is included in $X$ and $A_2$ is included in $Y$, then $A_1$ and $A_2$ contain pairwise entanglement~\cite{Dong2024pra}.
The pairwise entanglement between $A$ and $B$ in the three-qubit pure state is quantified by~\cite{Dong2024pra} 
\beax \label{pwc}
C_{A'B'}=\sqrt{C^2(AB)+\tau_{ABC}},
\eeax
and called pairwise concurrence. It is clear that $C_{A'B'}>0$ whenever $|\psi\ra^{ABC}$ is the GHZ state.

Taking motivation from the pairwise entanglement in Ref.~\cite{Dong2024pra}, Guo and Yang~\cite{Guo2025pra}
proposed the concept of the $k$-partitewise entanglement which is reduced to the pairwise entanglement whenever $k=2$. 
Let $|\psi\ra$ be a pure state in $\mH^{A_1A_2\cdots A_n}$, $2\leqslant k <n$. If
\beax\label{k-pwsp}
|\psi\ra=|\psi\ra^{X_1}|\psi\ra^{X_2}\cdots|\psi\ra^{X_k}
\eeax
for some $k$-partition $X_1|X_2|\cdots|X_k$ such that $A_{i}$ belongs to $X_i$, $1\leqslant i\leqslant k$,
then we call $|\psi\ra$ is $k$-partitewise separable up to $A_{1}A_{2}\cdots A_{k}$(or $A_{1}A_{2}\cdots A_{k}$ is $k$-partitewise separable in $|\psi\ra$). Or else, it is $k$-partitewise entangled ($k$-PWE) up to $A_{1}A_{2}\cdots A_{k}$ (or $A_{1}A_{2}\cdots A_{k}$ is $k$-PWE in $|\psi\ra$). Especially, if $\rho^{A_{1}A_{2}\cdots A_{k}}=\tr_{A_{k+1}\cdots A_n}|\psi\ra\la\psi|$ is genuinely entangled, $k\geqslant 3$, we call $|\psi\ra$ is genuinely $k$-partitewise entangled (G$k$-PWE) up to $A_{1}A_{2}\cdots A_{k}$ (or $A_{1}A_{2}\cdots A_{k}$ is G$k$-PWE in $|\psi\ra$).

For example, the subsystem $A_{1}A_{2}\cdots A_{k}$ is $k$-PWE but not G$k$-PWE in $|{\GHZ}_{n,2}\ra$, while the subsystem $A_{1}A_{2}\cdots A_{k}$ is not only $k$-PWE but also G$k$-PWE in $|{\W}_{n,2}\ra$, $3\leqslant k<n$. If 
\bea\label{case3}
|\phi\ra=|\phi\ra^{A_{1}A_{2}\cdots A_{k}}|\phi\ra^{A_{k+1}\cdots A_n},
\eea  
then $|\phi\ra$ is $k$-PWE (resp. G$k$-PWE) with respect to $A_{1}A_{2}\cdots A_{k}$ iff $|\phi\ra^{A_{1}A_{2}\cdots A_{k}}$ is $k$-entangled (resp. genuinely entangled).

Let $\rho$ be a mixed state in $\mS^{A_1A_2\cdots A_n}$, $2\leqslant k <n$. If
\bea\label{k-pwsm}
\rho=\sum_ip_i|\psi_i\ra\la\psi_i|
\eea
for some ensemble $\{p_i, |\psi_i\ra\}$ with $|\psi_i\ra$ are $k$-partitewise separable up to $A_{1}A_{2}\cdots A_{k}$, we call $\rho$ is $k$-partitewise separable up to $A_{1}A_{2}\cdots A_{k}$ (or $A_{1}A_{2}\cdots A_{k}$ is $k$-partitewise separable in $\rho$). Or else, it is $k$-PWE up to $A_{1}A_{2}\cdots A_{k}$ (or $A_{1}A_{2}\cdots A_{k}$ is $k$-PWE in $\rho$).
The G$k$-PWE mixed state follows directly. Namely, if $\rho^{A_{1}A_{2}\cdots A_{k}}=\tr_{A_{k+1}\cdots A_n}\rho$ is genuinely entangled, we call $\rho$ is G$k$-PWE up to $A_{1}A_{2}\cdots A_{k}$ (or $A_{1}A_{2}\cdots A_{k}$ is G$k$-PWE in $\rho$). Note here that, in Eq.~\eqref{k-pwsm}, the $k$-partition corresponding to $|\psi_i\ra$ might be different from that of $|\psi_j\ra$ when $i\neq j$. It was shown in Ref.~\cite{Guo2025pra} that:
If $|\psi\ra\in\mH^{A_1A_2\cdots A_n}$ is $k$-partitewise separable with respect to $A_{1}A_{2}\cdots A_{k}$, $2\leqslant k <n$, then $\rho^{A_{1}A_{2}\cdots A_{k}}$ is a product state. If $\rho\in\mS^{A_1A_2\cdots A_n}$ is $k$-partitewise separable with respect to $A_{1}A_{2}\cdots A_{k}$, then $\rho^{A_{1}A_{2}\cdots A_{k}}$ is fully separable. In particular, if $\dim\mH^{A_i}\leqslant3$ for any $i$,  then $|\psi\ra\in\mH^{A_1A_2\cdots A_n}$ is $k$-partitewise separable with respect to $A_{1}A_{2}\cdots A_{k}$ iff $\rho^{A_{1}A_{2}\cdots A_{k}}$ is a product state, and $\rho\in\mS^{A_1A_2\cdots A_n}$ is $k$-partitewise separable with respect to $A_{1}A_{2}\cdots A_{k}$ iff $\rho^{A_{1}A_{2}\cdots A_{k}}$ is fully separable.

If $|\psi\ra^{A_1A_2\cdots A_k}$ in Eq.~\eqref{case3} is neither fully separable nor genuinely entangled, the $k$-partitewise entanglement therein is different from both that of $|{\GHZ}_{n,2}\ra$ and $|{\W}_{n,2}\ra$, i.e., there is no a larger genuine entangled system contains the considered reference subsystem $A_{1}A_{2}\cdots A_{k}$. In order to distinguish these different $k$-partitewise entanglement more distinctly, Guo and Yang gave the following definition~\cite{Guo2025pra}:
Let $|\psi\ra$ be a pure state in $\mH^{A_1A_2\cdots A_n}$, $2\leqslant k <n$. If it is $k$-PWE with respect to $A_{1}A_{2}\cdots A_{k}$ and, moreover, there exists a partition $X|Y$ such that 
\beax \label{sk-pwe}
|\psi\ra=|\psi\ra^X|\psi\ra^Y
\eeax 
with $A_1A_2\cdots A_k\subseteq X$ and $|\psi\ra^X$ is genuinely entangled, we call $|\psi\ra$ is strongly $k$-partitewise entangled (S$k$-PWE) with respect to $A_{1}A_{2}\cdots A_{k}$ (or $A_{1}A_{2}\cdots A_{k}$ is S$k$-PWE in $|\psi\ra$).
The S$k$-PWE mixed state can be defined straightforwardly. Hereafter, we only say a state is $k$-partitewise separable/entangled (or $A_{1}A_{2}\cdots A_{k}$ is $k$-partitewise separable/entangled) if the reference subsystem $A_{1}A_{2}\cdots A_{k}$ and the global state $|\psi\ra$ are clear from the context. The G$k$-PWE state and S$k$-PWE state follow in the same way.

By definitions, the pairwise entangled $|\psi\ra\in\mH^{ABC}$ with respect to $AB$ has three different cases: (i) $\rho^{AB}$ is entangled but $|\psi\ra$ is not strongly pairwise entangled, (ii) $\rho^{AB}$ is entangled and $|\psi\ra$ is strongly pairwise entangled, (iii) $\rho^{AB}$ is separable but $|\psi\ra$ is strongly pairwise entangled. But the $k$-partitewise entangled states can be divided into four classes ($k\geqslant3$): (i) Neither genuinely $k$-partitewise entangled nor strongly $k$-partitewise entangled; (ii) Genuinely $k$-partitewise entangled but not strongly $k$-partitewise entangled; (iii) Strongly $k$-partitewise entangled but not genuinely $k$-partitewise entangled; (iv) Not only genuinely $k$-partitewise entangled but also strongly $k$-partitewise entangled.


\subsection{Coarsening relation of multipartite partition} \label{sec-6.2}


Let $X_1|X_2| \cdots |X_{k}$ and $Y_1|Y_2| \cdots |Y_{l}$ be two partitions of $A_1A_2\cdots A_n$ or subsystem of $A_1A_2\cdots A_n$, $k\leq  n$, $l\leq  n$. We denote by~\cite{Guo2022entropy,Guo2024pra}
\bea
X_1|X_2| \cdots| X_{k}\succ^a Y_1|Y_2| \cdots |Y_{l}, \label{succ-a}\\
X_1|X_2| \cdots| X_{k}\succ^b Y_1|Y_2| \cdots |Y_{l},\label{succ-b}\\
X_1|X_2| \cdots| X_{k}\succ^c Y_1|Y_2| \cdots |Y_{l}~\label{succ-c}
\eea 
if $Y_1|Y_2| \cdots |Y_{l}$ can be obtained from $X_1|X_2| \cdots| X_{k}$
by 
\begin{itemize}
	\item[(a)] discarding some subsystem(s) of $X_1|X_2| \cdots| X_{k}$,
	\item[(b)] combining some subsystems of $X_1|X_2| \cdots| X_{k}$,
	\item[(c)] discarding some subsystem(s) of some subsystem(s) $X_t$ provided that $X_{t}=A_{t(1)}A_{t(2)}\cdots A_{t(f(t))}$ with $f(t)\geq2$, $1\leq  t\leq  k$,
\end{itemize}
respectively. For example,
\beax 
&&A|B|C|D\succ^a A|B|D\succ^a B|D,\\
&&A|B|C|D\succ^b AC|B|D\succ^b AC|BD, \\
&&A|BC\succ^c A|B.
\eeax
We call $Y_1|Y_2| \cdots |Y_{l}$ is coarser than $X_1|X_2| \cdots| X_{k}$ if 
$Y_1|Y_2| \cdots |Y_{l}$ can be obtained from $X_1|X_2| \cdots| X_{k}$
by one or some of the ways in item (a)-item (c), and we denote it by 
\bea \label{succ}
X_1|X_2| \cdots| X_{k}\succ Y_1|Y_2| \cdots |Y_{l}
\eea 
uniformally.

Furthermore, if $X_1|X_2| \cdots| X_{k}\succ Y_1|Y_2| \cdots |Y_{l}$,
we denote by 
\bea \label{Xi}
\Xi(X_1|X_2| \cdots| X_{k}- Y_1|Y_2| \cdots |Y_{l})
\eea
the set of
all the partitions that are coarser than $X_1|X_2| \cdots| X_{k}$ but (i) neither coarser than $Y_1|Y_2| \cdots |Y_{l}$ nor the one from which one can derive $Y_1|Y_2| \cdots |Y_{l}$ by the coarsening means, and (ii) if it includes some or all subsystems of $Y_1|Y_2| \cdots |Y_{l}$, then all the subsystems $Y_j$'s included are regarded as one subsystem and (iii) if $Y_1|Y_2| \cdots |Y_{l}=X_1|X_2| \cdots|X_{l-1}|X_{l}\cdots X_{k}$, $\Xi(X_1|X_2| \cdots| X_{k}- Y_1|Y_2| \cdots |Y_{l})$ contains only $X_{l}|\cdots| X_{k}$ and the one coarser than it.
We call $\Xi(X_1|X_2| \cdots| X_{k}- Y_1|Y_2| \cdots |Y_{l})$ is the complementarity of $Y_1|Y_2| \cdots |Y_{l}$ up to $X_1|X_2| \cdots| X_{k}$. 
For example~\cite{explain1}, 
\beax 
\fl \qquad \Xi(A|B|CD|E-A|B)=&\{ A|CD|E, A|CDE, ACD|E, AE|CD, A|C|E, A|D|E,\\  
&AE|C, A|CE, AC|E, A|DE, AE|D, AD|E, A|CD,\\ 
&CD|E, A|C, A|D, A|E, C|E, D|E,  
B|CD|E, B|CDE,\\
& BCD|E, BE|CD, B|D|E, B|C|E,
BE|C,  B|CE, \\
& BC|E,  B|DE, BE|D, BD|E, B|CD,  B|C, B|D, B|E, \\
&  AB|CD|E, AB|C|E, AB|D|E, 
AB|CDE, ABCD|E, \\
& ABE|CD, AB|CD, AB|CE, AB|DE, 
ABE|C,\\
& ABE|D, ABC|E, ABD|E, 
AB|C, AB|D,  AB|E\},
\eeax 
\vspace{-6.5mm}
\beax 
\fl \qquad \Xi(A|B|C|D|E-A|B|C)
=&\{A|D|E, AD|E, AE|D, A|DE, A|D, A|E, D|E,\\
&  B|D|E, B|DE, BD|E, BE|D, B|D, B|E, C|D|E,\\
& C|DE, CD|E, CE|D, C|D, C|E, AB|D|E, ABD|E, \\
&ABE|D, AB|DE, AB|D, AB|E, AC|D|E, ACD|E, \\
& ACE|D, AC|DE, AC|D, AC|E,  BC|D|E, BCD|E,  \\
&BCE|D, BC|DE, BC|D, BC|E, ABC|D|E, \\
& ABCD|E, ABCD|E, ABCE|D, ABC|D, ABC|E\},
\eeax 
\vspace{-6.5mm}
\beax 
\fl \qquad \Xi(A|B|C|D-A|BCD)=\{B|C|D, B|CD, BC|D, C|BD, B|C,
C|D, B|D\}.
\eeax

Note here that, the coarser relation ``$\succ^b$'' is just the coarser relation ``$\preceq$'' in Ref.~\cite{Szalay2015pra}, namely, ``$\preceq$'' therein only refers to the coarser relation of type (b) and with the opposite direction.


\subsection{MEM} \label{sec-6.3}


A function 
$E^{(n)}: \mS^{A_1A_2\cdots A_n}\to\bb{R}^{+}$ 
is called an $n$-partite entanglement measure if
it satisfies (E1) $E^{(n)}(\rho)=0$ if $\rho$ is fully separable, and
(E2) $E^{(n)}$ cannot increase under $n$-partite LOCC, i.e., $E^{(n)}(\rho)\geq E^{(n)}(\rho')$ for any $n$-partite LOCC $\varepsilon$, $\varepsilon(\rho)=\rho'$. Item (E2) implies that MEM is locally unitary invariant, i.e., $E^{(n)}(\rho)=E^{(n)}(U_1\ot U_2\ot\cdots\ot U_n\rho U_1^\dagger\ot U_2^\dagger\ot\cdots\ot U_n^\dagger)$ for any local unitary operator $U_i$ acting on $\mH^{A_i}$, $i=1$, 2, $\dots$, $n$. An $n$-partite entanglement measure $E^{(n)}$ is said to be 
an $n$-partite entanglement monotone [or multipartite entanglement monotone (MEMo)] if it is convex and 
does not increase on average under $n$-partite stochastic LOCC. Let $\mS_{*\text{-}sep}$ be the set of all states that are separable with respect to some given kind of multipartite entanglement (for example, for $k$-entanglement and $k$-partite entanglement, they are $k$-separable and $k$-partite separable, respectively, also see below). If $E^{(n)}: \mS^{A_1A_2\cdots A_n}\to\bb{R}^{+}$ satisfies (E2) and
(E1$'$) $E^{(n)}(\rho)=0$ iff $\rho\in\mS_{*\text{-}sep}$ is separable 
it is called a faithful MEM (with respect to the corresponding kind of entanglement). 

Throughout this paper, for measure $E$ of any kind of entanglement, if it is convex, i.e., $E(\sum_ip_i\rho_i)\leq  \sum_ip_iE(\rho_i)$ for any $\{p_i,\rho_i\}$, $\rho_i\in\mathcal{S}$, $p_i>0$, $\sum_ip_i=1$, and does not increase on average under LOCC, we call it a monotone of this kind of entanglement. We denote by *Mo the later one, if the former one by *M (* denotes the associated kind of entanglement), e.g., MEMo \& MEM is such a case. In addition, as that of (E1) and (E1$'$), the later one corresponding to the measure is faithful while the former one is not necessarily faithful. Straightforwardly, any *Mo must be a *M. 
$E^{(k)}$'s are collectively called multipartite entanglement measures (MEMs) for any $2\leq k\geq n$. 
$E^{(n)}$ is called a global MEM if it quantifies all the entanglement (namely, the global entanglement) contained in the state.

An MEM $E^{(n)}$ is called a ``proper'' MEM if 
\beax 
E^{(n)}(|{\GHZ}_{n,2}\ra)>E^{(n)}(|{\W}_{n,2}\ra)
\eeax  
according to Ref.~\cite{Xie2021prl}, where 
\bea\label{GHZn2}
|{\GHZ}_{n,2}\ra=\frac{1}{\sqrt2}(|0\ra^{\ot n}+|1\ra^{\ot n})
\eea
is the $n$-qubit GHZ state, due to the fact that the GHZ state can faithfully teleport an
arbitrary single-qubit quantum state while the W state is
relatively less capable~\cite{Joo2003njp}.

Throughout this paper, for any $\rho$ and any given $k$-partition $X_1|X_2|\cdots|X_k$
	of $A_1A_2\cdots A_n$ or some subsystem of $A_1A_2\cdots A_n$,
	we denote by $\rho^{X_1|X_2|\cdots|X_k}$ the state for which we consider it as a $k$-partite state
	with respect to the partition $X_1|X_2|\cdots|X_k$. For example, in the four partite case, $\rho^{A|CD}$ denotes state which we regard it as a bipartite state with respect to the partition $A|CD$, where $\rho^{ACD}=\tr_B\rho^{ABCD}$ for some 
	$\rho^{ABCD}\in\mS^{ABCD}$. In general, one can clearly know which is the associated $\rho^{ABCD}$ from the context. For any kind of MEM $\left\lbrace E^{(k)}\right\rbrace$, hereafter, $E^{(k)}(X_1|X_2|\cdots|X_{p})$ for $\rho\in\mS^{A_1A_2\cdots A_n}$ means that $E^{(k)}(X_1|X_2| \cdots| X_{p})=E^{(k)}(\rho^{X_1|X_2| \cdots| X_{p}})$ with $\rho^{X_1|X_2| \cdots| X_{p}}$
	is the state with respect to the $k$-partition $X_1|X_2| \cdots| X_{p}$ of $A_1A_2\cdots A_n$ or some subsystem of $A_1A_2\cdots A_n$, the vertical bar indicates the split 
	across which the entanglement is measured. For example, $E^{(3)}(A|CD|E|F)$ for $\rho^{ABCDEF}$ means $E^{(3)}(\rho^{A|CD|E|F})$ with $\rho^{ACDEF}=\tr_B(\rho^{ABCDEF})$.

\subsubsection{Complete global MEM}

An MEM $E^{(n)}$ is called a {unified}
global MEM (U-g-MEM) if it satisfies the unification condition~\cite{Guo2020pra,Guo2024rip} (it is called unified MEM in Ref.~\cite{Guo2020pra,Guo2024rip}, but the term ``unified global MEM'' is more precise indeed):
\begin{itemize}
	\item (additivity): 
	\bea \label{additivity}
	&&E^{(n)}(A_1A_2\cdots A_k\ot{A_{k+1}\cdots A_n})\nonumber
	\\
	&=&E^{(k)}({A_1A_2\cdots A_k})+E^{(n-k)}({A_{k+1}\cdots A_n}),
	\eea 
	holds for all $\rho^{A_1A_2\cdots A_k}\ot\rho^{A_{k+1}\cdots A_n}\in\mS^{A_1A_2\cdots A_n}$, hereafter $E^{(n)}(X)$ refers to $E^{(n)}(\rho^X)$ and $E^{(1)}=0$;
	\item (symmetry): $E^{(n)}({A_1A_2\cdots A_n})=E^{(n)}({A_{\pi(1)}A_{\pi(2)}\cdots A_{\pi(n)}})$,
	for all $\rho\in\mS^{A_1A_2\cdots A_n}$ and any permutation $\pi$;
	\item (coarsening monotone): 
	\bea\label{coarsen}
	E^{(k)}(X_1|X_2| \cdots| X_{k})\geq E^{(l)}(Y_1|Y_2| \cdots |Y_{l})
	\eea
	holds for all $\rho\in\mS^{A_1A_2\cdots A_n}$ whenever $X_1|X_2| \cdots| X_{k}\succ^a Y_1|Y_2| \cdots |Y_{l}$,
	where $X_1|X_2| \cdots |X_{k}$ and $Y_1|Y_2| \cdots |Y_{l}$ are two partitions of $A_1A_2\cdots A_n$ or subsystem of $A_1A_2\cdots A_n$.
\end{itemize}
$E^{(n)}$ is called a {complete}
global MEM (C-g-MEM) if it satisfies both the unification condition above and the hierarchy condition~\cite{Guo2020pra}: 
\begin{itemize}
	\item (tight coarsening monotone): Eq.~\eref{coarsen} 
	holds for all $\rho\in\mS^{A_1A_2\cdots A_n}$ whenever $X_1|X_2|\cdots|X_{k}\succ^b Y_1|Y_2|\cdots |Y_{l}$.
\end{itemize}

\begin{figure}
\begin{center}
	\includegraphics[width=55mm]{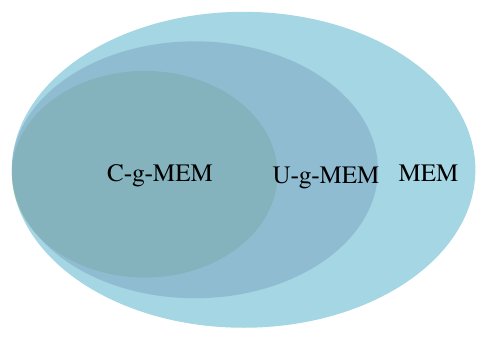}
\end{center}
\caption{\label{hierachyofmeasures} Schematic picture of the sets of MEMs, U-g-MEMs, and C-g-MEMs.
}
\end{figure}

By definition, unified MEM $E^{(n)}$ in fact refers to a family of measures $\{E^{(k)}: 2\leq k\leq n\}$.
There are MEMs that are not unified, and unified global MEMs that are not complete (see in Sec.~\ref{sec-9}-Sec.~\ref{sec-14}).
The relation between these 
MEMs are depicted in Fig.~\ref{hierachyofmeasures}.

\subsubsection{What is the additivity of an entanglement measure?}

We always expect intuitively that a measure of bipartite
entanglement should be additive in the sense of~\cite{Vollbrecht}
\bea\label{suppl-additive}
E\left( \rho^{AB}\otimes\sigma^{A'B'}\right) =E\left( \rho^{AB}\right) +E\left( \sigma^{A'B'}\right) ,
\eea
where $E(\rho^{AB}\otimes\sigma^{A'B'})$ is quantified under the partition $AA'|BB'$.
Eq.~\eref{suppl-additive} means that, 
from the resource-based point of view, sharing two particles from the
same preparing device is exactly ``twice as useful'' to Alice
and Bob as having just one. By now, we know that
the squashed entanglement~\cite{Christandl2004jmp} and the
conditional entanglement of mutual information~\cite{Yang2008prl}
are additive. 
Recall that, the additivity of the entanglement formation $E_f^{(2)}$ is a long
standing open problem which is conjectured to be
true~\cite{Plenio2007qic} and then disproved 
by Hastings in 2009~\cite{Hastings}.
Although EoF is not additive for all states, construction of additive states for
EoF is highly expected~\cite{Zhao2019pra}.
In~Ref.~\cite{Guo2020pra}, a new class of states such that $E_{f}^{(2)}$ 
is additive (and thus for such class of states, $E_f^{(2)}=E_c$~\cite{Plenio2007qic}) was presented.
Let $\rho^{AB}\otimes\sigma^{A'B'}$ be a state in $\mS^{AA'BB'}$.
If there exists an optimal ensemble $\{p_i,|\psi_i\ra^{AA'BB'}\}$ for $E_f$ [i.e., $E_f(\rho^{AB}\otimes\sigma^{A'B'} )=\sum_ip_iE(|\psi_i\ra^{AA'BB'} )  $]
such that any pure state $|\psi_i\ra^{AA'BB'}$ is a product state, i.e.,
$|\psi_i\ra^{AA'BB'}=|\phi_i\ra^{AB}|\varphi_i\ra^{A'B'}$ for some pure state
$|\phi_i\ra^{AB}\in\mH^{AB}$ and $|\varphi_i\ra^{A'B'}\in\mH^{A'B'}$,
then we have~\cite{Guo2020pra}
\beax\label{additivityEoF}
E_{f}\left( AB\ot A'B'\right) =E_{f}\left( AB\right) 
+E_{f}\left( {A'B'}\right) .
\eeax
Particularly, if $\rho^{AB}$ or $\sigma^{A'B'}$ 
is pure, then $\rho^{AB}\otimes\sigma^{A'B'}$ is additive under $E_f$.
Together with the result of Hastings in Ref.~\cite{Hastings},
we conclude that, the state $\rho^{AB}\otimes\sigma^{A'B'}$
that violates the additivity~\eref{suppl-additive} definitely have an optimal pure-state decomposition
in which some pure states are not product state up to the partition $AB|A'B'$.
The approach in Ref.~\cite{Guo2020pra} is far different from that of Ref.~\cite{Zhao2019pra}, 
in which it is shown that, if a state with range in the entanglement-breaking space is always additive.

We can now conclude that the additivity of bipartite entanglement measure is not universal. By the definition of the U-g-MEM, the additivity in Eq.~\eref{additivity} is an intrinsic attribute of U-g-MEM. So we proposed the following definition of additivity of an MEM $E^{(n)}$.

\begin{definition}
An MEM $E^{(n)}$ is called additive if it satisfies the additive condition in Eq.~\eref{additivity}.
\end{definition}

That is, any unified global MEM is additive.
But there are many MEMs are not additive (see in Sec.~\ref{sec-9}-Sec.~\ref{sec-14}).

\subsubsection{Complete GEM}

An MEM $E_g^{(n)}: \mS^{A_1A_2\cdots A_n}\to\bb{R}^{+}$ is defined to be a GEM if it admits the following conditions~\cite{Ma2011pra}:
\begin{itemize}
\item (GE1) $E_g^{(n)}(\rho)=0$ if $\rho\in\mS_2$.
\item (GE2) $E_g^{(n)}$ is convex.
\end{itemize}
In Ref.~\cite{Ma2011pra}, GEM need to be faithful. But there exist non-faithful GEMs, so we do not require $E_g^{(n)}(\rho)>0$ if $\rho\notin\mS_2$ necessarily. For example, the ``three tangle''~\cite{Coffman2000pra} is not faithful since it is vanished for the $W$ state.

Let $E_g^{(n)}$ be a GEM. It is defined to be a {unified} GEM if it satisfies the {unification condition of GEM}~\cite{Guo2022entropy}, i.e., 
\begin{itemize}
\item (symmetry): $E_g^{(n)}({A_1A_2\cdots A_n})=E_g^{(n)}({A_{\pi(1)}A_{\pi(2)}\cdots A_{\pi(n)}})$,
for any $\rho\in\mS$ and any permutation $\pi$;
\item (coarsening monotone): 
\bea\label{gcoarsen}
E_g^{(k)}(X_1|X_2|\cdots|X_{k})> E_g^{(l)}(Y_1|Y_2|\cdots|Y_{l})
\eea
holds for all $\rho\in\mS_g$ whenever $X_1|X_2| \cdots| X_{k}\succ^a Y_1|Y_2| \cdots |Y_{l}$.
\end{itemize}
A unified GEM $E_g^{(n)}$ is call a {complete} GEM 
if $E^{(n)}_{g}$ admits the {hierarchy condition of GEM}~\cite{Guo2022entropy}, i.e.,
\begin{itemize}
\item (tight coarsening monotone): 
\bea\label{ghierarchy}
E_g^{(k)}(X_1|X_2|\cdots|X_{k})\geq E_g^{(l)}(Y_1|Y_2| \cdots |Y_{l})
\eea 
holds for all $\rho\in\mS_g$ whenever $X_1|X_2|\cdots|X_{k}\succ^b Y_1|Y_2|\cdots |Y_{l}$.
\end{itemize}

\subsubsection{Complete $k$-entanglement measure}

An MEM $E_{k}: \mS^{A_1A_2\cdots A_n}\to\bb{R}^{+}$ is called a $k$-entanglement measure if $E_{k}(\rho)=0$ for any $k$-separable $\rho\in\mS_k$. As the CMEM and CGEM above, when we deal with the multipartite entanglement,
there are two steps to reveal such a completeness of a given measure. The first step is the unification condition, and the second one is the hierarchy condition. The unification condition is mainly related to the coarsen relation of type (a) while the hierarchy condition is corresponding to the coarsen relation of type (b). 
With the same spirit, the definitions of the unified $k$-EM and the complete $k$-EM were proposed in Ref.~\cite{Guo2024pra}.
A $k$-EM $E_k$ is called {unified} if it satisfies the {{unification condition of $k$-EM}}~\cite{Guo2024pra}:
\begin{itemize}
\item (symmetry) $E_k({A_1A_2\cdots A_n})=E_k({A_{\pi(1)}A_{\pi(2)}\cdots A_{\pi(n)}})$
for all $\rho\in\mS^{A_1A_2\cdots A_n}$ and any permutation $\pi$ of $\{1, 2, \cdots, n\}$;
\item ($k$-monotone)
\bea\label{k-monotone}
E_k(A_1A_2\cdots A_n)\geq E_{k-1}(A_1A_2\cdots A_n)
\eea
holds for all $\rho\in\mS^{A_1A_2\cdots A_n}$, $k\geq3$;
\item (coarsening monotone)
\bea\label{k-coarsen}
E_k(X_1|X_2| \cdots| X_{p})\geq E_l(Y_1|Y_2| \cdots |Y_{q})
\eea
holds for all $k$-entangled state $\rho\in\mS^{X_1X_2\cdots X_p}$ whenever $X_1|X_2| \cdots| X_{p}\succ^a Y_1|Y_2| \cdots |Y_{q}$ with $l\leq  q\leq  p$ and $l\leq  k\leq  p$.
\end{itemize}
Eq.~\eref{k-coarsen} is always true for any $k$-EM. If a $k$-EM $E_k$ obeys Eq.~\eref{k-monotone} and Eq.~\eref{k-coarsen}, it is called {$k$-monotonic} and {coarsening monotonic}, respectively. Note that, Eq.~\eref{k-coarsen} does not hold for $k$-separable state $\rho\in\mS_k$
in general. For example, $E_3\left(|\psi\ra^{AB}|\psi\ra^{C}|\psi\ra^D\right)=0$, but $E_3\left( |\psi\ra^{AB}|\psi\ra^{C}\ra\right)>0$ if $|\psi\ra^{AB}$ is entangled generally.

A unified $k$-EM $E_k$ is called {complete} if it satisfies the {hierarchy condition of $k$-EM} additionally~\cite{Guo2024pra}: 
\begin{itemize}
\item (tight coarsening monotone)
\bea\label{k-t-coarsen}
E_k(X_1|X_2| \cdots| X_{p})\geq E_l(Y_1|Y_2| \cdots |Y_{q})
\eea
holds for all $k$-entangled state $\rho\in\mS^{X_1X_2\cdots X_p}$ whenever $X_1|X_2|\cdots|X_{p}\succ^b Y_1|Y_2|\cdots |Y_{q}$ with $l\leq  q\leq  p$ and $l\leq  k\leq  p$.
\end{itemize}
For $k$-separable state $\rho\in\mS_k$, Eq.~\eref{k-t-coarsen} fails in general.
If a $k$-EM $E_k$ satisfies Eq.~\eref{k-t-coarsen}, we call it is {tightly coarsening monotonic}.
For any given $k$-EM $E_k$,   
\bea\label{hierachy3}
E_k(X_1|X_2|\cdots|X_{p})\geq E_k(X'_1|X'_2| \cdots |X'_{p})
\eea 
holds for any $\rho\in\mS^{A_1A_2\cdots A_n}$ whenever $X_1|X_2| \cdots| X_{p}\succ^c X'_1|X'_2| \cdots |X'_{p}$ since $\rho^{X'_1|X'_2| \cdots |X'_{p}}$ is obtained from $\rho^{X_1|X_2| \cdots| X_{p}}$
by partial trace and such a partial trace is indeed a $p$-partite LOCC, $2\leq  k\leq  p< n$.

\subsubsection{Complete $k$-partite entanglement measure}

An MEM $E_{(k)}:\mS^{A_1A_2\cdots A_n}\longrightarrow\bb{R}_+$ is called a $k$-partite entanglement measure if $E_{(k)}(\rho)=0$ for any $\rho\in \mS_{P(k-1)}$. A $k$-PEM/$k$-PEMo $E_{(k)}$ is called a genuine $k$-PEM/$k$-PEMo if $E_{(k)}(\rho)>0$ but $E_{(k+1)}(\rho)=0$ for any genuinely $k$-partite entangled state $\rho$.

In Ref.~\cite{Guo2025qip}, we presented the definitions of the unified $k$-PEM and the complete $k$-PEM based on the coarsening relation of the partitions of the system.
A $k$-PEM $E_{(k)}$ is called {unified} if it satisfies the unification condition: 
\begin{itemize}
	\item (i) (symmetry) $E_{(k)}({A_1A_2\cdots A_n})=E_{(k)}({A_{\pi(1)}A_{\pi(2)}\cdots A_{\pi(n)}})$ for all $\rho\in\mS^{A_1A_2\cdots A_n}$ and any permutation $\pi$ of $\{1, 2, \cdots, n\}$; 
	\item (ii) (additivity) $E_{(k)}(A_1A_2\cdots A_r\ot A_{r+1}A_{r+2}\cdots A_n)=E_{(k)}(A_1A_2\cdots A_r)+E_{(k)}(A_{r+1}A_{r+2}\cdots A_n)$ holds for all $\rho^{A_1A_2\cdots A_r}\ot\rho^{A_{r+1}A_{r+2}\cdots A_n}$; 
	\item (iii) ($k$-monotone)
	\bea\label{k-p-monotone}
	E_{(k)}(A_1A_2\cdots A_n)\leqslant E_{(k-1)}(A_1A_2\cdots A_n)
	\eea
	holds for all $\rho\in\mS^{A_1A_2\cdots A_n}$, $k\geqslant3$; 
	\item (iv) (coarsening monotone)
	\bea\label{k-p-coarsen}
	E_{(k)}(X_1|X_2| \cdots| X_{p})\geqslant E_{(k)}(Y_1|Y_2| \cdots |Y_{q})
	\eea
	holds for all states $\rho\in\mS^{A_1A_2\cdots A_n}$ whenever $X_1|X_2| \cdots| X_{p}\succ^a Y_1|Y_2| \cdots |Y_{q}$ with $k\leqslant q\leqslant p$. 
\end{itemize}
If a $k$-PEM $E_{(k)}$ obeys Eq.~\eqref{k-p-monotone} and Eq.~\eqref{k-p-coarsen}, we call it is {$k$-monotonic} and {coarsening monotonic}, respectively.

A unified $k$-PEM $E_{(k)}$ is called {complete} if it satisfies the hierarchy condition additionally: (v) (tight coarsening monotone) 
\bea\label{w-k-t-coarsen}
E_{(k)}(A_1A_2\cdots A_n)\geqslant E_{(k)}(Y_1|Y_2| \cdots |Y_{q})
\eea
holds for all state $\rho\in\mS^{A_1A_2\cdots A_n}$ whenever $A_1A_2\cdots A_n\succ^b Y_1|Y_2|\cdots |Y_{q}$ such that, for any $i$, either $\rho^{Y_i}$ is pure or $\rho^{Y_i}$ is the reduced state of some genuinely entangled pure state (or entangled bipartite pure state), $1\leqslant i\leqslant q<n$. If a $k$-PEM $E_{(k)}$ satisfies Eq.~\eref{w-k-t-coarsen}, we call it is {tightly coarsening monotonic}. One need note here that, for any given $k$-PEM $E_{(k)}$,   
\beax\label{p-hierachy3}
E_{(k)}(X_1|X_2|\cdots|X_{p})\geqslant E_{(k)}(X'_1|X'_2| \cdots |X'_{p})
\eeax 
holds for any $\rho\in\mS^{A_1A_2\cdots A_n}$ whenever $X_1|X_2| \cdots| X_{p}\succ^c X'_1|X'_2| \cdots |X'_{p}$ since $\rho^{X'_1|X'_2| \cdots |X'_{p}}$ is obtained from $\rho^{X_1|X_2| \cdots| X_{p}}$ by a partial trace and such a partial trace is indeed a $p$-partite LOCC, $1\leqslant k\leqslant p< n$.

\subsubsection{Partitewise entanglement measure}

An MEM $\check{E}^{A_1A_2}:\mS^{A_1A_2\cdots A_n}\longrightarrow\bb{R}^+$ is called a pairwise entanglement measure (PWEM) up to $A_1A_2$ if $\check{E}^{A_1A_2}(\rho)=0$ whenever $\rho$ is pairwise separable up to $A_1A_2$~\cite{Dong2024pra}.
An MEM $\check{E}^{A_{1}A_{2}\cdots A_{k}}:\mS^{A_1A_2\cdots A_n}\longrightarrow\bb{R}^+$ ($k\geqslant 2$) is called a $k$-partitewise entanglement measure ($k$-PWEM) with respect to $A_{1}A_{2}\cdots A_{k}$ if $\check{E}^{A_{1}A_{2}\cdots A_{k}}(\rho)=0$ whenever $\rho$ is $k$-partitewise separable up to $A_{1}A_{2}\cdots A_{k}$~\cite{Guo2025pra}.

The case of $k=2$ is just the PWEM. If a state $\rho$ is $k$-PWE, it is neither necessarily G$k$-PWE, nor necessarily S$k$-PWE in general. Consequently, Guo and yang gave the measures of genuine $k$-partitewise entanglement and strong $k$-partitewise entanglement respectively. An MEM $\check{E}_g^{A_{1}A_{2}\cdots A_{k}}:\mS^{A_1A_2\cdots A_n}\longrightarrow\bb{R}^+$ ($k\geqslant 3$) is called a genuine $k$-partitewise entanglement measure (G$k$-PWEM) with respect to $A_{1}A_{2}\cdots A_{k}$ if $\check{E}_g^{A_{1}A_{2}\cdots A_{k}}(\rho)=0$ whenever $\rho^{A_1A_2\cdots A_k}$ is not genuinely entangled~\cite{Guo2025pra}.
An MEM $\check{E}_s^{A_{1}A_{2}\cdots A_{k}}:\mS^{A_1A_2\cdots A_n}\longrightarrow\bb{R}^+$ is called a strong $k$-partitewise entanglement measure (S$k$-PWEM) with respect to $A_{1}A_{2}\cdots A_{k}$ if $\check{E}_s^{A_{1}A_{2}\cdots A_{k}}(\rho)=0$ provided that $\rho$ is not S$k$-PWE~\cite{Guo2025pra}.


\section{Complete monogamy of the MEM: basic concepts}\label{sec-7}


In this section, we review the complete monogamy of the ``global'' multipartite entanglement, the genuine entanglement, and the $k$-entanglement under the framework of complete monogamy of quantum correlation
established recently in~\cite{Guo2020pra,Guo2021qst,Guo2022entropy,Guo2023pra,Guo2024pra,Guo2024rip}, respectively. The monogamy relation of $k$-entanglement would also be discussed. This is the first kind of MEM that is involved with both monogamy and the complete monogamy.


\subsection{Complete monogamy of the unified global MEM}\label{sec-7.1}


In Ref.~\cite{Guo2020pra}, in order to characterize the distribution of entanglement in a ``complete'' sense, the term ``complete monogamy'' of the unified global MEM was proposed. For a unified global MEM $E^{(n)}$, it is said to be {{completely monogamous}} if for any
$\rho\in\mathcal{S}^{A_1A_2\cdots A_n}$ that satisfies~\cite{Guo2020pra}
\beax\label{cond2}
E^{(k)}(X_1|X_2| \cdots| X_{k})= E^{(l)}(Y_1|Y_2| \cdots |Y_{l})
\eeax
with $X_1|X_2| \cdots| X_{k}\succ^a Y_1|Y_2| \cdots |Y_{l}$ we have that
\beax\label{cond2x}
E^{(\ast)}({\gamma}) =0
\eeax
holds for all $\gamma\in \Xi(X_1|X_2| \cdots| X_{k}- Y_1|Y_2| \cdots |Y_{l})$, hereafter the superscript $(\ast)$ is associated with the partition $\gamma$, e.g., if $\gamma$ is an $m$-partite partition, then $(\ast)=(m)$. 
For example, $E^{(3)}$ is completely monogamous if for any $\rho^{ABC}$ that admits $E^{(3)}(ABC)=E^{(2)}(AB)$ we get $E^{(2)}(AC)=E^{(2)}(BC)=0$. Let $E^{(n)}$ be a complete global MEM. $E^{(n)}$ is defined to be
{tightly complete monogamous} if for any $\rho\in\mathcal{S}^{A_1A_2\cdots A_n}$ that satisfies~\cite{Guo2020pra}
\beax\label{cond3}
E^{(k)}(X_1|X_2| \cdots| X_{k})= E^{(l)}(Y_1|Y_2| \cdots |Y_{l})
\eeax
with $X_1|X_2| \cdots| X_{k}\succ^b Y_1|Y_2| \cdots |Y_{l}$ we have that
\beax\label{cond3x}
E^{(\ast)}({\gamma}) =0
\eeax
holds for all $\gamma\in \Xi(X_1|X_2| \cdots| X_{k}- Y_1|Y_2| \cdots |Y_{l})$. For instance, $E^{(3)}$ is tightly complete monogamous if for any $\rho^{ABC}$ that admits $E^{(3)}(ABC)=E^{(2)}(A|BC)$ we have $E^{(2)}(BC)=0$.

\begin{figure}
	\begin{center}
		\includegraphics[width=84mm]{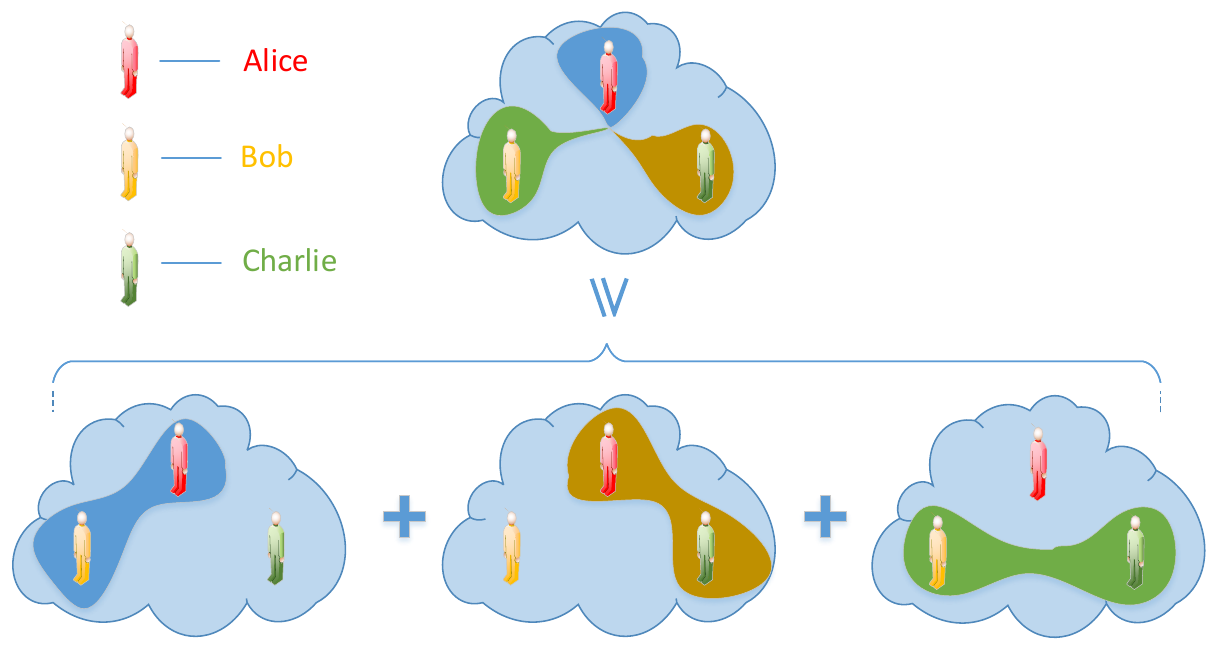}
		
		{\footnotesize (a)}
		
		\includegraphics[width=85mm]{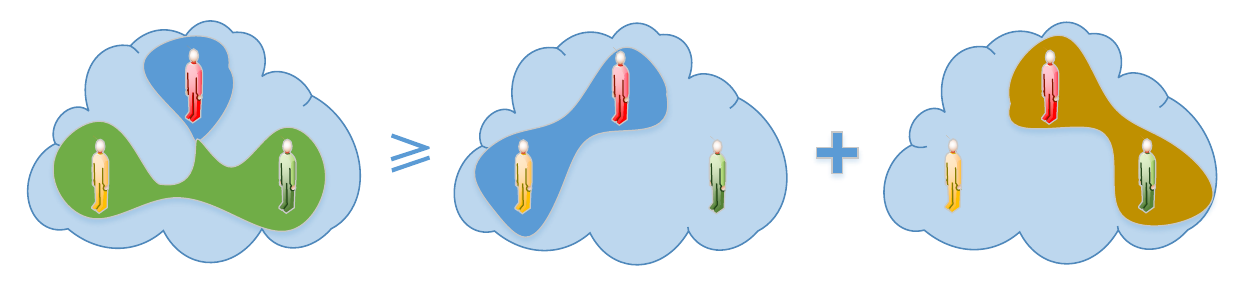}
		
		{\footnotesize (b)}
		\vspace{2mm}
	\end{center}
	\caption{\label{fig7-1-1}(color online). Schematic picture of 
		the monogamy relation under (a) 
		the unified tripartite entanglement measure and (b)
		the bipartite entanglement measure, respectively.}
\end{figure}

\begin{figure}
	\begin{center}
		\includegraphics[width=85mm]{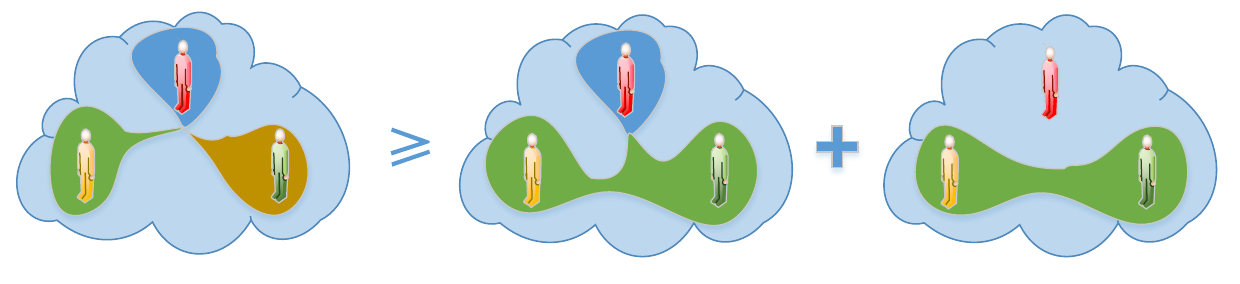}
	\end{center}
	\caption{\label{fig7-1-2}(color online). Schematic picture of the tight monogamy relation.}
\end{figure}

The difference between the monogamy relation and the complete monogamy relation is illustrated in Fig.~\ref{fig7-1-1} for the tripartite system. The tightly complete monogamy relation connects these two different kinds of monogamy relations (i.e., monogamy relation up to bipartite measure and the complete one) together (see Fig.~\ref{fig7-1-2} for the tripartite case).

\subsection{Complete monogamy of the unified GEM}\label{sec-7.2}


Let $E_g^{(n)}$ be a GEM. $E_g^{(n)}$ is completely monogamous if it obeys Eq.~\eref{gcoarsen}~\cite{Guo2022entropy}.	A complete GEM $E_g^{(n)}$ is
tightly complete monogamous if it satisfies the {genuine disentangling condition}, i.e., either for any $\rho\in\mS_g^{A_1A_2\cdots A_n}$ that satisfies~\cite{Guo2022entropy}  
\beax\label{g-tight}
E_g^{(k)}({X_1|X_2|\cdots|X_{k}})=E_g^{(l)}({Y_1|Y_2|\cdots|Y_{l}}) 
\eeax
with $X_1|X_2| \cdots| X_{k}\succ^b Y_1|Y_2| \cdots |Y_{l}$ we have that
\bea\label{g-tight1}
E_g^{(\ast)}({\gamma})=0
\eea
holds for all $\gamma\in \Xi(X_1|X_2| \cdots| X_{k}- Y_1|Y_2| \cdots |Y_{l})$, or 
\bea\label{g-tight2}
E_g^{(k)}({X_1|X_2|\cdots|X_{k}})>E_g^{(l)}({Y_1|Y_2|\cdots|Y_{l}}) 
\eea
holds for any $\rho\in\mS_g^{A_1A_2\cdots A_n}$.

In Ref.~\cite{Guo2020pra}, it was showed that the tightly complete monogamy is stronger than the complete monogamy for the complete global MEMs that defined by the convex-roof extension. One can easily find that it is also true for any complete GEM defined by the convex-roof extension.


\subsection{Complete monogamy of the unified $k$-entanglement measure}\label{sec-7.3}


The complete monogamy is related to the unification condition while the tightly complete monogamy to the hierarchy condition~\cite{Guo2020pra,Guo2021qst,Guo2022entropy,Guo2023pra}.  
Accordingly, the following two concepts are given in Ref.~\cite{Guo2024pra}.
A unified $k$-EM $E_k$ is {{completely monogamous}} if for any $k$-entangled state
$\rho\in\mathcal{S}\setminus \mS_k$ that satisfies
\bea\label{k-cm}
E_k(X_1|X_2| \cdots| X_{p})= E_l(Y_1|Y_2| \cdots |Y_{q})
\eea
with $X_1|X_2| \cdots| X_{p}\succ^a Y_1|Y_2| \cdots |Y_{q}$ we have that
\bea\label{k-cm2}
E_{\ast}({\gamma}) =0
\eea
holds for all $\gamma\in \Xi(X_1|X_2| \cdots| X_{p}- Y_1|Y_2| \cdots |Y_{q})$, $k\leq  p$, $l\leq  q$, $l\leq  k$. A complete $E_k$ is defined to be
{tightly complete monogamous} if Eqs.~\eref{k-cm} and~\eref{k-cm2} hold by replacing $X_1|X_2| \cdots| X_{p}\succ^a Y_1|Y_2| \cdots |Y_{q}$ with $X_1|X_2| \cdots| X_{p}\succ^b Y_1|Y_2| \cdots |Y_{q}$. 
For more clarity, we compare these three kinds of monogamy relations in Table~\ref{tab:table0}.

\begin{table}
	\caption{\label{tab:table0} Comparing of monogamy, complete monogamy, and tightly complete monogamy. Abbreviations in table are as follow:
		bipartite (B), unified (U), complete (C), monogamy (M), complete monogamy (CM), tightly complete monogamy (TCM), entanglement measure (EM), multipartite entanglement measure (MEM), genuine entanglement measure (GEM). }	
	\begin{indented}
		\lineup
		\item[]
		\begin{tabular}{ccc}\br
			& Coarser relation & Compatible EM\\ 
			\mr
			M        & $\succ^c$ & $k$-EM, BEM  \\			
			CM      &  $\succ^a$ & U$k$-EM, U-g-MEM, UGEM\\			
			TCM    & $\succ^b$ & C$k$-EM, C-g-MEM, CGEM\\
			\br
		\end{tabular}
	\end{indented}
\end{table}

As an illustrated example, we consider the 4-partite case $ABCD$.
If $E_4$ is completely monogamous, then
\beax
&&E_4(ABCD)= E_3(ABC)\Rightarrow E_2(ABC|D)=0,\\
&&E_4(ABCD)= E_2(AC)\Rightarrow E_3(AC|B|D)=E_2(BD)=0
\eeax
for any $4$-entangled state $\rho^{ABCD}\in\mS^{ABCD}$  (the other cases can be easily followed);
if $E_3$ is completely monogamous, then
\beax
&&E_3(A|B|CD)= E_2(A|CD)\Rightarrow E_2(ACD|B)=0
\eeax
for any $3$-entangled state $\rho^{ABCD}\in\mS^{ABCD}$ (the other cases can be easily followed).
If $E_4$ is tightly complete monogamous, then
\beax
&&E_4(ABCD)= E_3(A|B|CD)\Rightarrow E_2(CD)=0,\\
&&E_4(ABCD)= E_2(AB|CD)\Rightarrow E_2(AB)=E_2(CD)=0
\eeax
for any $4$-entangled state $\rho^{ABCD}\in\mS^{ABCD}$  (the other cases can be easily followed);
if $E_3$ is tightly completely monogamous, then
\beax
&&E_3(A|B|CD)= E_2(AB|CD)\Rightarrow E_2(AB)=0
\eeax
for any $3$-entangled state $\rho^{ABCD}\in\mS^{ABCD}$ (the other cases can be easily followed).


\subsection{Monogamy of the $k$-entanglement measure}\label{sec-7.4}


Observing the monogamy relations~\eref{monogamy1}, ~\eref{cond}, ~\eref{power}, one can find that the monogamy can be determined by the relation between states under the coarser relation of type (c), i.e., discarding the subsystem of the subsystems, together with the associated complementary relations. With this principle in mind, the monogamy of $k$-EM was established in Ref.~\cite{Guo2024pra}.

$E_k$ is said to be {{monogamous}} if for any
$\rho\in\mathcal{S}^{A_1A_2\cdots A_n}$ that satisfies~\cite{Guo2024pra}
\bea\label{k-monogamy}
E_k(X_1|X_2| \cdots| X_{p})
= E_k(X_1|X_2| \cdots|X_{s-1}|X'_s|X'_{s+1}|\cdots |X'_{p})
\eea
with $X_s|X_{s+1}| \cdots| X_{p}\succ^c X'_s|X'_{s+1}|\cdots |X'_{p}$, $1\leq  s\leq  p$, $k\leq  p$, we have that~\cite{explain2}
\bea \label{k-monogamy2}
E_*(\gamma)=0
\eea 
holds for any $\gamma\in\Xi(X_1|X_2| \cdots| X_{p}-X_1|X_2| \cdots|X_{s-1}|X'_s|X'_{s+1}|\cdots |X'_{p})$.
In such a sense, e.g., if $E_3$ is monogamous, then 
\beax \fl\quad \qquad E_3(A|B|CD)=E_3(A|B|C) \Rightarrow E_2(AB|D)=0, \\
\fl\quad\qquad E_3(A|B|CD|EF)=E_3(A|B|C|E) 
\Rightarrow \begin{cases}
	E_3(AB|D|F)=0, \\
	E_2(ABEF|D)=0, \\
	E_2(ABCD|F)=0, 
\end{cases}\\
\fl\quad\qquad E_3(AB|CD|EF)=E_3(A|C|E)
\Rightarrow \begin{cases}
	E_3(B|D|F)=0,\\
E_2(ABCD|F)=0,\\
	E_2(ABEF|D)=0,\\
E_2(B|CDEF)=0,
\end{cases}
\eeax  
which is similar to that of $E(A|BC)=E(AB)$ leads to $E(AC)=0$ for the monogamy of the bipartite entanglement measure $E$.
$E_k$ is {{weakly monogamous}} if for any
$\rho\in\mathcal{S}^{A_1A_2\cdots A_n}$ that satisfies Eq.~\eref{k-monogamy} with $p=k$ we have Eq.~\eref{k-monogamy2} holds. Note here that in Eq.~\eref{k-monogamy} only the case of $X_1|X_2| \cdots| X_{p}\succ^cX_1|X_2| \cdots|X_{s-1}|X'_s|X'_{s+1}|\cdots |X'_{p}$ is presented with no loss of generality since all other cases [i.e., any partitions with coarser relation of type (c)] can be followed easily due to the symmetry of $E_k$.

We need remark here that, in Ref.~\cite{Hou2022pra}, Hou \etal proposed the strong monogamy relation of the $k$-partite quantum correlation measure, which is in fact the same as the weakly monogamy of $E_k$. They also put forward a special case of strong monogamy relation of the $k$-partite quantum correlation measure in a more strict sense which is not valid in general.


\section{Polygamy relation of multipartite entanglement}\label{sec-8}


It is shown in Ref.~\cite{Jin2023rip,Qian2018njp,Zhu2015pra} that
\bea \label{polygamy-of-C}
C^2\left(|\psi\ra^{X|\overline{X}}\right)\leq\sum\limits_{Y\neq X}C^2\left(|\psi\ra^{Y|\overline{Y}}\right)
\eea
and
\bea \label{polygamy-of-C'}
C\left(|\psi\ra^{X|\overline{X}}\right)\leq  \sum\limits_{Y\neq X}C\left(|\psi\ra^{Y|\overline{Y}}\right)
\eea
holds for any $|\psi\ra\in\mH^{A_1A_2\cdots A_n}$, where $X|\overline{X}$ and $Y|\overline{Y}$ denote the bipartitions of $A_1A_2\cdots A_n$ with $X$ and $Y$ contain only one subsystem, i.e., $X=A_i$ and $Y=A_j$ for some $1\leq  i, j\leq  n$ (the 3-qubit case is proved in Refs.~\cite{Zhu2015pra} and~\cite{Qian2018njp}, respectively).

Guo \etal showed in Ref.~\cite{Guo2022jpa} that Eq.~\eref{polygamy-of-C} and Eq.~\eref{polygamy-of-C'} are valid for any bipartite continuous entanglement measure $E$:
For any given bipartite continuous entanglement measure $E$, 
there exists
$0<\alpha<\infty$ such that
\bea\label{triangle''}
E^{\alpha}\left(|\phi\ra^{A|BC}\right)\leq   E^{\alpha}\left(|\phi\ra^{B|AC}\right)
+ E^{\alpha}\left(|\phi\ra^{AB|C}\right)
\eea
for all pure states $|\phi\ra^{ABC}\in\mathcal{H}^{ABC}$ with fixed $\dim\mH^{ABC}=d<\infty$. We denote the supremum of exponent $\alpha$ in Eq.~\eref{triangle''} that associated with $E$ by $\acute{\alpha}(E)$, it is the largest one such that all the pure states satisfying Eq.~\eref{triangle''}. In addition, Eq.~\eref{triangle''} implies $E^{\eta}(|\phi\ra^{A|BC})\leq   E^{\eta}(|\phi\ra^{B|AC})
+ E^{\eta}(|\phi\ra^{AB|C})$ for any $\eta<\alpha$. 
In Ref.~\cite{Ge2024pra}, it is proved that, if the reduced function of a bipartite entanglement measure $E$ is subadditive, $E$ admits Eq.~\eref{triangle''} with $0\leq \alpha\leq  1$ but fails whenever $\alpha>1$, namely $\acute{\alpha}(E)\leq 1$ for these measures. So far, except for $E_2$ and $E_G$, all the bipartite entanglement measure $E$ wiht subadditive reduced function are continuous, so the latter one in Ref.~\cite{Ge2024pra}
coincides with Eq.~\eref{triangle''} in general. Eq.~\eref{triangle''} can be straightforwardly extended to $n$-partite case, $n\geq 4$. It includes the entanglement polygon inequalities explored in Refs.~\cite{Jin2023rip,Qian2018njp,Shi2023adp,Yang2022pra,Zhu2015pra}.
For example, for any continuous bipartite entanglement measure $E$ that is determined by the eigenvalues of the reduced state, there exists
$0<\alpha<\infty$ such that	~\cite{Guo2022jpa}
\bea\label{tetra-condition-2}
E^{\alpha}\left( |\psi\ra^{AB|CD}\right)\leq   E^{\alpha}\left(|\psi\ra^{AC|BD}\right)
+ E^{\alpha}\left(|\psi\ra^{AD|BC}\right)
\eea
for all $|\psi\ra^{ABCD}\in\mathcal{H}^{ABCD}$ with fixed $\dim\mH^{ABCD}=d<\infty$.

Moreover, if $E^{(3)}$ is a continuous unified tripartite entanglement measure, then there exists $0<\alpha<\infty$ such that~\cite{Guo2022jpa}
\bea\label{tetra-condition-1}
E^{\alpha}\left(|\psi\ra^{A|B|CD}\right)\leq   E^{\alpha}\left(|\psi\ra^{A|BD|C}\right)
+ E^{\alpha}\left( |\psi\ra^{AD|B|C}\right)
\eea
for all $|\psi\ra^{ABCD}\in\mathcal{H}^{ABCD}$ with fixed $\dim\mH^{ABCD}=d<\infty$,
Here we omit the superscript $^{(3)}$ of $E^{(3)}$ for brevity. Yang \etal also proposed the following entanglement polygon inequalities~\cite{Yang2022pra}: for any  $m+n$-partite pure entangled state $|\psi\rangle^{\bf{AB}}$ on Hilbert space $\mH^{\bf{A}}\ot\mH^{\bf{B}}$ with $\mH^{\bf{A}}=\otimes_{i=1}^m\mathcal{H}^{A_i}$ and $\mH^{\bf{B}}=\otimes_{j=1}^n \mathcal {H}^{B_{j}}$, the $q$-concurrence satisfies 
\bea\label{polygamy4}
C_{q}({\bf {A|B}})\leq \sum^m_{i=1}C_{q}(A_i|\overline{A}_i),
\eea
where $\overline{A}_i$ denotes all the particles except $A_i$. 
Similarly, the unified-$(q, s)$ entropy entanglement $E_{q,s}$ satisfies 
$E_{q,s}({\bf {A|B}})\leq \sum^m_{i=1}E_{q,s}({A_i|\overline{A}_i})$
for $q\geq1$ and $s\geq0$.

We note that, on one hand, the entanglement is not polygamous although the entanglement of assistance is, and on the other hand the triangle inequality always holds true [see Eqs.~\eref{triangle''}, \eref{tetra-condition-2}, \eref{tetra-condition-1}].
So when we say polygamy of entanglement, the polygamy inequalities as Eqs.~\eref{triangle''}, \eref{tetra-condition-2}, \eref{tetra-condition-2} and \eref{polygamy4} exhibit this feature well rather than the polygamy relation in Eqs.~\eref{polygamydefinition}, \eref{defcond}, and \eref{polygamy-power}. We thus give the following definition.

\begin{definition}
	Let $E^{(k)}$ be an entanglement measure of $k$-partite system. We call $E^{(k)}$ satisfies the entanglement polygamy inequality if for any $\gamma_i\in\Gamma_k$,
	\bea \label{polygam-relation}
	E^\alpha(|\psi\ra^{\gamma_i})\leq  \sum_{\gamma_j\neq \gamma_i}E^\alpha(|\psi\ra^{\gamma_j})
	\eea
	holds for any $|\psi\ra\in\mH^{A_1A_2\cdots A_n}$ for some $\alpha>0$.
\end{definition}

That is, when we say polygamy of entanglement, it refers to that the associated measure obeys Eq.~\eref{polygam-relation}. So we can discuss both the monogamy and polygamy of entanglement simultaneously.
We take concurrence for example. Together with the monogamy relation of concurrence~\cite{Yu2005pra}, we have
\beax
C^2(AB)+C^2(AC)\leq C^2(A|BC)\leq C^2(B|AC)+C^2(C|AB)
\eeax 
holds for any $|\psi\ra\in\mH^{ABC}$, where the left inequality is monogamy relation and the right one is the polygamy relation. However, whether the polygamy relation is valid for mixed states still remains open.


\section{Examples of the complete global MEMs}\label{sec-9}


In this section, we mainly review which global MEMs so far are complete in the sense of MEM 2.0. We fix some notations at first as in Ref.~\cite{Guo2024pra}. The set of all the $k$-partitions of $A_1|A_2|\cdots|A_n$ is denoted by $\Gamma_k$, $2\leq  k<n$, i.e., $\Gamma_k=\{\gamma_i\}$, where $\gamma_i=X_{1(i)}|X_{2(i)}|\cdots|X_{k(i)}$.


\subsection{Complete global MEM from sum of the reduced functions}\label{sec-9.1}


A natural candidate for the complete MEMs is the one defined by the sum of the reduced functions on all the single subsystems~\cite{Guo2020pra,Guo2024rip}, i.e.,
\bea\label{sum1}
E^{(n)}(|\psi\ra)=\frac{1}{2}\sum_ih(\rho^{A_i}).
\eea
It is a global MEM~\cite{Guo2020pra,Guo2024rip} (In Ref.~\cite{Guo2020pra,Guo2024rip} it is called MEM. We call it golbal MEM hereafter more precisely). We denote $E^{(n)}$ by $E_f^{(n)}$, $C^{(n)}$, $\tau^{(n)}$, $E_q^{(n)}$, $E_\alpha^{(n)}$, $N_F^{(n)}$, $E_{\mathcal{F}}^{(n)}$, $E_{\mathcal{F}'}^{(n)}$, $E_{A\mathcal{F}}^{(n)}$, $E_2^{(n)}$, $E_{\min}^{(n)}$, ${E}_{\min'}^{(n)}$, and $\hat{N}^{(n)}$ whenever
$h=S$, $h_C$, $h_\tau$, $h_q$, $h_\alpha$, $h_N$, $h_{\mathcal{F}}$, $h_{\mathcal{F}'}$, $h_{A\mathcal{F}}$, $h_2$, $h_{\min}$, $h_{\min'}$, and $\hat{h}$, respectively. Here, $E_f^{(n)}$, $C^{(n)}$, $\tau^{(n)}$, $E_q^{(n)}$, $E_\alpha^{(n)}$, and $N_F^{(n)}$ have been discussed in Ref.~\cite{Guo2020pra} for the first time. The coefficient ``1/2'' is fixed by the unification condition when $E^{(n)}$ is regarded as a unified MEM. One need note here that $E_{\mathcal{F}}^{(n)}$, $E_{\mathcal{F}'}^{(n)}$, and $E_{A\mathcal{F}}^{(n)}$ are different from $E_{\mathcal{F},F}^{(n)}$, $E_{\mathcal{F}',F}^{(n)}$, and $E_{A\mathcal{F},F}^{(n)}$ respectively in Ref.~\cite{Guo2020qip}. Also note that, $E_f^{(n)}$ coincides with the $\alpha$-entanglement entropy discussed in Ref.~\cite{Szalay2015pra}.

Let $E^{(n)}$ be a non-negative function defined as in Eq.~\eref{sum1}. Then the following statements hold true~\cite{Guo2024rip}:
\begin{itemize}
	\item $E^{(n)}$ is a unified global MEM and is completely monogamous;
	\item $E^{(n)}$ is a complete global MEM iff $h$ is subadditive;
	\item $E^{(n)}$ is tightly complete monogamous iff $h$ is subadditive with 
	\bea\label{sa-h}
	h(\rho^{AB})=h(\rho^A)+h(\rho^B)\Rightarrow \rho^{AB}~\mbox{is separable.}
	\eea
\end{itemize}

Together with the monogamy criterion, we obtain that, for these MEMs, both monogamy and tightly complete monogamy are stronger than the complete monogamy under the frame work of the complete MEM, and that monogamy is stronger than both complete monogamy and tightly complete monogamy (e.g., $E_2^{(n)}$).

\Table{\label{tab:table6-1} Comparing of $E^{(n)}$ with different different reduced functions, and $\mE^{(n)}$.
}	
	\br
	MEM     & Unified & Complete & CM & TCM \\ 
	\mr
	$E_f^{(n)}$         &$\checkmark$&$\checkmark$ &$\checkmark$&$\checkmark$\\
	$C^{(n)}$           &$\checkmark$&$\checkmark$ &$\checkmark$&$\checkmark$\\
	$\tau^{(n)}$        &$\checkmark$&$\checkmark$ &$\checkmark$&$\checkmark$  \\
	$E_q^{(n)}$         &$\checkmark$&$\checkmark$ &$\checkmark$&$\checkmark{\color{red}^{\rm a}}$\\
	$E_\alpha^{(n)}$    &$\checkmark$&$\times$     &$\checkmark$&$\times$   \\
	$N_F^{(n)}$         &$\checkmark$&$\times$     &$\checkmark$&$\times$\\
	$E_{\mF}^{(n)}$     &$\checkmark$&$\checkmark$ &$\checkmark$&$\checkmark{\color{red}^{\rm a}}$\\
	$E_{{\mF}'}^{(n)}$    &$\checkmark$&$\checkmark{\color{red}^{\rm b}}$&$\checkmark$&$\checkmark{\color{red}^{\rm a}}$\\
	$E_{A\mF}^{(n)}$    &$\checkmark$&$\checkmark{\color{red}^{\rm b}}$&$\checkmark$&$\checkmark{\color{red}^{\rm a}}$\\
	$E_{2}^{(n)}$       &$\checkmark$&$\checkmark$ &$\checkmark$&$\checkmark$\\
	$E_{\min}^{(n)}$    &$\checkmark$&$\times$     &$\checkmark$&$\times$\\
	${E}_{\min'}^{(n)}$ &$\checkmark$&$\times$     &$\checkmark$&$\times$\\
	$\hat{N}^{(n)}$     &$\checkmark$&$\checkmark{\color{red}^{\rm b}}$&$\checkmark$&$\checkmark{\color{red}^{\rm a}}$\\   \mr
	${\mE}^{(n)}~(n\geq4)$          &$\checkmark$&$\checkmark$ &$\checkmark$&$\checkmark$\\
	\br
\end{tabular}
	\item[] ${\color{red}^{\rm a}}$ It is tightly complete monogamous under the assumption that $h$ is subadditive and Eq.~\eref{sa-h} holds.
	\item[] ${\color{red}^{\rm b}}$ It is complete under the assumption that $h$ is subadditive.
\end{indented}
\end{table} 

Another candidate for global MEM is the one defined by the sum of all bipartite entanglement~\cite{Guo2022jpa}, i.e.,
\bea\label{sum2}
\mE^{(n)}(|\psi\ra)
=\frac12\sum\limits_{\gamma_i\in\Gamma_2} h(\rho^{X_{1(i)}}).
\eea
Note that $\mE^{(n)}$ is just $\frac12\mE_{12\cdots n(2)}$ in Ref.~\cite{Guo2022jpa} provided that the corresponding bipartite entanglement measure is an entanglement monotone.
This method has been presented in 2004 by Yu and He in Ref.~\cite{Yuchangshui2004pla} wherein the measure was called multipartite free entanglement measure which is denoted by $\bar{E}$ ($\bar{E}$ coincides with $\mE_{12\cdots n(2)}$).
Clearly, 
\beax
E^{(n)}\leq\mE^{(n)},
\eeax
and in general, $E^{(n)}<\mE^{(n)}$ whenever $n\geq4$. Indeed, $E^{(n)}(|\psi\ra)<\mE^{(n)}(|\psi\ra)$ iff $|\psi\ra$ is not fully separable~\cite{Guo2024rip}. $\mE^{(3)}$ coincides with $E^{(3)}$ but $\mE^{(n)}$ is different from $E^{(n)}$ whenever $n\geq4$. The properties of all these global MEMs are listed in Table~\ref{tab:table6-1} for convenience.

In Ref.~\cite{Cornelio2013pra}, it was proved that
\beax
\tau^{(3)}(\rho^{ABC})\geq \tau(\rho^{AB})+\tau(\rho^{AC})+\tau(\rho^{BC})
\eeax 
holds for any three-qubit state $\rho^{ABC}$ (Note that $\mC_3$ therein coincides with $\frac12\tau^{(3)}$ indeed).


\subsection{Maximally entangled state \& completely monogamous global MEM}\label{sec-9.2}


\subsubsection{The original definition of maximally entangled state}

The {maximally entangled state} (MES), as 
a crucial quantum resource in quantum information processing tasks such as quantum teleportation~\cite{Bennett1993teleporting,noh2009quantum,Zhang2006np},
superdense coding~\cite{barreiro2008beating,Bennett1992communication},
quantum computation~\cite{bennett2000quantum}
and quantum cryptography~\cite{Ekert1991prl},
has been explored considerably~\cite{aulbach2010maximally,du2007experimental,facchi2008maximally,gerry2002nonlinear,gerry2003generation,gilbert2008use,Guo2016pra,ishizaka2000maximally,Li2012qic,lougovski2005generation,peters2004maximally,revzen2010maximally,rubin2007loss,salavrakos2017bell,verstraete2001maximally,Vicente2013prl}.
For a bipartite system with state space $\mH^{AB}$, 
$\dim\mH^A=m$, $\dim\mH^B=n$ ($m\leq n$), a pure state $|\psi\rangle^{AB}$ 
is called a maximally entangled state if and only if 
$\rho^A=\frac{1}{m}I^A$~\cite{horodecki2001distillation}, 
where $\rho^A$ is the reduced state of $\rho^{AB}=|\psi\rangle\langle\psi|^{AB}$ 
with respect to subsystem $A$. Equivalently, $|\psi\rangle^{AB}$ is an MES if and only if
\beax\label{mes1}
|\psi\rangle^{AB}=\frac{1}{\sqrt{m}}\sum_{i=1}^m|i\rangle^A|i\rangle^B,\label{1}
\eeax
where $\{|i\rangle^A\}$ is an orthonormal basis
of $\mH^A$ and $\{|i\rangle^B\}$ is an orthonormal set of $\mH^B$.
An MES $|\psi\ra^{AB}$ always archives the maximal amount of entanglement for a certain
entanglement measure~\cite{Li2012qic} (such as entanglement of 
formation~\cite{Bennett1996pra3824,Horodecki01},
and concurrence~\cite{Hill,Rungta2001pra,Wootters}).
For example, the well-known EPR states are maximally entangled pure states.

It is proved in Ref.~\cite{Cavalcanti2005pra} that any MES in a $d\otimes d$ system is pure. Later,
Li \etal showed in Ref.~\cite{Li2012qic}
that the maximal entanglement can also exist in mixed states for $m\otimes n$ 
systems with $n\geq 2m$ (or $m\geq 2n$).
A necessary and sufficient condition of mixed maximally entangled state (MMES) is proposed \cite{Li2012qic}:
An $m\otimes n$ ($n\geq 2m$) bipartite mixed state $\rho^{AB}$ is maximally 
entangled if and only if
\begin{eqnarray}\label{mes2}
\rho^{AB}=\sum\limits_{k=1}^rp_k|\psi_k\rangle\langle\psi_k|^{AB},~~\sum\limits_kp_k=1,~p_k\geq0,
\end{eqnarray}
where $|\psi_k\rangle^{AB}$'s are maximally entangled pure states with
\beax
|\psi_k\rangle^{AB}=\frac{1}{\sqrt{m}}\sum_{i=0}^{m-1}|i\rangle^A|i_k\rangle^B,
\eeax
$\{|i\rangle^A\}$ is an orthonormal basis
of $\mH^A$ and $\{|i_k\rangle^B\}$
is an orthonormal set of $\mH^B$, satisfying $\langle i_s|j_t\rangle^B=\delta_{ij}\delta_{st}$.
Then the MMES $\rho^{AB}$ can be rewritten as~\cite{Guo2020pra}
\beax\label{mes3}
\rho^{AB}=|\Phi^+\ra\la\Phi^+|^{AB_1}
\otimes\left( \sum\limits_{k=1}^rp_k|k\rangle\langle k|^{B_2}\right),
\eeax
up to some local unitary on the subspace that spanned by $\{|i_k\rangle^B:i=0,1,\dots,m-1, k=1,2,\dots,r\}$,
$\sum_kp_k=1$, $p_k\geq0$.

\subsubsection{The incompatibility of MMES and the monogamy law}

Let $\rho^{ABC}$ be a state acting on
$\mH^{ABC}$ with $2\dim \mH^A\leq  \dim \mH^{B}$.
If $\rho^{AB}={\rm Tr}_{C}\rho^{ABC}$ is a mixed state as in Eq.~\eref{mes2},
then
$\rho^{AC}$ is a product state but $\rho^{BC}$ is not necessarily separable~\cite{Guo2020pra}. 
It seems that entanglement can be freely shared.
Moreover, we let
\beax\label{bac2}
\rho^{ABC}= \sum_{s=1}^l\frac1l |\phi_s\ra\la\phi_s|^{ABC} 
\eeax
be a state in $\mS^{ABC}$ with $\dim\mH^A=m$, $\dim\mH^B\geq rm$, $\dim\mH^C\geq rl$, and 
\beax
|\phi_s\ra^{ABC}=\sum_{i=0}^{m-1}\sum_{k=0}^{r-1} \frac{1}{rm}|i\ra^A|i\ra^{B_1}|k\ra^{B_2}|k\ra^{C_1}|s\ra^{C_2},
\eeax
where $\mH^{B_1}\ot\mH^{B_2}$ is a subspace of $\mH^B$ and $\mH^{C_1}\ot\mH^{C_2}$ is a subspace of $\mH^C$, 
then $\rho^{ABC}$ is an MES with respect to the cutting $B|AC$ according to Ref.~\cite{Li2012qic}.
Namely, $B$ can maximally entangle with $A$ and $C$ simultaneously~\cite{Guo2020pra}, which is in contradiction with the monogamy law.
This phenomenon indicates that the maximally entangled state defined in Ref.~\cite{Li2012qic} need to improved.
Consequently, the improved definition of MES was given in Ref.~\cite{Guo2020pra}: Let $\rho^{AB}$ be a state in $\mS^{AB}$ with $\dim\mH^A=m\leq \dim\mH^B$. Then
$\rho^{AB}$ is an MES if and only if i) 
$E_f^{(2)}(\rho^{AB})=\ln m$
and ii) for any extension
$\rho^{ABC}$ of $\rho^{AB}$ (i.e., $\rho^{AB}=\tr_C\rho^{ABC}$) we have 
$E_f^{(3)}(\rho^{ABC})=E_f^{(2)}(\rho^{AB})$.
We thus can conclude, to be more exact, the following definition.

\begin{definition}
Let $\rho^{AB}$ be a state in $\mS^{AB}$ with $\dim\mH^A=m\leq \dim\mH^B$. Then
$\rho^{AB}$ is an MES if and only if i) $\rho^A=\frac{1}{m}I^A$
and ii) for any extension
$\rho^{ABC}$ of $\rho^{AB}$ (i.e., $\rho^{AB}=\tr_C\rho^{ABC}$) we have 
\bea 
E^{(3)}(\rho^{ABC})=E^{(2)}(\rho^{AB})
\eea
holds for any completely monogamous global MEM $E^{(3)}$.
\end{definition}

This definition of MES is compatible with the monogamy law and
makes the concept of MES more clearly: If $\rho^{AB}$ is an MES, then by the complete monogamy 
of $E^{(3)}$, we immediately obtain that both $\rho^{AC}$ and $\rho^{BC}$ are separable.
This also indicates that the complete monogamy relation can reflects 
the monogamy law more effectively in some sense.
Together with the arguments above, we can conclude that there is no MMES in any bipartite quantum system~\cite{Guo2020pra}.


\subsection{Subadditivity of the reduced function}\label{sec-9.3}


As the subadditivity of the reduced function is closely related to the tightly complete monogamy of the global MEM $E^{(n)}$ above,
we summarize the subadditivity of the reduced functions in literature as following: 
\begin{itemize}
\item $S$ is additive and subadditive~\cite{Wehrl1978}, i.e.,
\beax
S(\rho\ot\sigma)=S(\rho)+S(\sigma)
\eeax 
and 
\beax 
S(\rho^{AB})\leq S(\rho^A)+S(\rho^B),
\eeax 
respectively.

\item $S_q$ is subadditive iff $q>1$, but not additive, and
for $0<q<1$, $S_q$ is neither subadditive nor superadditive~\cite{Raggio}
(superadditivity refers to $S_q(\rho^{AB})\geq S_q(\rho^A)+S_q(\rho^B)$).
In addition, 
\beax
S_q(\rho^{A}\ot\rho^B)= S_q(\rho^A)+S_q(\rho^B)
\eeax
iff $\rho^A$ or $\rho^B$ is pure~\cite{Raggio}.

\item $S_{q,s}$ is subadditive for $q>1$ and $s>1/q$~\cite{Rastegin2011jsp}.

\item $h_\alpha$ is additive but not subadditive~\cite{Beck,Aczel}.

\item $h_\tau$ is subadditive~\cite{Audenaerta2007jmp}, i.e.,
\bea\label{subadditive}
1+\tr\rho_{AB}^2\geq \tr\rho_A^2+\tr\rho_B^2.
\eea
In particular, the equality holds iff $\rho^A$ or $\rho^B$ is pure~\cite{Guo2020pra}.

\item $h_N$ is neither subadditive nor supperadditive~\cite{Guo2020pra}.

\item $h_2$ is subadditive~\cite{Guo2024rip}.

\item $h_2$, $h_{\min}$, $h_{\min'}$, and $\hat{h}$ are not additive, $h_{\min}$ and $h_{\min'}$ are subadditive on the states that satisfies $r(\rho^{AB})=r(\rho^{A})=r(\rho^{B})=2$~\cite{Guo2024rip}. 
\end{itemize}
Eq.~\eref{subadditive} implies $h_C$ is subadditive and the equality holds iff $\rho^A$ or $\rho^B$ is pure. $h_{\mathcal{F}}$ is subadditive since it coincides with $S_q/2$ ($q=3$). We conjecture that $h_{\mathcal{F}'}$ and $h_{A\mathcal{F}}$ are subadditive, and that $\hat{h}$ is subadditive~\cite{Guo2024rip}. In what follows, we always assume that $h_{\mathcal{F}'}$, $h_{A\mathcal{F}}$, and $\hat{h}$ are subadditive, and that $\hat{h}$ is concave. 
We summarize the properties of these reduced functions in Table~\ref{tab:table6-3} for more convenience.

\begin{table}[htbp]
	\caption{\label{tab:table6-3} Comparing of the properties of the reduced functions.
		C, SC, SA, and A signify the function is concave, strictly concave, subadditive, and additive, respectively.}
	\footnotesize
\hspace{5mm}	\begin{tabular}{@{}llllll}\br
$E$  &$h$&     C        & SC   &  SA & A  \\ \mr
$E_f$  &$S$                          &$\checkmark$         &$\checkmark$       &$\checkmark$&$\checkmark$\\
$C$    &$\sqrt{2(1-\tr\rho^2)}$      &$\checkmark$         &$\checkmark$       &$\checkmark$&$\times$\\
$\tau$ &$2(1-\tr\rho^2)$             &$\checkmark$         &$\checkmark$       &$\checkmark$&$\times$  \\
$E_q$  &$\frac{1-\tr\rho^q}{q-1}$    &$\checkmark$($q>0$)  &$\checkmark$($q>1$)&$\checkmark$($q>1$) &$\times$\\
$E_{q,s}$  &$\frac{1}{(1-q)s}\left[(\mathrm{Tr}\rho^q)^s-1\right]$    &$\checkmark$($0<q<1$, $s\leq1$)  &?&$\checkmark$($q>1$, $s>1/q$) &?\\
$E_\alpha$&$\frac{\ln(\tr\rho^\alpha)}{1-\alpha}$, $\alpha\in(0,1)$
&$\checkmark$       &$\checkmark$       &$\times$    &$\checkmark$ \\
$N_F$  &$\frac{(\tr\sqrt{\rho})^2-1}{2}$   &$\checkmark$         &$\checkmark$       &$\times$    &$\times$\\
$E_{\mF}$&$1-\tr\rho^3$                &$\checkmark$         &$\checkmark$       &$\checkmark$&$\times$\\
$E_{{\mF}'}$&$1-(\tr\rho^2)^2$            &$\checkmark$         &$\checkmark$       &$\checkmark{\color{red}^{\rm a}}$           &$\times$\\
$E_{A\mF}$&$1-\sqrt{\tr\rho^3}$         &$\checkmark$         &$\checkmark$       &$\checkmark{\color{red}^{\rm a}}$           &$\times$\\
$E_2$, $E_G$	   &$1-\|\rho\|$                 &$\checkmark$         &$\times$           &$\checkmark$&$\times$\\
$E_{\min}$&$\|\rho\|_{\min}$            &$\checkmark$         &$\times$           &$\times$           &$\times$\\
$E_{\min'}$&$r(\rho)\|\rho\|_{\min}$     &$\checkmark$         &$\times$           &$\times$            &$\times$\\
\br
\end{tabular}\\
\indent\quad ~~~${\color{red}^{\rm a}}$ We conjecture that they are subadditive. 
\end{table} 
\normalsize


\subsection{MEM from the maximal reduced function}\label{sec-9.4}


The quantity
\bea\label{max1}
{E'}^{(n)}(|\psi\ra)
=\max\limits_ih(\rho^{A_i})
\eea
can reports the amount of entanglement contained in $|\psi\ra$ to some extent~\cite{Guo2024rip}: If it is nonincreasing on average under LOCC, then (i) ${E'}^{(3)}$ is a complete global MEM but not tightly complete monogamous, and if $h$ is strictly concave, ${E'}^{(3)}$ is completely monogamous, and
(ii) ${E'}^{(n)}$ is not complete whenever $n\geq4$.
By definition, ${E'}^{(n)}\leq {E}^{(n)}$ if $h$ is subadditive.

We denote the corresponding ${E'}^{(n)}$ in the previous subsection by ${E'}_f^{(n)}$, ${C'}^{(n)}$, ${\tau'}^{(n)}$, ${E'}_q^{(n)}$, ${E'}_\alpha^{(n)}$, ${N'}_F^{(n)}$, ${E'}_{\mathcal{F}}^{(n)}$, ${E'}_{\mathcal{F}'}^{(n)}$, ${E'}_{A\mathcal{F}}^{(n)}$,  ${E'}_2^{(n)}$, ${E'}_{\min}^{(n)}$, ${E'}_{\min'}^{(n)}$, and ${\hat{N'}}^{(n)}$, respectively. Note here that we can not prove here ${E'}^{(n)}$ is nonincreasing on average under LOCC, but we conjecture that ${E'}^{(n)}$ does not increase on average under LOCC for these cases mentioned above. Hereafter we always assume the conjecture is true.

If $h$ is not strictly concave, then ${E'}^{(3)}$ is not completely monogamous~\cite{Guo2024rip}. 
Namely, ${E'}_2^{(3)}$, ${E'}_{\min}^{(3)}$, ${E'}_{\min'}^{(3)}$, and ${\hat{N'}}^{(3)}$ are not completely monogamous. Namely, for these four complete MEMs, monogamy coincides with complete monogamy, tightly complete monogamy seems stronger than both monogamy and complete monogamy.

It is worthy mentioning here that ${E'}^{(n)}$ may not a unified MEM if $n\geq 4$ since it may occur that
${E'}^{(k)}(X_1|X_2| \cdots| X_{k})<{E'}^{(l)}(Y_1|Y_2| \cdots |Y_{l})$ for some state $\rho\in\mathcal{S}^{A_1A_2\cdots A_n}$ whenever $X_1|X_2| \cdots| X_{k}\succ^a Y_1|Y_2| \cdots |Y_{l}$.

Moreover, we define~\cite{Guo2024rip}
\beax\label{max2}
{\mE'}^{(n)}(|\psi\ra)
=\max_{\gamma_i\in\Gamma_2}h(\rho^{X_{1(i)}})\quad\quad
\eeax
for pure states and for mixed states by the convex-roof structure, where the maximum is taken over all the bipartitions in $\Gamma_2$. By definition, 
\beax
{E'}^{(n)}\leq{\mE'}^{(n)},
\eeax
${\mE'}^{(3)}$ coincides with ${E'}^{(3)}$, ${\mE'}^{(n)}$ satisfies the hierarchy condition, but it may violate
the unification condition~\cite{Guo2024rip}. ${E'}^{(n)}<{\mE'}^{(n)}$ occurs whenever $n\geq4$~\cite{Guo2024rip}. We give a comparison for these MEMs in Table~\ref{tab:table6.4-3} for more clarity.

\begin{table}[htbp]
\caption{\label{tab:table6.4-3} Comparing of ${E'}^{(3)}$ with different reduced functions, ${E'}^{(4)}$~($n\geq4$), and ${\mE'}^{(4)}$~($n\geq4$).}	
\begin{indented}
\lineup
\item[]
\begin{tabular}{@{}lllll}\br
MEM     & Unified & Complete & CM & TCM \\ 
\mr
${E'}_f^{(3)}$          &$\checkmark$&$\checkmark$ &$\checkmark$&$\times$\\
${C'}^{(3)}$            &$\checkmark$&$\checkmark$ &$\checkmark$&$\times$\\
${\tau'}^{(3)}$         &$\checkmark$&$\checkmark$ &$\checkmark$&$\times$\\
${E'}_q^{(3)}$          &$\checkmark$&$\checkmark$ &$\checkmark$&$\times$\\
${E'}_\alpha^{(3)}$     &$\checkmark$&$\checkmark$ &$\checkmark$&$\times$\\
${N'}_F^{(3)}$          &$\checkmark$&$\checkmark$ &$\checkmark$&$\times$\\
${E'}_{F}^{(3)}$        &$\checkmark$&$\checkmark$ &$\checkmark$&$\times$\\
${E'}_{F'}^{(3)}$       &$\checkmark$&$\checkmark$ &$\checkmark$&$\times$\\
${E'}_{AF}^{(3)}$       &$\checkmark$&$\checkmark$ &$\checkmark$&$\times$\\
${E'}_{2}^{(3)}$        &$\checkmark$&$\checkmark$ &$\times$    &$\times$\\
${E'}_{\min}^{(3)}$     &$\checkmark$&$\checkmark$ &$\times$    &$\times$\\
${E'}_{\min'}^{(3)}$    &$\checkmark$&$\checkmark$ &$\times$    &$\times$\\
${\hat{N'}}^{(3)}$      &$\checkmark$&$\checkmark$ &$\times$    &$\times$\\  		
${E'}^{(4)}$~($n\geq4$)         &?&$\times$ &?&$\times$\\	
${\mE'}^{(n)}$~($n\geq4$)       &?&$\times$ &?&$\times$\\
\br
\end{tabular}
\end{indented}
\end{table} 


\section{Examples of GEMs}\label{sec-10}


In this section, we review the GEMs in literature so far. As we list below, there are two classes of GEMs, the complete one and the other one that are not complete. Some of them is not defined in a unified way for arbitrary $n$-party system, such as the triangle measure, and some of them is still not clear whether they are complete or not (e.g., the geometric measure).


\subsection{Three tangle}\label{sec-10.0}


In 2000, when Valerie Coffman, Joydip Kundu, and William K.
Wootters~\cite{Coffman2000pra} discussed the distribution of entanglement, they
discovered an interesting quantity for a tripartite two-level
system, referred to as the three tangle (or called residual tangle, or residual entanglement) that was defined by
\bea\label{3-tangle}
\tilde{\tau}(|\psi\ra^{ABC})=C^2(|\psi\ra^{A|BC})-C^2(\rho^{AB})-C^2(\rho^{AC}),
\eea
where $\rho^{AB}$ and $\rho^{AC}$ are the reductions of $|\psi\ra^{ABC}$, namely, $\tilde{\tau}$ coincides with $\tau^{(3)}_{ABC}$ in Eq.~\eref{n-residual} for the three-qubit state. It is the first GEM. For the three-qubit state $|\psi\ra^{ABC}=\sum_{i,j,k}a_{ijk}|i\ra^A|j\ra^B|k\ra^C$, its three tangle is given by
\beax
\tilde{\tau}(|\psi\ra^{ABC})=2\left|\sum a_{ijk}a_{i'j'm}a_{np}a_{n'p'm'}\epsilon_{ii'}\epsilon_{jj'}\epsilon_{kk'}\epsilon_{mm'}\epsilon_{nn'}\epsilon_{pp'} \right|,
\eeax 
where the summation is over all the indices, 
$\epsilon_{01}=-\epsilon_{10}=1$, and $\epsilon_{00}=\epsilon_{11}=0$. $\tilde{\tau}$ is invariant under
permutations of $A$, $B$, and $C$, so it can be regarded
as representing a collective property of three qubit and can
be used to quantify the three-way entanglement~\cite{Wong2001pra}.
For the three-qubit state $|\psi\rangle$ in the generalized Schmidt decomposition form \cite{Acin2000prl},
\begin{eqnarray}\label{ex1}
	|\psi\rangle=\lambda_0|000\rangle+\lambda_1e^{{\rm i}{\varphi}}|100\rangle+\lambda_2|101\rangle +\lambda_3|110\rangle+\lambda_4|111\rangle,
\end{eqnarray}
where $\lambda_i\geq0,~i=0,1,2,3,4$ and $\sum\limits_{i=0}\limits^4\lambda_i^2=1$, 
$\tilde{\tau}(|\psi\rangle)=4\lambda^2_{0}\lambda^2_{4}$.


\subsection{Complete GEM from sum of the reduced functions}\label{sec-10.1}


In Ref.~\cite{Guo2022entropy,Guo2024rip}, a class of GEMs defined by sum of all reduced functions of the single subsystems was introduced, i.e.,
\bea\label{gsum1}
E_g^{(n)}(|\psi\ra)
=\left\lbrace \begin{array}{ll}
	\!\!\frac{1}{2}\sum_jh(\rho^{A_j}), & h(\rho^{X_{1(i)}})>0~\forall~\gamma_i\in\Gamma_2,\\
	\!\!0, &h(\rho^{X_{1(i)}})=0~\exists~\gamma_i\in\Gamma_2.
\end{array}\right. 
\eea
If the reduced function $h$ is concave, then~\cite{Guo2024rip}
(i) $E_g^{(n)}$ is a unified GEM and is completely monogamous;
(ii) $E_g^{(n)}$ is a complete GEM iff $h$ is subadditive;
(iii) $E_g^{(n)}$ is tightly complete monogamous iff $h$ is subadditive with Eq.~\eref{sa-h} holds.
Therefore, for these GEMs, tightly complete monogamy are stronger than the complete monogamy under the frame work of the complete GEM.
In addition, it is obvious that $E_g^{(n)}$ is complete, completely monogamous, tightly complete monogamous, if and only if $E^{(n)}$ is complete, completely monogamous, tightly complete monogamous, respectively.

Ref.~\cite{Guo2022jpa,Guo2024rip} also proposed the following GEM
\bea\label{gsum2}
\mE_g^{(n)}(|\psi\ra)
=\left\lbrace \begin{array}{ll}
	\!\!\mE^{(n)}(|\psi\ra), & h(\rho^{X_{1(i)}})>0~\forall~\gamma_i\in\Gamma_2,\\
	\!\!0, & h(\rho^{X_{1(i)}})=0~\exists~\gamma_i\in\Gamma_2.
\end{array}\right. 
\eea
Clearly, 
\beax
{E}_g^{(n)}\leq{\mE}_g^{(n)},
\eeax
and ${\mE}_g^{(3)}$ coincides with ${E}_g^{(3)}$ but ${\mE}_g^{(n)}$ is different from ${E}_g^{(n)}$ whenever $n\geq4$. ${\mE}_g^{(n)}$ is just ${\mE}_{g-12\cdots n(2)}$ in Ref.~\cite{Guo2022jpa} if the corresponding bipartite entanglement measure is an entanglement monotone.

\Table{\label{tab:table7-1} Comparing of $E_{g}^{(n)}$ with different reduced functions and $\mE_{g}^{(n)}$ ($n\geq4$).}
\br
GEM                  & Unified    & Complete    & CM         & TCM   \\ \mr
$E_{g,f}^{(n)}$      &$\checkmark$&$\checkmark$ &$\checkmark$&$\checkmark$\\
$C_g^{(n)}$          &$\checkmark$&$\checkmark$ &$\checkmark$&$\checkmark$\\
$\tau_g^{(n)}$       &$\checkmark$&$\checkmark$ &$\checkmark$&$\checkmark$\\
$E_{g,q}^{(n)}$      &$\checkmark$&$\checkmark$ &$\checkmark$&$\checkmark{\color{red}^{\rm a}}$ \\
$E_{g,\alpha}^{(n)}$ &$\checkmark$&$\times$     &$\checkmark$&$\times$     \\
$N_{g,F}^{(n)}$      &$\checkmark$&$\times$     &$\checkmark$&$\times$     \\
$E_{g,F}^{(n)}$      &$\checkmark$&$\checkmark$ &$\checkmark$&$\checkmark{\color{red}^{\rm a}}$\\
$E_{g,F'}^{(n)}$     &$\checkmark$&$\checkmark{\color{red}^{\rm b}}$ &$\checkmark$&$\checkmark{\color{red}^{\rm a}}$\\
$E_{g,AF}^{(n)}$     &$\checkmark$&$\checkmark{\color{red}^{\rm b}}$&$\checkmark$&$\checkmark{\color{red}^{\rm a}}$\\
$E_{g,2}^{(n)}$      &$\checkmark$&$\checkmark$ &$\checkmark$&$\checkmark$\\
$E_{g,\min}^{(n)}$   &$\checkmark$&$\times$     &$\checkmark$&$\times$    \\
${E}_{g,\min'}^{(n)}$&$\checkmark$&$\times$     &$\checkmark$&$\times$    \\
${\hat{N}}_g^{(n)}$  &$\checkmark$&$\checkmark{\color{red}^{\rm b}}$&$\checkmark$&$\checkmark{\color{red}^{\rm a}}$\\ 
$\mE_{g}^{(n)}$~$(n\geq4)$     &$\checkmark$&$\checkmark$ &$\checkmark$&$\checkmark$\\
\br
\end{tabular}
\item[] ${\color{red}^{\rm a}}$ Assume that $h$ is subadditive and Eq.~\eref{sa-h} holds.
\item[] ${\color{red}^{\rm b}}$ Assume that $h$ is subadditive.
\end{indented}
\end{table} 

For the case of $n\geq4$, all these MEMs $\mE_{g}^{(n)}$ with the reduced functions we discussed in Sec.~\ref{sec-3.8} (i.e., if the reduced function is concave) are complete GEMs, and are not only completely monogamous but also tightly complete monogamous~\cite{Guo2024rip}. We denote $E_g^{(n)}$ in the previous Section by $E_{g,f}^{(n)}$, $C_g^{(n)}$, $\tau_g^{(n)}$, $E_{g,q}^{(n)}$, $E_{g,\alpha}^{(n)}$, $N_{g,F}^{(n)}$, $E_{g,\mathcal{F}}^{(n)}$, $E_{g,\mathcal{F}'}^{(n)}$, $E_{g,A\mathcal{F}}^{(n)}$, $E_{g,2}^{(n)}$, $E_{g,\min}^{(n)}$, ${E}_{g,\min'}^{(n)}$, and $\hat{N}_g^{(n)}$, respectively. For convenience, we list all these MEMs in Table~\ref{tab:table7-1}. In addition, it is obvious that $E_g^{(n)}<\mE_g^{(n)}$ whenever $n\geq4$ for any $E_g^{(4)}$ and $\mE_g^{(4)}$ mentioned above.


\subsection{Complete GEM from the maximal reduced function}\label{sec-10.2}


If
\bea\label{gmax1}
E_{g'}^{(n)}(|\psi\ra)
=\left\lbrace \begin{array}{ll}
\!\!\max\limits_jh(\rho^{A_j}),&  h(\rho^{X_{1(i)}})>0~\forall~\gamma_i\in\Gamma_2,\\
\!\!0, & h(\rho^{X_{1(i)}})=0~\exists~\gamma_i\in\Gamma_2 
\end{array} \right.  
\eea
is nonincreasing on average under LOCC, then~\cite{Guo2024rip} (i) ${E}_{g'}^{(3)}$ is a complete GEM but not tightly complete monogamous, and if $h$ is strictly concave, ${E}_{g'}^{(3)}$ is completely monogamous, and
(ii) ${E}_{g'}^{(n)}$ is not complete whenever $n\geq4$.

One need note here that, when $h$ is not strictly concave, $E_{g'}^{(n)}$ is not a unified GEM since it may happen that ${E}_{g'}^{(k)}(X_1|X_2|\cdots|X_{k})= {E}_{g'}^{(l)}(Y_1|Y_2|\cdots|Y_{l})$ for some $\rho\in\mS_g^{A_1A_2\cdots A_n}$ with $X_1|X_2| \cdots| X_{k}\succ^a Y_1|Y_2| \cdots |Y_{l}$, namely, it violates Eq.~\eref{gcoarsen}~\cite{Guo2024rip}. In addition, $E_{g'}^{(n)}$ also violates Eq.~\eref{g-tight1} or Eq.~\eref{g-tight2}~\cite{Guo2024rip}.
That is, whenever $h$ is strictly concave, $E_{g'}^{(n)}$ is complete, completely monogamous, tightly complete monogamous, if and only if ${E'}^{(n)}$ is complete, completely monogamous, tightly complete monogamous, respectively.

We assume in this subsection that $E_{g'}^{(n)}$ is nonincreasing on average under LOCC. 
If $h$ is strictly concave and $h(\rho\ot\sigma)\geq h(\rho)$ and $h(\rho\ot\sigma)\geq h(\sigma)$ for any $\rho$ and $\sigma$, $E_{g'}^{(3)}$ defined as in Eq.~\eref{gmax1} is tightly complete monogamous on the states that admit the form~\cite{Guo2024rip}
\beax\label{eta}
|\eta\ra^{ABC}=|\eta\ra^{AB_1}|\eta\ra^{B_2C}.
\eeax 
In fact, we always have $h(\rho\ot\sigma)\geq h(\rho)$ and $h(\rho\ot\sigma)\geq h(\rho)$ if $h\in\{S$, $h_C$, $h_\tau$, $h_q$, $h_\alpha$, $h_N$, $h_{\mathcal{F}}$, $h_{\mathcal{F}'}$, $h_{A\mathcal{F}}\}$.

If
\bea\label{gmax2}
{\mE}_{g'}^{(n)}(|\psi\ra)
=\left\lbrace \begin{array}{ll}
\!\!{\mE'}^{(n)}(|\psi\ra),& h(\rho^{X_{1(i)}})>0~\forall~\gamma_i\in\Gamma_2,\\
\!\!0, & h(\rho^{X_{1(i)}})=0~\exists~\gamma_i\in\Gamma_2
\end{array}\right. 
\eea
is nonincreasing on average under LOCC, it is a GEM~\cite{Guo2024rip}. We assume in this subsection that $\mE_{g'}^{(n)}$ is nonincreasing on average under LOCC. By definition, 
\beax
{E}_{g'}^{(n)}\leq{\mE}_{g'}^{(n)},
\eeax
${\mE}_{g'}^{(3)}$ coincides with ${E'}_g^{(3)}$, and ${\mE}_{g'}^{(n)}$ satisfies the hierarchy condition, but it violates the unification condition if $n\geq4$. For the case of $n\geq4$, it is possible that ${E}_{g'}^{(n)}<{\mE}_{g'}^{(n)}$. 
It is easy to see that all these GEMs ${{\mE}}_{g'}^{(n)}$ with the reduced function we discussed are not complete GEMs whenever $n\geq4$. 
We denote $E_{g'}^{(n)}$ the corresponding GEMs mentioned in the previous Subsection by $E_{g',f}^{(n)}$, $C_{g'}^{(n)}$, $\tau_{g'}^{(n)}$, $E_{g',q}^{(n)}$, $E_{g',\alpha}^{(n)}$, $N_{g',F}^{(n)}$,
$E_{g',\mathcal{F}}^{(n)}$, $E_{g',\mathcal{F}'}^{(n)}$, $E_{g',A\mathcal{F}}^{(n)}$, $E_{g',2}^{(n)}$, $E_{g',\min}^{(n)}$, ${E}_{g',\min'}^{(n)}$, and $\hat{N}_{g'}^{(n)}$, respectively. 
We give comparison for these GEMs in Table~\ref{tab:table5}.

\begin{table}
\caption{\label{tab:table5} Comparing of $E_{g'}^{(3)}$ with different reduced functions, $E_{g'}^{(4)}$, and $\mE_{g'}^{(n)}$.}	
\begin{indented}
\lineup
\item[]
\begin{tabular}{@{}lllll}\br
GEM                 & Unified    & Complete    & CM         & TCM   \\ 
\mr			
$E_{g',f}^{(3)}$     &$\checkmark$&$\checkmark$ &$\checkmark$&$\times$\\
$C_{g'}^{(3)}$       &$\checkmark$&$\checkmark$ &$\checkmark$&$\times$\\
$\tau_{g'}^{(3)}$    &$\checkmark$&$\checkmark$ &$\checkmark$ &$\times$\\
$E_{g',q}^{(3)}$     &$\checkmark$&$\checkmark$ &$\checkmark$ &$\times$\\
$E_{g',\alpha}^{(3)}$&$\checkmark$&$\checkmark$ &$\checkmark$ &$\times$\\
$N_{g',F}^{(3)}$     &$\checkmark$&$\checkmark$ &$\checkmark$ &$\times$\\
$E_{g',F}^{(3)}$   &$\checkmark$&$\checkmark$ &$\checkmark$ &$\times$\\
$E_{g',F'}^{(3)}$  &$\checkmark$&$\checkmark$ &$\checkmark$ &$\times$\\
$E_{g',AF}^{(3)}$  &$\checkmark$&$\checkmark$ &$\checkmark$ &$\times$\\
$E_{g',2}^{(3)}$     &$\times$&$\times$&$\times$&$\times$\\
$E_{g',\min}^{(3)}$  &$\times$&$\times$&$\times$&$\times$\\
${E}_{g',\min'}^{(3)}$&$\times$&$\times$&$\times$&$\times$\\
${\hat{N}}_{g'}^{(3)}$&$\times$&$\times$&$\times$&$\times$\\ \hline		
$E_{g'}^{(n)}$~$(n\geq4)$     &?&$\times$ &?&$\times$\\						
$\mE_{g'}^{(n)}$~$(n\geq4)$    &?&$\times$ &?&$\times$\\
\br
\end{tabular}
\end{indented}
\end{table} 


\subsection{GEM from the minimal reduced function}


For any reduced function $h$, 
\bea\label{gmin1}
E_{g''}^{(n)}(|\psi\ra)
=\left\lbrace \begin{array}{ll}
\!\!\min\limits_jh(\rho^{A_j}),& h(\rho^{X_{1(i)}})>0~\forall~\gamma_i\in\Gamma_2,\\
\!\!0, & h(\rho^{X_{1(i)}})=0~\exists~\gamma_i\in\Gamma_2
\end{array}\right. 
\eea
is a GEM~\cite{Guo2024rip}. Moreover, we can define
\bea\label{gmin2}
\mE_{g''}^{(n)}(|\psi\ra)
=\min_{\gamma_i\in\Gamma_2}h(\rho^{{X_{1(i)}}}),
\eea
it is also a GEM~\cite{Guo2024rip}. For example, GMC, denoted by $C_{gme}$~\cite{Ma2011pra}, is defined as in Eq.~\eref{gmin2}. Recall that,
\bea\label{Cgme}
C_{gme}(|\psi\ra)=\min\limits_{\gamma_i \in \Gamma_2} \sqrt{2\left[ 1-\tr(\rho^{X_{1(i)}})^{2}\right] }
\eea
for pure state $|\psi\ra\in\mH^{A_1A_2\cdots A_n}$.

By definition,
\beax
\mE_{g''}^{(n)}\leq E_{g''}^{(n)}\leq E_{g'}^{(n)}\leq E_{g}^{(n)}
\eeax
for any $h$, and $\mE_{g''}^{(3)}=E_{g''}^{(3)}$. If $n\geq4$, there does exist state such that $\mE_{g''}^{(n)}< E_{g''}^{(n)}$~\cite{Guo2024rip}.

$C_{gme}$ is not a complete GEM since it does not satisfy the hierarchy condition~~\eref{ghierarchy}~\cite{Guo2022entropy}. 
In general, $E_{g''}^{(n)}$ and $\mE_{g''}^{(n)}$ do not obey the unification condition~\eref{gcoarsen} and the hierarchy condition~\eref{ghierarchy}~\cite{Guo2024rip}.

For the generalized GHZ state $|\widetilde{\GHZ}_{n,d}\ra$,
$E_{g''}^{(n)}$ and $\mE_{g''}^{(n)}$ are complete monogamous and tightly complete monogamous. For this state, $E_{g''}^{(n)}=E_{g'}^{(n)}=\mE_{g''}^{(n)}=\mE_{g'}^{(n)}$, and $nE_{g''}^{(n)}=nE_{g'}^{(n)}=2E_{g}^{(n)}$. Moreover, for such a state, all the entanglement are shared between all of the particles. We thus regard this state as the maximal genuinely entangled state, and it reaches the maximal value whenever $\lambda_0=\lambda_1=\cdots=\lambda_{d-1}={1}/{\sqrt{d}}$ for the multi-qudit case.

Comparing $E_{g''}^{(3)}$ with $E_{g'}^{(3)}$ and $E_{g}^{(3)}$, $E_{g'}^{(3)}$ seems the best one since (i) it is complete and completely monogamous whenever the reduced function is strictly concave, (ii) it can be easily calculated, and (iii) it is monogamous iff it is completely monogamous. For the case of $n\geq4$, $E^{(n)}$, $\mE^{(n)}$, $E_g^{(n)}$, and $\mE_g^{(n)}$ seems better than the other cases as a g-MEM/GEM since these measures admit the postulates of a complete g-MEM/GEM.

Among these measures, $C_{gme}$ is extensively explored. An $n$-qubit X-state refers to the matrix in the form
\begin{align*}
\left(  \begin{array}{cccccccc}
a_{1} &  &  &  &  &  &  & z_{1} \\ 
& a_{2} &   &  &  &  & z_{2} &  \\ 
&  & \ddots &  &  & \iddots &  &  \\ 
&  &  & a_{m} & z_{m} &  &  &  \\ 
&  &  & z_{m}^{*} & b_{m} &  &  &  \\ 
&  & \iddots &  &  & \ddots &  &  \\ 
& z_{2}^{*} &  &  &  &  & b_{2} &  \\ 
z_{1}^{*} &  &  &  &  &  &  & b_{1} \\ 
\end{array}
\right),
\end{align*}
where $m=2^{n-1}$, and we require $|z_{i}|\le \sqrt{a_{i}b_{i}}$ and $\sum_{i}(a_{i}+b_{i})=1$ to ensure that the state is positive and normalized. It was shown that the GM concurrence of an $n$-qubit X-state $\rho$ is given by~\cite{Hashemi2012pra}
\beax
C_{gme}(\rho)= 2\max\{0,|z_{i}|-w_{i}\}, ~ i=0,1,\dots,m
\eeax
where $w_{i}=\sum_{j\neq i}^{m}\sqrt{a_{j} b_{j}}$.

Let $\rho$ be any given biseparable state in $\mS^{A_1A_2\cdots A_n}$, then~\cite{Huber2010pra,Ma2011pra}
\beax\label{ineqII}
\underbrace{\sqrt{\la\Phi|\rho^{\otimes 2}\Pi_{\{1,2,\cdots,n\}}|\Phi\ra}-\sum\limits_{\gamma}\sqrt{\la\Phi|\Pi_{\gamma}\rho^{\otimes 2}\Pi_{\gamma}|\Phi\ra}}_{=: I[\rho,|\Phi\rangle]}\leq 0\, ,
\eeax
where $|\Phi\ra$ is any fully separable pure state in $\mH^{A_1A_2\cdots A_n}\ot\mH^{A_1A_2\cdots A_n}$, the sum runs over all
inequivalent bipartitions $\gamma$ on  and $\Pi_{\{\alpha\}}$ is the cyclic permutation operator acting on the twofold copy Hilbert space in the subsystems defined by $\{\alpha\}$, i.e. exchanging the vectors of the subsystems $\{\alpha\}$ of the first copy with the vectors of the subsystems $\{\alpha\}$ of the second copy (e..g.,
$\Pi_{\{1\}}|\phi_1\phi_2\rangle\otimes|\psi_1\psi_2\rangle=|\psi_1\phi_2\rangle\otimes|\phi_1\psi_2\rangle$).
It turns out that~\cite{Ma2011pra}
\beax 
C_{gme}(\rho)\geq \max_{|\Phi\rangle}2I[\rho,|\Phi\rangle].\label{lowerbound}
\eeax
holds for all $n$-qudit state $\rho$. Later, in Ref.~\cite{Chenzhihua2012pra}, they derived improved lower bounds of $C_{gme}$ for $n$-qudit states.

If $\rho\in\mS^{ABC}$ with $\dim\mH^{A,B,C}=d$, then~\cite{Liming2024qip}
\bea\label{9-3-1}
C_{gme}(\rho)\geq \frac{C(\rho^{A|BC})+C(\rho^{AB|C})+C(\rho^{B|AC})}{\sqrt2}-2\sqrt{\frac{d-1}{d}}.
\eea 
If $\rho\in\mS^{A_1A_2\cdots A_n}$ with $\dim\mH^{A_i}=d$, $1\leq i\leq n$, then~\cite{Liming2024qip}
\beax 
C_{gme}(\rho)\geq\frac{1}{\sqrt2}\sum_{\gamma_i\in\Gamma_2}h_C(\rho^{X_{1(i)}})-(2^{n-1}-2)\sqrt{\frac{d-1}{d}}.
\eeax 
This is tighter than
\beax 
C_{gme}(\rho)\geq\frac{1}{\sqrt{d(d-1)}}\left[\max\{M_1(\rho), M_2(\rho)\}-\frac{1+2d}{3}\right]
\eeax 
given in~\cite{Liming2017srep}, where $M_1(\rho)=\frac13\left(\|\rho^{T_a}\|_\tr+ \|\rho^{T_b}\|_\tr+\|\rho^{T_c}\|_\tr\right)$, $M_2(\rho)=\frac13\left(\|\rho^{R_a}\|_\tr+ \|\rho^{R_b}\|_\tr+\|\rho^{R_c}\|_\tr\right)$, $\rho^{R_x}$ denote the realignment of $\rho$ under the split $x|\overline{x}$.
The lower bound in Eq.~\eref{9-3-1} is also better than~\cite{Liming2024qip,Qi2024rp}
\beax 
\fl \quad \quad\qquad C_{gme}(\rho)\geq\frac{1}{d(d-1)}\left[ M_{\alpha,\beta}(\rho)-\sqrt{(1+\alpha^2)(1+\beta^2)}-\frac{2(d-1)}{3}\right]
\eeax
for any $\alpha$, $\beta\in\mbR$, where
$$\eqalign{
M_{\alpha,\beta}(\rho)=\frac13\left[\|M_{\alpha,\beta}(\rho^{A|BC})\|_\tr+ \|M_{\alpha,\beta}(\rho^{AB|C})\|_\tr+\|M_{\alpha,\beta}(\rho^{B|AC})\|_\tr\right],\cr
M_{\alpha,\beta}(\rho^{AB})=\left(\begin{array}{cc}\alpha\beta&\alpha{\rm vec}(\rho_B)^T\\
\beta{\rm vec}(\rho_A)&\rho_{AB}^{R}\end{array}  \right),\cr
{\rm vec}([a_{ij}])=(a_{11}, \dots, a_{m1}, a_{12}, \dots, a_{m2}, \dots a_{1n},\dots a_{mn})^T\cr
}
$$
for $m$ by $n$ matrix $[a_{ij}]$.

Define
\begin{eqnarray*}\label{M:1} \mathscr{Q}(\rho)=\frac{1}{3}\left(\left\|\mathscr{Q}_{A|BC}
\left(\rho\right)\right\|_{\mathrm{T r}}+\left\|\mathscr{Q}_{B|AC}\left(\rho\right)\right\|_{\mathrm{T r}}+\left\|\mathscr{Q}_{C|AB}\!\left(\rho\right)\right\|_{\mathrm{T r}}\right),
\end{eqnarray*}
then for any tripartite state $\rho\in\mS^{ABC}$ with $\dim\mH^{A,B,C}=d$, we have~\cite{Luyu2025qip}
\begin{eqnarray*}
\fl \qquad \qquad C_{gme}(\rho)\geq \frac{1}{\sqrt{d(d-1)}}{\left( \mathscr{Q}(\rho)-\sqrt{\left(| \mu|^{2}+1\right)\left(|\nu|^{2}+1\right)}-\frac{2(d-1)}{3}\right) }.	
\end{eqnarray*}

In Ref.~\cite{Wang2021lpl}, a lower bound was obtained, which is more tighter than the ones in Ref.~\cite{Liming2017pra,Xu2021epjp}, i.e.,
\beax 
\fl \qquad  C_{gme}(\rho)\geq \frac{C(\rho^{A|BC})+C(\rho^{B|AC})+C(\rho^{AB|C})}{3\sqrt2}+\frac23+LT(\rho)-\frac23\sqrt{\frac{d-1}{d}},
\eeax 
where
\beax 
LT(\rho)=\max\left\lbrace \frac{1}{2\sqrt2}\|T^{ABC}\|_2-\frac{d-1}{d}\sqrt{\frac{d+1}{d}}, 0 \right\rbrace,
\eeax 
$T^{ABC}$ is the correlation vector (tensors) of the generalized Bloch representation of $\rho$.


\subsection{Genuine geometric measure}


The geometric approach~\cite{Shimony95} can also be used for quantifying the genuine multiparty entanglement.
In Ref.~\cite{Das2016pra,Sen2010pra}, a computable geometric measure that can quantify genuine multiparty entanglement was presented, i.e.,
\bea\label{m-geometric-measure}
E_{g,G}(|\psi\ra)=1-\max\limits_{|\phi\ra\in\mP^{\Gamma_2}} | \langle \phi|\psi\rangle |^2
\eea    
with the maximization being over all biseparable $|\phi\rangle$ in $\mP^{\Gamma_2}$, where $\mP^{\Gamma_2}$ denotes all the biseparable pure states in $\mH^{A_1A_2\cdots A_n}$. It can be calculated as~\cite{Das2016pra,Sen2010pra}
\beax
E_{g,G}(|\psi\rangle)= 1-\max\limits_{\gamma_i\in\Gamma_2}\lambda^2_{\gamma_i},
\label{m-GGM}
\eeax
where $\lambda_{\gamma_i}$ is the maximal Schmidt coefficient of $|\psi\ra^{\gamma_i}$ up to the bipartition $\gamma_i$, and the maximum runs over all possible bipartitions in $\Gamma_2$.
$E_{g,G}$ is not a proper GEM since $E_{g,G}(|\GHZ\ra)=\frac12<E_{g,G}(|\W\ra)=\frac59$~\cite{Weinbrenner2025}.


\subsection{Genuine negativity}


The genuine multiparite negativity (GMN) was defined as~\cite{Jungnitsch2011prl}
\bea
\label{eqn:sdpgennegold}
\tilde{N}_g(\rho) = 
-\min\limits_{W\in\mW} \tr\left( \rho W \right),
\eea
where the minimum runs over all witness operators in 
$\mW=\left\lbrace W = P_\alpha +Q_\alpha^{T_\alpha}\big| 0\le P_\alpha \le I\right.$, $0\le Q_\alpha \le I\left.\right\rbrace$ for all partitions $\alpha\vert \overline{\alpha}$ of $A_1A_2\cdots A_n$, $I$ is the identity operator on $\mH^{A_1A_2\cdots A_n}$.

For the three-particle case the witness operator $W\in\mW$ has to be decomposable into 
$W=P_1+Q_1^{T_a}$, $W=P_2+Q_2^{T_b}$ and $W=P_3+Q_3^{T_c}$ 
with $0 \le P_m,Q_m \le I$, $1\leq m\leq3$. By defintion, $\tilde{N}_g(\rho)$ is closely related to the approach which termed witnessed entanglement introduced in Ref.~\cite{Brandao2005pra}, and can be directly computed
using SDP~\cite{pptmixerprogram}, since calculation of $\tilde{N}_g(\rho)$ is an optimization problem of 
this class~\cite{Hofmann2014jpa}.

Later, by relaxing the constraints on the positive 
operators $P_m$, i.e., it is not bounded by $I$ anymore, Hofmann \etal modified GMN as~\cite{Hofmann2014jpa} 
\bea
\label{eqn:sdpgenneg}
N_g(\rho) = 
-\inf\limits_{W\in\mW} \tr\left( \rho W \right) 
\eea 
where the infimum runs over all witness operators in 
$\mW=\left\lbrace W = P_\alpha +Q_\alpha^{T_\alpha}\big| 0\le P_\alpha \right.$, $0\le Q_\alpha \le I\left.\right\rbrace$  for all partitions  $\alpha\vert \overline{\alpha}$ of $A_1A_2\cdots A_n$.
Note that this definition was already used to quantify genuine multiparticle entanglement in a device independent manner~\cite{Moroder2013prl}.
It was shown in Refs.~\cite{Jungnitsch2011prl,Jungnitsch2011pra} that $\tilde{N}_g$ satisfies the following:
(i) $\tilde{N}_g(\sigma)=0$ for any biseparable state
$\sigma$, and if $\rho$ is no PPT mixture, then $\tilde{N}_g(\rho) > 0$;
(ii) $\tilde{N}_g$ is non-increasing under any $n$-partite LOCC operation;
(iii) $\tilde{N}_g$ is invariant under local unitary operation;
(iv) $\tilde{N}_g$ is convex, i.e. $\tilde{N}_g(\rho) \le \sum_i p_i \tilde{N}_g(\rho_i)$ holds for all convex decompositions $\rho=\sum_i p_i \rho_i$;
(v) $\tilde{N}_g$ is bounded by $\tilde{N}_g(\rho) \le \frac 1 2 (d_{\mathrm{min}}-1)$, where $d_{\mathrm{min}}$ is the lowest dimension of any particle in the system \cite{Jungnitsch2011pra};
(vi) If $n=2$, $\tilde{N}_g$ reduces to the bipartite negativity \cite{VidalWerner}.
Items (i)-(vi) are also valid for ${N}_g$~\cite{Hofmann2014jpa}. Moreover, $\tilde{N}_g(\rho)\leq{N}_g(\rho)$ for any $\rho$ and $\tilde{N}_g(\rho)=0\Leftrightarrow{N}_g(\rho)=0$.
Sometimes ${N}_g$ coincides with $\tilde{N}_g$ for pure states, and sometimes not. For example~\cite{Hofmann2014jpa}, the 
three-qubit GHZ state $|\GHZ\ra$, 
where $N_g(|{\GHZ}\ra)=\tilde{N}_g(|{\GHZ}\ra)=1/2$. On the other hand, 
for the three-qubit W state, $N_g(|\W\ra) = \sqrt{2}/3 \approx 0.47$ and $\tilde{N}_g(|\W\ra)\approx 0.43$.

We denote by $N_\gamma(\rho)$ the negativity of $\rho$ with respect to the bipartition $\gamma$ and let $N_{\min}(\rho)=\min\limits_{\gamma}N_\gamma(\rho)$, where the minimum runs over all possible bipartitions of $A_1A_2\cdots A_n$. It turns out that~\cite{Hofmann2014jpa}
\bea \label{mixed-convex-roof}
{N}_g(\rho)=\min\limits_{\rho=\sum_kp_k\rho_k}\sum_kp_kN_{\min}(\rho_k),
\eea
where the minimum is taken over all decomposition $\rho=\sum_kp_k\rho_k$ in to mixed states. 
Here $N_g$ in Eq.~\eref{mixed-convex-roof} is the mixed convex-roof of $N_g$ in Eq.~\eref{eqn:sdpgenneg}. In general, the mixed convex-roof extension of a measure $E$ means the decompositions into mixed states (see e.g. \cite{Synak2006jpa,Yangdong2005prl,Yang2009ieee}), 
\beax 
E(\rho)=\inf_{\sum_i p_i\rho_i=\rho} \sum_i p_i E(\rho_i)
\eeax 
with $\rho_i$'s are mixed states. This approach works in both bipartite as well as the multipartite scenario. 
For pure state $|\psi\ra$, $N_g(|\psi\ra)=N_{\min}(|\psi\ra)$.

Analytic formulae of ${N}_g$ for the $n$-qubit GHZ-diagonal and four-qubit cluster-diagonal states were also given in Ref.~\cite{Hofmann2014jpa}. The $n$-qubit GHZ-diagonal state refers to the state which is diagonal in the $n$-qubit GHZ-basis consisting of $2^n$ states $|\psi_i\ra = 1 / \sqrt{2} \left( |x_1x_2\dots x_n\ra \pm|{\overline{x_1} \,\overline{x_2}\dots \overline{x_n}}\ra \right)$, where $x_j, \overline{x_j}\in \left\{ 0,1 \right\}$ and $x_j\not = \overline{x_j}$.
For example, in the three-qubit case, this 
basis consists of the states 
$1/\sqrt{2}\left( \vert 000 \rangle \pm \vert 111 \rangle \right)$, 
$1/\sqrt{2}\left( \vert 001 \rangle \pm \vert 110 \rangle \right)$,  
$1/\sqrt{2}\left( \vert 010 \rangle \pm \vert 101 \rangle \right)$ and  
$1/\sqrt{2}\left( \vert 011 \rangle \pm \vert 100 \rangle \right)$. 
A three-qubit state diagonal in the GHZ basis is of the form
\begin{equation*}
\label{eqn:ghzdiag}
\rho =
\begin{pmatrix}
\lambda_0 & & & & & & & \mu_0 \\
& \lambda_1 & & & & & \mu_1 & \\
& & \lambda_2 & & & \mu_2 & & \\
& & & \lambda_3 & \mu_3 & & &\\
& & & \mu_3 & \lambda_3 & & &\\
& & \mu_2 & & & \lambda_2 & & \\
& \mu_1 & & & & & \lambda_1 & \\
\mu_0 & & & & & & & \lambda_0,
\end{pmatrix},
\end{equation*}
with $\lambda_i, \mu_i \in \mathbb R$ (GHZ-diagonal states have real $\mu_i$, but all of the results in Ref.~\cite{Hofmann2014jpa} hold true for states with complex $\mu_i$ as well). A general $n$-qubit GHZ-diagonal state  would have the same shape, with $2^{n-1}$ independent real $\lambda_i$ on the diagonal and corresponding real $\mu_i$ on the anti-diagonal. 
For any GHZ-diagonal $n$-qubit state $\rho$,
\beax 
N_g(\rho)=\max\limits_i\{0, |\mu_i|-w_i\}=\max\limits_i\left\lbrace 0, F_i-\frac 1 2 \right\rbrace ,
\eeax
where $F_i = \la\psi_i|\rho|\psi_i\ra$ denotes the fidelity with the GHZ-basis state $|\psi_i\ra$. $N_g(\rho) = \max_i {0,|\mu_i| -w_i}$ holds also true for the slightly more general case with complex $\mu_i$ on the anti-diagonal.
For $n$-qubit state, we always have~\cite{Hofmann2014jpa}
\beax 
N_g(\rho)\leq C_{gme}(\rho).
\eeax 

Let $\rho = \sum_{\alpha\beta\gamma\delta} a_{\alpha\beta\gamma\delta} |\alpha\beta\gamma\delta\ra\la\alpha\beta\gamma\delta|$ be diagonal in the cluster graph basis $\{|++++\ra, |+++-\ra, \dots, |----\ra\}$, $|\pm\ra=\frac{1}{\sqrt2}(|0\ra+|1\ra)$, then the GMN of that state is given by~\cite{Hofmann2014jpa}
\beax
N_g(\rho) = - \min_{\alpha,\beta,\gamma,\delta,\mu,\nu}  \left\{ \tr( W_{\alpha\beta\gamma\delta}\rho)  \right\} \cup \left\{ \tr(W_{\alpha\beta\gamma\delta\mu\nu}\rho) \right\} \cup \left\{ 0 \right\},
\label{eqn:clustergenneg}
\eeax
where 
\beax
\label{eqn:clusterwone} &W_{\alpha\beta\gamma\delta} = \frac{I}{2} - |\alpha\beta\gamma\delta\ra\la\alpha\beta\gamma\delta| -\frac 1 2 \sum_{i,j} |\overline{\alpha} i j \overline{\delta}\ra\la \overline{\alpha} i j \overline{\delta}|,\\
\label{eqn:clusterwtwo} &W_{\alpha\beta\gamma\delta\mu\nu} =  \frac{I}{2} - |\alpha\beta\gamma\delta\ra\la\alpha\beta\gamma\delta| - |\overline{\alpha} \mu\nu \overline{\delta}\ra\la \overline{\alpha} \mu\nu \overline{\delta}|.
\eeax


\subsection{Geometric mean of bipartite entanglement measure}


For any $\gamma_i\in\Gamma_k$, let
\bea\label{mQ_k^{gamma_i}}
\mQ_k^{\gamma_i}(|\psi\ra)=\left[ \prod_{t=1}^kh(\rho^{X_{t(i)}})\right]^{1/k},~~2\leq  k<n.
\eea
Li and Shang proposed the geometric mean of bipartite concurrence (GBC) in Ref.~\cite{Liyinfei2022prr} and showed 
that it was a genuine entanglement monotone. Recall that, the GBC is defined by
\bea\label{CG}
C_G(|\psi\ra)=\left[ \prod_{\gamma_i\in\Gamma_2}\mQ_2^{\gamma_i}(|\psi\ra)\right]^{1/|\Gamma_2|},
\eea 
where $h(\rho)=\sqrt{\frac{d}{d-1}(1-\tr\rho_A^2)}$, $d$ is the dimension of the smaller subsystem of the corresponding system. The the methods of geometric mean and the sume of the reduced functions have been used in Ref.~\cite{Szalay2015pra} for defining the measure of the $\alpha$-entanglement entropy.

The geometric mean of $q$-concurrence was defined in Ref.~\cite{Shi2023adp}
\bea\label{geometric mean of $q$-concurrence}
C_{G\text{-}q}(|\psi\ra)=\left[ \prod_{\gamma_i\in\Gamma_2}\mQ_2^{\gamma_i}(|\psi\ra)\right]^{1/|\Gamma_2|},
\eea 
where $h(\rho)=1-\tr\rho_A^q$, $q\geq 2$.
Replacing $h(\rho)=1-\tr\rho_A^q$ with $h(\rho)=\tr\rho_A^{\alpha}-1$, $0\leqslant\alpha\leqslant\frac12$, it is just the geometric mean of $\alpha$-concurrence~\cite{Wang2025epjp}. In fact, this approach is valid for any concave reduced function $h$ (also see in Sec.~\ref{sec-11.2.4}).


\subsection{Triangle measure}


According to Eq.~\eref{polygamy-of-C} and Eq.~\eref{polygamy-of-C'}, for the tripartite case $|\phi\ra\in\mH^{ABC}$, both ($C_{A|BC}$, $C_{B|AC}$, $C_{C|AB}$) and ($C_{A|BC}^2$, $C_{B|AC}^2$, $C_{C|AB}^2$) can represent the lengths of the three edges of a triangle, which we call it concurrence triangle.
Consequently, in Ref.~\cite{Xie2021prl}, a new genuine three-qubit entanglement measure, called {concurrence fill}, was put forward. It is quantified as the square root of the area of concurrence triangle (up to some constant factor).
Under this scenario, the GHZ state is more entangled than the W state.
Recall that, the concurrence fill is defined as~\cite{Guo2024pra}
\bea\label{triangle}
F_{ABC}(|\phi\ra)=\left[\frac{16}{3}Q(Q-C^2_{A|BC})(Q-C^2_{B|AC})(Q-C^2_{C|AB})\right]^{1/4},
\eea
where $Q=\frac12(C^2_{1|23}+C^2_{2|13}+C^2_{3|12})$, the factor $\frac{16}{3}$ ensures the normalization $0\leq  F_{ABC}(|\phi\rangle)\leq 1$.
Unfortunately, it
has been shown that this measure can increase under LOCC~\cite{Ge2023pra},
which means it is not a proper entanglement measure.
Later, it is shown in Ref.~\cite{Jin2023rip} that 
\bea\label{triangle'}
F'_{ABC}(|\phi\rangle)=\Big[\frac{16}{3}Q'(Q'-C_{A|BC})(Q'-C_{B|AC})(Q'-C_{C|AB})\Big ]^{\frac{1}{2}}
\eea
defines a well defined genuine tripartite entanglement measure, 
where $Q'=\frac12(C_{A|BC}+C_{B|AC}+C_{C|AB})$.

\begin{figure}
\hspace{30mm}\includegraphics[width=38mm]{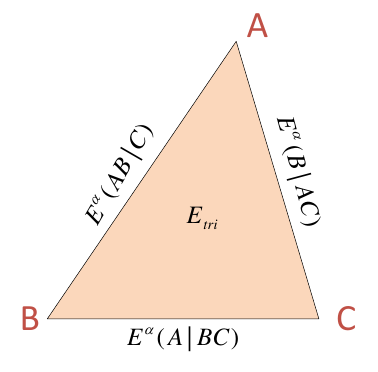}
\caption{\label{fig7-4}(color online). The $E$-triangle for a tripartite system. The
lengths of the three edges are set equal to the $\alpha$-th power of the three
one-to-other bipartite entanglement measured by $E$. }
\end{figure}

By Eq.~\eref{triangle''}, $E^{\alpha}(|\phi\ra^{A|BC})$, $E^{\alpha}(|\phi\ra^{B|AC})$ and $E^{\alpha}(|\phi\ra^{AB|C})$ can represent the lengths of the 
three edges of a triangle (the case of $E(|\phi\ra^{A|BC})=E(|\phi\ra^{B|AC})$, $E(|\phi\ra^{AB|C})=0$ is reduced to a line segment which is regarded as a trivial triangle hereafter), which is called $E$-triangle (See in Fig.~\ref{fig7-4}). Note that almost all entanglement measures so far are continuous, so we can always obtain the corresponding $E$-triangle.

For any bipartite entanglement measure with the reduced function is subadditive, for the case of $0<\eta\leq \frac12$, the triangle area
\bea\label{triangle area}
F_\eta(|\phi\ra)=\sqrt{Q(Q-E_{A|BC}^{\eta})(Q-E_{B|AC}^{\eta})(Q-E_{C|AB}^{\eta})}
\eea
define a genuine entanglement monotone but fails when $\eta>\frac12$~\cite{Ge2024pra}, where $Q=\frac12(E_{A|BC}^{\eta}+E_{A|BC}^{\eta}+E_{A|BC}^{\eta})$.

For any $|\phi\ra\in\mH^{ABC}$,
we write $E^{\alpha}_{A|BC}=x$, $E^{\alpha}_{AB|C}=y$, $E^{\alpha}_{B|AC}=z$ and denote the area of the $E$-triangle of $|\phi\ra$ by $E_{tri}(|\phi\ra)$.
Then $E_{tri}(|\phi\ra)>0$ if and only if $|\phi\ra$ is genuinely entangled, but we can not know whether $E_{tri}$
is nonincreasing under LOCC.
Guo introduced another quantity~\cite{Guo2022jpa}, 
\bea\label{gmem3}
\mF_{123}(\rho)=\left\lbrace \begin{array}{ll}
\frac{xyz}{x+y+z},&~x+y+z>0,\\
0,&~x+y+z=0
\end{array}\right. ~{\rm or}~\mF_{123}(\rho)={xyz}
\eea
for any $\rho\in\mS^{ABC}$
with $E^{A|BC}(\rho)=x$, $E^{AB|C}(\rho)=y$, $E^{B|AC}(\rho)=z$. 
Then $\mF_{123}$ is shown to be a 3-entanglement measure but not a faithful 3-entanglement measure for any bipartite entanglement measure $E$~\cite{Guo2022jpa}, and is not a GEM in general since there exists
2-separable but not fully separable mixed state $\sigma$ such that
$\mF_{123}(\sigma)>0$ in high-dimension system.	
In addition,
$\mF_{123}^F$ is not an entanglement measure any more in general
since we can not guarantee that it is nonincreasing under LOCC even for pure state.
This leads us to establish the GEM as in Eq.~\eref{gsum1}.

\begin{figure}	
\vspace{1mm}
\hspace{30mm}\includegraphics[width=35mm]{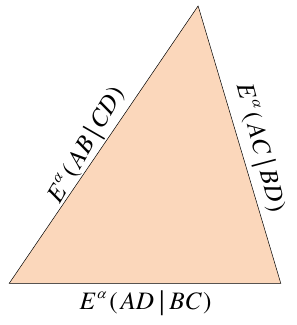}
\vspace{-1mm}
\caption{\label{fig7-4-2}(color online). The triangle structure for a 4-partite system via bipartite entanglement.}
\end{figure}

\begin{figure}	
\vspace{1mm}
\hspace{30mm}\includegraphics[width=37mm]{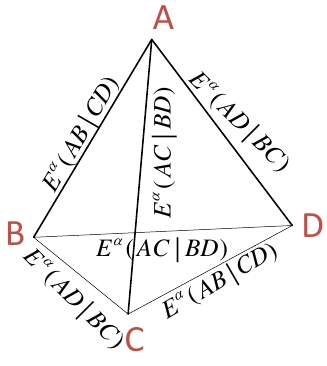}
\vspace{-1mm}
\caption{\label{fig7-4-3}(color online). The tetrahedron structure for a 4-partite system via bipartite entanglement. The opposite edges are equal.}
\end{figure}

It follows from Eq.~\eref{tetra-condition-2} that any pure state in $\mH^{ABCD}$
can induce a triangle for any bipartite entanglement measure that determined by the reduced state for the pure sates (see in Fig.~\ref{fig7-4-2}). In addition,
$E({AB|CD})$, $E({AC|BD})$, $E({AD|BC})$ can be regarded as
6 quantities
$\{E({AB|CD})$, $E({AC|BD})$, $E({AD|BC}), E({CD|AB})$, $E({BD|AC})$, $E({BC|AD})\}$, 
and therefore we can consider whether these 6 quantities can build a tetrahedron.
It depends obviously since in such a case it is equivalent to the fact that whether four same triangles can constitute a tetrahedron (namely, four copies of the same triangle can not necessarily constitute a tetrahedron).
When it does, the tetrahedron is illustrated in Fig.~\ref{fig7-4-3}.

\begin{figure}	
\hspace{30mm}\includegraphics[width=38mm]{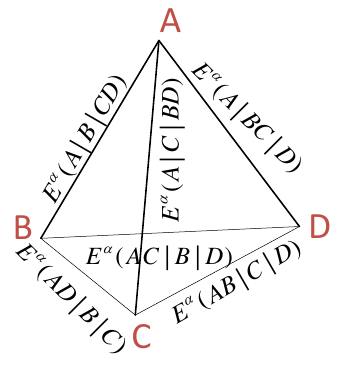}
\vspace{-4mm}
\begin{center}
{\footnotesize Case~(a)}
\end{center}
\vspace{-1mm}
\caption{\label{fig7-4-4}(color online). The tetrahedron for a 4-partite system for the case (a). The
lengths of the 6 edges are set equal to the $\alpha$-th power of $E^{(3)}(|\psi\ra^{X|Y|ZW})$, $\{X,Y,Z,W\}=\{A,B,C,D\}$. }
\end{figure}

\begin{figure}
\hspace{30mm}\includegraphics[width=45mm]{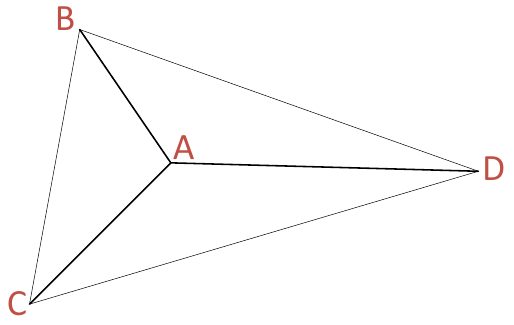}
\vspace{-2mm}
\begin{center}
{\footnotesize Case~(b1)}
\end{center}
\vspace{-1mm}
\caption{\label{fig7-4-5}(color online). The triangle structure for a 4-partite system in the case of (b1) (Assume that $S_{BCD}=S_{ABC}+S_{ACD}+S_{ABD}$).}
\end{figure}

\begin{figure}	
\hspace{30mm}\includegraphics[width=20mm]{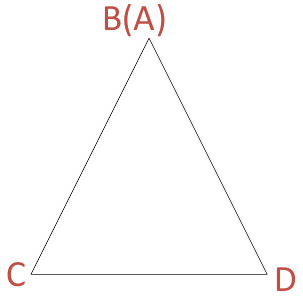}
\begin{center}
{\footnotesize Case~(b2)}
\end{center}
\vspace{-1mm}
\caption{\label{fig7-4-6}(color online). The regular triangle structure for a 4-partite system in the case of (b2) (Assume that $E^{(3)}(|\psi\ra^{A|B|CD})=0$ and $E^{(2)}(|\psi\ra^{CD})>0$).}
\end{figure}

\begin{figure}	
\vspace{3mm}
\hspace{30mm}\includegraphics[width=37mm]{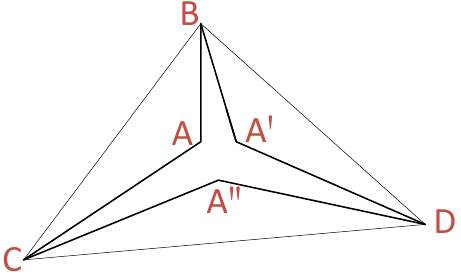}
\begin{center}
{\footnotesize Case (c)}
\end{center}
\vspace{-1mm}
\caption{\label{fig7-4-7}(color online). The triangle structure for a 4-partite system in the case of (c) (Assume that $E^{(3)}(|\psi\ra^{ABCD})>0$ for any tripartite partition of $ABCD$ and $S_{BCD}>S_{ABC}+S_{ACD}+S_{ABD}$).}
\end{figure}

Let $E^{(3)}$ be a continuous unified tripartite entanglement measure.
For any three-partition $R|S|T$ of $ABCD$ and any given pure state $|\psi\ra\in\mH^{ABCD}$,
we let $S_{RST}$ denotes the area of the triangle $\bigtriangleup_{RST}$ induced by $E^{\alpha}$ as in Eq.~\eref{tetra-condition-1}. 
With the same spirit as above, we can obtain three different triangle structures~\cite{Guo2022jpa}:
\begin{itemize}
\item [(a)] $S_{XYZ}+S_{XYW}+S_{XZW}> S_{YZW}$, for any $\{X,Y,Z,W\}=\{A,B,C,D\}$ (See in Fig.~\ref{fig7-4-4}).
\item [(b)] $S_{XYZ}+S_{XYW}+S_{XZW}= S_{YZW}$ for some $\{X,Y,Z,W\}=\{A,B,C,D\}$ (See in Figs.~\ref{fig7-4-5} and ~\ref{fig7-4-6}).
\item [(c)] $S_{XYZ}+S_{XYW}+S_{XZW}< S_{YZW}$ 
for some $\{X,Y,Z,W\}=\{A,B,C,D\}$ (See in Fig.~\ref{fig7-4-7}).
\end{itemize}

Going further, it is hard to find a symmetric geometric structure for $m$-partite system whenever $m>4$~\cite{Guo2022jpa}.
That is, the method in Ref.~\cite{Xie2021prl} is hard to be extended into $m$-partite system whenever $m>3$.


\section{Examples of $k$-entanglement measures}\label{sec-11}


We list in this section the $k$-entanglement measures proposed in Ref.~\cite{Guo2024pra,Hong2012pra,Li2024pra,Li2025cjp}. All these measures can be divided into two classes, they are either based on the sum of the reduced functions or from the product of the reduced functions.
For all these measures, we discussed whether they are complete on not and whether they are monogamous. Unfortunately, all of them may not even be unified. By comparing, the sum based class seems better than the product based one in general.


\subsection{$k$-EM from sum of the reduced functions}


In Ref.~\cite{Guo2024pra}, four classes of $k$-EMs in terms of the sum of the reduced functions are established.
For any $\gamma_i\in\Gamma_k$, let~\cite{Guo2024pra}
\bea\label{mP_k^{gamma_i}}
\mP_k^{\gamma_i}(|\psi\ra)=\frac12\sum_{t=1}^kh(\rho^{X_{t(i)}}),~2\leq  k<n,
\eea
where $\rho^{X}=\tr_{\overline{X}}|\psi\ra\la\psi|$. The coefficient ``1/2'' is fixed by the unification condition when the measures defined via $\mP_k^{\gamma_i}$ are regarded as unified $k$-EMs.

\subsubsection{The minimal sum}

The minimal $\mP_k^{\gamma_i}$, i.e.,
\bea\label{sum}
E_k(|\psi\ra)=
\min\limits_{\Gamma_k}\mP_k^{\gamma_i}(|\psi\ra),
\eea
is a well-defined $k$-EMo~\cite{Guo2023pra}, where the minimum is taken over all feasible $k$-partitions in $\Gamma_k$. If we take $k=n$, then $E_k$ defined in Eq.~\eref{sum} is reduced to the MEM $E^{(n)}$ in Ref.~\cite{Guo2024rip}.
When $k=2$ with $h$ is the reduced function of the concurrence, it reduces to the GEM $C_{gme}$ in Ref.~\cite{Ma2011pra}.

It is shown that~\cite{Guo2024pra}: (i) $E_k$ is $k$-monotonic whenever the reduced function is subadditive, (ii) if $h$ is strictly concave, $E_k$ is monogamous, (iii) $E_{k}$ is not coarsening monotonic (even for the genuine entangled states), (iv) $E_k$ is not tightly coarsening monotonic, moreover,
\beax\label{k-n-t-coarsen}
E_k(X_1|X_2| \cdots| X_{p})\leq  E_k(Y_1|Y_2| \cdots |Y_{q})
\eeax
holds for all $k$-entangled state $\rho\in\mS^{A_1A_2\cdots A_n}$ whenever $X_1|X_2|\cdots|X_{p}\succ^b Y_1|Y_2|\cdots |Y_{q}$ with $k\leq  q\leq  p$.

Particularly, for the case of $k=2$ (we denote it by $E_{2'}$ in order to distinguish it from the partial-norm of entanglement $E_2$), $E_{2'}$ is a GEM.
By the monogamy criterion, if the reduced function is strictly concave, $E_{2'}$ is monogamous. 
In such a case, $E_{2'}$ is monogamous means it is monogamous as a $2$-entanglement, but not regarded as genuine entanglement since genuine entanglement under some given partition is meaningless~\cite{Guo2024pra}. The same is true in other cases we proposed below where we do not repeat any more.
In addition, {\color{red}$E_{2'}$} as a GEM is neither completely monogamous nor tightly complete monogamous.
The same is true for other cases below which we do not restate again either.

\subsubsection{The maximal sum}

The maximal $\mP_k^{\gamma_i}(|\psi\ra)$, i.e.,
\bea\label{sum3}
E'_k(|\psi\ra)
=\left\lbrace \begin{array}{ll}
	\max\limits_{\Gamma_k}\mP_k^{\gamma_i}(|\psi\ra), & \min\limits_{\Gamma_k}\mP_k^{\gamma_i}(|\psi\ra)>0,\\
	0, & \min\limits_{\Gamma_k}\mP_k^{\gamma_i}(|\psi\ra)=0
\end{array}\right. 
\eea
is proposed in Ref.~\cite{Guo2024pra}, where the maximum/minimum is taken over all feasible $k$-partitions in $\Gamma_k$. $E'_k$ may be not a well-defined entanglement monotone since we can not guarantee that it is non-increasing on average under LOCC, and it is not even a well-defined entanglement measure~\cite{Guo2024pra}. 
If the reduced function $h$ is subadditive, it is shown that~\cite{Guo2024pra} (i) $E'_k$ is $k$-monotonic,
(ii) $E'_k$ is tightly coarsening monotonic, and (iii) 
\bea\label{m-coarsening} 
E'_k(X_1X_2\cdots X_p)\geq E'_l(Y_1Y_2\cdots Y_s)
\eea 
for any state $\rho\in\mS^{A_1A_2\cdots A_n}$, where $X_1X_2\cdots X_p\succ^a Y_1Y_2\cdots Y_s$, $l\leq  s\leq  k\leq  p$.

Although Eq.~\eref{m-coarsening} is valid for $E'_k$, we can not derive it is coarsening monotonic.
In general, tightly coarsening monotonicity if stronger than the coarsening monotonicity, we thus conjecture that
$E'_k$ is coarsening monotonic. In such a sense, if $E'_k$ is non-increasing under LOCC (resp. non-increasing on average under LOCC), it is a complete $k$-EM (resp. $k$-EMo) provided that $h$ is subadditive.

$E'_k$ is weakly monogamous if $h$ is strictly concave
since $E'_k=E_k$ for the state under any $k$-partition. But Eq.~\eref{hierachy3} may be not valid for $E'_k$ since it may not be a $k$-EMo. Even if it is a $k$-EMo/$k$-EM,
we can not derive that it is monogamous in general~\cite{Guo2024pra}.

\subsubsection{The arithmetric mean}

The following quantity
\bea\label{arithmetric mean of k-partition}
\bar{E}_k(|\psi\ra)
=\left\lbrace \begin{array}{ll}
	\frac{\sum_{\gamma_i}\mathcal{P}_k^{\gamma_i}(|\psi\ra)}{|\Gamma_k|},&\min\limits_{\Gamma_i}\mathcal{P}_k^{\gamma_i}(|\psi\ra)>0,\\
	0,&\min\limits_{\Gamma_i}\mathcal{P}_k^{\gamma_i}(|\psi\ra)=0
\end{array}\right. 
\eea
defines a $k$-EMo~\cite{Guo2024pra},
where the minimum is taken over all feasible $k$-partitions in $\Gamma_k$.
It is $k$-monotonic if the reduced function $h$ is subadditive, and is monogamous whenever $h$ is strictly concave~\cite{Guo2024pra}.
In general, $\bar{E}_k$ is neither coarsening monotonic nor tightly coarsening monotonic~\cite{Guo2024pra}.

\subsubsection{The geometric mean}

In Ref.~\cite{Li2025cjp}, two $k$-EMs called $k$-geometric multipartite concurrence and $q$-$k$-geometric multipartite concurrence respectively are proposed. 
In fact, the way of defining $k$-EM therein is valid for any non-negative concave function $h$.
The geometric mean
\bea\label{geoetric mean of k-parition}
E^G_k(|\psi\ra)=\left[ \prod_{\gamma_i\in\Gamma_k}\mathcal{P}_k^{\gamma_i}(|\psi\ra)\right]^ {1/|\Gamma_k|}
\eea
is a $k$-EMo~\cite{Guo2024pra}.
$\mathcal{P}_k^{\gamma_i}$ is different from that of $\mathcal{P}_k$ associated with concurrence and the $q$-concurrence in Ref.~\cite{Li2025cjp}: $2\mathcal{P}_k^{\gamma_i}(|\psi\ra)=k\left[ \mathcal{P}_k(|\psi\ra)\right] ^2$ when we take $h=\sqrt{2(1-\tr\rho^2)}$ or $h=\sqrt{2(1-\tr\rho^q)}$, $q>1$. 
It is checked in Ref.~\cite{Guo2024pra} that (i) ${E}^G_k$ is $k$-monotonic if $h$ is subadditive, and it
is also monogamous if the associated reduced function is strictly concave, (ii) $E^G_{k}$ is neither coarsening monotonic nor tightly coarsening monotonic.
It is clear that
\beax
E_k\leq  E^G_k\leq  \bar{E}_k\leq  E'_k,~~
E_n= E^G_n= \bar{E}_n= E'_n
\eeax
hold for any $\rho\in\mS^{A_1A_2\cdots A_n}$, 
and all these measures are faithful.

We summarize these properties in Table~\ref{tab:table8-1} for more clarity.
So far we have known that the von Neumann entropy, the Tsallis $q$-entropy for $q>1$, and the reduced function of concurrence are both strictly concave and subadditive~\cite{GG2018q,Guo2024rip,Raggio,Vidal2000,Wehrl1978}, so the $k$-EMs defined in this way with these reduced functions are better than the other ones.

\Table{\label{tab:table8-1} Comparing of $E_k$, $E'_k$, $\bar{E}_k$, and $E^G_k$ with the assumption that the reduced functions are strictly concave and subadditive.
	$k$-M, CoM, and TCoM signify the measure is $k$-monotonic, coarsening monotonic, and tightly coarsening monotonic, respectively.}	
	\br
	$k$-EM     & $k$-M&CoM&TCoM & Complete & M &CM & TCM \\ 
	\mr
	$E_k$         &$\checkmark$&$\times$ &$\times$&$\times$&$\checkmark$&$\times$&$\times$\\			
	$E'_k$${\color{red}^{\rm a}}$         &$\checkmark$&$\checkmark{\color{red}^{\rm b}}$ &$\checkmark$&$\checkmark{\color{red}^{\rm b}}$&$\times{\color{red}^{\rm c}}$&$\times$&$\times$\\			
	$\bar{E}_k$        &$\checkmark$&$\times$& $\times$&$\times$&$\checkmark$&$\times$&$\times$  \\
	$E^G_k$           &$\checkmark$&$\times$& $\times$&$\times$&$\checkmark$&$\times$&$\times$\\
	\br
\end{tabular}
\item[] ${\color{red}^{\rm a}}$ We assume here that $E'_k$ is a $k$-EMo.
\item[] ${\color{red}^{\rm b}}$ We only prove Eq.~\eref{m-coarsening} in this paper and we conjecture that it is CoM.
\item[] ${\color{red}^{\rm c}}$ It is weakly monogamous.
\end{indented}
\end{table}

\subsubsection{$k$-ME concurrence}

Hong \etal put forward the so-called $k$-ME concurrence, which was defined by~\cite{Hong2012pra}
\bea\label{CkME}
C_{k\text{-}ME}(|\psi\ra)=\min\limits_{\Gamma_k}\sqrt{\frac{2\sum_{t=1}^k\left[  1-\tr\left( \rho^{X_{t(i)}}\right)^2\right]  }{k}}
\eea
It is in fact a faithful $k$-entanglement monotone.

Very recently, according to the $k$-ME concurrence together with the $q$-concurrence and the $\alpha$-concurrence,  Li \etal proposed $q$-$k$-ME concurrence and $\alpha$-$k$-ME concurrence~\cite{Li2024pra}, i.e.,
\bea\label{Cqk}
C_{q\text{-}k}(|\psi\ra)=\min\limits_{\Gamma_k}\frac{\sum_{t=1}^k\left[  1-\tr\left( \rho^{X_{t(i)}}\right)^q\right]}{k},\quad q>1
\eea 
and
\bea \label{Calphak}
C_{\alpha\text{-}k}(|\psi\ra)=\min\limits_{\Gamma_k}\frac{\sum_{t=1}^k\left[ \tr\left( \rho^{X_{t(i)}}\right)^{\alpha}- 1\right]}{k},\quad 0\leq  \alpha<1,
\eea 
respectively. They were shown to be faithful $k$-entanglement monotones but they are note even $k$-monotonic.
. They are lower bounded by that of the permutational invariant part,
i.e.,
\beax
C_{q\text{-}k}(\rho)&\geq& \max\limits_{U}C_{q\text{-}k}(\rho_U^{\rm PI}),\\
C_{\alpha \text{-}k}(\rho)&\geq& \max\limits_{U}C_{\alpha\text{-}k}(\rho_U^{\rm PI}),
\eeax   
where the permutation invariant (PI) part of $\rho$, is defined as~\cite{Gao2014prl,Moroder2012njp,Toth2010prl}
\bea\label{PI}
\rho^{\rm PI}=\frac{1}{n!}\sum\limits_{i=1}^{n!}\Pi_i\rho\Pi_i^\dag,
\eea
where $\{\Pi_i\}$ denotes the set of all $n!$ permutations of the $n$ particles.
Permutation invariant (PI) part of the state can report the structure of the multipartite entanglement~\cite{Gao2014prl}. It was shown in Ref.~\cite{Gao2014prl} that for a series of MEMs, the maximum of theses MEMs of the PI part of the state was a lower bound on that of the state, namely,
\beax\label{permutation-ivariant}
E(\rho)\geq \max\limits_{U}E(\rho_U^{\rm PI}),
\eeax
where $\rho_U=U_1\ot U_2\ot\cdots \ot U_n\rho U_1^\dag \ot U_2^\dag \ot\cdots \ot U_n^\dag$ and the maximum runs over all local unitary operations $U$. Therefore, if $\rho^{\rm PI}$ is $k$-entangled, so is $\rho$, and conversely, $\rho^{\rm PI}$ is $k$-separable whenever $\rho$ is $k$-separable.

\subsubsection{$k$-GM concurrence}

In Ref.~\cite{Li2025cjp}, the $k$-GM concurrence was suggested,
\bea\label{k-GM}
C_{k\text{-}GM}(|\psi\ra)=\frac{\sqrt2}{\sqrt{k}}\left[ \prod\limits_{\gamma_i \in \Gamma_k}\sqrt{\mP_k^{\gamma_i}(|\psi\ra)}\right]^{1/|\Gamma_k|},\quad 2\leq  k\leq  n
\eea
with $h$ takes the reduced function of the tangle.
By replacing $h$ with the reduced function of the $q$-concurrence and the $\alpha$-concurrence in~\Eref{k-GM}, respectively, they are just the $q$-$k$-GM concurrence and the $\alpha$-$k$-GM concurrence~\cite{Li2025cjp}.
All these measures are shown to be well-defined faithful $k$-entanglement monotones~\cite{Li2025cjp}. But these measures are note even $k$-monotonic. For example, for $|\psi\ra^{ABC}=\frac12(|000\ra+|101\ra+|210\ra+|311\ra)$,
$C_{3\text{-}GM}(|\psi\ra^{ABC})<C_{2\text{-}GM}(|\psi\ra^{ABC})$.


\subsection{$k$-EM from the product of the reduced functions}


\subsubsection{The minimal product}\label{A}

The minimal $\mQ_k^{\gamma_i}$ which is defined by
\bea\label{product}
E_{G\text{-}k}(|\psi\ra)=
\min\limits_{\Gamma_k}\mQ_k^{\gamma_i}(|\psi\ra)
\eea
is a $k$-EMo~\cite{Guo2024pra}, where the minimum is taken over all feasible $k$-partition $\Gamma_k$.
$E_{G\text{-}k}$ is not faithful, and whenever
\beax
2\leq  k\leq \left\lbrace \begin{array}{ll}
\!\!\frac{n}{2}+1,& \mbox{if $n$ is even},\\
\!\!\frac{n+1}{2}+1,& \mbox{if $n$ is odd},
\end{array}\right. 
\eeax
$E_{G\text{-}k}$ admits the following properties~\cite{Guo2024pra}:
(i) ${E}_{G\text{-}k}(\rho)=0$ if and only if $\rho$ is genuinely entangled,
(ii) $E_{G\text{-}k}\geq E_{G\text{-}2}$ for any concave function $h$, 
But $E_{G\text{-}k}$ is not $k$-monotonic in general,
(iii) $E_{G\text{-}k}$ is not coarsening monotonic,
(iv) If $h$ is strictly concave, $E_{G\text{-}k}$ is monogamous,
(v) $E_{G\text{-}k}$ is not completely monogamous as a GEM since it is not coarsening monotonic on genuine entangled states,
and it is not tightly complete monogamous as a GEM since it is not tightly coarsening monotonic on genuine entangled states.

\subsubsection{The maximal product}

The maximal $\mQ_k^{\gamma_i}$ that defined by
\bea\label{product3}
E'_{G\text{-}k}(|\psi\ra)
=\left\lbrace \begin{array}{ll}
\!\!\max\limits_{\Gamma_k}\mQ_k^{\gamma_i}(|\psi\ra), & \min\limits_{\Gamma_k}\mP_k^{\gamma_i}(|\psi\ra)>0,\\
\!\!0, & \min\limits_{\Gamma_k}\mP_k^{\gamma_i}(|\psi\ra)=0,
\end{array}\right. 
\eea
is considered in Ref.~\cite{Guo2024pra},
where the maximum/minimum is taken over all feasible $k$-partitions in $\Gamma_k$.
$E'_{G\text{-}k}$ may be not a well-defined entanglement monotone since we can not guarantee that it non-increasing on average under stochastic LOCC, and it is not even a well-defined entanglement measure. 
$E'_{G\text{-}k}$ satisfies~\cite{Guo2024pra}: (i)
\beax\label{k-n-t-coarsen-3}
E'_{G\text{-}k}(X_1|X_2| \cdots| X_{p})\geq E'_{G\text{-}k}(Y_1|Y_2| \cdots |Y_{q})
\eeax
holds for all $k$-entangled state $\rho\in\mS^{X_1X_2\cdots X_{p}}$ whenever $X_1|X_2|\cdots|X_{p}\succ^b Y_1|Y_2|\cdots |Y_{q}$ with $k\leq  q\leq  p$, but it is not tightly coarsening monotonic,  
(ii) It is also weakly monogamous if the associated reduced function is strictly concave, (iii) It is not $k$-monotonic,
(iv) $E'_{G\text{-}k}$ is not coarsening monotonic.

\subsubsection{The arithmetric mean}

The arithmetric mean of $\mathcal{Q}_k^{\gamma_i}$, i.e.,
\bea\label{arithmetric mean}
\bar{E}_{G\text{-}k}(|\psi\ra)
=\left\lbrace \begin{array}{ll}
\!\!\frac{\sum_{\gamma_i}\mathcal{Q}_k^{\gamma_i}(|\psi\ra)}{|\Gamma_k|},&\min\limits_{\Gamma_i}\mathcal{P}_k^{\gamma_i}(|\psi\ra)>0,\\
\!\!0,&\min\limits_{\Gamma_i}\mathcal{P}_k^{\gamma_i}(|\psi\ra)=0,
\end{array}\right. 
\eea
define a $k$-EMo~\cite{Guo2024pra}, where the minimum is taken over all feasible $k$-partitions in $\Gamma_k$.
If a $k$-entanglement measure is $k$-monotonic, it is faithful, but not vice versa. Although $\bar{E}_{G\text{-}k}$ is not $k$-monotonic, it is faithful~\cite{Guo2024pra}. In addition, one has~\cite{Guo2024pra}
(i) $\bar{E}_{G\text{-}k}$ is monogamous if the associated reduced function is strictly concave, (ii)$E'_{G\text{-}k}$ is not $k$-monotonic, (iii) $\bar{E}_{G\text{-}k}$ is neither coarsening monotonic nor tightly coarsening monotonic.

\Table{\label{tab:table8-2} Comparing of $E_{G\text{-}k}$, $E'_{G\text{-}k}$, and $\bar{E}_{G\text{-}k}$ with the assumption that the reduced functions is strictly concave.}	
\br
$k$-EM     & $k$-M&CoM&TCoM & Complete & M &CM & TCM \\ 
\mr
$E_{G\text{-}k}$         &$\times$&$\times$ &$\times$&$\times$&$\checkmark$&$\times$&$\times$\\			
$E'_{G\text{-}k}$${\color{red}^{\rm a}}$         &$\times$&$\times$ &$\times$&$\times$&$\times$${\color{red}^{\rm b}}$&$\times$&$\times$\\			
$\bar{E}_{G\text{-}k}$        &$\times$&$\times$& $\times$&$\times$&$\checkmark$&$\times$&$\times$  \\
\br
\end{tabular}
\item[] ${\color{red}^{\rm a}}$ We assume here that $E'_{G\text{-}k}$ is a $k$-EMo.
\item[] ${\color{red}^{\rm b}}$ It is weakly monogamous.
\end{indented}
\end{table}

\subsubsection{The geometric mean}\label{sec-11.2.4}

The geometric mean of $\mathcal{Q}_k^{\gamma_i}$, i.e.,
\bea\label{geometric mean}
E^G_{G\text{-}k}(|\psi\ra)
=\left[\prod_{\gamma_i}\mathcal{Q}_k^{\gamma_i}(|\psi\ra)\right]^{1/|\Gamma_k|},
\eea
define a $k$-EMo, where $\gamma_i$ is taken over all feasible $k$-partitions in $\Gamma_k$.

By definition,
\beax 
E_{G\text{-}k}\leq \bar{E}_{G\text{-}k}\leq  E'_{G\text{-}k},~
E_{G\text{-}n}=\bar{E}_{G\text{-}n}= E'_{G\text{-}n},~~~
\eeax
$\bar{E}_{G\text{-}k}$ is faithful, but ${E}'_{G\text{-}k}$ and ${E}_{G\text{-}k}$ are not faithful.
For the case of $k=2$, $E_2=E_{G\text{-}2}$ coincide with the GEM $\mE^{(n)}_{g''}$ in Ref.~\cite{Guo2024rip}, and
$E'_2=E'_{G\text{-}2}$ coincide with the GEM $\mE^{(n)}_{g'}$ in Ref.~\cite{Guo2024rip}.
We compare these product-based measures in Table~\ref{tab:table8-2}.
Compared to the sum-based $k$-entanglement monotones investigated in the previous subsection, the former class is better than
the product-based three ones here as they are even not $k$-monotonic. 
In both classes, the monogamy is related to whether the reduced function is strictly concave, which is the same as the bipartite entanglement measures~\cite{GG2018q}.

Before the end of this section, we compare these different monogamy relations in light of various entanglement measures and measures of other quantum correlations. Consequently, we find out that the monogamy seems independent of the complete ones, and that the tightly complete monogamy is stronger than the complete monogamy in general.  For more clear, we list them in the following Table~\ref{tab:table10-2-4}.

\Table{\label{tab:table10-2-4} Comparing of the monogamy, the complete monogamy and the tightly complete monogamy by different entanglement measures and measure of quantum correlations. $I$ denotes the multipartite mutual information (see in Ref.~\cite{Guo2023pra}).}
\br
Entanglement measure     & M  &CM &TCM  \\ 
\mr
$E^{(n)}$ in Sec.~\ref{sec-9.1} & $-$ & $\checkmark$& $\checkmark$ (most of them)\\
$E_g^{(n)}$ in Sec.~\ref{sec-10.1}& $-$ & $\checkmark$& $\checkmark$ (most of them)\\
$E'^{(n)}$ in Sec.~\ref{sec-9.4}& $-$ & $\checkmark$ (most of them)& $\times$\\
$E_{g'}^{(n)}$ in Sec.~\ref{sec-10.2}& $-$ & $\checkmark$ (most of them)& $\times$\\
$E_k$, $\bar{E}_k$, $E^G_k$, $E_{G\text{-}k}$, $\bar{E}_{G\text{-}k}$         &$\checkmark$&$\times$ &$\times$\\			
$E'_k$${\color{red}^{\rm a}}$,$E'_{G\text{-}k}$${\color{red}^{\rm a}}$         &$\checkmark$${\color{red}^{\rm b}}$&$\times$ &$\times$\\	
$I$	&$\times$  & $\checkmark$ &	$\checkmark$\\
\br
\end{tabular}
\item[] ${\color{red}^{\rm a}}$ We assume that $E'_k$ and $E'_{G\text{-}k}$ are $k$-EMos.
\item[] ${\color{red}^{\rm b}}$ They are weakly monogamous.
\end{indented}
\end{table}


\section{Eaxmples of $k$-PEMs}\label{sec-12}


$k$-PEMs were only discussed in Ref.~\cite{Guo2024pra,Hong2023epjp,Lihui2024aqt} wherein Ref.~\cite{Guo2024pra} 
was devoted to the complete measure of $k$-partite entanglement and presented some general ways of quantifying $k$-partite entanglement.


\subsection{$k$-PEMo from unified MEM}


For pure state $|\psi\ra$ in $\mH^{A_1A_2\cdots A_n}$, we assume it is not $(k-1)$-producible. Then there exists an $l$-fitness partition $X_1|X_2|\cdots|X_m$, $l\geqslant k$, such that 
\bea\label{condition of X_t}
\left\lbrace \begin{array}{l}\Delta(X_t)=s(t)\geq k, \\
	\rho^{X_t}~\textrm{is a genuinely entangled pure state}
\end{array}\right. ~~~~
\eea
for some subsystem $X_t$ in the partition $X_1|X_2|\cdots|X_m$. Let $t_1$, $t_2$, $\dots$, $t_l$ be all of the subscripts such that $X_{t_i}$ satisfies the condition~\eqref{condition of X_t} corresponding to all possible $l$-fitness partitions with $l\geqslant k$. It turns out that
\beax 
|\psi\ra=|\psi\ra^{X_{t_1}}|\psi\ra^{X_{t_2}}\cdots|\psi\ra^{X_{t_l}}|\phi\ra^{X_*}
\eeax
under some permutation of the subsystems, where $X_*$ denotes the subsystem complementary to $X_{t_1}X_{t_2}\cdots X_{t_l}$. In Ref.~\cite{Guo2025qip}, Zhao \etal gave the following $k$-PEMo which was defined by
\bea\label{E-k2}
E_{(k)}(|\psi\ra)=\begin{cases}
	\sum\limits_{j=1}^lE^{(s(t_j))}\left(|\psi\ra^{X_{t_j}}\right),&s(t_j)\geq k~\text{for some}~j,\\
	0,& \text{otherwise}.
\end{cases}\quad ~
\eea
They proved that $E_{(k)}$ is a unified $k$-PEMo, and is a complete $k$-PEMo whenever the reduced function is subadditive. 

Another candidate for the unified global MEM is $\mE^{(n)}$. Let $\mE_{(k)}$ be the quantity that is defined as in Eq.~\eqref{E-k2} just with $\mE^{(s(t_j))}$ replacing $E^{(s(t_j))}$. By definition,  
$\mE_{(k)}$ is a unified $k$-PEMo and is a complete $k$-PEMo if the reduced function is subadditive.

In Eq.~\eqref{E-k2}, $|\psi\ra^{X_{t_j}}$ is genuinely entangled, so $E^{(s(t_j))}$ can choose any GEM such as $E_{g''}^{(n)}$ and $\mE_{g''}^{(n)}$ instead~\cite{Guo2025qip}. Then the corresponding $k$-PEMo is not unified in general since $E_{g''}^{(n)}$ and $\mE_{g''}^{(n)}$ may increase under the coarsening relation of type (a)~\cite{Guo2024rip}.


\subsection{$k$-PEMo from the minimal sum}   


Hong \etal presented a $k$-PEMo in Ref.~\cite{Hong2023epjp}.
For any pure state $|\psi\ra\in\mH^{A_1A_2\cdots A_n}$, the $k$-PEM via concurrence was defined as~\cite{Hong2023epjp}
\bea\label{k-part-entanglement-C0}
C_{(k)}(|\psi\rangle)=\min\limits_{\Gamma_{\!k-\!1}^f}\frac{\sum_{t=1}^{m}\sqrt{2[1-{\rm Tr}(\rho_{X_t}^{2})]}}{m},
\eea
where $\rho_{X_t}={\rm Tr}_{{\overline X}_t}(|\psi\rangle\langle\psi|)$, ${\overline X}_t$ is the complement of subsystem $X_t$, the minimum is taken over all the $(k\!-\!1)$-fineness partitions in $\Gamma_{\!k-\!1}^f$.

Very recently, Li \etal proposed two $k$-PEMos in Ref.~\cite{Lihui2024aqt}. For any pure state $|\psi\ra$, the $k$-PEMo via the $q$-concurrence was defined as~\cite{Lihui2024aqt}
\bea\label{k-part-entanglement-C1}
C_{q(k)}(|\psi\rangle)=\min\limits_{\Gamma_{\!k-\!1}^f}\sqrt{\frac{\sum_{t=1}^{m}[1-{\rm Tr}(\rho_{X_t}^{q})]}{m}},
\eea
and the $k$-PEMo via the $\alpha$-concurrence was by~\cite{Lihui2024aqt}
\bea\label{k-part-entanglement-C2}
C_{\alpha(k)}(|\psi\rangle)=\min\limits_{\Gamma_{\!k-\!1}^f}\sqrt{\frac{\sum_{t=1}^{m}[{\rm Tr}(\rho_{X_t}^{\alpha})-1]}{m}},
\eea
where the minimum is taken over all the $(k\!-\!1)$-fineness partitions in $\Gamma_{\!k-\!1}^f$. Note here that, the notations here are different from $E_{q\text{-}k}$ and $E_{\alpha\text{-}k}$ in Ref.~\cite{Lihui2024aqt}: $C_{q(k+1)}=E_{q\text{-}k}$, $C_{\alpha(k+1)}=E_{\alpha\text{-}k}$.

Let $h$ be some reduced function. For any $\gamma_i^f\in\Gamma_k^f$, we write 
\bea\label{P_k}
\mP_k^{\gamma_i^f}(|\psi\ra)=\frac12\sum_{t=1}^{m}h(\rho^{X_{t(i)}}),~1\leqslant k<n,
\eea
where $X_{1(i)}|X_{2(i)}|\cdots|X_{m(i)}$ corresponds to $\gamma_i^f$, $\rho^{X}=\tr_{\overline{X}}|\psi\ra\la\psi|$. The coefficient ``1/2'' is fixed by the unification condition when the measures defined via $\mP_k^{\gamma_i^f}$ are regarded as unified $k$-PEMs. A general way of quantifying the $k$-partite entanglement was defined as~\cite{Guo2025qip}
\bea\label{p-sum}
E'_{(k)}(|\psi\ra)=\min\limits_{\Gamma_{\!k-\!1}^f}\mP_{\!k-\!1}^{\gamma_i^f}(|\psi\ra),
\eea
where the minimum is taken over all feasible $(k\!\!-\!\!1)$-fineness partitions in $\Gamma_{\!k-\!1}^f$. For mixed states, we define it by the convex-roof structure. In what follows, we give only the measures for pure states, for the case of mixed states they are all defined by the convex-roof extension with no further statement. 
If was shown in Ref.~\cite{Guo2025qip} that (i) $E'_{(k)}$ is a unified $k$-PEMo, (ii) $E'_{(2)}$ is complete if the reduced function $h$ is subadditive, (iii) If $h$ is subadditive, then the minimal partition is the ones that contained in $\Gamma_{\!k-\!1}^f\!\!\setminus\!\Gamma_{\!k-\!2}^f$, and (iv) $E'_{(k)}$ is not complete in general if $k\geqslant 3$, (iv) if the reduced function is subadditive, it can be easily checked that~\cite{Guo2025qip}
\beax 
E'_{(k)}(\rho)\leqslant E_{(k)}(\rho)\leqslant\mE_{(k)}(\rho).
\eeax (v) $C_{(k)}$, $C_{q(k)}$, and $C_{\alpha(k)}$ do not satisfy both the unification condition (except for the symmetry) and the hierarchy condition.

Ref.~\cite{Lihui2024aqt} also gave the following two $k$-PEMos:
\bea\label{k-part-entanglement-GC1}
C_{G,q(k)}(|\phi\rangle)=\left\lbrace \frac{\prod\limits_{\gamma_i\in\Gamma_{\!k-\!1}^f}\left[ \sum\limits_{t=1}^{m_i}(1-{\rm Tr}\rho_{X_{t(i)}}^q)\right] }{\prod_{i=1}^{\left| \Gamma_{\!k-\!1}^f\right| }m_{i}}\right\rbrace^{\frac{1}{2\left| \Gamma_{\!k-\!1}^f\right| }},	~~~
\eea
and 
\bea\label{k-part-entanglement-GC2}
C_{G,\alpha(k)}(|\phi\rangle)=\left\lbrace\frac{\prod\limits_{\gamma_i\in\Gamma_{\!k-\!1}^f}\left[ \sum\limits_{t=1}^{m_i}({\rm Tr}\rho_{X_{t(i)}}^{\alpha}-1)\right] }{\prod_{i=1}^{\left| \Gamma_{\!k-\!1}^f\right| }m_{i}}\right\rbrace^{\frac{1}{2\left| \Gamma_{\!k-\!1}^f\right| }},~~~
\eea
where $\rho_{X_{t(i)}}$ is the reduced density operator with respect to subsystem ${X_{t(i)}}$, and $m_i$ refers to $\gamma_i$ is an $m_i$-partition, $\left| \Gamma_{\!k-\!1}^f\right|$ is the cardinal number of $\Gamma_{\!k-\!1}^f$. The notations in Eqs.~\eqref{k-part-entanglement-GC1},~\eqref{k-part-entanglement-GC2} are different from $\varepsilon_{q\text{-}k}$ and $\varepsilon_{\alpha\text{-}k}$ in Ref.~\cite{Lihui2024aqt}: $C_{G,q(k+1)}=\varepsilon_{q\text{-}k}$, $C_{G,\alpha(k+1)}=\varepsilon_{\alpha\text{-}k}$. 
It was proved in Ref.~\cite{Guo2025qip} that $C_{G,q(k)}$ and $C_{G,\alpha(k)}$ in Eqs.~\eqref{k-part-entanglement-GC1}-\eqref{k-part-entanglement-GC2} do not satisfy both the unification condition (except for the symmetry) and the hierarchy condition. By definitions, both $E_{(k)}$ and $E'_{(k)}$ are not genuine $k$-partite entanglement measures.


\section{Examples of partitewise entanglement measure}\label{sec-13}


The partitewise entanglement measure proposed in Ref.~\cite{Guo2025pra}, wherein they gave three classes of measures defined in a wide-ranging manner via any suitable reduced function. They also provided a quantity called measure of partitewise entanglement extensibility, which is not a partitewise entanglement measure indeed, but has a close relation to it.  


\subsection{$k$-PWEMo from GEMo}


For any reduced function $h$, let $E$ be the associated entanglement measure, and $E_g^{(3)}$ be a genuine entanglement measure induced from $h$~\cite{Suppl}. Then
\bea\label{pem}
\check{E}^{AB}(|\psi\ra^{ABC})=E(AB)+E_g^{(3)}(ABC).
\eea
is a faithful PWEMo if both $E$ and $E_g^{(3)}$ are faithful~\cite{Guo2025pra}. In nature, the pairwise concurrence is such a case since $\mC_{A'B'}^2$ is just a special case of $\check{E}^{AB}$. But $\tau_{ABC}$ vanished on W state, so $\mC_{A'B'}$ can not measure all the pairwise entanglement faithfully.
For $|\psi\ra=|\psi\ra^{A_1A_2\cdots A_n}\in\mH^{A_1A_2\cdots A_n}$,
Eq.~\eref{pem} can be extended as
\bea\label{pwem}
\fl	~~\check{E}^{A_1A_2}(|\psi\ra)=
	\begin{cases}
	\!	E(A_1A_2)+ E_g^{(l)}(A_1A_2A_{3'}\cdots A_{l'}), 3\le l\le n, & \\
		\qquad\text{if $\exists$ $|\psi\ra^{A_1A_2A_{3'}\cdots A_{l'}}
			$ is a genuinely entangled state} \\
	\!	E(A_1A_2),\qquad\qquad\text{otherwise},
	\end{cases}
	\eea
where $E_g^{(l)}$ is a GEM, $E$ and $E_g^{(l)}$ are induced from the same reduced function $h$, $\{3'$, $4'$, $\dots$, $l'\}\subseteq\{3, 4, \dots, n\}$.

In what follows, if $|\psi\ra=|\psi\ra^{X_1}|\psi\ra^{X_2}\cdots|\psi\ra^{X_l}$ with $|\psi\ra^{X_i}$ is either a single state or a composite non-biseparable state, $1\leqslant l<k$, $1\leqslant i\leqslant l$, we let
$\varUpsilon^{X_1X_2\cdots X_l}=\{X_s: X_s$~contains at least two subsystems of $A_1A_2\cdots A_k$ and also at least one subsystem of $A_{k+1}\cdots A_n\}$
and denote by $t_s$ the number of subsystems contained in $X_s$, $1\leqslant s\leqslant l$. {Hereafter, if we write $|\psi\ra=|\psi\ra^{X_1}|\psi\ra^{X_2}\cdots|\psi\ra^{X_l}$, it always refers to $|\psi\ra^{X_i}$ is either a single state or a composite non-biseparable state. For any $|\psi\ra\in\mH^{A_1A_2\cdots A_n}$, $k\geqslant 3$, Guo and Yang proposed the following measures~\cite{Guo2025pra}:
\bea \label{kpwem}
\fl~~\check{E}^{A_1A_2\cdots A_k}(|\psi\ra)=
		\begin{cases}
			\!E^{(k)}(A_1A_2\cdots A_k)+\sum\limits_{X_s\in\varUpsilon^{X_1X_2\cdots X_l}}E_g^{(t_s)}\left(|\psi\ra^{X_s}\right), &\\ \qquad \qquad\text{if $|\psi\ra=|\psi\ra^{X_1}|\psi\ra^{X_2}\cdots|\psi\ra^{X_l}|\psi\ra^Y$, $l<k$}\\
			\!E^{(k)}(A_1A_2\cdots A_k), \qquad\text{otherwise},
		\end{cases}
		\eea
	\bea \label{gkpwem}
	\check{E}_g^{A_1A_2\cdots A_k}(|\psi\ra)=
	E_g^{(k)}\left(\rho^{A_1A_2\cdots A_k}\right),
	\eea
	and
	\bea \label{skpwem}
	\check{E}_s^{A_1A_2\cdots A_k}(|\psi\ra)=
	\begin{cases}
		\!E_g^{(t_X)}\left(|\psi\ra^{X}\right), & |\psi\ra=|\psi\ra^{X}|\psi\ra^Y ~~~~~\\
		\!0, &\text{otherwise},
	\end{cases}
	\eea
	where $X_i$ contains at least one subsystem of $A_1A_2\cdots A_k$, $A_1A_2\cdots A_k\subseteq X$, $|\psi\ra^X$ is genuinely entangled, $t_X$ denotes the number of subsystems contained in $X$, $E^{(k)}$ is some multipartite entanglement measure as in Sec.~\ref{sec-9} or some $k$-entanglement measure as in Sec.~\ref{sec-11}, $E^{(k)}$ and $E_g^{(l)}$ are induced from the same reduced function $h$. It is easy to check that $\check{E}^{A_1A_2\cdots A_k}$ is a $k$-PWEM, $\check{E}_g^{A_1A_2\cdots A_k}$ is a G$k$-PWEM, and $\check{E}_s^{A_1A_2\cdots A_k}$ is a S$k$-PWEM (they are all $k$-PWEMos if both $E^{(k)}$ and $E_g^{(l)}$ therein are entanglement monotones). For the case of $k=2$, Eq.~\eref{kpwem} turns back to Eq.~\eref{pwem}.


\subsection{$k$-PWEMo from 2-partitions}


Let $h$ be a reduced function. Guo and Yang defined~\cite{Guo2025pra}
\bea\label{min-h-kpwem}
\check{E}_{\min}^{A_1A_2\cdots A_k}(|\psi\ra)=\min_{1\leqslant i\leqslant k}\min_{Z_i} h(A_iZ_i),
\eea 
where $Z_i$ is contained in $\overline{A_i}$ but $Z_i\neq\overline{A_i}$ and $Z_i$ exclude at least one $A_j$, $j\neq i$, and the second minimum is taken over all possible $Z_i$ ($Z_i$ can be the empty set as well). In addition, they defined
	\bea\label{E2} 
	\check{\bar{E}}^{A_1A_2\cdots A_k}(|\psi\ra)=\begin{cases}
		\frac12\sum\limits_{i=1}^kh(A_i), &\min\limits_{1\leqslant i\leqslant k}\min\limits_{Z_i} h(A_iZ_i)>0\\
		0,&\min\limits_{1\leqslant i\leqslant k} \min\limits_{Z_i}h(A_iZ_i)=0,
	\end{cases}
	\eea
	\bea\label{kEG}
	\check{E}_{G}^{A_1A_2\cdots A_k}(|\psi\ra)=\begin{cases}
		\left[ \prod_{i=1}^kh(A_i)\right] ^{\frac1k}, &\min\limits_{1\leqslant i\leqslant k}\min\limits_{Z_i} h(A_iZ_i)>0\\
		0,&\min\limits_{1\leqslant i\leqslant k} \min\limits_{Z_i}h(A_iZ_i)=0, 
	\end{cases}
	\eea 
where $Z_i$ is taken as in Eq.~\eqref{min-h-kpwem}. It is shown in Ref.~\cite{Guo2025pra} that $\check{E}_{\min}^{A_1A_2\cdots A_k}$, $\check{\bar{E}}^{A_1A_2\cdots A_k}$, and $\check{E}_{G}^{A_1A_2\cdots A_k}$ are faithful S$k$-PWEMos, $k\geqslant3$.
But they are not G$k$-PWEMos since there exist non-G$k$-PWE state $\sigma$ such that
$\check{E}_{\min}^{A_1A_2\cdots A_k}(\sigma)>0$, $\check{\bar{E}}^{A_1A_2\cdots A_k}(\sigma)>0$, and $\check{E}_{G}^{A_1A_2\cdots A_k}(\sigma)>0$.


\subsection{Distance based $k$-PWEMo}


We denote the set of all $k$-partitewise separable states in $\mS^{A_1A_2\cdots A_n}$ by $\mS_{pws}^{A_1A_2\cdots A_k}$, and the set of $k$-partitewise separable pure states in $\mH^{A_1A_2\cdots A_n}$ by $\mP_{pws}^{A_1A_2\cdots A_k}$.
Guo and Yang defined~\cite{Guo2025pra}
\bea
\check{S}_r^{A_1A_2\cdots A_k}(\rho)&=&\min\limits_{\sigma\in\mS_{pws}^{A_1A_2\cdots A_k}}S(\rho\|\sigma),\label{kpwemo}\\
\check{E}_{\rm G}^{A_1A_2\cdots A_k}(|\psi\ra)&=&1-\max\limits_{|\phi\ra\in\mP_{pws}^{A_1A_2\cdots A_k}}|\la\psi|\phi\ra|^2,\label{kpwemo2}
\eea
and proved that $\check{S}_r^{A_1A_2\cdots A_k}$ is a $k$-PWEMo and $\check{E}_{\rm G}^{A_1A_2\cdots A_k}$ is a $k$-PWEM.


\subsection{Partitewise entanglement extension}


The $k$-partitewise entanglement discuss the shared entanglement of fixed subsystem in a given global system. Guo and Yang  considered also another issue: For a given state that we do not know the global system where it lived in, whether there exists a larger global state such that it is $k$-PWE with respect to the given state, and moreover, how can we quantify such a capability and what is the relation between this capability and the $k$-PWE of the global state.

A state $\rho\in\mS^{A_1A_2\cdots A_k}$ is $k$-partitewise entanglement extendable ($k$-PWEE)~\cite{Guo2025pra} if there exists a $k$-PWE state $\rho^{A_1A_2\cdots A_l}$ with respect to $A_1A_2\cdots A_k$ such that $\rho=\tr_{A_{k+1}\cdots A_l}\rho^{A_1A_2\cdots A_l}$ and $\rho^{A_1A_2\cdots A_l}\neq \rho^{A_1A_2\cdots A_k}\ot\rho^{A_{k+1}\cdots A_l}$, $k<l$. In such a case, $\rho^{A_1A_2\cdots A_l}$ is called a $k$-partitewise entanglement extension of $\rho$. In particular, $\rho$ is  strongly $k$-partitewise entanglement extendable (S$k$-PWEE)~\cite{Guo2025pra} if there exists a genuinely entangled state $\rho^{A_1A_2\cdots A_l}$ such that $\rho=\tr_{A_{k+1}\cdots A_l}\rho^{A_1A_2\cdots A_l}$, $k<l$, and $\rho^{A_1A_2\cdots A_l}$ is called a strong $k$-partitewise entanglement extension of $\rho$.

Clearly, any pure state $|\psi\ra^{A_1A_2\cdots A_k}$ is not $k$-partitewise entanglement extendable. In fact,
$\rho\in\mS^{A_1A_2\cdots A_k}$ is $k$-PWEE iff $\rho$ has at least one bipartite mixed reduced state that contains two mixed reduced states (if $k=2$, $\rho$ is a mixed state that contains two mixed reduced states), and it is S$k$-PWEE iff it is a mixed state that contains no pure reduced state~\cite{Guo2025pra}.

\subsubsection{Measure of partitewise entanglement extensibility}

Ref.~\cite{Guo2025pra} provided a measure of the pairwise entanglement extensibility
 \bea \label{extensiblity}
\fl ~~~~\quad ~~~E_{ext}(\rho^{AB})=
\begin{cases}
	E_g^{(3)}(|\Phi\ra^{ABC}),& \!\!\!\text{$\rho$ is pairwise entanglement extendable}~~~\\
	0, &\!\!\!\text{otherwise}, 
\end{cases}
\eea
where $|\Phi\ra^{ABC}$ is defined as 
\bea\label{purification2}
|\Phi\ra^{ABC}=\sum_j\sqrt{q_j}|\psi_j\ra^{AB}|j\ra^C,
\eea 
with $\{|i\ra^C\}$ is an orthogonal set of $\mH^C$ if $\rho^{AB}$ is a mixed state and $\rho^{AB}=\sum_jq_j|\psi_j\ra\la\psi_j|^{AB}$ is the spectral decomposition,
$E_g^{(3)}$ is some genuine entanglement measure.

For any given state $\rho^{A_1A_2\cdots A_k}$, we denote by $\mS_{ext}(\rho^{A_1A_2\cdots A_k})$ the set of all the $k$-paritewise entanglement extensions of $\rho^{A_1A_2\cdots A_k}$ and by $\mS_{sext}(\rho^{A_1A_2\cdots A_k})$ the set of all the strong $k$-paritewise entanglement extensions of $\rho^{A_1A_2\cdots A_k}$. Then
\beax 
\mS_{sext}(\rho^{AB})=\mS_{ext}(\rho^{AB}),
\eeax
i.e., $\rho^{AB}$ is pairwise entanglement extendable iff it is strongly pairwise entanglement extendable. But 
\beax
\mS_{sgext}(\rho^{A_1A_2\cdots A_k})\subsetneq\mS_{ext}(\rho^{A_1A_2\cdots A_k})
\eeax 
whenever $k\geqslant3$. It turns out that both $\mS_{ext}(\rho^{A_1A_2\cdots A_k})$ and $\mS_{sext}(\rho^{A_1A_2\cdots A_k})$ are convex sets with all the purification as in Eq.~\eqref{purification2} are the extreme points. But $\mS_{ext}(\rho^{A_1A_2\cdots A_k})$ is not the convex hull of these points. For example, $\rho_p=p\rho^{AB}\ot\rho^C+(1-p)|\Phi\ra\la\Phi|\in \mS_{ext}(\rho^{AB})$, $0<p<1$, but $\rho_p$ is not a convex combination of these extreme points. 
Let $\rho^{AB}$ be a mixed state in $\mS^{AB}$, and $r(\rho^{A,B})>1$. Then~\cite{Guo2025pra}
	\bea 
	E_{ext}(\rho^{AB})=\max\limits_{\rho\in\mS_{sext}(\rho^{AB})} E_g^{(3)}(\rho). 
	\eea 
Namely, although the global system is unknown, up to local unitary operation, the maximal pairwise entanglement extension is exactly decided by the given state. This also manifests that the state supposition is the nature of the entanglement.

With the notations as in Eq.~\eqref{kpwem}, the $k$-paritewise entanglement extensibility is defined by~\cite{Guo2025pra}
\bea \label{k-ent-ext0}
E_{ext}^{\psi}(\rho^{A_1A_2\cdots A_k})=
\sum\limits_{X_s\in\varUpsilon^{X_1X_2\cdots X_l}}E_g^{(t_s)}(|\psi\ra^{X_s})
\eea
if there exits a $k$-partitewise entanglement extension $|\psi\ra=|\psi\ra^{X_1}|\psi\ra^{X_2}\cdots|\psi\ra^{X_l}$, $l<k$.
Namely, 
	\bea \label{k-ent-ext}
	E_{ext}(\rho^{A_1A_2\cdots A_k})=
	\begin{cases}
		\!E_{ext}^{\psi}(\rho^{A_1A_2\cdots A_k}), &\text{$\rho^{A_1A_2\cdots A_k}$ is $k$-PWEE}\\
		\!0, & \text{otherwise}
	\end{cases}
	\eea 
with some abuse of notation.
And in particular, they defined
\bea\label{m-k-ent}
E_{ext}^{\Phi}(\rho^{A_1A_2\cdots A_k})=E_g^{(k+1)}(|\Phi\ra^{A_1A_2\cdots A_kA_{k+1}})
\eea 
if $|\Phi\ra^{A_1A_2\cdots A_kA_{k+1}}$ is a strong $k$-partitewise entanglement extension of $\rho^{A_1A_2\cdots A_k}$. They also showed that
\bea E_{ext}^{\Phi}(\rho^{A_1A_2\cdots A_k})=\max\limits_{\rho\in\mS_{sext}(\rho^{A_1A_2\cdots A_k})}E_g^{(k+1)}(\rho).
\eea

\subsubsection{PWE versus PWE extensibility}

For any S$k$-PWE state $|\psi\ra$, with the notations as in Eqs.~\eqref{kpwem} and~\eqref{skpwem}, it is clear that~\cite{Guo2025pra} 
\beax 
&&\check{E}^{A_1A_2\cdots A_k}(|\psi\ra)\\
&=&E^{(k)}(A_1A_2\cdots A_k)+E_g^{(t_X)}(|\psi\ra^{X})\\
&=&\!E^{(k)}(A_1A_2\cdots A_k)+E_{ext}(A_1A_2\cdots A_k).
\eeax 
In particular, for any tripartite pure state $|\psi\ra^{ABC}$, 
\beax 
\check{E}^{AB}(|\psi\ra^{ABC})=E(AB)+E_{ext}(AB)
\eeax since $\check{E}_s^{AB}(|\psi\ra^{ABC})=E_{ext}(\rho^{AB})$ whenever one choose the same $E_g^{(3)}$ in $\check{E}_s^{AB}$ and $E_{ext}$. Together with Eqs.~\eqref{pem}, \eqref{extensiblity}, \eqref{k-ent-ext0}, \eqref{k-ent-ext} and \eqref{m-k-ent}, we have~\cite{Guo2025pra}
\bea 
\check{E}^{A_1A_2\cdots A_k}(|\psi\ra)\!=\!E^{(k)}(A_1A_2\cdots A_k)+E_{ext}^{\psi}(A_1A_2\cdots A_k)
\eea 
with some abuse of notation (here $E_{ext}^{\psi}=0$ if $\rho^{A_1A_2\cdots A_k}$ is not $k$-partitewise entanglement extendable). In other words, $E_{ext}$ is just the ``global entanglement'' shared by $A_1A_2\cdots A_k$ in its extension $|\psi\ra$ under the measure $\check{E}^{A_1A_2\cdots A_k}$.

 If $\rho^{AB}$ is maximally entangled, then it is pure~\cite{Guo2020pra}, this results in $E_{ext}(AB)=0$. Conversely, if a three-qubit state $|\psi\ra^{ABC}$ is absolutely maximally entangled, i.e., $|\psi\ra^{ABC}$ is the GHZ state, then $E(AB)=0$. In addition, the GHZ state is regarded more entangled than the W state~\cite{Xie2021prl}. So this seemingly indicates that, $E(AB)$ and $E_{ext}$ are complementary to each other. But it is shown in Ref.~\cite{Guo2025pra} that there exist states such that both $E(AB)$ and $E_{ext}$ can increase simultaneously, states such that $E(AB)$ creases with $E_{ext}$ decreases. and states such that $E(AB)$ decreases with $E_{ext}$ increases, and states such that $E(AB)$ remain invariant with $E_{ext}$ increases.

Another issue is: what is the supremum of $E(AB)+E_{ext}(AB)$? It seems a difficult task but~\cite{Guo2025pra}  
\bea 
E(AB)+E_{ext}(AB)<2
\eea 
for any bipartite state with any normalized measures $E(AB)$ and $E_{ext}(AB)$ since they can not be maximally entangled at the same time. Here, an entanglement measure $E$ is called normalized if $|E|\leqslant1$ for any state and $E=1$ for the maximally entangled states.

Let $\varepsilon^{AB}$ be an LOCC on $\mS^{AB}$. For any given $\rho\in\mS^{ABC}$, let $\varepsilon(\rho)=\varepsilon^{AB}\ot\mathbbm{1}^C(\rho)=\sigma\in\mS^{ABC}$. Then~\cite{Guo2025pra}
\bea \label{ext-var}
E_{ext}(\sigma^{AB})-E_{ext}(\rho^{AB})\leqslant E(\rho^{AB})-E(\sigma^{AB})
\eea 
since $\check{E}^{AB}(\sigma)\leqslant\check{E}^{AB}(\rho)$. In other words, if $E_{ext}(\sigma^{AB})\geqslant E_{ext}(\rho^{AB})$, the increment of the entanglement extensibility of $AB$ is not exceeding the decrement of entanglement in $AB$ under LOCC. Note that it may happen that $E_{ext}(\sigma^{AB})<E_{ext}(\rho^{AB})$~\cite{Guo2025pra}.


\section{Other examples of MEMs}\label{sec-14}


This section lists the general MEMs proposed in literature that, may not complete or not belong to the MEMs in previous sections. They were defined from various perspectives of those involved with the project and thus provided us different understanding of multipartite entanglement.


\subsection{The geometric measure of multipartite entanglement}\label{sec-14.1}


We denote by $\mP^{A_1A_2\cdots A_n}$ the set of all fully separable pure states in $\mH^{A_1A_2\cdots A_n}$, i.e., 
\bea\label{full separable pure states}
 \mP^{A_1A_2\cdots A_n}=\left\lbrace |\phi\ra=\bigotimes\limits_{i=1}^n|\phi\ra^{A_i}\bigg|   |\phi\ra^{A_i}\in\mH^{A_i}\right\rbrace.
\eea
The geometric measure of entanglement for multipartite cases is defined by~\cite{Shimony95}:
\beax \label{GEM1}
E_G^{(n)}(|\psi\ra)=\frac12\left(  1-\max _{|\phi\rangle \in \mP^{A_1A_2\cdots A_n}}|\langle\phi \vert \psi\rangle|^2\right),
\eeax
where the maximum runs over all the fully separable pure states in $\mP^{A_1A_2\cdots A_n}$.
Later, the factor $\frac12$ was omitted in literaure~\cite{Cao2007jpa,Das2016pra,Sen2010pra,Wei2003pra,Wei2003arxiv,Wei2004pra}, i.e.,
\bea\label{geometric-measure}
E_G^{(n)}(|\psi\ra)=  1-\max_{|\phi\rangle \in \mP^{A_1A_2\cdots A_n}}|\langle\phi \vert \psi\rangle|^2.
\eea
$E_G^{(n)}$ is shown to be an entanglement monotone~~\cite{Wei2003pra}.

It was checked in~\cite{Weinbrenner2025} that $E_G^{(3)}(|\W\ra)=5/9$ and $E_G^{(3)}(|\GHZ\ra)=1/2$. So $E_G^{(n)}$ is not proper. In addition, any state $|\psi\ra$ in the subspace spanned by $|\W\ra$ and $\frac{1}{\sqrt3}(|011\ra+e^{\frac{2\pi}{3}{\rm i}}|110\ra+e^{\frac{4\pi}{3}{\rm i}}|101\ra)$, we have $E_G^{(3)}(|\psi\ra)=5/9$~\cite{Weinbrenner2025}. Let $\Lambda(|\psi\ra)=\max_{|\phi\rangle \in \mP^{A_1A_2\cdots A_n}}|\langle\phi \vert \psi\rangle|$. It follows that~\cite{Weinbrenner2025} 
\beax
\Lambda^2(|\psi\ra)=\sup\limits_{|a\ra, |b\ra}\la ab|\rho^{AB}|ab\ra,
\eeax 
where $\rho^{AB}=\tr_C|\psi\ra\la\psi|$ for any $|\psi\ra\in\mH^{ABC}$.

There is an operational interpretation for $E_G^{(n)}$~\cite{Hayashi2006prl,Weinbrenner2025}: If the discrimination of the orthogonal states
$\{|\psi_i\ra\in\mH^{A_1A_2\cdots A_n}| \dim\mH^{A_j}=d, 1\leq j\leq n, i = 1, 2, \dots, m\}$ via LOCC is possible,
then $\sum\limits_{i=1}^m1/\Lambda^2(|\psi_i\ra)\leq d^n$.

The revised geometric measure of multipartite entanglement was defined by\cite{Cao2007jpa}
\begin{eqnarray}\label{m-revised geometric-measure}
	\check{E}_{G}^{(n)}(\rho) = \min\limits_{\sigma\in \mathcal{S}_n}[{1-F({\rho,\sigma})}].
\end{eqnarray}
$\check{E}_{G}^{(n)}(\rho)$ is non-increasing under local operation and classical communication, thus it is an MEM.
For any state $\rho$ and $\sigma$, the RGME and the Fidelity have the following relation
\beax
1 - \sqrt{F}\left( {\rho ,\sigma } \right) \le \sqrt{\check{E}_{G}^{(n)} \left( \rho \right)}.
\eeax
Streltsov \etal proved in Ref.~\cite{Streltsov2010njp} that
\beax 
E_G^{(n)}= \min\limits_{\sigma\in \mathcal{S}_{n}} \left[  {1 -
	F( {\rho ,\sigma } )} \right],
\eeax 
namely $\check{E}_{G}^{(n)}(\rho)=E_G^{(n)}(\rho)$ for any state and thus $\check{E}_{G}^{(n)}$ is also an entanglement monotone.
They also proved that
\beax
S(\rho\lVert\sigma)\leq \tr(\rho\log_2\rho)-\log_2F(\rho, \sigma).
\eeax

In Ref.~\cite{Eisert2001pra,Zhu-arXiv-2311.10353}, the tensor rank of $|\psi\ra$ was defined, which is the minimum number $r$ such that there exist states $|\phi_i\rangle^{A_k}\in \mathcal{H}^{A_k}$, $1\leq i\leq r$, $1\leq k\leq n$, satisfying $|\psi\rangle=\sum_{i=1}^r\lambda_i |\phi_i\rangle^{A_1}|\phi_i\rangle^{A_2}\cdots|\phi_i\rangle^{A_n}$, where the coefficients $\lambda_i$ are positive and the different product terms may not be orthogonal to each other. We denote by $\mT_r$ the set of all states of tensor rank at most $r$. A generalization of geometric measure of entanglement
was defined by~\cite{Zhu-arXiv-2311.10353}
\bea\label{E_{tr}^{(n)}}
E_{tr}^{(n)}(|\psi\rangle)=1-\max _{|\varphi\rangle \in \mT_{r-1}}|\langle\varphi \vert \psi\rangle|^2.
\eea

The geometric measure is also valid for the $k$-entanglement~\cite{Cavalcanti2006pra}
\bea \label{geometric-k-entanglement}
E_{G,k}^{(n)}=1-\max_{|\phi\ra\la\phi| \in \mS_k}|\langle\phi \vert \psi\rangle|^2.
\eea


\subsection{The relative entropy of multipartite entanglement}\label{sec-14.2}


The relative entropy of entanglement is also valid for multiparty system~\cite{Vedral1997,Vedral1998pra}:
\bea\label{E_r^n}
E_r^{(n)}(\rho)= \min\limits_{\sigma}S\left( \rho||\sigma\right) ,
\eea
where the minimum is taken over all the corresponding separable states $\sigma$ in $\mS_X$. $\mS_X$ is some compact and convex subset in $\mS^{A_1A_2\cdots A_n}$. In such a case, there exists an optimal reference state $\sigma^*\in\mS_X$ such that $E_r^{(n)}(\rho)=S(\rho\|\sigma^*)$. For example if the reference set $\mS_X$ is consisted of all biseparable states in $\mS^{A_1A_2\cdots A_n}$, $E_r^{(n)}$ is a GEM; if $\mS_X=\mS_n$, it is a global MEMo; if $\mS_X=\mS_k$, it is a $k$-entanglement monotone, etc. Namely, $E_r^{(n)}$ can define different kinds of entanglement by taking distance from the different set of states.
But $E_r^{(n)}$ denotes the one whose reference set is $\mS_n$ in general unless otherwise specified.

We denote by $\mathbb{M}_{\rm LOCC}$ the set of all LOCCs acting on $A_1A_2\cdots A_n$ and by $\mathbb{M}_{\rm SEP}$ the
set of all separable measurements. 
Piani defined~\cite{Piani2009prl}
\beax 
\mathbb{M}S(\rho\|\sigma)=\sup\limits_{\varepsilon\in\mathbb{M}}S(\varepsilon(\rho)\|\varepsilon(\sigma)), \quad \mathbb{M}=\mathbb{M}_{\rm LOCC}~\text{or}~\mathbb{M}_{\rm SEP};
\eeax 
and the $\mathbb{M}$-relative entropy of $\rho$ with respect to $\mS_X$ was defined by
\bea\label{M-relative entropy}
\mathbb{M}E_r^{X(n)}(\rho)=\min\limits_{\sigma\in\mS_X}\mathbb{M}S(\rho\|\sigma).
\eea 
By definition, $\mathbb{M}E_r^{X(n)}(\rho)\leq E_r^{X(n)}(\rho)$ since the relative entropy is non-increasing under CPTP measurement. Piani proved that~\cite{Piani2009prl}, for any combination of $\mathbb{M}=\mathbb{M}_{\rm LOCC}$, $\mathbb{M}_{\rm SEP}$, $\mS_X=\mS_n$, $\mS_{\rm PPT}$ (the set of all PPT states), $\mathbb{M}E_r^{X(n)}$ was a faithful entanglement monotone and strongly superadditive, i.e., $\mathbb{M}E_r^{X(n)}(\rho^{XY})\geq \mathbb{M}E_r^{X(\Delta(X))}(\rho^{X})+\mathbb{M}E_r^{X(\Delta(Y))}(\rho^{Y})$ for any possible bipartition $X|Y$ of $A_1A_2\cdots A_n$.


\subsection{$n$-tangle}


Motivated from the three tangle, Wong and Christensen defined a generalization of the three tangle, $n$-tangle, for the $n$-qubit state, which was formulated by~\cite{Wong2001pra}
\bea\label{taun}
\acute{\tau}^{(n)}(|\psi\ra)&=&2\left| \sum a_{\alpha_1\cdots\alpha_n}a_{\beta_1\cdots\beta_n}a_{\gamma_1\cdots\gamma_n}a_{\delta_1\cdots\delta_n}\epsilon_{\alpha_1\beta_1}\epsilon_{\alpha_2\beta_2}\nonumber\right. \\
&&\left. \quad \cdots \epsilon_{\alpha_{n-1}\beta_{n-1}}\epsilon_{\gamma_1\delta_1}\epsilon_{\gamma_2\delta_2}\epsilon_{\gamma_{n-1}\delta_{n-1}}\epsilon_{\alpha_n\gamma_n}\epsilon_{\beta_n\delta_n}\right| 
\eea
for all even $n$ and $n=3$, where
$|\psi\ra=\sum_{i_1\cdots i_n}a_{i_1\cdots i_n}|i_1i_2\cdots i_n\ra$.
It is also an multipartite entanglement monotone and is invariant under permutations of the qubits~\cite{Wong2001pra}.

In 2005, Yu and Song~\cite{Yu2005pra} proved 
\beax
C^2(\rho^{AB_1})+C^2(\rho^{AB_2})+\cdots C^2(\rho^{AB_n})\leq  C^2(\rho^{A|B_1B_2\cdots B_n}),
\eeax
for any $\rho^{AB_1B_2\cdots B_n}$
and consequently generalized the residual entanglement into the $n$-particle case as
\bea\label{ntangle}
\check{\tau}^{(n)}(|\psi\ra^{AB_1B_2\cdots B_n})=\min\limits_{X}\left\lbrace\tau_{X}(|\psi\ra^{AB_1B_2\cdots B_n})\right\rbrace,
\eea
where $X$ corresponds to all possible foci, e.g., $\tau_{B_1}(|\psi\ra^{AB_1B_2\cdots B_n})=C^2(\rho^{B_1|AB_2\cdots B_n})-C^2(\rho^{B_1A})-C^2(\rho^{B_1B_2})-\cdots C^2(\rho^{B_2B_n})$,
$\tau_{AB_1}(|\psi\ra^{AB_1B_2\cdots B_n})=C^2(\rho^{AB_1|B_2\cdots B_n})-C^2(\rho^{AB_1|B_2})-C^2(\rho^{AB_1|B_3})-\cdots C^2(\rho^{AB_1|B_n})$.
It is not invariant under permutations of the particles but it provides a good measure for $n$-way
entanglement in arbitrary dimensions.


\subsection{Global multipartite concurrence}


Ref.~\cite{Carvalho2004prl} put forward a generalization of bipartite concurrence $C$ as
\be\label{multi}
C_n(|\psi\ra)=2^{1-\frac{n}{2}}
\sqrt{2^n-2- \sum\limits_{\gamma_i}\tr\rho_{X_{1(i)}}^2},
\label{multi}
\ee
where $\gamma_i$ runs over all bipartitions in $\Gamma_2$, $d=|\Gamma_2|$. It is an entanglement monotone and $C_n$ adopts its maximal value for the $n$-qudit GHZ states
\beax 
|{\GHZ}_{n,d}\ra=\frac{1}{\sqrt{d}}(|0\ra^{\ot n}+|1\ra^{\ot n}+\cdots|d-1\ra^{\ot n}).
\eeax

Yu and Song presented the following global multipartite concurrence in Ref.~\cite{Yu2006pra}
\bea\label{dcn}
C'_n(|\psi\ra)=\sqrt{2\left(n-\sum\limits_{i=1}^n\tr\rho_i^2 \right). }
\eea
It is shown to be an entanglement monotone. 

Another one is~\cite{Gao2008epj}
\bea\label{ddcn}
C''_n(|\psi\ra)=\sqrt{d-\sum\limits_{\gamma_i}\tr\rho_{X_{1(i)}}^2},
\eea
where $\gamma_i$ runs over all bipartitions in $\Gamma_2$, $d=|\Gamma_2|$.


\subsection{Robustness of multipartite entanglement}\label{Robustness of multipartite entanglement}


For a given state $\rho$ and a fully separable
state $\sigma\in\mS_n$. Let 
\beax 
R(\rho|\sigma)=\min\left\lbrace t \Big|  \frac{\rho+t\sigma}{1+t}\in\mS_n\right\rbrace. 
\eeax 
The robustness of multipartite entanglement is defined
as~\cite{Vidal1999pra} 
\bea\label{multipartite robustness of entanglement}
R^{(n)}(\rho)=\inf\limits_{\sigma\in\mS_n}R(\rho|\sigma).
\eea 
$R^{(n)}(\rho)$ reports what minimal level of mixing of $\rho$ with fully separable state will produce fully separable state again. $R^{(n)}$ was shown to be convex and nonincreasing on average under LOCC, so it is a global multipartite entanglement monotone. Similarly, one can define the generalized robustness of multipartite entanglement $\check{R}^{(n)}$ in which the reference set is consisted of all states.

Let
\beax 
R_k(\rho|\sigma)=\min\left\lbrace t \Big|  \frac{\rho+t\sigma}{1+t}\in\mS_k\right\rbrace, 
\eeax 
then
\bea \label{robustness-k-ent}
\check{R}_k^{(n)}(\rho)=\inf\limits_{\sigma\in\mS}R_k(\rho|\sigma)
\eea 
define a $k$-entanglement monotone~\cite{Cavalcanti2006pra}. In addition~\cite{Cavalcanti2006pra},
\beax 
\check{R}_k^{(n)}(|\psi\ra)\geq E_{G,k}^{(n)}(|\psi\ra).
\eeax 

The logarithmic robustness was defined by~\cite{Wei2008pra}
\bea \label{logarithmic robustness}
\check{R}_L^{(n)}(\rho)=\log_2[1+\check{R}^{(n)}(\rho)],
\eea 
which was shown to be a upper bound of the relative entropy~\cite{Acin2003qic55,Hayashi2006prl,Wei2004qic252,Wei2008pra}
\beax 
\check{R}_L^{(n)}(|\psi\ra)\geq E_r^{(n)}(|\psi\ra)\geq-2\log_2\Lambda_{\max}(|\psi\ra),\\
\check{R}_L^{(n)}(\rho)\geq E_r^{(n)}(\rho)\geq E_{LG}^{(n)}(\rho)-S(\rho),
\eeax
where $\Lambda_{\max}(|\psi\ra)=\max\limits_{|\phi\ra\in\mP^{A_1A_2\cdots A_n}}|\la\phi|\psi\ra|$, and
\beax 
E_{LG}^{(n)}(\rho)=-\min\limits_{p_i,|\psi_i\ra}\sum\limits_ip_i[2\log_2\Lambda_{\max}(|\psi_i\ra)],
\eeax
the minimum runs over all pure state ensembles of $\rho$.


\subsection{MEMs induced by the fidelity distance}


There is yet another way of defining tripartite entanglement measure~\cite{Guo2020qip}:
\bea
E_{\mF}^{(3)}\left( |\psi\rangle\right)
&=&1-F\left( |\psi\rangle\la\psi|,\rho^A\otimes\rho^B\otimes\rho^C\right),
~~\label{multipartite1}\\
E_{{\mF}'}^{(3)}\left( |\psi\rangle\right)
&=&1-\sqrt{F}\left( |\psi\rangle\la\psi|,\rho^A\otimes\rho^B\otimes\rho^C\right),
~~\label{multipartite2}\\
E_{A\mF}^{(3)}\left( |\psi\rangle\right)
&=&1-F_A\left( |\psi\rangle\la\psi|,\rho^A\otimes\rho^B\otimes\rho^C\right),
~~\label{multipartite3}
\eea
for any pure state $|\psi\ra$ in $\mH^{ABC}$,
where
$\rho^{x}=\tr_{\overline{x}}|\psi\rangle\la\psi|$.
It is proved that
$E_{Fi,F}^{(3)}$,
$E_{Fi',F}^{(3)}$ and
$E_{AFi,F}^{(3)}$ are unified continuous tripartite
entanglement monotones but they
violate the hierarchy condition~\cite{Guo2020qip}.

There is example which indicates
that the relative entropy (as a distance) and
the fidelity induced distance (e.g., the sine distance $C(\rho,\sigma)=\sqrt{1-F(\rho,\sigma)}$)
are not equivalent~\cite{Guo2020qip}. That is, there exist $\rho_{i}$
and $\sigma_i$, $i=1,2$, such that
$S(\rho_1\|\sigma_1)\geq S(\rho_2\|\sigma_2)$
while
$C(\rho_1,\sigma_1)\leq  C(\rho_2,\sigma_2)$.


\subsection{Multipartite conditional entanglement of mutual information}
\label{sec-Conditional entanglement for multipartite mutual information}


The multipartite version of $E_I$ is defined by~\cite{Yang2008prl}
\bea\label{multipartite EI}
E_I^{(n)}(\rho)=\inf\frac12\{I_n(A_1A_1':\cdots:A_nA_n')-I_n(A_1':\cdots A_n')\},
\eea
where $I_n=\sum_iS(A_i)-S(A_1A_2\cdots A_n)$, the infimum is taken over all extensions of $\rho^{A_1A_2\cdots A_n}$, i.e.,
over all states satisfying the equation $\tr_{A'_1A'_2\cdots A'_n}\rho^{A_1A_1'\cdots A_nA_n'}=\rho^{A_1A_2\cdots A_n}$.
It is also additive in the sense that
\beax
E_I^{(n)}(\rho^{A_1A_2\cdots A_n}\ot \sigma^{B_1B_2\cdots B_n})=E_I^{(n)}(\rho^{A_1A_2\cdots A_n})+E_I^{(n)}(\sigma^{B_1B_2\cdots B_n}).
\eeax


\subsection{Concentratable entanglement}


Let $\hat{N}=\{1, 2, \dots, n\}$ be the set of labels for each qubit, and $\mP(\hat{N})$ be its power set (i.e., the set of subsets, with cardinality $|\mP(\hat{N})|=2^n$). 
In Ref.~\cite{Beckey2021prl}, a multi-qubit entanglement monotone called concentratable entanglement was introduced.
For any set of qubit labels $S\in\mP(\hat{N})\setminus\emptyset$, the concentratable entanglement was defined as
\bea\label{Concentratable entanglement}
C_S^{(n)}(|\psi\ra)-1-\frac{1}{2^{|S|}}\sum\limits_{t\in\mP(S)}\tr\rho_t^2,
\eea
where $|S|$ is the cardinality of the set $S$, and $\mP(S)$ its power set, $\rho_t$ the corresponding reduced state of $|\psi\ra$.


\subsection{Ergotropic-gap concentratable entanglement}


The ergotropic-gap concentratable
entanglement measures and the battery capacity-gap concentratable
entanglement measures were introduced in Ref.~\cite{Bai2025pra}. 
For an $n$-qubit pure state $|\psi\ra$ and a subset $s\subseteq\{1, 2, \dots, n\}$, 
the ergotropic-gap concentratable
entanglement is defined by
\bea \label{E_s}
E_{s\text{-}g,s}^{(n)}(|\psi\ra)=\frac{1}{2^{|s|}}\sum\limits_{X\in\mP(s)}\Delta_{X|\overline{X}}(|\psi\ra),
\eea
where $\mP(s)$ is the power set of $s$, and $\Delta_{X|\overline{X}}(|\psi\ra)=0$ when $X=\emptyset$.
It was shown in Ref.~\cite{Bai2025pra} that $E_{s\text{-}g,s}^{(n)}$ is a multipartite entanglement monotone.


\subsection{Global entanglement measure of multi-qubit state}


In Ref.~\cite{Meyer2002jmp}, Meyer and Wallach proposed an $n$-qubit entanglement
	measure which was derived as a formal function on many qubit states but has a simple
	physical interpretation. They define
\bea\label{n-qubit entanglement}
E_Q^{(n)}(|\psi\ra)=\frac4n\sum\limits_{k=1}^{n}D(|\widetilde{u}^k\ra,|\widetilde{u}^k\ra),
\eea
where $|\widetilde{u}^k\ra$ and $|\widetilde{u}^k\ra$ are vectors in $\left( \C^2\right) ^{\ot(n-1)}$ which are non-normalized and obtained by projecting on state with local projectors on the $k$th qubit,
\beax
|\psi\ra=|0\ra_k\ot |\widetilde{u}^k\ra+|1\ra_k\ot |\widetilde{v}^k\ra,
\eeax
the function $D(|\widetilde{u}^k\ra,|\widetilde{u}^k\ra)$ refers a ``distance'' between the two vectors $|\widetilde{u}^k\ra$ and $|\widetilde{v}^k\ra$, which is defined by the generalized cross product
\beax 
D(|\widetilde{u}^k\ra,|\widetilde{v}^k\ra)=\sum\limits_{i<j}|\widetilde{u}_i^k\widetilde{v}_j^k-\widetilde{u}_j^k\widetilde{v}_i^k|^2
\eeax 
with $|\widetilde{u}^k\ra=\sum_i\widetilde{u}_i^k|i\ra$ and $|\widetilde{v}^k\ra=\sum_j\widetilde{v}_j^j|k\ra$,
$0\leq i, j<2^{n-1}$ are ($n-1$)-bit strings.
Brennen showed in Ref.~\cite{Brennen2003qic} that, 
\beax 
D(|\widetilde{u}^k\ra,|\widetilde{u}^k\ra)=\frac12(1-\tr\rho_k^2),
\eeax
and thus
\beax
E_Q^{(n)}(|\psi\ra)=2\left( 1-\frac{1}{n}\sum\limits_{k=1}^{n}\tr\rho_k^2\right).
\eeax
This quantity can be obtained in the laboratory either by performing state tomography on each qubit or by a more efficient technique using bitwise interactions with a second identically prepared register.

Later, Scott~\cite{Scott2004pra} generalized $E_Q^{(n)}$ into the multiqudit system ${\C^d}^{\ot n}$, and all possible bipartite partitions were considered:
\bea\label{Generalized EQn}
E_{Q,m}^{(n)}(|\psi\ra)=\frac{d^m}{d^m-1}\left[1-\frac{m!(n-m)!}{n!}\sum\limits_{|S|=m}\tr\rho_S^2\right],
\eea
where $\{1, 2, \dots, n\}$ is the set of labels for each qudit, $S\subset\{1, 2, \dots, n\}$, $m=1$, $2$, $\dots$, $\lfloor n/2\rfloor$, and $\rho_S=\tr_{\overline{S}}|\psi\ra\la\psi|$ is the reduced state after tracing out the rest $\overline{S}$.
$E_{Q,m}^{(n)}$ reduces to $E_Q^{(n)}$ whenever $m=1$ and $d=2$.
Clearly, $E_{Q,m}^{(n)}$ is an entanglement monotone and $0\leq  E_{Q,m}^{(n)}\leq  1$.


\subsection{Entanglement monotone via relative purity}


Somma \etal~\cite{Somma2004pra} proposed a measure of entanglement via projection onto the associated Lie algebra. For the given state space $\mH$, let $\mathfrak{h}$ be the real Lie algebra consisting of Hermitian operators on $\mH$, with the Lie bracket 
\beax 
[X,Y]={\rm i}(XY-YX).
\eeax
In this way, operators in $\fh$ can be directly associated with physical observables.
Since $\fh$ is closed under Hermitian conjugation, the projection of a quantum state $\rho$ onto $\fh$ with respect to the trace inner product is uniquely defined. Let $\mP_{\fh}$ denote the projection map, $\rho\mapsto\mP_{\fh}(\rho)$.
For pure state $|\psi\ra\in\mH$, the so-called $\fh$-purity was defined as~\cite{Barnum2003pra,Barnum2004prl,Somma2004pra}
\bea \label{fh purity}
P_{\fh}(|\psi\ra)=\tr[\mP_{\fh}(|\psi\ra\la\psi|)]^2.
\eea 
For any Hermitian orthogonal basis of $\fh$, $\{E_i\}$, $E_i^\dag=E_i$, $\tr (E_iE_j)=\delta_{ij}$, and $E_i$ have a common, rescaled norm chosen to ensure that the maximal value is 1, $P_{\fh}(|\psi\ra)=\sum_i|\la\psi|A_i|\psi\ra|^2$.
For composed system, the basis is a collection of local form, i.e., $E_i^{(j)}=I^{A_1}\ot \cdots\ot I^{A_{j-1}}\ot E_i^{A_j}\ot I^{A_{j+1}}\ot\cdots\ot I^{A_n}$.
It turns out that
\beax 
P_{\fh}^{(n)}(|\psi\ra)=p\sum\limits_{j=1}^n\left( \tr\rho_{A_j}^2-\frac{1}{d_j}\right),
\eeax 
where $d_j=\dim\mH^{A_j}$, and
\beax 
p=\frac{1}{n(1-\frac{1}{n}\sum_j\frac{1}{d_j})}.
\eeax
It is clear that 
\bea \label{Entanglement monotone}
E_{\fh}^{(n)}(|\psi\ra)=1-P_{\fh}^{(n)}(|\psi\ra)
\eea
define an entanglement monotone.
Interestingly, for the $n$-qubit pure state $|\psi\ra$, 
\beax 
E_Q^{(n)}(|\psi\ra)=E_{\fh}^{(n)}(|\psi\ra).
\eeax


\subsection{Witnessed entanglement}


Brand\~{a}o presented a general expression for the quantification of entanglement via entanglement witness (EW), which was defined as~\cite{Brandao2005pra}
\bea\label{E_W}
E_W^{(n)}(\rho) = \max \left\lbrace 0, -\min_{W \in \mathcal {M}} \mathrm{Tr}(W\rho) \right\rbrace ,
\eea
where $\mathcal {W}$ denotes the set of entanglement witnesses, $\mathcal{M} = \mathcal{W} \cap \mathcal {C}$, the set $\mathcal {C}$ is what distinguish the quantities such that $\mathcal {M} $ is compact. Any measure expressed by equation above was called witnessed entanglement. Here, $\mathcal{W}$ could be any type of entanglement witness. For example, when we deal with the genuine entanglement, $\mathcal{W}$ refers to the set of all $W$'s such that $\tr(W\sigma)\geq0$ for any biseparable state $\sigma$, and there exists genuine entangled state $\rho$ with $\tr(W\rho)<0$.

For the bipartite case, some well-known entanglement measure can be expressed as above equation, such as 
negativity, 
\beax  
N(\rho) = \max \left\lbrace  0, - \min_{0 \leq W \leq I}\mathrm{Tr}(W^{T_{a}}\rho) \right\rbrace 
\eeax 
and concurrence of two-qubit state, according to Verstraete~\cite{Verstraete2002thesis},
\beax 
C(\rho)=\max\left\lbrace 0, -\min\limits_{A\in\mS L_2(\C)}\tr\left( |A\ra\la A|^{T_b}\rho\right) \right\rbrace, 
\eeax 
where $|A\ra$ denotes the unnormalized state $A\ot I|I\ra$ with $|I\ra=\sum_i|ii\ra$, $\det A=1$.

In addition,
\bea \label{Bn}
B^{(n)}(\rho) = \max_{||W - I||_{2} \leq 1}\left\lbrace \min_{\sigma \in \mathcal {S}_n}\tr(W\sigma) - \tr(W\rho)\right\rbrace, 
\eea 
where $B^{(n)}(\rho)$ was introduced by Bertlmann \etal~\cite{Bertlmann2002pra} wherein it was shown that $B^{(n)}(\rho)=\min\limits_{\sigma\in\mS_n}\|\rho-\sigma\|_2$, $W \in  \mathcal {W}$. The robustness of $k$-entanglement can be expressed as
\beax 
\check{R}_k^{(n)}(\rho)=\max\{0, -\min\limits_{W_k}\tr(W_k\rho)\},
\eeax 
where $W_k\leq I$ denotes the $k$-entanglement witness~\cite{Brandao2005pra,Cavalcanti2006pra}.

Let $\mC$ be the set of all Hermitian operators $W$ that satisfying $-kI\leq W\leq lI$, where $k$, $l\geq 0$, then we denote the witnessed entanglement derived from Eq.~\eref{E_W} by $E^{(n)}_{k:l}$.
It was shown in~\cite{Brandao2005pra} that $E_{k:l}$ are entanglement monotones for any $k$, $l\geq 0$.
It is difficult to compute in general. But for the $d\ot d$ isotropic state
\beax 
\rho_t=tP^++(1-t)\frac{I}{d^2},
\eeax 
it was found that~\cite{Brandao2005pra}
\beax 
E_{k:1}^{(2)}(\rho_t)=\begin{cases}
	\!(k+1)t+\frac{(1-t)(k+1)}{d^2}-1,&k\leq d-1,\\
\!	dt+\frac{1-t}{d}-1,&k\geq d.
\end{cases}
\eeax 
For any bipartite state $\rho$, 
\beax 
E_d(\rho)\leq\log_2[1+E_{k:1}^{(2)}(\rho)],\quad k\geq 1.
\eeax


\subsection{Invariance under determinant 1 SLOCC operations}
	

Recall that $B_1\otimes \cdots\otimes B_n|\psi\rangle$ with $B_i$ being arbitrary local operator on $\mH^{A_i}$ is called a SLOCC operation on $|\psi\ra$, and determinant 1 SLOCC operations mean that $\{B_i\}$ is the group of full-rank local operators having a determinant equal to 1 up to some given orthonormal basis~\cite{Verstraete2003pra}.

In Ref.~\cite{Verstraete2003pra}, Verstraete \etal proposed an entanglement monotone that is defined as a linearly homogeneous positive function of a pure state, $E_D^{(n)}(|\psi\rangle\langle\psi|)$, that remains invariant under determinant 1 SLOCC operations. Namely, if for any $|\psi\ra$ and any $c>0$,
\bea \label{SLOCC-invariant}\eqalign{
	E_D^{(n)}(c|\psi\ra)\geq 0, \cr
	E_D^{(n)}(c|\psi\ra)=cE_D^{(n)}(|\psi\ra),\cr
	E_D^{(n)}(|\psi\ra)=E_D^{(n)}(B_1\otimes \cdots\otimes B_n|\psi\ra)
}\eea
for any determinant 1 SLOCC operation $B_1\otimes \cdots\otimes B_n$,
then $E_D^{(n)}$ is an entanglement monotone.

Later, a series of determinant 1 SLOCC invariant entanglement monotones were investigated~\cite{Eltschka2012pra,Gour2010prl,Leifer2004pra,Lidafa2009qic,Luque2003pra,Luquejpa2006,Osterloh2005pra,Osterloh2009jmp,Sharma2010pra,Viehmann2011pra}. For example,
the quantity defined as in Eq.~\eref{SLOCC-invariant} with relacing $E_D^{(n)}(c|\psi\ra)=cE_D^{(n)}(|\psi\ra)$ by
$E_D^{(n)}(c|\psi\ra)=c^tE_D^{(n)}(|\psi\ra)$ with $0<t\leq 4$ was shown to be an entanglement monotone~\cite{Eltschka2012pra}.


\subsection{Barycentric measure}


Consider an $n$-qubit pure state $|\psi\rangle$
symmetric with respect to permutation of subsystems:
\bea
|\psi\rangle=\frac{1}{\sqrt{K}}\sum\limits_\pi|\phi_{\pi(1)}
\rangle|\phi_{\pi(2)}\rangle\cdots|\phi_{\pi(n)}\rangle.
\label{eq:maj}
\eea 
The sum is taken over all permutations $\pi$,  and $K=n!\sum\limits_\pi\prod\limits_{i=1}^n|\langle
\phi_i| \phi_{\pi(i)}\rangle|$ is the normalization constant.

For any permutation symmetric state~\eref{eq:maj} with
$|\phi_{i}\rangle=\cos\frac{\theta_i}{2}|0\rangle+
e^{i\Phi_i}\sin\frac{\theta_i}{2}|1\rangle$ ($i=1,2,...,n$). The barycentric measure of entanglement was defined as~\cite{Ganczarek2012pra}
\bea\label{Barycentric measure of entanglement}
E_B^{(n)}(|\psi\ra)=1-d^2,
\eea
where $d$ is the distance between the barycenter of the Majorana points (see Fig.~\ref{fig:2q} borrowed from~\cite{Ganczarek2012pra})~\cite{Majorana1932}
representing the state and the center of the Bloch ball:
\begin{eqnarray*} 
	d= \left| \frac{1}{n}\sum\limits_{i=1}^n (\sin\theta_i\cos\Phi_i ,
	\,\sin\theta_i\sin\Phi_i , \,\cos\theta_i)\right|,
\end{eqnarray*} 
where $| x |$ denotes the length of a vector $x$.
By definitioin, $0\leq E_B^{(n)}\leq 1$.

\begin{figure}
	\centering
	\includegraphics[height=4cm]{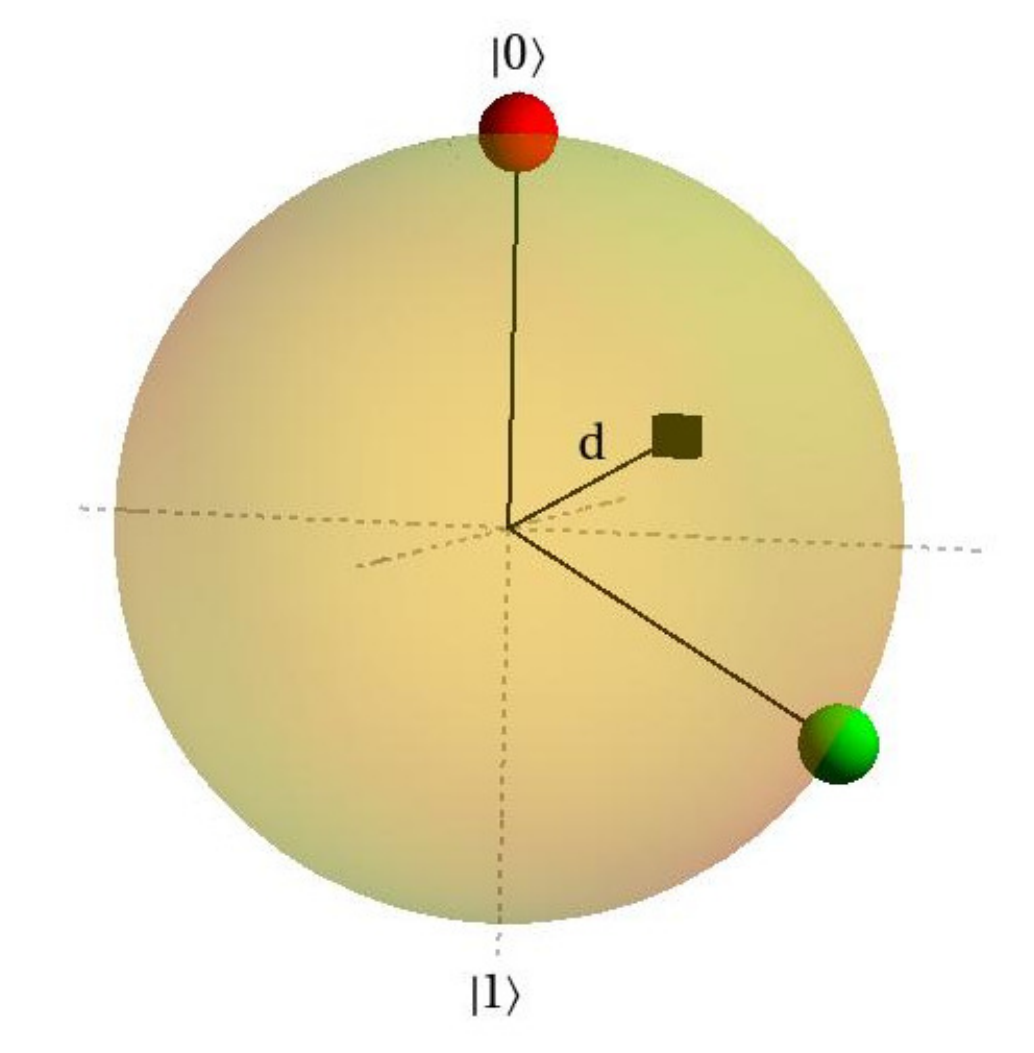}
	\caption{ Majorana representation of a symmetric 2-qubit state
		of the form $|\phi_1\rangle=|0\rangle$ and
		$|\phi_2\rangle=\cos\frac{\theta}{2}|0\rangle+\sin\frac{\theta}{2}|1\rangle$
		with $\theta=\frac{2}{3}\pi$. Majorana points
		are shown as red and green balls. The black
		cube denotes the barycenter of Majorana points located inside the Bloch ball
		and its distance to the center of the ball is equal to $d$.}
	\label{fig:2q}
\end{figure}

A useful entanglement measure should not increase under
local operations. The barycentric measure $E_B^{(n)}$ is defined only
for symmetric states so monotonicity can be checked only
for local unitary operations which preserve the permutation
symmetry. It was proved in Ref.~\cite{Ganczarek2012pra}
that $E_B^{(n)}$ is invariant under local operations preserving the permutation symmetry.
Namely, in such a sense $E_B^{(n)}$ can be regarded as an entanglement measure.


\subsection{MEM via the product of the reduced function}


In Ref.~\cite{Guo2022jpa}, for the case of $n=4$, the quantity 
\bea\label{1234(2)}
F_{1234(2)}(\rho)=\left\lbrace \begin{array}{ll}
	\!\!\frac{\prod_ih(\rho^{A_i})}{\sum_ih(\rho^{A_i})},&~\sum_ih(\rho^{A_i})>0,\\
	\!\!0,&~\sum_ih(\rho^{A_i})=0
\end{array}\right. 
\eea
was shown to be an entanglement monotone therein.

Going further, we can also define MEM by the product of all the reduced function of the single reduced states, i.e.,
\bea\label{MEM2}
\check{\mE}^{(n)}(|\psi)=\sqrt[n]{\prod\limits_ih(\rho^{A_i})}.
\eea
Moreover, it can be defined by the product of all possible reduced functions, i.e.,
\bea\label{prod2}
\hat{\mE}^{(n)}(|\psi\ra)
=\sqrt[\hat{n}]{\prod\limits_{\gamma_i\in\Gamma_2}h(\rho^{X_{1(i)}})}.
\eea
They are  multipartite entanglement monotones due to the geometric mean function $f(x_1, x_2, \cdots, x_n)=(\prod_i^nx_i)^{1/n}$ is concave~\cite{Boyd2004book}, where $\hat{n}=|\Gamma_2|$ denotes the number of all possible bipartite partition.
The case of $h=h_c$ (up to some factor) and $h=h_N$ were firstly discussed in Ref.~\cite{Love2007qip} and Ref~\cite{Sabin2008epjd}, respectively.	$\hat{\mE}^{(n)}$ is indeed a GEM, and $\check{\mE}^{(n)}$	is also a GEM whenever $n=3$.


\subsection{Schmidt measure}


The Schmidt decomposition is not valid for multipartite case since only
rare pure states in the multipartite case admit the
Schmidt decomposition form\cite{Peres1995,Thapliyal}
\begin{eqnarray}\label{n-Schmidt-decmoposition}
	|\psi\rangle=\sum_{k=1}^{r}
	\lambda_k|e_k\rangle^{A_1}|e_k\rangle^{A_2}\otimes\cdots\otimes|e_k\rangle^{A_n},
\end{eqnarray}
where $r\leq \min\{d_1,d_2,\dots,d_n\}$, $d_i=\dim\mH^{A_i}$, $\{|e_k\rangle^{A_i}\}$ is an orthonormal set of $\mH^{A_i}$.
For the simplest three-partite case, any
three-qubit pure state can be written as Eq.~\eref{ex1}, which is not a form of Schmidt
decomposition as Eq.~\eref{n-Schmidt-decmoposition}. That is, there is no correspondence of Schmidt number
for multipartite case in general.

Eisert and Briegel introduced the so-called Schmidt measure in Ref.~\cite{Briegel2001prl,Eisert2001pra}.
Any $|\psi\rangle\in \mH^{A_1A_2\cdots A_n}$ can be written in the form
\beax
|\psi\rangle=\sum_{i=1}^s p_i|\psi_i\rangle^{A_1}\otimes \cdots \otimes |\psi_i\rangle^{A_n}
\eeax
for some $s$. 
Let $r_t(|\psi\ra)$ be the minimal number of product terms $s$ in such a decomposition of $|\psi\ra$.
The Schmidt measure was defined by~\cite{Eisert2001pra}
\bea\label{Schmidt measure}
E_{\rm Sr}(|\psi\rangle)=\log_2r_t(|\psi\ra),
\eea
and by the convex-roof extension for the mixed state, which is different from the $S_r$ for mixed state.
In case of a bipartite system $r_t(|\psi\ra)$ reduces to the Schmidt rank of the state $|\psi\ra$. Thus, the Schmidt measure could be conceived as a generalization of the Schmidt rank to multipartite systems.
$E_{\rm Sr}$ was shown to be a well-defined entanglement monotone~\cite{Eisert2001pra}.
The Schmidt measure cannot distinguish genuine entanglement from bipartite entanglement. But it may be useful in many contexts, e.g., see in~\cite{Mora2005prl}.
By definition, $E_{\rm Sr}$ is difficult to calculate, but for some symmetric state, it can be computed.
For example, for the two-qubit Werner state 
\beax 
\rho(x)=x|\psi^{-}\ra\la\psi^{-}|+(1-x)I/4
\eeax 
with $|\psi^{-}\ra=\frac{1}{\sqrt2}(|01\ra-|10\ra)$, $0\leq x\leq 1$,
\beax 
E_{\rm Sr}(\rho(x))=\left\lbrace \begin{array}{ll}
	\!\!\frac{3x}{2}-\frac12, &\frac13<x\leq 1,\\
	\!\!0, & 0\leq x\leq \frac13.
\end{array}\right. 
\eeax

In Ref.~\cite{Briegel2001prl}, the authors introduced the persistency of entanglement of an entangled state for the $n$-qubit system, which is defined as the minimum number of local measurements such that, for all measurement outcomes, the state is completely disentangled. For a number of states one can show that the Schmidt measure coincides with the persistency of entanglement introduced in Ref.~\cite{Briegel2001prl}, such as the $n$-qubits cluster states.


\subsection{A generalization of the Schmidt number}


Not that the scenario in the above subsection is not based on the Schmidt decomposition. In Ref.~\cite{Guo2015ijqi}, Guo and Fan extended the
Schmidt number to multipartite systems according to the Schmidt decomposition of the subsystems.
As desired, the generalized Schmidt number, denoted by $\check{S}_r^{(n)}$, can be used to quantify entanglement for multipartite states as well.

\begin{table}[h]
	\caption{\label{tab:2}The Schmidt number of the
		three qubit pure state $|\psi\ra$ ($r_i$ denotes the rank
		of $\rho_i$, $`i-j$' means $\rho_{ij}$ is not a
		pure separable state, $`i~~~j-k$' means
		$\rho_{jk}=|\psi_{jk}\rangle\langle\psi_{jk}|$ is an
		entangled pure state, etc.)}
	
	\begin{indented}
		\lineup
		\item[] 
		\begin{tabular}{@{}llll} 
			\br 
			Type & Model            &  Reductions                                         & $\check{S}_r^{(3)}(|\psi\ra)$\\
			\mr
			$1|2|3$      &$1~~~ 2~~~3$      & $r_i=1$                                             & 1 \\
			$1|23$       &$1~~~2-3$         & $r_1=1$, $r_2=r_3=2$                                & 2\\
			$12|3$       &$1-2~~~3$         & $r_3=1$, $r_1=r_2=2$                                & 2\\
			$2|13$       &$2~~~1-3$         & $r_2=1$, $r_1=r_3=2$                                & 2\\
			GE           &figure (a) & $r_i=2$, $\rho_{\overline{i}}$ is separable, $i=1,2,3$   & 3\\
			GE           &figure (a) & $r_i=2$, $\rho_{\overline{i}}$ is entangled for some $i$ & 4\\
			\br
		\end{tabular}
	\end{indented}
\end{table}

According to the scenario in Ref.~\cite{Guo2015ijqi}, there are four types
of entanglement indeed for the three qubit case (see
Table~\ref{tab:2}). In this subsection, the part $A_i$ is abbreviated as (the subscript) $i$ for simplicity.  Similarly, there are at most
$d_1+d_3$ types of entanglement for the $d_1\otimes d_2\otimes d_3$ case (see
Table~\ref{tab:3}), $d_1\leq d_2\leq d_3$. $\check{S}_r^{(2)}$ reduces to the bipartite Schmidt number $S_r$ as Eq.~\eref{Schmidt number}.

\begin{table}[h]
	\caption{\label{tab:3}The Schmidt number of $|\psi\rangle$ in
		the three-partite system.}
	\begin{indented}
		\lineup
		\item[]
		\begin{tabular}{@{}lll} 
			\br
			Type    &  Reductions                      & $\check{S}_r^{(3)}(|\psi\ra)$\\
			\mr
			$1|2|3$         & $r_i=1$                          & 1\\
			$1|23$          & $r_1=1$, $r_2=r_3=\check{S}_r^{(2)}(|\psi_{23}\ra)$ & $\check{S}_r^{(2)}(|{\psi_{23}}\ra)$\\
			$12|3$          & $r_3=1$, $r_1=r_2=\check{S}_r^{(2)}(|\psi_{12}\ra)$ & $\check{S}_r^{(2)}(|{\psi_{12}}\ra)$\\
			$2|13$          & $r_2=1$, $r_1=r_3=\check{S}_r^{(2)}(|\psi_{13}\ra)$ & $\check{S}_r^{(2)}(|{\psi_{13}}\ra)$\\
			GE              & $r_i\geq 2$, $i=1$,2,3               & $\max\limits_i(r_i+\check{S}_r^{(3)}({\rho_{\overline{i}}}))$\\
			\br
		\end{tabular}
	\end{indented}
\end{table}

\begin{table}[pht]
	\caption{\label{tab:4}The Schmidt number of
		$|\psi\rangle$ in the four-partite system. }
	\begin{indented}
		\lineup
		\item[]
		\begin{tabular}{@{}llll} \br
			Type  &Model          &  Local rank   & $\check{S}_r^{(4)}(|\psi\ra)$\\
			\mr
			$1|2|3|4$   & $1~~~2~~~3~~~4$ & $r_i=1$       & 1\\
			$1|23|4$    & $1~~~2-3~~~4$   &$r_1=r_4=1$    & $\check{S}_r^{(2)}(|\psi_{23}\ra)$\\
			$12|3|4$    & $1-2~~~3~~~4$   &$r_3=r_4=1$    & $\check{S}_r^{(2)}(|\psi_{12}\ra)$\\
			$2|3|14$    & $2~~~3~~~1-4$   &$r_2=r_3=1$    & $\check{S}_r^{(2)}(|\psi_{14}\ra)$\\
			$2|13|4$    & $2~~~1-3~~~4$   &$r_2=r_4=1$    & $\check{S}_r^{(2)}(|\psi_{13}\ra)$\\
			$1|3|24$    & $1~~~3~~~2-4$   &$r_1=r_3=1$    & $\check{S}_r^{(2)}(|\psi_{24}\ra)$\\
			$1|2|34$    & $1~~~2~~~3-4$   &$r_1=r_2=1$    & $\check{S}_r^{(2)}(|\psi_{34}\ra)$\\
			$1|234$     & $1~~~2-3-4$     & $r_1=1$       & $\check{S}_r^{(3)}(|\psi_{234}\ra)$\\
			$2|134$     & $2~~~1-3-4$     &$r_2=1$        & $\check{S}_r^{(3)}(|\psi_{134}\ra)$\\
			$3|124$     & $3~~~1-2-4$     &$r_3=1$        & $\check{S}_r^{(3)}(|\psi_{124}\ra)$\\
			$4|123$     & $4~~~1-2-3$     &$r_4=1$        & $\check{S}_r^{(3)}(|\psi_{123}\ra)$\\
			$12|34$     & $1-2~~~3-4$     &$r_i\geq2$     & $\check{S}_r^{(2)}(|\psi_{12}\ra)+\check{S}_r^{(2)}(|\psi_{34}\ra)$\\
			$13|24$     & $1-3~~~2-4$     &$r_i\geq2$     & $\check{S}_r^{(2)}(|\psi_{13}\ra)+\check{S}_r^{(2)}(|\psi_{24}\ra)$\\
			$14|23$     & $1-4~~~2-3$     &$r_i\geq2$     & $\check{S}_r^{(2)}(|\psi_{14}\ra)+\check{S}_r^{(2)}(|\psi_{23}\ra)$\\
			GE          & figure (b) &$r_i\geq2$ & $\max\limits_i(r_i+\check{S}_r^{(3)}(\rho_{\overline{i}})$\\
			\br
		\end{tabular}
	\end{indented}
\end{table}

The Schmidt number for the four-partite system is
as Table~\ref{tab:4}. Moreover,
this approach can be extended to $n$-partite case step by step for
both pure and mixed states. That is, for
$|\psi\rangle\in\mathbb{C}^{d_1}\otimes\mathbb{C}^{d_2}\otimes\cdots\otimes\mathbb{C}^{d_n}$,
the $\check{S}_r^{(n)}(|{\psi}\ra)$ can be defined as the program discussed above:
if $|\psi\rangle$ is not genuinely entangled and assume with on loss of generality that it is $12|34|5\cdots n$ separable (resp. $12|3|4|5\cdots n$ separable),
then $\check{S}_r^{(n)}(|\psi\ra)=\check{S}_r^{(2)}(|\psi_{12}\ra)+\check{S}_r^{(2)}(|\psi_{34}\ra)+\check{S}_r^{(n-4)}(|\psi_{5\cdots n}\ra)$ [resp. $\check{S}_r^{(n)}(|\psi\ra)=\check{S}_r^{(2)}(|\psi_{12}\ra)+\check{S}_r^{(n-2)}(|\psi_{5\cdots n}\ra)$]; if $|\psi\rangle$ is genuinely entangled, then
$\check{S}_r^{(n)}(|\psi\ra)=\max\limits_{i}\left[ r_i+\check{S}_r^{(n-1)}({\rho_{\overline{i}}})\right]$. For
mixed state $\rho$,
it can be defined by the convex-roof structure
\begin{eqnarray}\label{n-Schmidt measure}
	\check{S}_r^{(n)}(\rho):
	=\min\limits_{\{p_i,|\psi_i\ra\}}\max\limits_{|\psi_i\ra}\check{S}_r^{(n)}(|\psi_i\ra),\label{a}
\end{eqnarray}
where the minimization is over all ensembles $\{p_i, |\psi_i\ra\}$ of $\rho$.

The ``structure'' of the Schmidt number for genuinely entanglement with the
following figures. $`r_i-R_{ij}-r_j$' means $\rho_{ij}$ is a mixed state
with Schmidt number $R_{ij}$. For the tripartite case, the Schmidt number
\begin{eqnarray*}
	&\begin{array}{c}\xymatrix{
			r_1 \ar@{-}[rd]|-{R_{{13}}}  \ar@{-}[r]
			&{~_{R_{{12}}}} \ar@{-}[r]&
			r_2 \ar@{-}[l]\ar@{-}[ld]|-{R_{{23}}}\\
			&r_3&}\\ \\
		(a)\end{array}\quad\quad\quad
	\begin{array}{c}\xymatrix{
			&r_1 \ar@{-}[ldd]|-{R_{13}}  \ar@{-}[d]|-{R_{{12}}} \ar@{-}[rdd]|-{R_{14}}\\
			&r_2 \ar@{-}[ld]|-{R_{23}} \ar@{-}[rd]|-{R_{24}}\\
			r_3\ar@{-}[r] &_{R_{34}}&\ar@{-}[l]r_4 }\\ \\
		(b)\end{array} &
\end{eqnarray*}
can be ``explained'' by Figure (a): $\check{S}_r^{(3)}(|\psi\ra)=\max\limits_{i}\{r_i+R_{\overline{i}}\}$. For the
four-partite pure state $|\psi\rangle$,
$\check{S}_r^{(4)}(|\psi\ra)$ is determined by $r_i$
and $R_{\overline{i}}$ while
$R_{\overline{i}}$ is decided by the
Schmidt number of the six bipartite
reductions and the four single
part reductions. It is clear that
$\check{S}_r^{(4)}(|\psi\ra)\leq\max\limits_{i\neq j}\{r_i+r_j+R_{\overline{ij}}\}$
[see the ``relation'' among $r_i$, $r_j$ and $R_{\overline{ij}}$ in Figure (b)].

In such a sense, if $n>3$, then there exist $|\psi\rangle$ and $|\phi\rangle$ in $n$-partite systems, such that $|\psi\rangle$ is $k$-separable, $|\phi\rangle$ is genuinely entangled but $\check{S}_r^{(n)}(|\psi\ra)>\check{S}_r^{(n)}(|\phi\ra)$. The Schmidt number defined as Eq.~\eref{a} is an entanglement monotone since it is reduced to the fact that the bipartite Schmidt number is an entanglement monotone, and it is invariant under the invertible SLOCC (stochastic local operations and classical communication) since invertible SLOCC preserves the rank of the reduction\cite{Dur2000pra2} (also see in Refs.~\cite{Cornelio,Lidafa,Wangshuhao}).


\subsection{Operational MEMs}


We say that a state $|\psi\rangle $ can reach a state $|\phi\rangle $ if there exists an LOCC protocol which transforms $|\psi\rangle $ into $|\phi\rangle $ (deterministically). In this case $|\phi\rangle $ is accessible from $|\psi\rangle $. For a given state, $|\psi\rangle$, we denote by $M_a(|\psi\rangle)$ the set of states which can be accessed via LOCC from $|\psi\rangle $ and by $M_s(|\psi\rangle )$ the set of states which can reach $|\psi\rangle $. 

The following two magnitudes occur then naturally in the context of possible LOCC transformations~\cite{Schwaiger2015prl}: the source volume, $V_s(|\psi\rangle )=\mu[M_s(|\psi\rangle ]$, which measures the amount of states that can be used to reach the state $|\psi\rangle $ and the accessible volume, $V_a(|\psi\rangle )=\mu[M_a(|\psi\rangle )]$, which measures the amount of states that can be accessed by $|\psi\rangle $ via LOCC. Here, $\mu$ denotes an arbitrary measure in the set of local unitary (LU) equivalence classes. 

Hence, any properly normalized and rescaled measure of these sets is indeed an entanglement measure, i.e. it does not increase under LOCC. A possible choice would be~\cite{Schwaiger2015prl}
\bea \label{Eac}
E_{ac}^{(n)}(|\psi\rangle )=V_a(|\psi\rangle )/V_a^{sup}
\eea 
and 
\bea \label{Eso}
E_{so}^{(n)}(|\psi\rangle)=1-V_s(|\psi\rangle )/V_s^{sup},
\eea  
where $V_a^{sup}$ ($V_s^{sup}$) denote the maximally accessible (source) volume according to the measure $\mu$. Note that these operational entanglement measures are applicable to arbitrary multipartite systems of any dimension. Moreover, these are valid entanglement measures for mixed states.


\subsection{Concurrence vector in multipartite quantum systems}


Let $|\psi\ra$ be a pure state in $\mH^{A_1A_2\cdots A_n}$, then 
\beax\label{mpartitepsi}
\left|\psi\right>=\sum_{i_1}^{d_1}\sum_{i_2}^{d_2}\cdots
\sum_{i_n}^{d_n} a_{i_1 i_2 \cdots i_n}|i_1i_2\cdots i_n\ra
\eeax
in the standard basis $|i_1i_2\cdots i_n\ra$, $\dim\mH^{A_i}=d_i$.
With $\rho^{T_{ij}}$, $\rho=|\psi\ra\la\psi|$, we denote the partial transposition of $\rho$ with respect to subsystems $i$ and $j$. The associated $\Pi_{i=1}^{n}d_i(d_i-1)/2$ dimensional concurrence vector ${\bf C}$ with components
$C^{\{ij\}}_{\alpha_i\alpha_j}$ is defined as~\cite{Akhtarshenas2005jpa}
\beax
C^{\{ij\}}_{\alpha_i\alpha_j}= \sqrt{ \langle
	\psi|\tilde{\rho}^{\{ij\}}_{\alpha_i,\alpha_j}|\psi\rangle},
\eeax
where
\beax
\tilde{\rho}^{\{ij\}}_{\alpha_i,\alpha_j}=M^{\{ij\}}_{\alpha_i\alpha_j}\rho^{T_{ij}}
M^{\{ij\}}_{\alpha_i\alpha_j}
\eeax
with
\beax
\fl \quad \quad M^{\{ij\}}_{\alpha_i\alpha_j}=&I_1\otimes \cdots \otimes I_{i-1}
\otimes L_{\alpha_i} \otimes I_{i+1} \otimes \cdots 
\otimes
I_{j-1} \otimes L_{\alpha_j} \otimes I_{j+1} \otimes \cdots
\otimes I_n,
\eeax
for $1\le i < j \le n$, $\alpha_i=1,\cdots, d_i(d_i-1)/2$ and $\alpha_j=1,\cdots, d_j(d_j-1)/2$. Here $I_k$ denotes the identity operator on $\mH^{A_k}$, and $L_{\alpha_i}$ represents the set of $d_i(d_i-1)/2$ generators of an $SO(d_i)$ group and $L_{\alpha_{j}}$ are generators of an $SO(d_j)$ group with similar a definition. The concurrence vector ${\bf C}$ is regarded as a direct sum of elementary subvectors ${\bf C}^{\{ij\}}$, i.e.,
\beax\label{direcsum}
{\bf C}=\sum_{\bigoplus ij}{\bf C}^{\{ij\}},
\eeax
such that each subvector ${\bf C}^{\{ij\}}$ corresponds to pair
$A_iA_j$. The entanglement
contribution of pair $A_iA_j$ in the entanglement of
$\left|\psi\right>$ is defined as the norm of the concurrence
subvector ${\bf C}^{\{ij\}}$~\cite{Akhtarshenas2005jpa}
\bea\label{Cij}
\fl \quad \begin{array}{ll}
	C^{\{ij\}}&=  \left\| {\bf C}^{\{ij\}}\right\| =
	\sqrt{\sum_{\alpha_i=1}^{d_i(d_i-1)/2}\quad\sum_{\alpha_j=1}^{d_j(d_j-1)/2}
		\langle
		\psi|\tilde{\rho}^{\{ij\}}_{\alpha_i,\alpha_j}|\psi\rangle}
	\vspace{3mm}  \\
	&=  \left\{\sum_{\{K\}}\sum_{\{L\}}\sum_{k_i <l_i}^{d_i} \sum_{k_j
		<l_j}^{d_j}\left|a_{\{k_i, k_j, K\}} a_{\{l_i, l_j, L\}}-\right. \right. \\
	&~~~~~\left. \left. a_{\{k_i,
		l_j, K\}}a_{\{l_i, k_j, L\}}-a_{\{l_i, k_j, K\}} a_{\{k_i, l_j,
		L\}}+a_{\{l_i, l_j, K\}}a_{\{k_i, k_j,
		L\}}\right|^2\right\}^{1/2}.
\end{array}
\eea
In Eq.~\eref{Cij}, $\{k_i, k_j, K\}$ stands for $n$ indices such
that $k_i$ and $k_j$ correspond to sybsystems $i$ and $j$
respectively, and $K$ denotes the set of $n-2$ indices for other
subsystems. Also $\sum_{\{K\}}$ stands for summation over indices
of all subsystems except subsystems $i$ and $j$.
The total concurrence of $\left|\psi\right>$ is
defined as the norm of the concurrence vector ${\bf C}$, i.e.~\cite{Akhtarshenas2005jpa}
\bea\label{Cmpartite}
\fl \quad \begin{array}{rl}
	\check{C}^{(n)}= & \|{\bf C}\|=\sqrt{\sum_{1\le i < j \le n} \left\| {\bf C}^{\{ij
			\}}\right\|^2}
	\vspace{3mm}  \\
	= & \left\{\sum_{1\le i < j \le n}
	\sum_{\left\{K\right\}}\sum_{\left\{L\right\}}
	\sum_{k_i<l_i}^{d_i} \sum_{k_j <l_j}^{d_j} \left|a_{\{k_i, k_j, K\}} a_{\{l_i, l_j,L\}}\right.\right.  \\
	&	~~~\left. \left. -a_{\{k_i,
		l_j,K\}}a_{\{l_i, k_j, L\}} - a_{\{l_i, k_j, K\}} a_{\{k_i,
		l_j,L\}}+a_{\{l_i, l_j,K\}}a_{\{l_i, l_j, L\}}
	\right|^2\right\}^{1/2}
\end{array}
\eea
Note that whether this total concurrence is an entanglement measure is unknown although it was conjectured to be true~\cite{Akhtarshenas2005jpa}.

There are other concurrence-type measures which are defined via the projectors onto the antisymmetric and symmetric subspace of state space~(see~\cite{Carvalho2004prl,Demkowicz2006pra,Mintert2005prl} for detail).


\subsection{Ent}


Ref.~\cite{Hedemann2018QIC} presented an entanglement monotone termed ent which was defined by
\begin{eqnarray}\label{Hedemann2018QIC}
	\Upsilon (|\psi\ra ) = \frac{1}{M}\left( {1 - \frac{1}{n}\sum\limits_{i = 1}^n {\frac{{d_i \tr\rho_{A_i}^2 - 1}}{{d_i  - 1}}} } \right)\!,
	\label{eq:2}
\end{eqnarray}
where  $\rho_{A_i}$ is the reduce state of $\rho=|\psi\ra\la\psi|$ in the $i$th subsystem with dimension $d_i$, and the proper normalization factor is
\begin{eqnarray*}%
	M= M(L_{*}) = 1 - \frac{1}{n}\sum\limits_{i = 1}^n {\frac{{d_i P_{\mathrm{MP}}^{(i)}(L_{*}) - 1}}{{d_i  - 1}}},
	\label{eq:3}
\end{eqnarray*}
where, given a pure state $\rho$, $ P_{\mathrm{MP}}^{(i)} (L_* ) $ is the minimum physical purity of $\rho_{A_i}$~\cite{Hedemann2018QIC},
\begin{eqnarray}\label{universal-measure}                
\fl\quad 	P_{\mathrm{MP}}^{(i)} (L_* ) =  &{\bmod (L_* ,d_i )\left( {\frac{{1 + \lfloor L_* /d_i\rfloor}}{{L_* }}} \right)^2 }   { + [d_i  - \bmod (L_* ,d_i )]\left( {\frac{{\lfloor L_* /d_i \rfloor}}{{L_* }}} \right)^2 ,} 
\end{eqnarray}
where $\bmod (a,b) = a - \lfloor a/b\rfloor b$, $L_*$ is any number of levels of $\rho$ with equal nonzero probabilities that can support maximal entanglement, and by convention we use the smallest of these in Eq. (\ref{universal-measure}), as $L_*  = \min \{ \mathbf{L}_* \}$, where $\mathbf{L}_*  = \{ L_* \}$ is the list of values of $L$ that satisfy
\beax     
\mathop {\min }\limits_{L \in 2, \ldots, d_{\overline{\max}} } [1 - M(L)],
\label{eq:5}
\eeax
where $d_{\overline{\max}} = \frac{d}{{d_{\max } }}$ is the product of all $d_m$ except $d_{\max }$, where $d_{\max } = \max d_i$, $d=d_1d_2\cdots d_n$.  See details in Ref.~\cite{Hedemann2018QIC} on how to compute $L_*$. 

One can also define some MEMs based on Eq. (\ref{Hedemann2018QIC}) such as full genuinely multipartite (FGM) ent, full simultaneously multipartite ent and full distinguishably multipartite ent~\cite{Hedemann2018qic}. For example, FGM entanglement is defined as ~\cite{Hedemann2018qic}
\begin{eqnarray}\label{FGM}
	\Upsilon _{\mathrm{FGM}} (\rho ) = \frac{1}{M_{\mathrm{FGM}}}\sum\limits_{k = 2}^n {\Upsilon_{\mathrm{GM}_k } (\rho )},
	\label{eq:6}
\end{eqnarray}
where \smash{$M_{\mathrm{FGM}}$} is a normalization factor, and
\begin{eqnarray}\label{GMk}
	\Upsilon _{\mathrm{GM}_{k}} (\rho ) = \min\limits_{\gamma_i^{(k)}} \left\lbrace \Upsilon_{\gamma_i^{(k)}(\rho ) }\right\rbrace ,
\end{eqnarray}
which is the $\mathrm{GM}_k$ ent that refers to the minimum ent over all $k$-partitions $\gamma_i^{(k)}$ in $\Gamma_k$, where \smash{$\{ \Upsilon _{\gamma_i^{(k)}}(\rho ) \}$} is the set of all $n$-partite $k$-partitional entanglements, each labeled by $\gamma_i^{(k)}$.


\subsection{Majorization-derived monotones}


Motivated by 
\beax 
\sum\limits_{i=1}^k\lambda_i(\rho^A)=\max\limits_{\text{rank-}k \,{\rm projectors}\, P}\left\| I\ot P|\psi\ra^{AB}\right\|^2,
\eeax
where $\rho^A=\tr_B|\psi\ra\la\psi|^{AB}$, $\lambda_i(\rho^A)$ denotes the $i$th decreasingly ordered eigenvalue of $\rho^A$, Barnum and Linden gave the majorization-derived entanglement as~\cite{Barnum}
\beax \label{themonotones} 
\max_{\Gamma_1,...,\Gamma_N} \|\Gamma_1 \otimes \cdots \Gamma_n
|\psi\rangle\|^2\;,
\eeax
where each of 
$\Gamma_i$ is a $k_i$-dimensional projector on $\mH^{A_i}$. 
It is indeed an increasing entanglement monotone in the sense that it is nondecreasing on average under LOCC~\cite{Barnum}. So we improve it as
\bea \label{majorization-derived entanglement}
E_{k_1,k_2,\cdots,k_n}(|\psi\rangle) = 1-
\max_{\Gamma_1,...,\Gamma_N} \|\Gamma_1 \otimes \cdots \Gamma_n
|\psi\rangle\|^2,
\eea
and it is an entanglement monotone. Specially, for a bipartite system, pure states $|\psi\rangle^{AB}$ satisfy:
\begin{eqnarray*}
	E_{k_1,k_2}(|\psi\rangle^{AB}) = E_{k_2,k_1}(|\psi\rangle^{AB}) 
	= 1-\sum_{i=1}^{\min\{k_1,k_2\}} \lambda_i(\rho^A),
\end{eqnarray*}
where $\lambda_i(\rho)$ is the $i$th decreasingly ordered eigenvalue of $\rho$.


\subsection{Hyperdeterminant of higher order}


In order to find the SLOCC classification of the multipartite pure states~\cite{Levay2006jpa,Miyake2003pra,Miyake2004ijqi}, the hyperdeterminant of the $n$-dimensional matrix $[a_{i_1i_2\cdots i_n}]$ for $|\psi\ra=\sum\limits_{a_{i_1i_2\cdots i_n}}a_{i_1i_2\cdots i_n}|i_1i_2\cdots i_n\ra\in\mH^{A_1A_2\cdots A_n}$ are shown to be necessary and efficient tools. 
For example, the ranks of the reduced states of both $|\psi_1\ra=\frac{1}{\sqrt3}(|000\ra+\frac{1}{\sqrt2}(|011\ra+|101\ra)+|112\ra)$ and $|\psi_2\ra=\frac{1}{\sqrt3}(|000\ra+|011\ra+|112\ra)$ of the $2\ot2\ot3$ system are the same, but the
hyperdeterminant of the format $2 \times 2 \times 3$~\cite{Miyake2004ijqi},
\begin{align}
	\label{eq:hdet223}
	{\rm Det}_{223}(|\psi\ra) =&
	\left|\begin{array}{ccc}
		a_{000} & a_{001} & a_{002} \\ a_{010} & a_{011} & a_{012} \\
		a_{100} & a_{101} & a_{102} \\
	\end{array}\right|
	\left|\begin{array}{ccc}
		a_{010} & a_{011} & a_{012} \\ a_{100} & a_{101} & a_{102} \\
		a_{110} & a_{111} & a_{112} \\
	\end{array}\right| \nonumber\\
	& - 
	\left|\begin{array}{ccc}
		a_{000} & a_{001} & a_{002} \\ a_{010} & a_{011} & a_{012} \\
		a_{110} & a_{111} & a_{112} \\
	\end{array}\right|
	\left|\begin{array}{ccc}
		a_{000} & a_{001} & a_{002} \\ a_{100} & a_{101} & a_{102} \\
		a_{110} & a_{111} & a_{112} \\
	\end{array}\right|,
\end{align}
are different from each other: ${\rm Det}(|\psi_1\ra)\neq0$ while ${\rm Det}(|\psi_2\ra)=0$.
Also note that 
the 3-tangle~\eref{3-tangle} can be expressed as the hyperdeterminant of the format $2 \times 2 \times 2$~\cite{Levay2006jpa,Miyake2003pra,Miyake2004ijqi}, 
\begin{align}
	\label{eq:hdet222}
	{\rm Det}_{222}(|\psi\ra) 
	=& \mbox{\;} a_{000}^2 a_{111}^2 + a_{001}^2 a_{110}^2
	+ a_{010}^2 a_{101}^2 + a_{100}^2 a_{011}^2 \nonumber\\ 
	& \mbox{} - 2(a_{000}a_{001}a_{110}a_{111}+  
	a_{000}a_{010}a_{101}a_{111}
	+ a_{000}a_{100}a_{011}a_{111} \nonumber\\
	& \mbox{} +a_{001}a_{010}a_{101}a_{110} 
	+ a_{001}a_{100}a_{011}a_{110}+  
	a_{010}a_{100}a_{011}a_{101})  \nonumber\\
	& \mbox{} + 4 (a_{000}a_{011}a_{101}a_{110} +
	a_{001}a_{010}a_{100}a_{111})
\end{align}
which distinguishes the GHZ class and the W class.

In general, the absolute value of the hyperdeterminant of format $k_1 \times \cdots \times k_l$ was shown to be an entanglement monotone~\cite{Levay2006jpa,Miyake2002qic,Miyake2003pra,Miyake2004ijqi}.


\subsection{$q$- and $c$-squashed entanglement }


Yang \etal proposed in Ref.~\cite{Yang2009ieee} two generalizations of the squashed entanglement, termed $q$- and $c$-squashed entanglement, based on two versions of the multiparty mutual information. For the $n$ party state $\rho$, the multipartite $q$-squashed entanglement was defined as follows
\bea \label{11-23-sq-q}
E_{sq}^q(\rho)=\inf I(A_1:A_2:\cdots :A_n|E), 
\eea 
where the infimum is taken over states $\sigma^{A_1A_2\cdots A_nE}$ that are extensions of $\rho$, i.e. $\tr_E \sigma= \rho$,
\beax 
I(A_1:A_2: \cdots :A_n|E) =&I(A_1:A_2|E)+I(A_3:A_1A_2|E) \\
&+ \cdots+I(A_n:A_1\cdots A_{n-1}|E). 
\eeax
with
\beax
I(A:B|E)= S(AE) + S(BE) - S(ABE) - S(E)
\eeax
is the mutual information conditioned on $E$.

There is yet another formula for mutual information in the multipartite case, as proposed in \cite{Cerf2002pra}, where multipartite secrecy monotones were considered. It is of the form 
\bea 
I'(A_1:\cdots:A_n)=\sum_{i=1}^n S(\rho^{\overline{A_i}})-(n-1) S(\rho), \label{eq:defsn} 
\eea
and the conditioned version is~\cite{Yang2009ieee}
\beax 
I'(A_1:\cdots:A_n|X)=	I'(XA_1:\cdots:A_n)-I(X:A_2\cdots A_n).
\eeax 
In the case of bipartite systems, both candidates are equal to the bipartite mutual information.
Alternatively, one can replace $I$ in Eq.~\eref{11-23-sq-q} with $I'$ and it gives rise to an independent definition of multipartite squashed entanglement~\cite{Yang2009ieee}. Both of them are entanglement monotones and they are additive, i.e., $E_{sq}^q(\rho\ot\sigma)=E_{sq}^q(\rho)+E_{sq}^q(\sigma)$ for any $\rho$, $\sigma$, and superadditive, i.e., $E_{sq}^q(\rho^{XY})=E_{sq}^q(\rho^X)+E_{sq}^q(\rho^Y)$, $\rho^{X,Y}\in\mS^{A_1A_2\cdots A_n}$.

Multipartite $c$-squashed entanglement is defined as follows~\cite{Yang2009ieee}.
For the $n$-party state $\rho$ 
\bea \label{Multipartite c-squashed entanglement}
E_{sq}^c(\rho)=\inf I(A_1:A_2:\cdots :A_n|E), 
\eea 
where infimum is taken over the extension states $\sigma^{A_1A_2\cdots A_nE}$ of the form $\sum p_i\rho_{i}\otimes |i\ra\la i|^E$. 
Replacing $I$ in Eq.~\eref{11-23-sq-q} with $I'$, it gives rise to an independent definition of $c$-squashed entanglement. $E_{sq}^c$ is nothing but the entanglement measure obtained by mixed convex-roof of multipartite quantum mutual information function. A mixed convex-roof of a function $g$ is given by~\cite{Yang2009ieee}
\bea \label{Mixed convex roof}
E_{\check{g}}(\rho)=\inf \sum_i p_i g(\rho_i), 
\eea 
where infimum is taken over all ensembles $\{ p_i, \rho_i\}$ satisfying $\sum_i p_i \rho_i=\rho$.
In Ref.~\cite{Synak2006jpa} it was shown that if $g$ is continuous, then there exists a finite decomposition that realizes the infimum. Moreover it was shown, that if $g$ is asymptotically continuous, then so is its mixed convex-roof $E_{\check{g}}$ (which is a generalization of~\cite{Nielsen2000pra}).
It was proved that, for any continuous function $g$ which is nonincreasing on average under local operations,
its mixed convex-roof $E_{\check{g}}$ is is nonincreasing on average under LOCC~\cite{Yang2009ieee}. Therefor both of the two $c$-squashed entanglement are entanglement monotones. In addition, these two classes of multipartite entanglement monotones, $E_{sq}^q$ and $E_{sq}^c$ are shown to be asymptotically continuous. But whether $E_{sq}^c$ is additive remains open.


\section{Outlook}


As we have seen, although a great deal of methods were used to quantify entanglement, there is still lack of the one that are both precise and easy to calculate especially for the higher dimensional case. The measure now is either hard to compute such as the convex-roof extended ones or not faithful or has other drawbacks. The monogamy of the measures that are not defined via the convex-roof extension such as the negativity and the relative entropy of entanglement still remains open, which is also an interesting mathematical problem. This would be worked out with the advent of the machine learning or the quantum computer.

Reviewing all these researches so far, one may conclude that: (i) The version 2.0 of MEM has more potential advantages than the version 1.0 when we exploring the multipartite quantum entanglement. (ii) We need the strict entanglement monotone instead of the original entanglement monotone or entanglement measure in quantifying bipartite entanglement. (iii) Monogamy, complete monogamy, and the tightly monogamy relation should be probed simultaneously to cope with the multiparty entanglement distribution. (iv) The disentangling condition is very effective in demonstrating the monogamy relation. (v) Three types of coarsening relations are corresponding to three kinds of monogamy relation, and in turn three progressive measures, the original measures, unified ones, and the complete one, respectively. (vi) The key point in revealing an entanglement monotone is the peculiarity of the reduced function for many cases. The entropy as a reduced function exhibits superior features, such as not only the strict concavity, but also the additivity and the sub-additivity. (vii) The origin additivity of the entanglement measure did not work well in general, but the unified MEM is additive automatically, from another perspective. (viii) The triangle relation~\eref{polygam-relation}, when it is regarded as the polygamy of entanglement, outperforms that of the entanglement assistance.

This topical review suggests several areas for further research. A promising one is to find new method/tool of quantifying entanglement, which is both faithful and convenient to figure up. Apart from verifying the monogamy of negativity and the relative entropy of entanglement, etc., evaluating of the monogamy exponent $\alpha(E)$, polygamy exponent $\beta(E_a)$, and the exponent $\acute{\alpha}(E)$ in the triangular relation seems another hard work. But these exponents could reflect the entanglement distribution more accurate. In addition, the relation among the monogamy, complete monogamy, and the tightly complete monogamy need further investigation.

Moreover, this review discussed neither 
the entanglement measure for the continuous-variable quantum system such as the Gaussian state~\cite{Adesso2007prl,Hiroshima25007prl,Onoe2020pra,Roy2020pra,Tserkis2017pra,Wolf2004pra} nor that of the quantum network entanglement~\cite{Glover2024prl,Liu2021np,Xuzhenpeng2025q,Yangxue2022prr}. In general, the method here would be failed in the measure of network. The relation between entanglement and the other resources such as coherence, imaginarity, asymmetry, etc., was not discussed either.

\ack{
This work is supported by the National Natural Science Foundation of China under Grant Nos.~12471434 and 11971277, the Program for Young Talents of Science and Technology in Universities of Inner Mongolia Autonomous Region under Grant No. NJYT25010, and the High-Level Talent Research Start-up Fund of Inner Mongolia University under Grant No. 10000-A23207007.
}	
\\

\noindent{{\bf Author Contributions}
	
	\vspace{3mm}
\noindent Z.J. mainly wrote Sections 4-5, Y.G. wrote all the other parts.
}


\begin{thebibliography} {99}
\bibitem{Suppl} We checked by many examples that it is true. Especially, if $[\rho, \sigma]=0$, $\hat{h}(\rho/2+\sigma/2)\geq\hat{\rho}/2+\hat{h}(\sigma)/2$ is always true.
\end{thebibliography}

\section*{Appendix A: List of symbols of bipartite entanglement measures/quantifiers}
\vspace{-3mm}

{\small
	\begin{align*}
		\begin{array}{lll}
			E_d & \text{Distillable entanglement}  & \text{page 15 Eq.~\eref{E_d}}\\
			E_c & \text{Entanglement cost}         & \text{page 16 Eq.~\eref{E_c}}\\
			E_f & \text{Entanglement of formation} & \text{page 16 Eq.~\eref{EoF}} \\
			C   & \text{Concurrence}               & \text{page 17 Eq.~\eref{concurrence}}\\
			\tau& \text{Tangle}                    & \text{page 20 Eq.~\eref{tangle}}\\
			C_k & \text{Concurrence hierarchy}     & \text{page 20 Eq.~\eref{Concurrence hierarchy}}\\
			G_d &\text{G-concurrence}              & \text{page 20 Eq.~\eref{G-concurrence}}\\
			C_q &    \text{$q$-concurrence}        &\text{page 21 Eq.~\eref{C_q}} \\
			C_\alpha   &    \text{$\alpha$-concurrence}  &\text{page 21 Eq.~\eref{C_alpha}} \\
			C_q^{to}     &    \text{Total concurrence}                  & \text{page 21 Eq.~\eref{C_q^{to}}}\\
			E_q      &    \text{Tsallis-$q$ entanglement}                  &\text{page 22 Eq.~\eref{E_q}} \\
			E_\alpha       &    \text{R\'{e}nyi-$\alpha$ entanglement}                  &\text{page 22 Eq.~\eref{E_alpha}} \\
			E_{q,s} &    \text{Unifired-($q,s$) entanglement}                  &\text{page 23 Eq.~\eref{uep}} \\
			E_t &    \text{Dual entropy of entanglement}                  &\text{page 23 Eq.~\eref{E_t}} \\
			N  &    \text{Negativity}                  &\text{page 24 Eq.~\eref{negativity}}\\
			E_R   &    \text{quantifier of entanglement based on realignment}                 &\text{page 25 Eq.~\eref{E_R}} \\
			E_N   &    \text{Logarithmic negativity}                 &\text{page 25 Eq.~\eref{E_N}} \\
			E_{N,G}   &    \text{Trace-norm group negativity}                  &\text{page 26 Eq.~\eref{eq:tngn}} \\
			E_{N\text{-}q} & \text{$q$-logarithm negativity}   &\text{page 26 Eq.~\eref{E_{N-q}}}\\
			E_r & \text{Relative entropy of entanglement} &\text{page 26 Eq.~\eref{E_r}} \\
			E_A & \text{{Variant of the relative entropy of entanglement}} &\text{page 26 Eq.~\eref{E_A}} \\
			E_M & \text{Variant of the entropy of entanglement} &\text{page 26 Eq.~\eref{E_M}} \\
			E_{sq} &\text{Squashed entanglement} &\text{page 27 Eq.~\eref{E_{sq}}} \\
			E_I & \text{Conditional entanglement of mutual information} &\text{page 27 Eq.~\eref{E_I}} \\
			E_G & \text{Geometric measure} &\text{page 27 Eq.~\eref{geometric-measure-1}}\\
			\check{E}_G &\text{Revised geometric measure} &\text{page 28 Eq.~\eref{b-revised geometric-measure}}\\
			E_{\rm HS} &\text{Hilbert-Schmidt entanglement}     &  \text{page 28 Eq.~\eref{H-S-entanglement}}    \\
			E_{\tr} &\text{Trace distance entanglement}     &  \text{page 28 Eq.~\eref{E_tr}}    \\
			E'_{\tr} &\text{Variation of the trace distance entanglement}     &  \text{page 28 Eq.~\eref{E'_tr}}  \\
			E_{\mF}& \text{Fidelity based entanglement measure}&\text{page 29 Eq.~\eref{definition-purestate1}} \\
			E_{{\mF}'} &  \text{Square-root fidelity based entanglement measure} &\text{page 29 Eq.~\eref{definition-purestate2}}\\
			E_{A\mF} & \text{A-fidelity based entanglement measure}&\text{page 29 Eq.~\eref{definition-purestate3}}\\  				
			E_{\B} & \text{Bures measure of entanglement} &\text{page 29 Eq.~\eref{Bures measure}}\\
			E_{\rm Gr} & \text{Groverian measure of entanglement} &\text{page 30 Eq.~\eref{f4}}\\
			E_{SG} & \text{Sharp geometric measure of entanglement} &\text{page 30 Eq.~\eref{E_SG}}\\
			R   & \text{Robustness of entanglement} &\text{page 30 Eq.~\eref{Robustness of entanglement}} \\
			\check{R} & \text{Generalized robustness of entanglement} &\text{page 30 Eq.~\eref{generalized robustness}} \\
			E_{V\text{-}k} & \text{Vidal's entanglement monotone} &\text{page 31 Eq.~\eref{E_k}} \\
			E_2 & \text{Partial-norm of entanglement} &\text{page 31 Eq.~\eref{E_2}}\\
				E_{\rm min} &\text{Minimal partial-norm of entanglement} &\text{page 31 Eq.~\eref{E_min}}\\
			E_{\rm min'}~~~~~ &\text{Reinforced minimal partial-norm of entanglement}~~ ~~~&\text{page 31 Eq.~\eref{E'_min}}~~~\\
			E_\kappa &\text{Kaniadakis entropy of entanglement}&\text{page 32 Eq.~\eref{E_kappa}}\\
			E_{\rm ic} &\text{Informationally complete entanglement measure} &\text{page 32 Eq.~\eref{ICEM}}\\
			S_r&  \text{Schmidt number} &\text{page 33 Eq.~\eref{Schmidt number}}\\
			E_{\max\text{-}r} &\text{Max-relative entropy of entanglement} &\text{page 33 Eq.~\eref{max-relative}}
		\end{array}
	\end{align*}
	
}



{\small
	\begin{align*}
		\begin{array}{lll}
		E_{\min\text{-}r} &\text{Min-relative entropy of entanglement} &\text{page 33 Eq.~\eref{min-relative}}\\
			E_{\check{R}}&\text{Logarithmic version of $\check{R}$} &\text{page 33 Eq.~\eref{E_{check{R}}}}\\
			E_N^\alpha &\text{$\alpha$-logarithmic negativity} & \text{page 35 Eq.~\eref{alpha-logarithmic negativity}}\\
			\check{E}_\kappa &\text{$\kappa$ entanglement} &\text{page 35 Eq.~\eref{Kappa-entanglement}}\\
			E_{cr} &\text{Entanglement monotone from complementary relation}&\text{page 35 Eq.~\eref{E_{cr}}} \\
			\check{E} & \text{Entanglement measure in terms of LOCC on pure states} &\text{page 36 Eq.~\eref{Ev}}\\
			E_p &\text{Passive-state energy} &\text{page 36 Eq.~\eref{pasivemonotone}}\\
			E_{e\text{-}g} &\text{Ergotropic gap entanglement} &\text{page 37 Eq.~\eref{E_{e-g}}}\\
		\mathcal{\tilde{E}} &\text{Entanglement parameter} &\text{page 38 Eq.~\eref{entanglement-parameter}} \\
			C_E &\text{Entanglement coherence} &\text{page 38 Eq.~\eref{ecm}}\\
			E_{oc} &\text{Entanglement measure based on observable correlations} &\text{page 39 Eq.~\eref{E_oc}}\\
			E_a&\text{Entanglement of assistance} &\text{page 39 Eq.~\eref{eoa}}\\
			E_S&\text{Negative entanglement measure} &\text{page 40 Eq.~\eref{E_S}}\\
			E_{co} &\text{Entanglement of collaboration} &\text{page 43 Eq.~\eref{EoC}}\\
				{\tau}^{(n)}_{A_1A_2\cdots A_n}  &    \text{Residual $n$-tangle }                             & \text{page 61 Eq.~\eref{n-residual}}  \\
			E^{(n)}& \text{Complete global MEM from sum of reduced functions} &\text{page 83 Eq.~\eqref{sum1}}\\
			\mE^{(n)}& \text{Complete global MEM from sum of bipartite entanglement} &\text{page 84 Eq.~\eqref{sum2}}\\
			\tilde{\tau} &\text{Three tangle} &\text{page 88 Eq.~\eref{3-tangle}}\\	
			E_{g}^{(n)}              &\text{GEM from sum of reduced functions} &\text{page 89 Eq.~\eref{gsum1}}
		\end{array}
	\end{align*}
	
}


\section*{Appendix B: List of symbols of multipartite entanglement measures/quantifiers I}

\vspace{-3mm}
{\small

	\begin{align*}
		\begin{array}{lll}
				\mE_{g}^{(n)}              &~~\quad\text{GEM from sum of bipartite entanglement} &\text{page 90 Eq.~\eref{gsum2}}\\	
			E_{g'}^{(n)}              &~~\quad\text{GEM from the maximal single reduced function} &\text{page 90 Eq.~\eref{gmax1}}\\
				\mE_{g'}^{(n)}              &~~\quad\text{GEM from the maximal reduced function} &\text{page 91 Eq.~\eref{gmax2}}\\	
					E_{g''}^{(n)}             &~~\quad\text{GEM from the minimal single reduced function} &\text{page 91 Eq.~\eref{gmin1}}\\		
				\mE_{g''}^{(n)}             &~~\quad\text{GEM from the minimal reduced function} &\text{page 91 Eq.~\eref{gmin2}}\\
				C_{gme}                   &~~\quad\text{Genuinely multipartite concurrence } &\text{page 91 Eq.~\eref{Cgme}} \\
				E_{g,G} & ~~\quad\text{Genuine geometric measure}                                                    & \text{page 94 Eq.~\eref{m-geometric-measure}} \\
				\tilde{N}_g & ~~\quad\text{Genuine multipartite negativity}                                               & \text{page 94 Eq.~\eref{eqn:sdpgennegold}} \\				
				N_g & ~~\quad\text{Modified genuine multipartite negativity}                                      & \text{page 95 Eq.~\eref{eqn:sdpgenneg}} \\
				C_G                   & ~~\quad\text{Geometric mean of bipartite concurrence}&\text{page 96 Eq.~\eref{CG}}\\
				C_{G\text{-}q}                   & ~~\quad\text{Geometric mean of $q$-concurrence}&\text{page 97 Eq.~\eref{geometric mean of $q$-concurrence}}\\
				F_{ABC}                   &~~\quad \text{Concurrence fill}&\text{page 97 Eq.~\eref{triangle}}\\
				F'_{ABC}                   &~~\quad \text{Improved concurrence fill}&\text{page 97 Eq.~\eref{triangle}}\\
				F_\eta                    & ~~\quad\text{Triangle area from subadditive reduced functions}&\text{page 97 Eq.~\eref{triangle area}}	\\
					E_k & ~~\quad\text{Minimal sum of the reduced functions} & \text{page 100 Eq.~\eref{sum}} \\
				E'_k&~~\quad\text{Maximal sum of the reduced functions} &\text{page 101 Eq.~\eref{sum3}}\\
				\bar{E}_k& ~~\quad\text{Arithmetric mean of $k$-partitions}& \text{page 101 Eq.~\eref{arithmetric mean of k-partition}}\\
				E_k^G & ~~\quad\text{geoetric mean of $k$-paritions}& \text{page 102 Eq.~\eref{arithmetric mean of k-partition}}\\
				C_{k\text{-}ME} &~~\quad\text{$k$-ME concurrence } &\text{page 102 Eq.~\eref{CkME}} \\
				C_{q\text{-}k} &~~\quad\text{$q$-$k$-ME concurrence } &\text{page 102 Eq.~\eref{Cqk}} \\
				C_{\alpha\text{-}k} &~~\quad\text{$\alpha$-$k$-ME concurrence } &\text{page 103 Eq.~\eref{Calphak}} \\
				C_{k\text{-}GM}                  & ~~\quad\text{$k$-GM concurrence} & \text{page 103 Eq.~\eref{k-GM}}\\
				E_{G\text{-}k}                   & ~~\quad\text{Minimal geometric mean of product of reduced functions} & \text{page 103 Eq.~\eref{product}} \\
				E'_{G\text{-}k}                  & ~~\quad\text{Maximal geometric mean of product of reduced functions}~~ & \text{page 104 Eq.~\eref{product3}}\\
				\bar{E}_{G\text{-}k}             &~~\quad \text{Arithmetric mean of product of reduced functions} & \text{page 104 Eq.~\eref{arithmetric mean}}\\
				E_{G\text{-}k}^{G}         & ~~\quad \text{Geometric mean of all $k$-partitions} & \text{page 104 Eq.~\eref{geometric mean}}						
				\end{array}
	\end{align*}

}			
{\small

	\begin{align*}
		\begin{array}{lll}
			E_{(k)} & \text{$k$-PEMo from unified MEM} &\text{page 106 Eq.~\eref{E-k2}}\\
			C_{(k)}                   &\text{$k$-PEM via concurrence } &\text{page 106 Eq.~\eref{k-part-entanglement-C0}} \\
			C_{q(k)}                  &\text{$k$-PEMo via the $q$-concurrence}& \text{page 106 Eq.~\eref{k-part-entanglement-C1}}\\
			C_{\alpha(k)}             &\text{$k$-PEMo via the $\alpha$-concurrence}& \text{page 106 Eq.~\eref{k-part-entanglement-C2}}\\
			E'_{(k)} & \text{$k$-PEMo from minimum partition} &\text{page 107 Eq.~\eref{p-sum}}\\
			C_{G,q(k)} & \text{$k$-PEMo from geometric mean of $q$-concurrence}  & \text{page 107 Eq.~\eref{k-part-entanglement-GC1}}\\
			C_{G,\alpha(k)}   &\text{$k$-PEMo from geometric mean of $\alpha$-concurrence}   &  \text{page 107 Eq.~\eref{k-part-entanglement-GC2}}\\
			\check{E}^{AB}            & \text{Pairwise entanglement measure  w.r.t $AB$  }&\text{page 108 Eq.~\eref{pem}}\\
			\check{E}^{A_1A_2}            & \text{Pairwise entanglement measure  w.r.t $A_1A_2$  }&\text{page 108 Eq.~\eref{pwem}}\\
			\check{E}^{A_1A_2\cdots A_k} &\text{$k$-PWEM  w.r.t $A_1A_2\cdots A_k$ } &\text{page 108 Eq.~\eref{kpwem}}\\
			\check{E}_g^{A_1A_2\cdots A_k} &\text{Genuine $k$-PWEM w.r.t $A_1A_2\cdots A_k$  } &\text{page 108 Eq.~\eref{gkpwem}}\\
			\check{E}_s^{A_1A_2\cdots A_k} &\text{Strong $k$-PWEM  w.r.t $A_1A_2\cdots A_k$  } &\text{page 108 Eq.~\eref{skpwem}}\\
			\check{E}_{\min}^{A_1A_2\cdots A_k}& \text{$k$-PWEMo from minimal reduced function}& \text{page 108 Eq.~\eref{min-h-kpwem}}\\
			\check{\bar{E}}^{A_1A_2\cdots A_k} &\text{$k$-PWEMo from sum of reduced functions } &\text{page 109 Eq.~\eref{E2}}\\
			\check{E}_{G}^{A_1A_2\cdots A_k} & \text{$k$-PWEMo from geometric mean of reduced functions}&\text{page 109 Eq.~\eref{kEG}}\\
			\check{S}_r^{A_1A_2\cdots A_k} & \text{$k$-PWEMo via relative entropy}& \text{page 109 Eq.~\eref{kpwemo}}\\
			\check{E}_{\rm G}^{A_1A_2\cdots A_k} &  \text{Geometric measure of $k$-PWE}& \text{page 109 Eq.~\eref{kpwemo2}}\\
			E_{ext} &\text{Pairwise entanglement extensibility measure} & \text{page 110 Eq.~\eref{extensiblity}}\\
			E_G^{(n)} & \text{Multipartite geometric measure}                                     & \text{page 112 Eq.~\eref{geometric-measure}} \\
			\check{E}_G^{(n)} &  \text{Revised multipartite geometric measure}                    & \text{page 112 Eq.~\eref{m-revised geometric-measure}} \\		
		\end{array}
	\end{align*}
	
}


\section*{Appendix C: List of symbols of multipartite entanglement measures/quantifiers II}

\vspace{-3mm}

{\small

	\begin{align*}
		\begin{array}{lll}
			E_{tr}^{(n)} \quad&  \text{Generalized geometric MEM}& \text{page 113 Eq.~\eref{E_{tr}^{(n)}}}\\
			E_{G,k}^{(n)} &  \text{Geometric measure of $k$-entanglement}& \text{page 113 Eq.~\eref{geometric-k-entanglement}}\\
			E_r^{(n)} &  \text{Relative entropy of multipartite entanglement} \qquad~~~~& \text{page 113 Eq.~\eref{E_r^n}}\\		
				\mathbb{M}E_r^{X(n)} & \text{$\mathbb{M}$-relative entropy of entanglement} & \text{page 113 Eq.~\eref{M-relative entropy}} \\
			\acute{\tau}^{(n)}  &    \text{$n$-qubit tangle}                                      & \text{page 113 Eq.~\eref{taun}}  \\
			\check{\tau}^{(n)}  &    \text{$n$-tangle}                                            &  \text{page 114 Eq.~\eref{ntangle}} \\	
			C_n  &    \text{Generalization of concurrence}                              &  \text{page 114 Eq.~\eref{multi}} \\		
			C'_n &\text{Global concurrence }                                         &  \text{page 114 Eq.~\eref{dcn} } \\
			C''_n &    \text{Another global concurrence}                                     &  \text{page 114 Eq.~\eref{ddcn} } \\
			R^{(n)}   &    \text{The robustness of multipartite entanglement }                    &  \text{page 114 Eq.~\eref{multipartite robustness of entanglement} } \\
			\check{R}^{(n)}   &    \text{The generalized robustness of multipartite entanglement }                    &  \text{page 114 } \\
			\check{R}_k^{(n)}& \text{Generalized robustness of $k$-entanglement} & \text{page 115 Eq.~\eref{robustness-k-ent}}\\
			\check{R}_L^{(n)}& \text{Logarithmic robustness} & \text{page 115 Eq.~\eref{logarithmic robustness}}\\
			E_{\mF}^{(3)}     &    \text{Fidelity based entanglement monotone}&  \text{page 115 Eq.~\eref{multipartite1}} \\
			E_{\mF'}^{(3)}      &    \text{Entanglement monotone via square root of fidelity} &  \text{page 115 Eq.~\eref{multipartite2}} \\
			E_{A\mF}^{(3)}  & \text{A-fidelity based tripartite entanglement monotone} &  \text{page 115 Eq.~\eref{multipartite3}} \\
			E_I^{(n)} &    \text{Multipartite version of $E_I$}                                 & \text{page 115 Eq.~\eref{multipartite EI}}  \\
			C_S^{(n)} &    \text{Concentratable entanglement}                                    &  \text{page 116 Eq.~\eref{Concentratable entanglement}} \\
			E_{s\text{-}g,s}^{(n)}&    \text{Ergotropic-gap concentratable entanglement}                             &  \text{page 116 Eq.~\eref{E_s}} \\
			E_Q^{(n)} &    \text{Global $n$-qubit entanglement}                                         & \text{page 116 Eq.~\eref{n-qubit entanglement}}  \\
			E_{Q,m}^{(n)}   &    \text{Generalized $E_Q^{(n)}$ into the multiqudit system}      &  \text{page 117 Eq.~\eref{Generalized EQn}} \\
			E_{\fh}^{(n)}   &    \text{Entanglement monotone via relative purity}                                  &  \text{page 117 Eq.~\eref{Entanglement monotone}} 	\\
			E_W^{(n)}&\text{Witnessed entanglement}&\text{page 118 Eq.~\eref{E_W}}	\\
				B^{(n)}&\text{Bertlmann' entanglement quantifier}&\text{page 118 Eq.~\eref{Bn}}
		\end{array}
\end{align*}
}		
{\small

	\begin{align*}
		\begin{array}{lll}		
				E_{k:l}^{(n)} & \text{Witnessed entanglement}                                     & \text{page 118} \\			
			E_D^{(n)} & \text{Invariant under determinant 1 SLOCC}                        & \text{page 119 Eq.~\eref{SLOCC-invariant}}  \\
			E_B^{(n)} &\text{Barycentric measure}                          &  \text{page 119 Eq.~\eref{Barycentric measure of entanglement}} \\
			F_{1234(2)} & \text{Entanglement monotone via product of reduced functions}                                &  \text{page 120 Eq.~\eref{1234(2)}} \\
			\check{\mE}^{(n)} & \text{Geometric mean of product of all single reduced function}   & \text{page 120 Eq.~\eref{MEM2}} \\
			\hat{\mE}^{(n)} &\text{Geometric mean of product of all possible bipartitions}  & \text{page 120 Eq.~\eref{prod2}} \\
			E_{\rm Sr}&   \text{Schmidt measure}   & \text{page 121 Eq.~\eref{Schmidt measure}}  \\
			\check{S}_r^{(n)} &  \text{$n$-partite Schmidt number} & \text{page 122 Eq.~\eref{n-Schmidt measure}} \\
				E_{ac}^{(n)} & \text{Operational MEM from accessible volume} & \text{page 123 Eq.~\eref{Eac}} \\
			E_{so}^{(n)} & \text{Operational MEM from source volume} & \text{page 123 Eq.~\eref{Eso}} \\
			\check{C}^{(n)}& \text{Total concurrence } & \text{page 124 Eq.~\eref{Cmpartite}} \\
			\Upsilon &  \text{Ent } &\text{page 125 Eq.~\eref{Hedemann2018QIC}}\\
			\Upsilon_{\rm FGM}   & \text{FGM entanglement} &  \text{page 125 Eq.~\eref{FGM}} \\
			\Upsilon_{\mathrm{GM}_k}& \text{GM-$k$ entanglement} &  \text{page 125 Eq.~\eref{GMk}} \\
			E_{k_1,k_2,\cdots,k_n} & \text{Majorization-derived entanglement} &  \text{page 126 Eq.~\eref{majorization-derived entanglement}} \\
			{\rm Det}_{d_1d_2\cdots d_n}&\text{Hyperdeterminant} & \text{page 126} \\
			E_{sq}^q &\text{Multipartite $q$-squashed entanglement} & \text{page 127 Eq.~\eref{11-23-sq-q}} \\
			E_{sq}^c &\text{Multipartite $c$-squashed entanglement}& \text{page 127 Eq.~\eref{Multipartite c-squashed entanglement}} \\
			E_{\check{g}} &\text{Mixed convex-roof} &	 \text{page 127 Eq.~\eref{Mixed convex roof}} 
		\end{array}
	\end{align*}
	
}

\vspace{-3mm}

\section*{Appendix D: List of symbols of several special states}


\vspace{-3mm}
{\small

	\begin{align*}
		\begin{array}{lll}
			|\Phi^+\ra &   \text{Maximally entangled state in $d\ot d$ system} \qquad~~~ \quad&   \text{page 14 }\\	
			P^+=|\Phi^+\ra\la\Phi^+| & \text{Maximally entangled state in $d\ot d$ system} &   \text{page 15 } \\
			|\phi^+\ra  & \text{Singlet state, i.e.,} |\phi^+\ra=\frac{1}{\sqrt2}(|00\ra+|11\ra) & \text{page 15} \\
			|\check{\W}\ra&    \text{Three-qubit W calss state}                   & \text{page 43 Eq.~\eref{wstate}} \\	
			|\W\ra  & \text{Three-qubit W state} & \text{page 45} \\   
			\text{$|\widetilde{\W}_{n,2}\ra$} &    \text{Generalized $n$-qubit W state}                   &  \text{page 50 Eq.~\eref{W-state}} \\				
			|\widetilde{\W}_{n,d}\ra  &    \text{Generalized $n$-qudit W state}                  &  \text{page 50 Eq.~\eref{GW}} \\
			|\GWV\ra &\text{$n$-qudit generalized W state and vacuum} &\text{page 54 Eq.~\eref{GWV} }\\
			|{\W}_{n,2}\ra &    \text{$n$-qubit W state}                   &  \text{page 62 Eq.~\eref{wn2}} \\
			|\widetilde{\GHZ}_{n,d}\ra  &    \text{Generalized $n$-qudit GHZ state}                  &  \text{page 63 Eq.~\eref{GHZnd}} \\
			|\GHZ\ra  &  \text{Three-qubit GHZ state}                    & \text{page 71} \\	  		
			|{\GHZ}_{n,2}\ra&    \text{$n$-qubit GHZ state}                   &  \text{page 74 Eq.~\eref{GHZn2}} \\
		\end{array}\\
		\footnotetext{$E_2$ is just the partial-norm entanglement}
	\end{align*}
	
}

\vspace{-8mm}


\section*{Appendix E: List of abbreviations}


\vspace{-3mm}

{\small

	\begin{align*}
		\begin{array}{lll}
			{\rm LOCC} &\qquad\text{Local operation and classical communication}\qquad\quad ~~&  \text{page 6}\qquad~~~~  \\
			\text{GEM} & \qquad\text{Genuine entanglement measure} &\text{page 7}  \\
			\text{GMC} &\qquad\text{Genuinely multipartite concurrence} &\text{page 7} \\
			\text{$k$-EM} & \qquad\text{$k$-entanglement measure} &\text{page 9} \\
			\text{MEM} &\qquad\text{Multipartite
				entanglement measure}&\text{page 9}\\
			\text{CPTP}&\qquad\text{Completely positive and trace preserving}  & \text{page 11} \\
			\text{QRT} &\qquad\text{Quantum resource theory} &\text{page 11} \\
			\text{POVM}~~&\qquad\text{Positive operator-valued measure} &  \text{page 19}\\
			\text{SIC} &\qquad\text{Symmetric informationally complete}  &  \text{page 19}\\
			\text{PPT} &\qquad\text{Positive partial transpose} &\text{page 24} \\
			\text{CFT} &\qquad\text{Conformal ffeld theory}&\text{page 25} \\
			\text{RGME} &\qquad\text{Revised geometric measure of entanglement} &\text{page 28} 
		\end{array}
		\end{align*}
	\begin{align*}			
			\begin{array}{lll}			
			\text{CM} &\text{Covariance matrice} &\text{page 37}\\
			\text{LSE} &\text{Locally symmetric extension}&\text{page 42} \\
			\text{SEM} &\text{Strict entanglement monotone}  &  \text{page 48} \\
			\text{S-LEM} &\text{Strict-LOCC entanglement monotone }&\text{page 49} \\
			\text{SM} &\text{Strong monogamy} &\text{page 61} \\
				\text{$k$-PEM} & \text{$k$-partite entanglement measure} & \text{page 68} \\
			\text{$k$-PWE} &\text{$k$-partitewise entangled} &\text{page 71} \\
			\text{G$k$-PWE} &\text{Genuinely $k$-partitewise entangled} &\text{page 71} \\
			\text{S$k$-PWE} &\text{Strongly $k$-partitewise entangled} &\text{page 72} \\
			\text{U-g-MEM} &\text{Uniffed global MEM} &\text{page 74} \\
			\text{C-g-MEM}&\text{Complete global MEM} &\text{page 75} \\
			\text{PWEM} & \text{Pairewise entanglement measure} & \text{page 78} \\
			\text{$k$-PWEM} & \text{$k$-partitewise entanglement measure} & \text{page 78} \\
			\text{G$k$-PWEM}~~~ & \text{Genuine $k$-partitewise entanglement measure}\quad & \text{page 78} \\
				\text{S$k$-PWEM} & \text{Strong $k$-partitewise entanglement measure} & \text{page 78} \\
			\text{MES} &\text{Maximally entangled state  } &\text{page 85} \\
			\text{MMES} & \text{Mixed maximally entangled state }&\text{page 85} \\
			\text{GMN}&\text{Genuine multipartite negativity} &\text{page 94} \\
			\text{GBC} &\text{Geometric mean of bipartite concurrence} &\text{page 96} \\
			\text{PI} &\text{Permutation invariant } &\text{page 103} \\
			\text{EW} &\text{Entanglement witness } &\text{page 118} \\
			\text{LU} &\text{Local unitary} &\text{page 123} \\
			\text{FGM} &\text{Full genuinely multipartite } &\text{page 125}\qquad~~~~ \\
		\end{array}
	\end{align*}
	
}


\section*{Appendix F: Other symbols}

\vspace{-3mm}

{\small

	\begin{align*}
		\begin{array}{lll}
			\preceq       &  \text{Majorization relation} \quad\quad  \qquad\qquad &  \text{page 13 Eq.~\eref{majorization}} \\
			S(\rho) &   \text{von Neumann entropy} &\text{page 14}\\	
			F &  \text{Fidelity} &\text{page 14 Eq.~\eref{fidelity}} \\	
				S(\rho\|\sigma) & \text{Relative entropy} &\text{page 14}\\
			\text{$\check{F}$} & \text{Flip operator or called swap operator} &\text{page 15}\\
			\rho_x        &  \text{Werner state}    &  \text{page 15 Eq.~\eref{rhox}} \\
			\rho_c        &  \text{Werner state}    &  \text{page 15 Eq.~\eref{rhoc}} \\		
			\rho_t        &  \text{Isotropic sate}    &  \text{page 15 Eq.~\eref{rhot}} \\
			\rho_f        &  \text{Isotropic sate}    &  \text{page 15 Eq.~\eref{rhof}} \\	
			C_{\Theta}    &    \text{$\Theta$-concurrence} &\text{page 17 Eq.~\eref{Ctheta}} ~~\\
			A^*&\text{Complex conjugation of $A$ under given basis} \quad~~&\text{page 17} \\
			\rho^{T_{a,b}}  &\text{Partial transpose of $\rho$} &  \text{page 19 Eq.~\eref{partial-transpose}} \\
			\rho^R         & \text{Realignment of $\rho$} & \text{page 19 Eq.~\eref{realignment}} \\
				S_q(\rho) & \text{Quantum Tsallis-$q$ entropy}&\text{page 22 Eq.~\eref{Quantum Tsallis-entropy}} \\
			S_\alpha(\rho) &\text{R\'{e}nyi-$\alpha$ entropy} & \text{page 22 Eq.~\eref{Salpha}}\\
			S_{q,s}(\rho) & \text{Unified entropy}&\text{page 23 Eq.~\eref{unified entropy}}\\
			S^{to}(\rho) &\text{Quantum dual entropy} &\text{page 24 Eq.~\eref{quantum dual entropy}}\\
			F_s  &\text{Fidelity of separability} &\text{page 29 Eq.~\eref{Fs}}\\
			\mS_{sep} &\text{The set of all bipartite separable states} &\text{page 29}\\
			\|\rho\|_{\min} &\text{Minimal partial-norm} &\text{page 31 Eq.~\eref{min norm}}  \\
			\ln_{\kappa} &\text{Kaniadakis' logarithm} &\text{page 32 Eq.~\eref{ln-kappa}} \\
			S_{\kappa} &\text{Quantum version of Kaniadakis entropy} &\text{page 32 Eq.~\eref{Kaniadakis entropy}} \\
			D_{\max}(\rho\| \sigma)~~ &\text{Max-relative entropy} &\text{page 33 Eq.~\eref{max-relative entropy}}\\
			D_{\min}(\rho\| \sigma) &\text{Min-relative entropy} &\text{page 33 Eq.~\eref{dmin}}\\
			\|X\|_{\alpha} &\text{$\alpha$-norm of an operator $X$} &\text{page 34 Eq.~\eref{alphanorm}} \\
			\widetilde{D}_{\alpha}(X\Vert\sigma) & \text{Sandwiched R\'{e}nyi relative entropy} & \text{page 34 Eq.~\eref{SR relative entropy}}\\
			I_{oc} & \text{Observable correlation} & \text{page 38 Eq.~\eref{Ioc}}		
		\end{array}
\end{align*}

}
{\small

	\begin{align*}
		\begin{array}{lll}
			\lfloor x\rfloor &\text{Floor function} &\text{page 42} \\
			\alpha(E) &\text{Monogamy exponent} &\text{page 50 } \\
			\beta(E_a) &\text{Polygamy exponent} &\text{page 63 } \\
			\mS_{k} &\text{The set of all $k$-separable states} &\text{page 68}\\
			\varDelta(X_t) &\text{Subsystems number contained in $X_t$} &\text{page 69}\\
			\mS_g &\text{The set of all genuinely entangled states} &\text{page 69}\\
			\Gamma_k^f &\text{The set of all $k$-fineness partitions} &\text{page 69}\\
			\mS_{P(k)} &\text{The set of all $k$-producible quantum states} &\text{page 70} \\
			\succ^a        & \text{Coarsening relation of type (a)} & \text{page 72 Eq.~\eref{succ-a}} \\
			\succ^b        & \text{Coarsening relation of type (b)} & \text{page 72 Eq.~\eref{succ-b}} \\
			\succ^c        & \text{Coarsening relation of type (c)} & \text{page 72 Eq.~\eref{succ-c}} \\
			\succ        & \text{Coarsening relation} & \text{page 72 Eq.~\eref{succ}} \\
			\Xi(\gamma_p-\gamma_q) &\text{Complementarity partitions of $\gamma_q$ w.r.t $\gamma_p$} &\text{page 73 Eq.~\eref{Xi}} \\
			\Gamma_k &\text{The set of all $k$-partitions} &\text{page 83} \\
			\mQ_k^{\gamma_i} & \text{Geometric mean of the reduced functions w.r.t. $\gamma_i$}&\text{page 96 Eq.~\eref{mQ_k^{gamma_i}}} \\
			\mP_k^{\gamma_i} &\text{Sum of reduced functions w.r.t. $\gamma_i$} &\text{page 100 Eq.~\eref{mP_k^{gamma_i}}} \\				
		\rho^{\rm PI} &\text{Permutation invariant part of $\rho$} &\text{page 103~Eq.~\eref{PI}} \\			
			\mP_k^{\gamma_i^f} & \text{Unified $k$-EMs}&\text{page 106 Eq.~\eref{P_k}} \\
			\mP^{A_1A_2\cdots A_n}&\text{The set of all fully separable pure states}&\text{page 112 Eq.~\eref{full separable pure states}}\\
			P_{\fh} & \text{$\fh$-purity} &\text{page 117 Eq.~\eref{fh purity}}		\\
			\mathcal{W} & \text{The set of all entanglement witnesses}&\text{page 118 }\\
		\end{array}
	\end{align*}
	
}


\end{document}